\documentclass[preprint,authoryear,12pt]{elsarticle}
\usepackage{amsfonts,amssymb,amsmath,amsopn}
\usepackage{rotating}
\usepackage{bigints}
\usepackage{booktabs}
\usepackage{subfigure}
\usepackage{graphicx}
\usepackage{url}
\usepackage{amssymb}
\usepackage{color}
\usepackage{multirow}
\usepackage{tabularx}
\usepackage{diagbox}
\usepackage{enumerate}
\usepackage{epstopdf}
\usepackage[colorlinks=true]{hyperref}
\newcommand{\ignore}[1]{}
\newtheorem{asu}{{\sc Assumption}}
\newtheorem{corollary}{{\sc Corollary}}
\newtheorem{pro}{{\sc Proposition}}
\newtheorem{thm}{Theorem}
\newtheorem{lem}{Lemma}
\newdefinition{rmk}{Remark}
\usepackage[margin=1in]{geometry}
\newproof{pf}{Proof}
\newproof{pot}{Proof of Theorem \ref{thm2}}

\usepackage{amssymb}
\usepackage{booktabs}
\newcommand{\E}{\mathbb{E}}
\newcommand{\red}{\textcolor{red}}

\usepackage{xr}

\allowdisplaybreaks

\begin{document}
	\begin{frontmatter}
		\title{
	Empirical Characteristic Function Method for Leverage Effect and Volatility of Volatility: Estimation and Feasible Inference
		}
		\author[SUFE]{Qiang LIU}
		\author[UM]{Zhi LIU}
		\author[JU]{Guangren YANG}
		\author[NUS]{Wang ZHOU}
	
		\address[SUFE]{School of Statistics and Data Science, Shanghai University of Finance and Economics, Shanghai, China.}
		\address[UM]{Department of Mathematics, University of Macau, Macau, China.}
		\address[JU]{Department of Statistics and Data Science, Jinan University, Guangzhou, China}
		\address[NUS]{Department of Statistics and Data Science, National University of Singapore, Singapore.}
		\begin{abstract}
We develop jump-robust estimators of the leverage effect and volatility of volatility using high-frequency data. Our construction begins with a spot volatility estimator based on the empirical characteristic function of high-frequency increments. This method can mitigate the contamination from jumps, which can be of infinite variation. We then construct estimators of the leverage effect and volatility of volatility and correct for the bias induced by spot volatility estimation. We establish consistency and central limit theorems under conditions that allow greater jump activity than existing methods. We also develop consistent estimators of the asymptotic variances, making the limiting results feasible for statistical inference. Simulation studies demonstrate the improved finite-sample performance of the proposed estimators, particularly in the presence of infinite variation jumps. An empirical application provides evidence of nonzero leverage effect and volatility of volatility, when the jump activity is intensive.
		\end{abstract}
	\end{frontmatter}
	\newpage
       \section{Introduction}
Semimartingale processes are widely used in finance. For example, the fundamental theorem of asset pricing states that, in an arbitrage-free and frictionless financial market, the logarithmic price process of an asset must be a semimartingale (\cite{DS1994}). We denote the log-price process of an asset over $[0,T]$ by $\{X_t\}_{0\leq t\leq T}$ and assume that it is an It$\hat{\text{o}}$ semimartingale on the filtered probability space $(\Omega,\mathcal{F},(\mathcal{F}_t)_{ 0\leq t\leq  T},\mathbb{P})$. In general, $X_t$ can be represented as
\begin{equation}\label{model-price}
	X_t =X_0 + \int_0^tb_sds + \int_0^t\sigma_sdB_s + J_t, \quad t\in[0,T], 
\end{equation}
where $b$ and $\sigma$ are adapted and locally bounded c$\grave{\text a}$dl$\grave{\text a}$g processes, $B$ is a standard Brownian motion, $\sigma^2$ is the volatility process, and $J$ is a pure-jump process.
In high-frequency financial econometrics, the path of $X$ over $[0,T]$ cannot be observed continuously and is available only at discrete time points. We assume that $X$ is observed at equally spaced times $t_i^n:=i \Delta_n$, for $i=0, 1,\cdots, n$, with $n=\lfloor\frac{T}{\Delta_n}\rfloor$, where $\Delta_n$ is constant for a given $n$. We consider the infill asymptotic regime in which $\Delta_n\rightarrow0$ for fixed $T$.

To quantify the intensity of the jump process, \cite{YJ2009} introduced the following jump activity index (JAI) for a semimartingale process $X$:
\begin{align}
	\text{JAI}:= \inf\{r>0: \sum_{t\leq T} |\Delta X_{t}|^{r} < \infty\},
\end{align}
where $\Delta X_{t} = X_{t} - X_{t-}$ is the jump size at time $t$. 
With this definition, $0\leq \text{JAI}< 2$ holds almost surely for a pure jump process, and small jumps tend to become more frequent as the JAI increases. A process with finite jump activity has $\text{JAI}=0$, whereas a process with $\text{JAI}>0$ has infinite jump activity. Moreover, when $\text{JAI}<1$, the jumps are locally summable and hence have finite variation, whereas they have infinite variation when $\text{JAI}>1$. For a L$\acute{\text{e}}$vy process, the JAI coincides with the Blumenthal--Getoor index. In the special case in which $X$ is a stable L$\acute{\text{e}}$vy process, the JAI is also the stable index of the process. We refer to \cite{YJ2009} for further details.  
Empirically, it was reported in \cite{YJ2009} and \cite{JKLM2012} that the jump activity of real high-frequency financial data can be intensive as of infinite variation. 

The increasing availability of high-frequency data has stimulated extensive research on volatility-related quantities, including integrated volatility $\int_{0}^{T}\sigma_t^2dt$, spot volatility $\sigma^2_t$ for any fixed $t\in[0,T]$, and the general volatility functional $\int_{0}^{T} g(\sigma_t^2)dt$ for a function $g$; see \cite{AJ2014} for a comprehensive introduction. In the absence of the jump component in \eqref{model-price}, integrated volatility can be estimated by the realized volatility estimator (see, e.g., 
\cite{ABDL2003}). The presence of jumps introduces bias into standard realized volatility estimator, and various methods have been proposed to mitigate their effect. These include the thresholding approach in 
\cite{M2011} and \cite{MR2011}, which filters increments through a truncation procedure, and the bi-power and multi-power estimators in 
\cite{BSW2006}, \cite{W2006} and \cite{J2008}, which use products of two or more consecutive increments to reduce the effect of jumps. When the jumps have infinite variation, these estimators remain consistent, but their central limit theorems are no longer valid; see \cite{JR2014} and \cite{JT2014}. To this end, by using the empirical characteristic function of high-frequency increments, \cite{JT2014} constructed an integrated volatility estimator that achieves both the optimal convergence rate and the efficient asymptotic variance. Spot volatility estimators have been studied by 
\cite{D2010}, \cite{JP2012}, \cite{YP2014}, \cite{LLL2018}, and some others.

In recent years, the dynamics of the volatility process and its relationship with the log-price process, including the leverage effect and volatility of volatility, have attracted considerable attention. The leverage effect is defined as the covariance between the asset price process and its volatility (\cite{C1982}). The estimation of the leverage effect using high-frequency data in a continuous setting was investigated by \cite{AJ2014} and \cite{WM2014}. In the presence of jumps, \cite{AFLWY2017} and \cite{KX2017} applied thresholding technique to remove jump-contaminated increments. They then constructed leverage effect estimators by replacing the true spot volatilities with estimates and appropriately correcting for the bias induced by the estimation procedure. These existing estimators impose restrictions on jump activity; for example, consistency holds only when the jump component has finite variation. Volatility of volatility quantifies the variability of the volatility process and thereby provides a more detailed description of asset price dynamics. Because the volatility process is unobservable and must itself be estimated, estimating volatility of volatility is particularly challenging. The estimation of volatility of volatility was considered by \cite{AJ2014} and \cite{M2015} without jumps and by \cite{BV2009} in the presence of jumps. In the latter case, only consistency was established, under the condition that the jumps in the log-price process have finite variation.
The primary objective of this paper is to investigate the effect of possible infinite-variation jumps on the estimation of the leverage effect and volatility of volatility. For comparison, Table \ref{tab:comp_the} summarizes the conditions on the jump activity index for the aforementioned estimators and our proposed estimators when all estimators achieve the optimal convergence rate of $n^{-1/4}$.
The finite-sample performance of these estimators is examined in Section \ref{sec:simu}.
Feasible inference of the leverage effect and volatility of volatility was also investigated in \cite{CT2024} by using high-frequency observations of short-dated options, when the jumps are assumed to have finite variation. 
\begin{table}[!htbp]
	\caption{Conditions on the jump activity index for the consistency and central limit theorems (CLTs) of different leverage effect and volatility of volatility estimators when the optimal convergence rate of $n^{-1/4}$ is achieved. The explicit forms of the estimators are given in Section \ref{sec:estimator}.}\label{tab:comp_the}
	\vspace{0.2cm}
	\centering
	\scriptsize
	\begin{tabular}{|c|c|c|c|c|c|c|c|}
		\hline
		&\multicolumn{4}{c|}{Leverage effect} &\multicolumn{3}{c|}{Volatility of volatility} \\
		\cline{2-8}
		&Lev-AJ14, -WM14& Lev-AFLWY17&Lev-KX17&Lev-our &Vov-AJ14, -V15& Vov-BV09& Vov-our\\
		\hline
		\text{Consistency}&no jump  & JAI$<1$  & not given  &JAI$\leq 4/3$& no jump& JAI$<1$& JAI$<4/3$ \\
		\hline
		\text{CLT}	&no jump  & JAI$<1/2$ & JAI$<1/3$ & JAI$\leq 1 $& no jump& not given & JAI$<1$ \\
		\hline
	\end{tabular}
\end{table}

The main challenge in estimating the leverage effect and volatility of volatility is that the volatility path cannot be observed directly and must be estimated. In addition, jumps introduce estimation bias, which may be more severe when jump activity is high. To mitigate the influence of jumps, we apply the jump-robust spot volatility estimator proposed by \cite{LLL2018}. This estimator is based on the empirical characteristic function of high-frequency increments and is more effective than thresholding and bipower estimators in reducing the effect of jumps, especially infinite-variation jumps. We then construct estimators of the leverage effect and volatility of volatility by substituting the estimated volatility path into their definitions and appropriately correcting for the bias induced by the estimation procedure. Under mild conditions, we establish the consistency and asymptotic normality of the proposed estimators. Consistent estimation of the relevant volatility functionals then yields feasible central limit theorems.

This paper makes several contributions. First, as shown in Table \ref{tab:comp_the}, our theoretical results accommodate a broader class of jumps than existing results. The conditions can be relaxed further when the jump component has the special structure specified in Assumption \ref{asu-cha}. Specifically, for the estimation of volatility of volatility, consistency holds for any jump activity index below 2, whereas asymptotic normality requires the index to be below $1.5$, as stated in Theorem \ref{thm-vov}. In Section \ref{sec:remark}, we provide intuitive explanations of how thresholding, bi-power estimator, and our method mitigate the effect of jumps and why our approach outperforms existing methods.
We also find that the spot volatility estimator proposed by \cite{LLL2018} is particularly suitable when increments of estimated volatility are used, as in the estimation of volatility of volatility, because the jump-induced bias terms cancel. Second, some studies, such as \cite{WM2014}, define the leverage effect as the quadratic covariation between the log-price process and a general function of the volatility process. We extend our theory to this setting and derive the corresponding central limit theorem. Although both our estimator and that of \cite{WM2014} achieve the optimal convergence rate, our estimator has a smaller asymptotic variance. Moreover, \cite{WM2014} did not consider jumps. Third, unlike traditional thresholding technique, our method does not require a parameter-tuning procedure, making it more convenient for empirical applications. Although the spot volatility estimator involves a parameter $u$, our simulation studies show that fixing it at a constant yields satisfactory finite-sample performance for estimating both the leverage effect and volatility of volatility.
Fourth, as a by-product, we show that the spot volatility estimator of \cite{LLL2018} achieves the optimal convergence rate of $n^{1/4}$, rather than the nearly optimal rate of $n^{1/4}/\log(n)$ established therein. The convergence rate is governed by two terms that must be balanced through the choice of bandwidth. Proposition \ref{pro-1} characterizes how the estimator's asymptotic properties vary with the bandwidth, paralleling the analysis of spot volatility estimation in \cite{AJ2014}. More importantly, \cite{AJ2014} established only an infeasible central limit theorem for the spot volatility estimator, whereas we provide a feasible version.

The remainder of this paper is organized as follows. Section \ref{sec:setup} presents the model and assumptions. Section \ref{sec:est} introduces the proposed estimators of spot volatility, the leverage effect, volatility of volatility, and volatility functionals, together with their asymptotic properties. 
We discuss several issues in Section \ref{sec:diss}, including alternative definitions of the leverage effect used in the existing literature and the effects of jumps. 
Simulation studies are conducted in Section \ref{sec:simu}.
We apply our estimators to a real high-frequency dataset in Section \ref{sec:emp}.
Section \ref{sec:con} concludes. All proofs of the theoretical results in the paper are given in Section \ref{sec:proofs}. 

\section{Setup and assumptions}\label{sec:setup}
For the jump component $J$ in the data-generating process \eqref{model-price}, we consider
\begin{equation}\label{model-price-jump}
	J_t = \int_0^t \gamma_sdL_s + J'_t,
\end{equation}
where $L$ and $J'$ are pure-jump processes and
$\gamma$ is an adapted and locally bounded c$\grave{\text a}$dl$\grave{\text a}$g process.
According to \citet{JS2003}, we can write $L$ and $J'$ as
\begin{align}\label{jump:dec}
	\begin{split}
		dL_t&= \int_{|x| \leq 1}x(\mu-\nu)(dt,dx)+\int_{|x| >1}x\mu(dt,dx),\\
		J'_t &= \int_0^t \int_{|x|\leq 1}\delta(s,x) (\mu'-\nu')(ds,dx) + \int_0^t \int_{|x|> 1}\delta(s,x) \mu'(ds,dx),
	\end{split}
\end{align}
where $\delta$ is a predictable process on $\Omega \times [0,T] \times \mathbb{R}$;
$\mu$ and $\mu'$ are Poisson random measures on $[0,T] \times \mathbb{R}$ with intensity measures $\nu(dt,dx)=dt \otimes \lambda(dx)$ and $\nu'(dt,dx) = dt \otimes \lambda'(dx)$, respectively. We further assume that $L$ and $J'$ are independent.
Under these definitions, $X$ takes the following form:
\begin{equation}\label{model2}
	X_t =X_0 + \int_0^tb'_sds + \int_0^t\sigma_sdB_s + \int_0^t\int_{\mathbb{R}} \gamma_s\cdot x (\mu-\nu)(ds,dx) + \int_0^t \int_{\mathbb{R}}\delta(s,x) (\mu'-\nu')(ds,dx),
\end{equation}
where $b'_t = b_t+\int_{\{|x| > 1\}}\gamma_t\cdot x\lambda(dx) + \int_{\{|x| > 1\}}\delta(t,x)\lambda'(dx)$. The same model was used by \cite{JT2014} to estimate integrated volatility $\int_{0}^{T} \sigma_t^2dt$ and by \cite{LLL2018} to estimate spot volatility $\sigma_t^2$ for $t\in[0,T]$.
In \eqref{model-price-jump}, we decompose the jump component $J$ into two pure-jump processes, $L$ and $J'$, for the following reason.
Both processes may have infinite variation, but we impose structural assumptions on $L$ and not on $J'$. We show that the conditions on the jump activity index required for the asymptotic properties of our leverage effect and volatility of volatility estimators differ depending on whether this structure is imposed, as summarized in Section \ref{sec:remark}.


We further assume that $\sigma$ is a continuous It$\hat{\text{o}}$ semimartingale and can be written as
\begin{align}\label{model-vol}
	\sigma_t =\sigma_0 + \int_0^t \tilde{b}_sds + \int_0^t\tilde{\sigma}_sdB_s + \int_0^t \tilde{\sigma}'_sdB'_s,
\end{align}
where $\tilde{b}$, $\tilde{\sigma}$, and $\tilde{\sigma}'$ are adapted and locally bounded c$\grave{\text a}$dl$\grave{\text a}$g processes, and $B'$ is a standard Brownian motion independent of $B$.
A direct application of It$\hat{\text{o}}$'s lemma implies that the volatility process $\sigma^2$ can be written as
\begin{align}\label{model-volvol}
	d\sigma_s^2 =  \left(2\sigma_s\tilde{b}_s+(\tilde{\sigma}_s)^2 + (\tilde{\sigma}'_s)^2\right) ds + 2\sigma_s\tilde{\sigma}_sdB_s +  2\sigma_s\tilde{\sigma}'_sdB'_s. 
\end{align}

\begin{asu}\label{asu-con}
	Let $\{\tau_n, n=1,2,\cdots\}$ be a sequence of stopping times increasing to infinity, let $a_n$ be a sequence of real numbers, let $G$ be a deterministic nonnegative Lebesgue-integrable function on $\mathbb{R}$ satisfying $\int G(x)\lambda(dx) < \infty$ and $\int G(x)\lambda'(dx) < \infty$, and let $0 \leq r < 2$. If $0 \leq t<s\leq \tau_n(\omega)$, then
	\begin{equation}
		|V_t|\leq a_n,  \qquad ~\hbox{for}~V=b, \gamma, \sigma, \delta, \tilde{b}, \tilde{\sigma}, \tilde{\sigma}',
	\end{equation}
	and 
	\begin{equation}\label{cond-vov}
		\begin{split}
			& \E[(V_{s}-V_{t})^2] \leq a_n (s-t), \qquad ~\text{for}~V=\tilde{\sigma}, \tilde{\sigma}',\delta,\gamma,
		\end{split}
	\end{equation}
	and
	\begin{equation}
		|\delta(\omega,t,x)|^r\wedge 1\leq a_nG(x), \quad |\gamma_t(\omega)x|^2\wedge 1\leq a_nG(x). 
	\end{equation}
	\ignore{
		Moreover, it holds that, for some $\alpha>0$, 
		\begin{align}\label{cond-gamma}
			{\red{\E[|\gamma_s - \gamma_t |^2] \leq a_n|s-t|^{\alpha}.??????}}
		\end{align}
	}
\end{asu}

\begin{asu}\label{asu-cha}
	The process $L$ is a symmetric L$\acute{\text{e}}$vy process with Blumenthal--Getoor index $\beta$ and is independent of both $B$ and $B'$.
	We decompose $L$ as $L= L^{+}-L^{-}$, where $L^{+}$ and $L^{-}$ are two independent L$\acute{\text{e}}$vy processes with the same index $\beta$ and positive jumps. We assume that the characteristics of $L^{\pm}$ are $(0,0,F^{\pm})$ and that, for $x\in (0,1]$, there exist a uniform constant $r\in[0,1)$ and a function $f$ such that the tail functions $\overline{F}^{\pm}(x)=F^{\pm}((x,+\infty))$ satisfy
	\begin{align}\label{con-beta}
		\left|\overline{F}^{\pm}(x) - \frac{1}{x^{\beta}}  \right|\leq f(x),
	\end{align}
	where $f$ is a decreasing function with $\int_{0}^{1} x^{r-1}f(x)dx<\infty.$
\end{asu} 

Assumption \ref{asu-con} imposes local boundedness and smoothness conditions on the processes driving $X$ and $\sigma$, which are standard in the high-frequency literature. Examples satisfying condition \eqref{cond-vov} include It$\hat{\text{o}}$ semimartingales. Assumption \ref{asu-cha} requires the jump process $L$ to behave locally like a stable process near zero by restricting the deviation between the tail functions $\overline{F}^{\pm}(x)$ and $1/x^{\beta}$. Under this condition, the characteristic function of $L_t$ can be approximated by $E[e^{iuL_t}]=e^{-C|u|^\beta t}$.
\cite{JT2014} provided a detailed analysis showing that this assumption accommodates tempered stable processes, including time-changed Brownian motion, the normal inverse Gaussian process, and the Carr--Geman--Madan--Yor (CGMY) model. These models are widely used in finance.
We also note that the symmetry assumption can, to some extent, be relaxed for volatility estimation by replacing the original increments with differences between consecutive increments. This point was also discussed in \cite{JT2014}. Under the two assumptions above, the jump activity indices of $L$ and $J'$ in \eqref{model-price-jump} are $\beta$ and $r$, respectively.

\section{Estimators and asymptotic results}\label{sec:est}
We begin by estimating spot volatility and then use this estimator to construct estimators of the leverage effect and volatility of volatility. We establish the consistency and asymptotic normality of these estimators. Furthermore, to obtain feasible central limit theorems, we propose a consistent estimator of a general volatility functional.
Throughout the paper, for a process $Z$, we define the increments $\Delta_i^nZ = Z_{t_i^n} - Z_{t_{i-1}^n}$ for $i=1,...,n$.
We use $\longrightarrow^{p}, \longrightarrow^{L}, \longrightarrow^{L_s}$ to denote convergence in probability, convergence in law, and stable convergence in law\footnote{A detailed definition and discussion are provided after Proposition \ref{pro-1} in Section \ref{sec-spot}.}, respectively.

\subsection{Estimation of spot volatility}\label{sec-spot}
For model \eqref{model-price}, \cite{LLL2018} proposed a kernel-based spot volatility estimator that uses the empirical characteristic function to separate volatility from infinite-variation jumps. We apply this estimator with the uniform kernel $K(x) = 1_{\{0< x\leq 1\}}$\footnote{For clarity of exposition, we do not consider a general kernel function. We use the uniform kernel because it has the minimum asymptotic variance and is widely used.}, namely,
\begin{align}\label{est-spot}
	\widehat{\sigma}^2_{t} =\frac{-2}{u^2} \log\left( \left( \frac{1}{k_n}\sum_{j= \lfloor t /\Delta_n \rfloor +1}^{ \lfloor t /\Delta_n \rfloor +k_n} \cos{ \left(\frac{u\Delta_j^nX}{\sqrt{\Delta_n}} \right)} \right) \vee \frac{1}{\sqrt{k_n}}\right),
\end{align}
where $k_n\in \mathbb{Z}^{+}$ is the number of increments and $u$ is a positive real number.
As shown by \cite{LLL2018}, the threshold $\frac{1}{\sqrt{k_n}}$ ensures that the logarithm is well-defined and plays no asymptotic role. Their analysis shows that the jump component $L$ in \eqref{model-price} induces the following bias in the spot volatility estimator at time $t \in [0,T]$:
\begin{align}\label{bias}
	b_{t,n}=2C|\gamma_t|^\beta |u|^{\beta-2}\Delta_n^{1-\frac{\beta}{2}}.
\end{align}
Using an infeasible debiasing procedure, we obtain the following limiting result.
\begin{pro}\label{pro-1}
	Under Assumptions \ref{asu-con} and \ref{asu-cha}, suppose that, as $n\rightarrow \infty$, $k_n \rightarrow \infty$, $k_n\Delta_n \rightarrow 0$, and $k_n \sqrt{\Delta_n} \rightarrow \kappa$, where $\kappa$ is nonnegative and may be infinite.
	If $\beta <2$ and $r<4/3$, then, for $t\in [0, T)$\footnote{We do not consider estimation near the endpoint $t=T$. One possible method for addressing this boundary problem is described in Remark 3.1 of \cite{LL2024}.}, as $n\rightarrow \infty$,
	\begin{align}\label{pro1-res1}
		\sqrt{k_n}\left( \widehat{\sigma}^2_{t} -\sigma^2_{t} - b_{ t,n}(u) \right) &\longrightarrow^{L_s} V_{t}, \qquad \text{if} \quad  \kappa = 0,\\\label{pro1-res2}
		\sqrt{k_n}\left( \widehat{\sigma}^2_{t} -\sigma^2_{t} - b_{ t,n}(u) \right) &\longrightarrow^{L_s} V_{t} + \kappa V'_{t}, \qquad \text{if} \quad 0< \kappa < \infty,\\	\label{pro1-res3}
		\frac{1}{\sqrt{k_n\Delta_n}}\left( \widehat{\sigma}^2_{t} -\sigma^2_{t} - b_{ t,n}(u) \right) &\longrightarrow^{L_s} V'_{t}, \qquad \text{if} \quad  \kappa = \infty,
	\end{align}
	where $(V_{t}, V'_{t})$ is a vector of normal random variables defined on an extension of the original probability space $(\Omega,\mathcal{F},(\mathcal{F}_t)_{0\leq t\leq T},\mathbb{P})$. Conditional on the $\sigma$-field $\mathcal{F}$, the vector has zero mean, $\mathcal{F}$-conditional covariance $Cov(V_{t},V'_{t}|\mathcal{F})= \E[V_{t}V'_{t}|\mathcal{F}] = 0$, and $\mathcal{F}$-conditional variances
	\begin{align}\label{pro1-var}
		Var(V_{t}|\mathcal{F}) = h_1(u,t, \sigma_{t}^2), \qquad 
		Var(V'_{t}|\mathcal{F})= h_2(t, \sigma_{t}^2,(\tilde{\sigma}_{t})^2, (\tilde{\sigma}'_{t})^2), 
	\end{align}
	with
	\begin{align}\label{def_h12}
		\begin{split}
			&h_1(u,t, \sigma_{t}^2)= \frac{2(\exp{(-2 u^2\sigma_t^2)} - 2\exp{(- u^2\sigma_t^2)} + 1)}{u^4\exp{(- u^2\sigma_t^2 )}}, \\
			&h_2(t, \sigma_{t}^2,(\tilde{\sigma}_{t})^2, (\tilde{\sigma}'_{t})^2) = \frac{4(\sigma_t^2 )((\tilde{\sigma}_t)^2+(\tilde{\sigma}'_t)^2)}{3}. 
		\end{split}
	\end{align}
\end{pro}

Stable convergence in law is stronger than convergence in law. 
Specifically, let $Z_n$ be a sequence of random variables defined on the probability space $(\Omega,\mathcal{F},\mathbb{P})$, and let $Z$ be a random variable defined on an arbitrary extension $(\tilde{\Omega},\tilde{\mathcal{F}},\tilde{\mathbb{P}})$ of $(\Omega,\mathcal{F},\mathbb{P})$. Stable convergence in law of $Z_n$ to $Z$ implies
\begin{align*}
	\E[Y f(Z_n)] \rightarrow \tilde{\E}[Y f(Z)],
\end{align*}
for any bounded continuous function $f$ and any bounded random variable $Y$ on $(\Omega,\mathcal{F})$, where $\E$ and $\tilde{\E}$ denote expectations with respect to $\mathbb{P}$ and $\tilde{\mathbb{P}}$, respectively; see \cite{JS2003}, \cite{JP2012}, and \cite{PV2010} for further details.

By taking $k_n = \lfloor \kappa n^{1/2} \rfloor$, \eqref{pro1-res2} can be written as
\begin{align*}
	n^{1/4}\left( \widehat{\sigma}^2_{t} -\sigma^2_{t} - b_{ t,n}(u_n) \right) &\longrightarrow^{L_s} \frac{1}{\sqrt{\kappa}}V_{t} + \sqrt{\kappa}V'_{t}.
\end{align*}
Our convergence rate is faster than the rate $n^{1/4}/\log{n}$ obtained by \cite{LLL2018}; indeed, it is optimal for spot volatility estimation.
The limiting variance is minimized by setting $\kappa = \sqrt{Var(V_{t}|\mathcal{F})/ Var(V'_{t}|\mathcal{F}) }$, yielding the minimum value $2\sqrt{Var(V_{t}|\mathcal{F})\cdot Var(V'_{t}|\mathcal{F}) }$.
The central limit theorem established by \cite{LLL2018} corresponds to \eqref{pro1-res1}, one component of our result.
For this component, the asymptotic variance of the spot volatility estimator in \cite{LLL2018} is $2 (\sigma_{t})^4$, whereas ours is $\E[(V_{t})^2|\mathcal{F}]$.
These results are not contradictory. \cite{LLL2018} require $u$ to be a sequence $u_n$ that converges to zero at an appropriate rate as $n \rightarrow \infty$, and we recover the same result under this specification. A Taylor expansion at zero gives $\E[(V_{t})^2|\mathcal{F}] \longrightarrow^{p} 2 (\sigma_{t})^4$ if $u \rightarrow 0$.
We do not require $u\rightarrow 0$ because doing so would complicate the asymptotic conditions.
In addition, this specification introduces further approximation error.
More importantly, from the proof of the following Theorem \ref{thm-lev-1} and \ref{thm-vov}, we find that allowing $u \rightarrow 0$ does not further relax the restriction on jump activity index for leverage effect and volatility of volatility estimation. 

The results in \eqref{pro1-res1}--\eqref{pro1-res3} are infeasible because the bias terms $ b_{ t,n}(u)$ are unobservable. However, these terms are asymptotically negligible under suitable conditions. Specifically, we require $\sqrt{k_n}b_{ t,n}(u) \longrightarrow^{p} 0$ for \eqref{pro1-res1} and \eqref{pro1-res2}, and $\frac{1}{\sqrt{k_n\Delta_n}}b_{ t,n}(u) \longrightarrow^{p} 0$ for \eqref{pro1-res3}; that is,
\begin{align*}
	\sqrt{k_n}|u|^{\beta-2}\Delta_n^{1-\frac{\beta}{2}} \rightarrow 0, \quad \frac{1}{\sqrt{k_n\Delta_n}}|u|^{\beta-2}\Delta_n^{1-\frac{\beta}{2}} \rightarrow 0. 
\end{align*}
At the optimal convergence rate, corresponding to $k_n = O(\sqrt{n})$, the bias is negligible whenever $\beta<3/2$. 
When the jump activity index is sufficiently large that the bias term $b_{ t,n}(u)$ is not asymptotically negligible, an additional procedure removing the bias as in \cite{LLL2018} can be considered. 
We do not consider this issue because it is not the focus of the paper.

\subsection{Estimation of leverage effect}\label{sec-lev}
The leverage effect over $[0,T]$, denoted by $\mathcal{L}_{[0,T]}$, is defined as the quadratic covariance between $X$ and its volatility process $\sigma^2$:
\begin{align}\label{leverage}
	\mathcal{L}_{[0,T]}:= \langle X, \sigma^2\rangle_T = \int_0^T 2\sigma_t^2\tilde{\sigma}_tdt.
\end{align}
By definition,
\begin{align*}
	\langle X, \sigma^2\rangle_T = \lim_{n \rightarrow \infty} [X, \sigma^2]^n_T,
\end{align*}
where $[X, \sigma^2]^n_T:=\sum_{i=1}^{n} (\Delta_i^n X \cdot \Delta_i^n \sigma^2)$. A natural estimator of the leverage effect is obtained by replacing the spot volatility $\sigma_{t_i^n}^2$ in $[X, \sigma^2]_T^n$ with its estimator $\widehat{\sigma}_{t_i^n}^2$ from \eqref{est-spot}.
This yields
\begin{align}\label{est-main}
	\widehat{\mathcal{L}}_{[0,T]} = \sum_{i=k_n+1}^{n-k_n} \left( \Delta_i^n X \cdot \left( \widehat{\sigma}^2_{t_{i+}^n} - \widehat{\sigma}^2_{t_{i-}^n} \right) \right),
\end{align}
with
\begin{align}\label{est-spot++}
	&\widehat{\sigma}^2_{t_{i+}^n} =\frac{-2}{u^2} \log\left( \left( \frac{1}{k_n}\sum_{j \in I^n_{i+} } \cos{ \left(\frac{u\Delta_j^nX}{\sqrt{\Delta_n}} \right)} \right) \vee \frac{1}{\sqrt{k_n}}\right),\\\label{est-spot-}
	&\widehat{\sigma}^2_{t_{i-}^n} =\frac{-2}{u^2} \log\left(\left( \frac{1}{k_n}\sum_{j \in I^n_{i-} } \cos{ \left(\frac{u\Delta_j^nX}{\sqrt{\Delta_n}} \right)} \right)\vee \frac{1}{\sqrt{k_n}}\right),
\end{align}
where $u \in \mathbb{R}^{+}$, $I^n_{i+} = \{i+1,...,i+k_n\}$, and $I^n_{i-} = \{i-k_n,...,i-1\}$\footnote{We use $\widehat{\sigma}^2_{t_{i-}^n}$, rather than $\sigma_{t_{(i-1)+}^n}^2$, to estimate $\sigma_{t_{i-1}^n}^2$. This choice avoids overlapping increments between $\widehat{\sigma}^2_{t_{i-}^n}$ and $\widehat{\sigma}_{t_{i+}^n}^2$. The same idea was adopted by \cite{AFLWY2017}.}. The sets $I^n_{i+}$ and $I^n_{i-}$ are two local windows of length $k_n\Delta_n$ immediately after and before time $t_i^n$, respectively. In fact, $\widehat{\sigma}^2_{t_{i-}^n}$ is the spot volatility estimator of \cite{LLL2018} at time $t_{i-1}^n$ with kernel $K(x) = 1_{\{-1\leq x < 0\}}$; that is, $t_{i-}^n = t_{i-1}^n$. And it is obvious that $\widehat{\sigma}^2_{t_{i-}^n} = \widehat{\sigma}^2_{t_{(i-k_n-1)+}^n} $.
\begin{thm}\label{thm-lev-1}
	Under Assumptions \ref{asu-con} and \ref{asu-cha}, suppose that, as $n\rightarrow \infty$, $k_n \rightarrow \infty$ and $k_n\Delta_n \rightarrow 0$.
	Let $k_n = \lfloor \kappa n^{b} \rfloor$ with $0<b<1$ and $\kappa$ a positive constant.\\
	(1). If $\max\{\beta,r\}\leq 1$, or $\max\{\beta,r\}\leq 4/3$ with $b=1/2$,
	we have
	\begin{align}\label{thm-lev-con}
		\widehat{\mathcal{L}}_{[0,T]} \longrightarrow^{p} \mathcal{L}_{[0,T]}.
	\end{align}
	(2). 
	For $\max\{\beta,r\}\leq 1$, we have
	\begin{align}\label{thm-lev-clt}
		\sqrt{n}^{b\wedge (1-b)} \cdot \left(\widehat{\mathcal{L}}_{[0,T]} - \mathcal{L}_{[0,T]} \right) \longrightarrow^{L_s} U,
	\end{align}
	where $U$ is a normal random variable defined on an extension of the original probability space $(\Omega,\mathcal{F},\mathcal{F}_{0\leq t\leq T},\mathbb{P})$. Conditional on the $\sigma$-field $\mathcal{F}$, it has zero mean and $\mathcal{F}$-conditional variance
	\begin{align}\label{lev-var}
		\begin{split}
			Var(U|\mathcal{F}) &= 
			\frac{2}{\kappa } \int_{0}^{T} \sigma_t^2  h_1(u,t, \sigma_{t}^2)  
			dt  \cdot 1_{\{0<b\leq \frac{1}{2}\}}
			+  2\kappa T	\int_{0}^{T} \sigma_{t}^2 h_2(t, \sigma_{t}^2,(\tilde{\sigma}_{t})^2, (\tilde{\sigma}'_{t})^2)  
			dt \cdot 1_{\{\frac{1}{2}\leq b < 1\}},
		\end{split}
	\end{align}
	where the functions $h_1(\cdot)$ and $h_2(\cdot)$ are defined in \eqref{def_h12}.
\end{thm}

A similar leverage effect estimator was proposed by \cite{AFLWY2017}, who used classical thresholding to address jumps. They require the strict condition $r<1$ for consistency, whereas our result allows $r\leq 1$. For the central limit theorem, a comparison of their Theorem 3 with part (2) of Theorem \ref{thm-lev-1} shows that our restriction on jump activity is weaker: we require $r \leq 1$, whereas they require $r < \frac{1}{2}$. Moreover, the performance of the thresholding method depends strongly on the appropriate selection of two tuning parameters ($\alpha$ and $\varpi$ in \cite{AFLWY2017}), which are themselves related to the jump activity index ($\varpi \in [\frac{3}{4(2-\beta)},\frac{1}{2})$). Selecting these parameters is therefore a critical and complicated problem, especially in empirical applications in which the jump activity index is unknown. Our estimator avoids this difficulty because its parameter $u$ does not depend on the jump activity indices $\beta$ and $r$, and the simulation studies in Section \ref{sec:simu} show that fixing $u=1$ yields satisfactory results.

When $b<\frac{1}{2}$, taking $u\rightarrow 0$ implies that $h_1(u,t)$ in \eqref{lev-var} converges to $2\sigma_t^4$, as discussed in Section \ref{sec-spot}. Consequently, the asymptotic variance $Var(U|\mathcal{F})$ becomes $\frac{4}{\kappa}\int_{0}^{T} \sigma_t^6 dt$, which coincides with that obtained by \cite{AFLWY2017}.
When $b=\frac{1}{2}$, the estimator achieves the convergence rate $n^{-1/4}$, which is optimal for estimating the leverage effect in the absence of microstructure noise. The limiting variance is minimized by setting
\begin{align}\label{opt_kk}
	\kappa_{opt} = \sqrt{ \frac{\int_{0}^{T} \sigma_t^2  h_1(u,t, \sigma_{t}^2)
			dt} {T	\int_{0}^{T} \sigma_{t}^2 h_2(t, \sigma_{t}^2,(\tilde{\sigma}_{t})^2, (\tilde{\sigma}'_{t})^2) dt  }}.
\end{align}
When $b>\frac{1}{2}$, the asymptotic variance $Var(U|\mathcal{F})$ is the same as that in \cite{AFLWY2017}.

\ignore{
	\begin{rmk}\label{rmk:beta}
		{\red{It may be weird that the same condition $\max\{\beta,r \} \leq 1$ is required for both the consistency \eqref{thm-lev-con} and the asymptotic normality \eqref{thm-lev-clt}, at the first glance. This is because we allow the parameter $b$ to be as close to $0$ or $1$, making the main term in $\left(\widehat{\mathcal{L}}_{[0,T]} - \mathcal{L}_{[0,T]} \right)$ as close to $O_p(1)$, while other error terms caused by the jumps can always be neglected, irrespective of the jump activity index. Moreover, if we fix $b=\frac{1}{2}$, then the requirement for the consistency result \eqref{thm-lev-con} can be relaxed to $\max\{ \beta,r \}\leq \frac{4}{3}$. 
				The above points are also explained with detailed theoretical derivations in the proof of Theorem \ref{thm-lev-1}.}}
	\end{rmk}
}
To make the central limit theorem in Theorem \ref{thm-lev-1} feasible, Section \ref{sec:fea-clt} proposes a consistent estimator of the asymptotic variance $Var(U|\mathcal{F})$.

\subsection{Estimation of volatility of volatility}\label{sec:vov}
We now consider the integrated volatility of volatility of the process $\sigma^2$. Under model \eqref{model-volvol}, it is defined as
\begin{align}\label{vov:form}
	VoV_{[0,T]} := \langle \sigma^2, \sigma^2\rangle_T  =\int_{0}^{T} 4(\sigma_{t}^2 )((\tilde{\sigma}_{t})^2+(\tilde{\sigma}'_{t})^2) dt.
\end{align}
To construct the estimator, we first discretize $VoV_{[0,T]}$ as
\begin{align}\label{vov-dis}
	\Delta_n\sum_{i=k_n+1}^{n-k_n} 4(\sigma_{t_i^n}^2 )((\tilde{\sigma}_{t_i^n})^2+(\tilde{\sigma}'_{t_i^n})^2).
\end{align}
We estimate this quantity by substituting the spot volatility estimates, an approach also used to estimate volatility functionals (see, e.g., \cite{JR2013}). The integrand in \eqref{vov:form} appears in the asymptotic variance of the spot volatility estimator $\widehat{\sigma}^2_{t_{i+}^n}$ in Proposition \ref{pro-1} when $\frac{1}{2} \leq b<1$\footnote{The integrand is contained in the exact variance for any $0<b<1$. However, when $0<b< \frac{1}{2}$, the asymptotic variance is dominated by $h_1(u,t, \sigma_{t}^2)$, which can be seen from the proof of Lemma \ref{lem-lev-1}.}. A similar result holds for $\widehat{\sigma}^2_{t_{i-}^n}$. In addition, Lemma \ref{lem-lev-1} in the Appendix establishes a joint central limit theorem for $(\widehat{\sigma}^2_{t_{i+}^n} , \widehat{\sigma}^2_{t_{i-}^n})$. Based on this result, we obtain
\begin{align}\label{vov-spotest}
	\begin{split}
		&\E \left[\left( \frac{1}{\sqrt{k_n\Delta_n}}\left( \widehat{\sigma}^2_{t_{i+}^n}  -(\sigma^2_{t_{i+}^n} + b_{ t_{i+}^n,n}(u) ) \right) - \frac{1}{\sqrt{k_n\Delta_n}}\left( \widehat{\sigma}^2_{t_{i-}^n}  -(\sigma^2_{t_{i-}^n} + b_{ t_{i-}^n,n}(u) ) \right) \right)^2  \right]  \\
		& - \frac{2}{k_n^2\Delta_n}h_1(u,t_{i+}^n,\sigma_{t_i^n}^2) - \frac{2}{3} \left( 4(\sigma_{t_i^n}^2 )((\tilde{\sigma}_{t_i^n})^2+(\tilde{\sigma}'_{t_i^n})^2) \right)\longrightarrow^{p} 0,
	\end{split}
\end{align}
where $h_1(\cdot)$ and $h_2(\cdot)$ are defined in \eqref{def_h12}.
This result motivates the following spot volatility of volatility estimator at $t \in [0,T]$:
\begin{align*}
	\frac{1}{\widetilde{m}k_n} \sum_{i=\lfloor t/\Delta_n \rfloor+1}^{\lfloor t/\Delta_n \rfloor + \widetilde{m}k_n}\left( \frac{ 3\left( \left( \widehat{\sigma}^2_{t_{i+}^n}  -(\sigma^2_{t_{i+}^n} + b_{ t_{i+}^n,n}(u) ) \right) -\left( \widehat{\sigma}^2_{t_{i-}^n}  -(\sigma^2_{t_{i-}^n} + b_{ t_{i-}^n,n}(u) ) \right) \right)^2}{2k_n\Delta_n} - \frac{3h_1(u,t_{i+}^n,\sigma_{t_i^n}^2)}{k_n^2\Delta_n} \right),
\end{align*}
where $\widetilde{m}$ is an integer sequence tending to infinity such that $\widetilde{m}k_n\Delta_n \rightarrow 0$. For the integrated version, we can equivalently partition $[0,T]$ into $\lfloor T/(\widetilde{m}k_n\Delta_n) \rfloor$ blocks, each of length $\widetilde{m}k_n\Delta_n$, and construct the estimator as
\begin{align*}
	&\sum_{j=0}^{\lfloor T/(\widetilde{m}k_n\Delta_n) \rfloor-1 } \Bigg( \widetilde{m} k_n\Delta_n \cdot  \\
	&\frac{1}{\widetilde{m}k_n} \sum_{i=j\widetilde{m}k_n+1}^{(j+1)\widetilde{m}k_n}\Bigg( \frac{ 3\left( \left( \widehat{\sigma}^2_{t_{i+}^n}  -(\sigma^2_{t_{i+}^n} + b_{ t_{i+}^n,n}(u) ) \right) -\left( \widehat{\sigma}^2_{t_{i-}^n}  -(\sigma^2_{t_{i-}^n} + b_{ t_{i-}^n,n}(u) ) \right) \right)^2}{2k_n\Delta_n} - \frac{3h_1(u,t_{i+}^n,\sigma_{t_i^n}^2)}{k_n^2\Delta_n} \Bigg)\Bigg).
\end{align*}
Although the processes $\sigma^2$ and $b$ are unobservable, Assumption \ref{asu-con} implies that $\sigma^2_{t_{i+}^n}  - \sigma^2_{t_{i-}^n} = O_p(\Delta_n^{1/2}) $ and $b_{ t_{i+}^n,n}(u)  - b_{ t_{i-}^n,n}(u) = O_p(\Delta_n^{(3-\beta)/2})$, both of which are asymptotically negligible. Furthermore, $h_1(u,t_i^n,\sigma_{t_i^n}^2)$ can be estimated by substituting the volatility estimator $\widehat{\sigma}_{t_{i+}^n}^2$. This yields the following integrated volatility of volatility estimator:
\begin{align}\label{vov-main}
	\widehat{VoV}_{[0,T]} = \sum_{i=k_n+1}^{n-k_n} \left( \frac{3}{2k_n}  \left( \widehat{\sigma}^2_{t_{i+}^n} - \widehat{\sigma}^2_{t_{i-}^n} \right)^2 - \frac{3}{k_n^2} h_1(u,t_{i+}^n,\widehat{\sigma}_{t_{i+}^n}^2) \right).
\end{align}

\begin{thm}\label{thm-vov}
	Under Assumptions \ref{asu-con} and \ref{asu-cha}, suppose that
	as $n\rightarrow \infty$, $k_n \rightarrow \infty$, $k_n\Delta_n \rightarrow 0$.
	Let $k_n = \lfloor \kappa n^{b} \rfloor$ with $ \frac{1}{2}\leq b<1$ and $\kappa$ a positive constant.\\
	(1). If $\beta<2$ and $r<4/3$, we have, as $n\rightarrow \infty$,
	\begin{align}\label{thm-vov-con}
		\widehat{VoV}_{[0,T]}  \longrightarrow^{p} VoV_{[0,T]}.
	\end{align}
	(2). Furthermore, if $\beta<3/2$ and $r<1$, we have, as $n\rightarrow \infty$,
	\begin{align}\label{thm-vov-clt}
		n^{\frac{1-b}{2}} \left(\widehat{VoV}_{[0,T]} - VoV_{[0,T]} \right) \longrightarrow^{L_s} W,
	\end{align}
	where $W$ is a normal random variable defined on an extension of the original probability space $(\Omega,\mathcal{F},\mathcal{F}_{0\leq t\leq T},\mathbb{P})$. Conditional on the $\sigma$-field $\mathcal{F}$, it has zero mean and $\mathcal{F}$-conditional variance
	\begin{align}\label{var_W}
		Var(W|\mathcal{F}) &= \int_{0}^{T} H(u,t, \sigma_{t}^2,(\tilde{\sigma}_{t})^2, (\tilde{\sigma}'_{t})^2) )dt,
	\end{align}
	with 
	\begin{align*}
		&H(u,t, \sigma_{t}^2,(\tilde{\sigma}_{t})^2, (\tilde{\sigma}'_{t})^2)\\
		&= \Bigg(9\frac{(h'_1(u,t,\sigma_{t}^2))^2h_1(u,t,\sigma_{t}^2)}{\kappa^5T^2} +  \frac{9}{2} \frac{(h'_1(u,t,\sigma_{t}^2))^2h_2(t,\sigma_{t}^2,(\tilde{\sigma}_{t})^2, (\tilde{\sigma}'_{t})^2)}{\kappa^3T}  + \frac{27}{2} \frac{(h_1(u,t,\sigma_{t}^2))^2}{\kappa^3T} \\
		&\quad + \frac{709}{40} \frac{h_1(u,t,\sigma_{t}^2)h_2(t,\sigma_{t}^2,(\tilde{\sigma}_{t})^2, (\tilde{\sigma}'_{t})^2) }{\kappa} \Bigg)\cdot 1_{\{b=\frac{1}{2}\}}+ \kappa T \cdot  \frac{837}{70} (h_2(t, \sigma_{t}^2,(\tilde{\sigma}_{t})^2, (\tilde{\sigma}'_{t})^2))^2,
	\end{align*}
	where $h_1(\cdot)$ and $h_2(\cdot)$ are defined in \eqref{def_h12}, and $h'_1(u,t,\sigma_{t}^2)$ denotes the first derivative of $h_1(u,t,\sigma_{t}^2)$ with respect to $\sigma_t^2$.
\end{thm}
The preceding results show that the choice $0<b<\frac{1}{2}$ is not applicable. In this case, the convergence rate is $\sqrt{k_n}$, and
\begin{align*}
	\sqrt{k_n}\left(\widehat{VoV}_{[0,T]} - VoV_{[0,T]} \right) = k_n\sqrt{\Delta_n} \cdot	\frac{1}{\sqrt{k_n\Delta_n}} \left(\widehat{VoV}_{[0,T]} - VoV_{[0,T]} \right).
\end{align*}
The asymptotic variance diverges when $b<\frac{1}{2}$, invalidating the central limit theorem.

The convergence rate in Theorem \ref{thm-vov} is the same as those obtained by \cite{AJ2014} and \cite{M2015} for the continuous case. Setting $b=\frac{1}{2}$ yields the optimal convergence rate of $1/n^{1/4}$, and our asymptotic variance is close to those in \cite{AJ2014} and \cite{M2015}.\ignore{\footnote{This can be seen after rewriting
		\begin{align*}
			H(u,t, \sigma_{t}^2,(\tilde{\sigma}_{t})^2, (\tilde{\sigma}'_{t})^2) )&= \left( \frac{54}{\kappa^3} \left(\frac{h_1(u,t,\sigma_{t}^2)}{2}\right)^2 + \frac{709}{60}\frac{1}{\kappa} \frac{h_1(u,t,\sigma_{t}^2)}{2}\cdot 3h_2(t,\sigma_{t}^2,(\tilde{\sigma}_{t})^2, (\tilde{\sigma}'_{t})^2) \right) \cdot 1_{\{b=\frac{1}{2}\}}\\
			&\quad +\kappa \cdot  \frac{93}{70} (3h_2(t, \sigma_{t}^2,(\tilde{\sigma}_{t})^2, (\tilde{\sigma}'_{t})^2)))^2,
		\end{align*} 
		and doing the comparison after applying $(h_1(u,t,\sigma_{t}^2))^2 \approx 2\sigma_t^4$, $3h_2(t, \sigma_{t}^2,(\tilde{\sigma}_{t})^2, (\tilde{\sigma}'_{t})^2) =  4(\sigma_{t}^2 )((\tilde{\sigma}_{t})^2+(\tilde{\sigma}'_{t})^2) $, the spot volatility of volatility, and matching corresponding notations. We notice that the three constant coefficients driving above three terms are 54, $\frac{709}{60}$, $\frac{93}{70}$, while the ones obtained by \cite{AJ2014} and \cite{M2015} are 48, 12, $\frac{151}{70}$. In the simulation study, we present both these theoretical results and compare them with the sample variances, the simulation results show that our asymptotic variance is closer to the sample variance. We also note that we provide a step-by-step calculation for the derivation of asymptotic variance, while most of the details are omitted in \cite{M2015}. And our proof is also different from the ones in \cite{AJ2014} and \cite{M2015}.}.}
Because the asymptotic variance depends on $\kappa$ in a complicated nonlinear manner, we do not consider its minimization.
As before, a feasible central limit theorem can be obtained using a consistent estimator of the asymptotic variance $Var(W|\mathcal{F})$, presented in Section \ref{sec:fea-clt}.

\subsection{Estimation of volatility functionals}\label{sec:vol-fun}
The central limit theorems in Proposition \ref{pro-1}, Theorem \ref{thm-lev-1}, and Theorem \ref{thm-vov} are infeasible in practice because the limiting variances $Var(V_{t}|\mathcal{F})$, $Var(V'_{t}|\mathcal{F})$, $Var(U|\mathcal{F})$, and $Var(W|\mathcal{F})$ are unknown. Consistent estimators of these quantities are therefore required. Because $Var(U|\mathcal{F})$ and $Var(W|\mathcal{F})$ are specific forms of integrated volatility functionals, we first consider the consistent estimation of such functionals. Local versions of these estimators can then be used for functions of spot volatility, yielding consistent estimators of $Var(V_{t}|\mathcal{F})$ and $Var(V'_{t}|\mathcal{F})$.

We consider the integral over $[0,T]$ of a given function $g$ of the volatility process $\sigma^2$:
\begin{align*}
	I(g) := \int_{0}^{T} g(\sigma_t^2)dt,
\end{align*}
where $g$ is continuous.
A natural estimator of $I(g)$ approximates the integral by a Riemann sum based on local estimators of pointwise volatility:
\begin{align}\label{est-vol-func}
	\widehat{I}(g)= \Delta_n \sum_{i=0}^{n-k_n} g( \widehat{\sigma}^2_{t_{i+}^n}).
\end{align}
The same approach was adopted by \cite{JP2012}, \cite{JR2013}, and \cite{JT2014}, among others.

\begin{thm}\label{thm:vovfunc}
	Under Assumption \ref{asu-con}, suppose that
	as $n\rightarrow \infty$, $k_n \rightarrow \infty$, $k_n\Delta_n \rightarrow 0$. For any continuous function $g$, as $n\rightarrow \infty$,
	\begin{align}
		\widehat{I}(g) \longrightarrow^{p} I(g).
	\end{align}
\end{thm}

\begin{rmk}
	The result above holds for any $r \in [0,2)$. Thus, the assumption that $L$ is a L$\acute{\text{e}}$vy process, as in Assumption \ref{asu-cha}, is unnecessary for the consistency of the volatility functional estimator.
\end{rmk}

We now provide consistent estimators of $Var(U|\mathcal{F})$ and $Var(W|\mathcal{F})$. Theorem \ref{thm:vovfunc}, together with the identity $VoV_{[0,T]} = 3 \int_{0}^{T} h_2(t, \sigma_{t}^2,(\tilde{\sigma}_{t})^2, (\tilde{\sigma}'_{t})^2)dt$, motivates the following estimator of $Var(U|\mathcal{F})$:
\begin{align}
	\begin{split}
		\widehat{Var(U|\mathcal{F})}&= \frac{2}{\kappa } \cdot \left(\Delta_n \sum_{i=0}^{n-k_n}  (\widehat{\sigma}^2_{t_{i+}^n} h_1(u,t_{i+}^n, \widehat{\sigma}^2_{t_{i+}^n}) ) \right) \cdot 1_{\{0<b\leq \frac{1}{2}\}}\\
		&\quad +  2\kappa T	\cdot \left( \sum_{i=k_n+1}^{n-k_n} \frac{\widehat{\sigma}^2_{t_{i+}^n} }{3} \left( \frac{3}{2k_n}  \left( \widehat{\sigma}^2_{t_{i+}^n} - \widehat{\sigma}^2_{t_{i-}^n} \right)^2 - \frac{3}{k_n^2} h_1(u,t_{i+}^n,\widehat{\sigma}_{t_{i+}^n}^2) \right) \right) \cdot 1_{\{\frac{1}{2}\leq b < 1\}}.
	\end{split}
\end{align}
Similarly, for $Var(W|\mathcal{F})$ in \eqref{var_W}, \ignore{we recall that
	\begin{align*}
		&H(u,t, \sigma_{t}^2,(\tilde{\sigma}_{t})^2, (\tilde{\sigma}'_{t})^2) )\\
		&=\kappa \cdot \left(\frac{27}{2\kappa^4} \int_{0}^{T}  (h_1(u,t,\sigma_{t}^2))^2dt + \frac{709}{40\kappa^2} \int_0^T ( h_1(u,t,\sigma_{t}^2)h_2(t,\sigma_{t}^2,(\tilde{\sigma}_{t})^2, (\tilde{\sigma}'_{t})^2)) dt \right)\cdot 1_{\{b=\frac{1}{2}\}} \\
		&\quad +\kappa \cdot  \frac{837}{70} (h_2(t, \sigma_{t}^2,(\tilde{\sigma}_{t})^2, (\tilde{\sigma}'_{t})^2)))^2 dt.
\end{align*}}
The terms $\int_{0}^{T} \frac{(h_1(u,t,\sigma_{t}^2))^2}{\kappa^4T^2}dt$, $\int_{0}^{T} \frac{h_1(u,t,\sigma_{t}^2)}{\kappa^2T} h_2(t,\sigma_{t}^2,(\tilde{\sigma}_{t})^2, (\tilde{\sigma}'_{t})^2) dt$, $\int_{0}^{T}\frac{(h'_1(u,t,\sigma_{t}^2))^2(h_1(u,t,\sigma_{t}^2))}{\kappa^6T^3} dt$, and $\int_{0}^{t}\frac{(h'_1(u,t,\sigma_{t}^2))^2h_2(t,\sigma_{t}^2,(\tilde{\sigma}_{t})^2, (\tilde{\sigma}'_{t})^2)}{\kappa^4T^2}dt$ can be estimated by $\widehat{H_n^1}$, $\widehat{{H_n}^2}$, $\widehat{{H'_n}^1}$, and $\widehat{{H'_n}^2}$, respectively:
\begin{align}
	\widehat{H_n^1} &=\Delta_n \sum_{i=0}^{n-k_n}  \frac{(h_1(u,t_{i+}^n, \widehat{\sigma}^2_{t_{i+}^n}) )^2}{(k_n^2\Delta_n)^2},\  \widehat{{H'_n}^1}=\Delta_n \sum_{i=0}^{n-k_n}  \frac{(h'_1(u,t_{i+}^n, \widehat{\sigma}^2_{t_{i+}^n}) )^2h_1(u,t_{i+}^n, \widehat{\sigma}^2_{t_{i+}^n})}{(k_n^2\Delta_n)^3}  \\
	\widehat{H_n^2} &=   \sum_{i=k_n+1}^{n-k_n} \frac{h_1(u,t_{i+}^n, \widehat{\sigma}^2_{t_{i+}^n}) }{3k_n^2\Delta_n} \left( \frac{3}{2k_n}  \left( \widehat{\sigma}^2_{t_{i+}^n} - \widehat{\sigma}^2_{t_{i-}^n} \right)^2 - \frac{3}{k_n^2} h_1(u,t_{i+}^n,\widehat{\sigma}_{t_{i+}^n}^2) \right),\\
	\widehat{{H'_n}^2} &=   \sum_{i=k_n+1}^{n-k_n} \frac{(h'_1(u,t_{i+}^n, \widehat{\sigma}^2_{t_{i+}^n}))^2 }{3(k_n^2\Delta_n)^2} \left( \frac{3}{2k_n}  \left( \widehat{\sigma}^2_{t_{i+}^n} - \widehat{\sigma}^2_{t_{i-}^n} \right)^2 - \frac{3}{k_n^2} h_1(u,t_{i+}^n,\widehat{\sigma}_{t_{i+}^n}^2) \right).
\end{align}
Theorem \ref{thm-uv-hat} shows that
\begin{align*}
	\widehat{H_n^3}  \longrightarrow^{p} &\int_{0}^{T} \Big( 30 \frac{(h_1(u,t,\sigma_{t}^2))^2 }{\kappa^4T^2} +24 \frac{h_1(u,t,\sigma_{t}^2)h_2(t,\sigma_{t}^2,(\tilde{\sigma}_{t})^2, (\tilde{\sigma}'_{t})^2))}{\kappa^2T}\\
	&\quad \quad \ + 30 (h_2(t, \sigma_{t}^2,(\tilde{\sigma}_{t})^2, (\tilde{\sigma}'_{t})^2)))^2 \Big)dt,
\end{align*}
where
\begin{align}
	\widehat{H_n^3} =\frac{1}{k_n^2\Delta_n}  \sum_{i=k_n+1}^{n-k_n}   \left( \widehat{\sigma}^2_{t_{i+}^n} - \widehat{\sigma}^2_{t_{i-}^n} \right)^4.
\end{align}
Using these results, we define
\begin{align}
	\widehat{Var(W|\mathcal{F})} =\kappa T \cdot \left( \left( \frac{891}{35}\widehat{H_n^1} + \frac{38207}{1400} \widehat{H_n^2} + 9\widehat{{H'_n}^1} + \frac{9}{2} \widehat{{H'_n}^2} \right) \cdot 1_{\{b=\frac{1}{2}\}} + \frac{279}{70}\widehat{H_n^3}\right).
\end{align}

\begin{thm}\label{thm-uv-hat}
	Under Assumption \ref{asu-con}, suppose that 
	as $n\rightarrow \infty$, $k_n \rightarrow \infty$, $k_n\Delta_n \rightarrow 0$. Then, as $n\rightarrow \infty$,
	\begin{align}
		\widehat{Var(U|\mathcal{F})} \longrightarrow^{p} Var(U|\mathcal{F}), \quad \widehat{Var(W|\mathcal{F})} \longrightarrow^{p} Var(W|\mathcal{F}).
	\end{align}
\end{thm}

Similarly, $Var(V_{t}|\mathcal{F})$ can be estimated by
\begin{align}
	\begin{split}
		&\widehat{Var(V_{t}|\mathcal{F}) }= h_1(u,t, \widehat{\sigma}_{t}^2).
	\end{split}
\end{align}
Because $h_2(t, \sigma_{t}^2,(\tilde{\sigma}_{t})^2, (\tilde{\sigma}'_{t})^2))$ is a local version of $VoV_{[0,T]} /3$, we estimate $Var(V'_{t}|\mathcal{F})$ using a local version of $\widehat{VoV}_{[0,T]}$:
\begin{align}
	\widehat{Var(V'_{t}|\mathcal{F}) }= \frac{1}{m_n\Delta_n} \sum_{i=\lfloor t/\Delta_n \rfloor +1}^{\lfloor t/\Delta_n \rfloor +m_n} \left( \frac{1}{2k_n}  \left( \widehat{\sigma}^2_{t_{i+}^n} - \widehat{\sigma}^2_{t_{i-}^n} \right)^2 - \frac{1}{k_n^2} h_1(u,t_{i+}^n,\widehat{\sigma}_{t_{i+}^n}^2) \right),
\end{align}
where $m_n$ is an integer sequence that tends to infinity as $n\rightarrow\infty$.

\begin{thm}\label{thm-vv-hat}
	Under Assumption \ref{asu-con}, suppose that
	as $n\rightarrow \infty$, $m_n \rightarrow \infty$, $m_n/k_n \rightarrow \infty$, $m_n\Delta_n \rightarrow 0$. Then, as $n\rightarrow \infty$,
	\begin{align}
		\widehat{Var(V_{t}|\mathcal{F})} \longrightarrow^{p} Var(V_{t}|\mathcal{F}), \quad \widehat{Var(V'_{t}|\mathcal{F})} \longrightarrow^{p} Var(V'_{t}|\mathcal{F}).
	\end{align}
\end{thm}

\subsection{Feasible central limit theorems and their applications}\label{sec:fea-clt}
Combining the results in Sections \ref{sec-spot}--\ref{sec:vol-fun} with Proposition 2.5 of \cite{PV2010} yields the following feasible versions of Proposition \ref{pro-1}, Theorem \ref{thm-lev-1}, and Theorem \ref{thm-vov}.
\begin{corollary}\label{cor-spot-fea}
	Let $\mathcal{N}(0,1)$ denote a standard normal random variable. \\
	(a). Under the same assumptions and conditions as in Proposition \ref{pro-1}, for $t\in[0,T)$, we have, as $n\rightarrow \infty$,
	\begin{align}
		\begin{split}
			\sqrt{k_n}\left( \widehat{\sigma}^2_{t} -\sigma^2_{t} - b_{t,n}(u) \right) / \sqrt{\widehat{Var(V_{t}|\mathcal{F})} }&\longrightarrow^{L} \mathcal{N}(0,1), \text{if} \ \kappa = 0,\\
			\sqrt{k_n}\left( \widehat{\sigma}^2_{t} -\sigma^2_{t} - b_{ t,n}(u) \right) /\sqrt{\widehat{Var(V_{t}|\mathcal{F})} + \kappa^2 \widehat{Var(V'_{t}|\mathcal{F})}}& \longrightarrow^{L} \mathcal{N}(0,1) ,\text{if} \ 0< \kappa < \infty,\\
			\frac{1}{\sqrt{k_n\Delta_n}}\left( \widehat{\sigma}^2_{t} -\sigma^2_{t} - b_{ t,n}(u) \right) / \sqrt{\widehat{Var(V'_{t}|\mathcal{F})} } &\longrightarrow^{L} \mathcal{N}(0,1),\text{if} \ \kappa = \infty.
		\end{split}
	\end{align}
	(b). Under the same assumptions and conditions as in Theorem \ref{thm-lev-1}, we have, as $n\rightarrow \infty$, 
	\begin{align}\label{fea:lev}
		\sqrt{n}^{b\wedge (1-b)} \cdot \frac{\left(\widehat{\mathcal{L}}_{[0,T]} - \mathcal{L}_{[0,T]} \right)}{\sqrt{\widehat{Var(U|\mathcal{F})}}} \longrightarrow^{L} \mathcal{N}(0,1).
	\end{align}
	(c). Under the same assumptions and conditions as in Theorem \ref{thm-vov}, we have, as $n\rightarrow \infty$,
	\begin{align}\label{fea:vov}
		n^{\frac{1-b}{2}} \frac{ \left(\widehat{VoV}_{[0,T]} - VoV_{[0,T]} \right) }{\sqrt{\widehat{Var(W|\mathcal{F})}}} \longrightarrow^{L} \mathcal{N}(0,1).
	\end{align}
\end{corollary}

The feasible central limit theorems in Corollary \ref{cor-spot-fea} can be used to construct confidence intervals for the leverage effect and volatility of volatility and to test hypotheses that these quantities equal specified values.
For example, to test whether the volatility of volatility $VoV_{[0,T]}$ is zero, we consider the null and alternative hypotheses
\begin{align}\label{vov_test}
	H_0: \int_{0}^{T} 4(\sigma_{t}^2 )((\tilde{\sigma}_{t})^2+(\tilde{\sigma}'_{t})^2) dt = 0, \quad \text{v.s.} \quad H_1: \int_{0}^{T} 4(\sigma_{t}^2 )((\tilde{\sigma}_{t})^2+(\tilde{\sigma}'_{t})^2) dt > 0.
\end{align}
According to \eqref{fea:vov}, we can use the test statistic
\begin{align}\label{vov_test_sta}
	\widetilde{T}_n := n^{\frac{1-b}{2}} \frac{ \widehat{VoV}_{[0,T]}  }{\sqrt{\widehat{Var(W|\mathcal{F})}}},
\end{align}
Under the null hypothesis, $\widetilde{T}_n  \longrightarrow^{L} \mathcal{N}(0,1)$; under the alternative hypothesis, $\widetilde{T}_n$ diverges to $+\infty$ at the rate $n^{\frac{1-b}{2}}$.
In a setting with both jumps and microstructure noise, \cite{LLZ2022} showed that the convergence rate can be improved further. Under the null hypothesis, the diffusion terms in the volatility process vanish, and spot volatility becomes a bounded-variation process. This additional smoothness should facilitate the estimation of spot volatility relative to the usual setting in which volatility is an It$\hat{\text{o}}$ process.
Future work could extend our analysis to this setting and develop a test statistic with a faster convergence rate.
 
 \section{Discussions}\label{sec:diss}
 In Section \ref{sec-lev}, we define the leverage effect as the quadratic covariance between $X$ and its volatility process $\sigma^2$, while several alternative definitions have been used in the existing literature. We discuss these definitions and their estimation. We then summarize the effect of jumps on the estimation of the leverage effect and volatility of volatility. 
 
 \subsection{Leverage effect: A correlation perspective}
 \cite{KX2017} defined the leverage effect as a time-varying correlation process with 
 \begin{align}
 	\rho_t:=\frac{\langle X, \sigma^2\rangle'_t}{\sqrt{\langle X,X \rangle'_t \cdot \langle \sigma^2,\sigma^2\rangle'_t}}, \ t \in [0,T],
 \end{align}
 where $'$ denotes the first derivative with respect to time. For simplicity, we assume that it is constant over $[0,T]$, with $\rho_t \equiv \rho$. This setting includes the popular Heston model, in which $\rho$ is the constant correlation between the two Brownian motions. 
 Under this constancy assumption, the leverage effect can alternatively be defined, from the correlation perspective, as 
 \begin{align}
 	\mathcal{L}^{cor}_{[0,T]}:= \frac{\langle X, \sigma^2\rangle_T}{\sqrt{\langle X,X \rangle_T \cdot \langle \sigma^2,\sigma^2\rangle_T}}.
 \end{align}
 It can be estimated by plugging in the corresponding consistent estimators for the numerator and denominator terms, which yields
 \begin{align}
 	\widehat{\mathcal{L}}^{cor}_{[0,T]}  = \frac{\widehat{\mathcal{L}}_{[0,T]} }{\sqrt{\widehat{IV}_{[0,T]}}\sqrt{\widehat{VoV}_{[0,T]}}}, 
 \end{align}
 where $\widehat{\mathcal{L}}_{[0,T]}$ and $\widehat{VoV}_{[0,T]}$ are given by \eqref{est-main} and \eqref{vov-main}, respectively, and the integrated volatility estimator $\widehat{IV}_{[0,T]}$ can be found in \cite{JT2014}. 
 
 \begin{thm}\label{thm:lev-cor}
 	Under the same assumptions as in Theorem \ref{thm-lev-1}, if $\max\{\beta, r\} < 1$, then, as $n\rightarrow \infty$, 
 	\begin{align}
 		\widehat{\mathcal{L}}^{cor}_{[0,T]}  \longrightarrow^{p} \mathcal{L}^{cor}_{[0,T]}.
 	\end{align}
 \end{thm}
 For a central limit theorem for $\widehat{\mathcal{L}}^{cor}_{[0,T]}$, we need to derive the joint distribution of $\langle X, \sigma^2\rangle_T$, $\langle X, X\rangle_T$, and $\langle \sigma^2, \sigma^2\rangle_T$. This is beyond the scope of this paper and will be considered in future work.  
 
 \subsection{Leverage effect: A volatility functional perspective}
 \cite{WM2014} defined the leverage effect as the quadratic covariance between $X$ and $F(\sigma^2)$, namely,
 \begin{align}
 	\mathcal{L}^{func}_{[0,T]}:= \langle X, F(\sigma^2)\rangle_T,
 \end{align}
 where $F(\cdot)$ is a twice continuously differentiable function that is monotone on $(0,\infty)$. Applying It$\hat{\text{o}}$'s lemma to \eqref{model-volvol}, we obtain
 \begin{align}
 	\begin{split}
 		dF(\sigma_s^2) &=  \left(2\sigma_s\tilde{b}_s+(\tilde{\sigma}_s)^2 + (\tilde{\sigma}'_s)^2 + 2F''(\sigma_s^2)\sigma_s^2((\tilde{\sigma}_s)^2+(\tilde{\sigma}')^2)\right) ds \\
 		&\quad + 2F'(\sigma_s^2)\sigma_s\tilde{\sigma}_sdB_s +  2F'(\sigma_s^2)\sigma_s\tilde{\sigma}'_sdB'_s,
 	\end{split}
 \end{align}
 where $F'$ and $F''$ are the first and second derivatives of $F$, respectively. As a result, we have 
 \begin{align}
 	\langle X, F(\sigma^2)\rangle_T = 2F'(\sigma_s^2)\sigma_s^2\tilde{\sigma}_s.
 \end{align}
 Similarly to the estimation of $\mathcal{L}_{[0,T]}$, $\mathcal{L}^{func}_{[0,T]}$ can be estimated by 
 \begin{align}
 	\widehat{\mathcal{L}}^{func}_{[0,T]} = \sum_{i=k_n+1}^{n-k_n} \left( \Delta_i^n X \cdot \left( F( \widehat{\sigma}^2_{t_{i+}^n} )- F( \widehat{\sigma}^2_{t_{i-}^n}) \right) \right).
 \end{align}
 We can then establish the following result. 
 \begin{thm}\label{thm:lev-fun}
 	Under the same assumptions as in Theorem \ref{thm-lev-1}, \\
 	(1). If $\max\{\beta,r\}\leq 1$, or $\max\{\beta,r\}\leq 4/3$ with $b=1/2$, we have 
 	\begin{align}
 		\widehat{\mathcal{L}}^{func}_{[0,T]}  \longrightarrow^{p} \mathcal{L}^{func}_{[0,T]}.
 	\end{align}
 	(2). For $\max\{\beta,r\}\leq 1$, we have
 	\begin{align}
 		\sqrt{n}^{b\wedge (1-b)} \cdot \left( 	\widehat{\mathcal{L}}^{func}_{[0,T]}  -  \mathcal{L}^{func}_{[0,T]} \right) \longrightarrow^{L_s} \widetilde{U},
 	\end{align}
 	where $\widetilde{U}$ is a normal random variable defined on an extension of the original probability space $(\Omega,\mathcal{F},\mathcal{F}_{0\leq t\leq T},\mathbb{P})$. Moreover, conditionally on the $\sigma$-field $\mathcal{F}$, it has zero mean and $\mathcal{F}$-conditional variance
 	\begin{align}
 		\begin{split}
 			Var(\widetilde{U}|\mathcal{F}) &= 
 			\frac{2}{\kappa } \int_{0}^{T} \sigma_t^2 (F'(\sigma_t^2))^2 h_1(u,t, \sigma_{t}^2)  
 			dt  \cdot 1_{\{0<b\leq \frac{1}{2}\}}\\
 			& \quad +  2\kappa T	\int_{0}^{T} \sigma_{t}^2 (F'(\sigma_t^2))^2 h_2(t, \sigma_{t}^2,(\tilde{\sigma}_{t})^2, (\tilde{\sigma}'_{t})^2)  
 			dt \cdot 1_{\{\frac{1}{2}\leq b < 1\}}. 
 		\end{split}
 	\end{align}
 \end{thm}
 Comparing the asymptotic variance in the above theorem with that in \cite{WM2014}, we conclude that our theoretical variance is always smaller, regardless of the convergence rate. 
 Similarly, a feasible version of the theorem can be obtained by estimating the asymptotic variance with the volatility functional estimator discussed in Section \ref{sec:vol-fun}. 
 
 \subsection{The effect of jumps}\label{sec:remark}
 
 We now summarize the effect of jumps on the estimation of the leverage effect and volatility of volatility. Recall that, in model \eqref{model-price-jump}, our jump process consists of two parts. The first part is a L$\acute{\text{e}}$vy process driven by $L$ satisfying Assumption \ref{asu-cha}. The second part has jump activity index $r$ and a general structure without further conditions. 
 Theorems \ref{thm-lev-1}--\ref{thm-vov} show that the restriction on the first jump component is more relaxed: $\beta\leq 1$ $(r\leq 1)$ and $\beta<2$ $(r<4/3)$ are required for the consistency of the leverage effect estimator and the volatility of volatility estimator, respectively. For asymptotic normality, the corresponding conditions are $\beta\leq 1$ $(r\leq 1)$ and $\beta<3/2$ $(r<1)$. The reason is that, under condition \eqref{cond-vov}, the bias $b_{t,n}$ in \eqref{bias} can be diminished by taking the difference between the spot volatility estimates at two consecutive time points, namely $ \left( \widehat{\sigma}^2_{t_{i+}^n} - \widehat{\sigma}^2_{t_{i-}^n} \right)$ in \eqref{est-main} and \eqref{vov-main}. To illustrate this point, consider the extreme case $\gamma_t \equiv \gamma_0$ for $t\in[0,T]$, under which $b_{t_{i+}^n,n} \equiv b_{t_{i-}^n,n}$; 
 therefore, the difference between $\widehat{\sigma}^2_{t_{i+}^n}$ and $\widehat{\sigma}^2_{t_{i-}^n}$ can remove the bias terms $b_{t_{i+}^n,n}$ and $b_{t_{i-}^n,n}$. 
 As for general $\gamma$ satisfying Assumption \ref{asu-con}, we can obtain $b_{t_{i+}^n,n} - b_{t_{i-}^n,n} = O_p(\Delta_n^{(3-\beta)/2})$ from \eqref{cont-sigma+gamma} in the Appendix of proof. 
 
 We next provide some intuition for how the bipower estimator, the thresholding technique, and our method reduce the effect of jumps. The volatility estimators based on these three methods are given in \eqref{est:bip}, \eqref{est:thr}, and \eqref{est-spot}, respectively. For illustration, consider the special case $\Delta_i^n X = \Delta_i^n B + \Delta_i^n J$. As is well known, $\Delta_i^n B = O_p(\sqrt{\Delta_n})$ and $\Delta_i^n J = O_p(1)$. 
 As in \eqref{jump:dec}, jumps can be classified into rare ``large" jumps, whose sizes are larger than 1, and more frequent ``small" jumps, whose sizes are no larger than 1. The bipower estimator is based on $|\Delta_i^n X||\Delta_{i+1}^n X|$, and as the sampling frequency increases ($n\rightarrow \infty$), at most one increment, say $\Delta_i^n X$, contains jumps. Thus 
 \begin{align*}
 	|\Delta_i^n X||\Delta_{i+1}^n X| = |\Delta_i^n B + \Delta_i^n J||\Delta_{i+1}^n B| &\approx |\Delta_i^n B| |\Delta_{i+1}^n B|+ |\Delta_i^n J||\Delta_{i+1}^n B| \\
 	&= O_p(\Delta_n) + O_p(\sqrt{\Delta_n}),
 \end{align*}
 which shows that the influence of $|\Delta_i^n J|$ is reduced to $O_p(\sqrt{\Delta_n})$ for both ``large" and ``small" jumps. The thresholding technique is based on $ (\Delta_i^n X)^2\cdot 1_{\{|\Delta_i^n X| \leq \alpha \Delta_n^{\omega} \}}$ with parameters $\alpha, \omega$, so that the indicator function removes ``large" jumps with probability one, but cannot detect ``small" jumps. In this sense, by the mean value theorem, there exists $\xi_i^n \in [\Delta_i^n B, \Delta_i^n B+ \Delta_i^n J]$ such that 
 \begin{align*}
 	(\Delta_i^n X)^2 = (\Delta_i^n B + \Delta_i^n J)^2  = (\Delta_i^n B)^2+ 2\xi_i^n \Delta_i^n J = O_p(\Delta_n) + O_p(1),
 \end{align*}
 which shows that the effect of ``small" jumps remains of order $O_p(1)$. 
 The above analysis suggests that thresholding is more effective for ``large" jumps, while the bipower estimator performs better for ``small" jumps. The numerical comparison in \cite{V2011} verifies this intuition. Our estimator is constructed from $\cos(\Delta_i^n X)$, and by the mean value theorem, there exists $\xi_i^{'n} \in [\Delta_i^n B, \Delta_i^n B+ \Delta_i^n J]$ such that 
 \begin{align*}
 	\cos(\Delta_i^n X) = \cos(\Delta_i^n B + \Delta_i^nJ) =\cos(\Delta_i^n B) -\sin(\xi_i^{'n}) \Delta_i^nJ.
 \end{align*}
 Since $\sin(\xi_i^{'n})$ is bounded, the influence of jumps is always controlled. Moreover, if $ \Delta_i^nJ$ is relatively small, our method can reduce the effect of small jumps more effectively than thresholding, since $|\sin(x)| \leq |x|$ for $x\in[-1,1]$. This observation also suggests that our method could be improved by using thresholded increments to remove ``large" jumps entirely.  
 
\section{Simulation studies}\label{sec:simu}
We now conduct simulation studies to compare the finite sample performance of our estimators with that of existing ones and verify the theoretical results established in the previous sections.

\subsection{Existing leverage effect and volatility of volatility estimators}\label{sec:estimator}

We first introduce several existing leverage effect and volatility of volatility estimators and present their explicit forms; these estimators will be compared with our proposed estimators through simulation studies. For the leverage effect, we denote the estimators in \cite{AJ2014}, \cite{WM2014}, \cite{AFLWY2017}, \cite{KX2017}, and this paper by Lev-AJ14, Lev-WM14, Lev-AFLWY17, Lev-KX17, and Lev-our, respectively. For volatility of volatility, we denote the estimators in \cite{AJ2014}, \cite{M2015}, \cite{BV2009}, and this paper by Vov-AJ14, Vov-V15, Vov-BV09, and Vov-our, respectively. The conditions on the jump activity index for these estimators are compared in Table \ref{tab:comp_the}. 

Several estimators of the leverage effect have been proposed under different settings. Without jumps, the spot volatility can be estimated by
\begin{align*}
	\widehat{\sigma}^{2,con}_{t_{i}^n} = \frac{1}{k_n\Delta_n}\sum_{j \in I^n_{i+} } \left((\Delta_j^n X)^2 \right),
\end{align*}
for $i=0,...,n-k_n$. 
Using increments over non-overlapping time intervals, \cite{WM2014} proposed the following leverage effect estimator:
\begin{align}\label{est-WM2014} 
	\text{``$\text{Lev-WM14}$" Estimator:} \ 2\sum_{j=0}^{\lfloor n/k_n \rfloor -2} \left( \left( X_{\tau_{j+1}^n}- X_{\tau_{j}^n}\right)  \cdot \left( \widehat{\sigma}^{2,con}_{\tau_{(j+1)+}^n} - \widehat{\sigma}^{2,con}_{\tau_{j+}^n} \right) \right),
\end{align}
where $\tau_{j}^n:= t_{jk_n}^n$ and $\tau_{j+}^n:=t_{jk_n+}^n$ for $j=0,...,\lfloor n/k_n \rfloor-1$.
In \cite{AJ2014}, the proposed estimator takes the following form:
\begin{align}\label{est-AJ2014} 
	\text{``Lev-AJ14" estimator:} \ \frac{1}{k_n}\sum_{i=1}^{n-2k_n+1} \left( \left( X_{t_{i+2k_n-1}^n}- X_{t_{i-1}^n}\right)  \cdot \left( \widehat{\sigma}^{2,con}_{t_{(i+k_n-1)}^n} - \widehat{\sigma}^{2,con}_{t_{i-1}^n} \right) \right).
\end{align}
When jumps may be present in $X$, \cite{AFLWY2017} applied the thresholding technique (see, e.g., \cite{Mancini2009}, \cite{MR2011}) to filter jumps and constructed the estimator 
\begin{align}\label{est-AFLWY(2017)}
	\text{``Lev-AFLWY17" Estimator:} \ \sum_{i=k_n+1}^{n-k_n} \left( \Delta_i^n X \cdot \left( \widehat{\sigma}^{2,thr}_{t_{i+}^n} - \widehat{\sigma}^{2,thr}_{t_{i-}^n} \right)\cdot 1_{\{
		|\Delta_i^n X| \leq \alpha \Delta_n^{\omega} \}} \right),
\end{align}
with 
\begin{align}\label{est:thr}
	&\widehat{\sigma}^{2,thr}_{t_{i+}^n} = \frac{1}{k_n\Delta_n}\sum_{j \in I^n_{i+} } \left((\Delta_j^n X)^2\cdot 1_{\{
		|\Delta_j^n X| \leq \alpha \Delta_n^{\omega} \}} \right), \ 
	&\widehat{\sigma}^{2,thr}_{t_{i-}^n} = \frac{1}{k_n\Delta_n}\sum_{j \in I^n_{i-} } \left((\Delta_j^n X)^2\cdot 1_{\{
		|\Delta_j^n X| \leq \alpha \Delta_n^{\omega} \}} \right),
\end{align}
where $\alpha>0$ and $\omega\in(0,\frac{1}{2})$ are constants. 
For \eqref{est-AFLWY(2017)}, following their simulation setting, we take $\omega=0.49$ and $\alpha=5\sqrt{\text{BV}_{n}}$, where $\text{BV}_{n}$ denotes the bipower variation estimator (see, e.g., \cite{BN2004}) and can be written as 
\begin{align}\label{est:bip}
	\text{BV}_{n} = \frac{\pi}{2} \sum_{i=1}^{n-1} \left( |\Delta_i^n X|\cdot |\Delta_{i+1}^n X| \right).
\end{align}
For volatility of volatility, the estimators in \cite{AJ2014} (``Vov-AJ14") and \cite{M2015} (``Vov-V15") were proposed for the continuous case, while the estimator in \cite{BV2009} (``Vov-BV09") allows for jumps. These estimators can be written as 
\begin{align}
	&\text{``Vov-AJ14" Estimator}: \ \frac{3}{2k_n} \sum_{i=0}^{n-2k_n}\left( \left( \widehat{\sigma}^{2,con}_{t_{(i+k_n)}^n} - \widehat{\sigma}^{2,con}_{t_{i}^n} \right)^2 - \frac{4}{k_n} \left( \widehat{\sigma}^{2,con}_{t_{i}^n} \right)^2  \right), \\
	& \text{``Vov-V15" Estimator:}\ \sum_{i=0}^{n-2k_n}\left( \frac{3}{2k_n} \left( \widehat{\sigma}^{2,con}_{t_{(i+k_n)}^n} - \widehat{\sigma}^{2,con}_{t_{i}^n} \right)^2 - \frac{6}{k_n^2} \widehat{\sigma}^{(4)}_{t_{i}^n}  \right), \\
	&\text{``Vov-BV09" Estimator:} \  \sum_{i=0}^{n-k_n}\left( \left( \widehat{\sigma}^{2,thr}_{t_{(i+1)}^n} - \widehat{\sigma}^{2,thr}_{t_{i}^n} \right)^2  \right),
\end{align}
where $\widehat{\sigma}^{(4)}_{t_{i}^n}  = \frac{1}{3k_n\Delta_n^2} \sum_{j=1}^{k_n} |\Delta_{i+j}^n X|^4$.
In addition, we reformulate the estimator ``Vov-V15" by further removing jumps through thresholding. This yields the estimator
\begin{align} 
	& \text{``Vov-V15-thr" Estimator:}\ \sum_{i=0}^{n-2k_n}\left( \frac{3}{2k_n} \left( \widehat{\sigma}^{2,thr}_{t_{(i+k_n)+}^n} - \widehat{\sigma}^{2,thr}_{t_{i+}^n} \right)^2 - \frac{6}{k_n^2} \widehat{\sigma}^{(4,thr)}_{t_{i}^n}  \right),
\end{align}
with $\widehat{\sigma}^{(4,thr)}_{t_{i}^n}  = \frac{1}{3k_n\Delta_n^2} \sum_{j=1}^{k_n} \left( |\Delta_{i+j}^n X|^4 \cdot 1_{\{
	|\Delta_j^n X| \leq \alpha \Delta_n^{\omega} \}}\right)$.
For comparison, we refer to our leverage estimator \eqref{est-main} and volatility of volatility estimator \eqref{vov-main} as ``Lev-our" and ``Vov-our", respectively. 

\subsection{Simulation design}
We consider the following models for the log-price process $X$ and the volatility process $\sigma^2$:
\begin{align}\label{sim:model}
	\begin{split}
		dX_t&= (v- \sigma_t^2/2)dt+\sigma_t(\rho dW_t + \sqrt{1-\rho^2} dV_t)+ \gamma (dL_t + dJ_t),\\
		d\sigma_t^2 &= \zeta (\theta-\sigma_t^2) dt + \eta \sigma_t  dW_t,
	\end{split}
\end{align}
where $W$ and $V$ are two independent standard Brownian motions; $L$ is a strictly symmetric stable L$\acute{\text{e}}$vy process with Blumenthal--Getoor index $\beta$; and $J_t=\sum_{i=1}^{N_t} Y_{i}$ is a compound Poisson process, where $N_t$ is a Poisson process with intensity $\lambda'$ and $Y_i\stackrel{i.i.d}\sim N(0, 0.01^2)$. The continuous part of $X$ in \eqref{sim:model} is a Heston model, while the discontinuous part consists of a possible infinite variation jump component $L$ and a finite jump component $J$. 
We fix $T=1$, with time measured in months. To match trading schemes in real financial markets and mimic high-frequency data, 
we assume that each month has 21 trading days and that each trading day has 130 observations, corresponding to sampling every 3 minutes during a 6.5-hour trading day. Given these considerations, we set $n=2730$ $(21\times 130)$ and repeat the simulation 1000 times. 
The same model was also considered by \cite{AFLWY2017} and \cite{LLL2018}, and we use the same parameter setting. 
Specifically, $X_0 = 0$, $\sigma_0^2 = 0.02$, $v=0.05$, $\lambda' = 3$, $\theta=0.02$, and $\zeta= 5$\footnote{The Feller condition $2\zeta\theta>\eta^2$ ensures that the volatility process is strictly positive.}. For $\rho, \eta$, and $\gamma$, we consider different choices for robustness checks; these choices are specified below. Under model \eqref{sim:model}, the true leverage effect is $\eta \rho \int_{0}^{1} \sigma_t^2dt$, and the true volatility of volatility is $\eta^2\int_{0}^{1} \sigma_t^2dt$. We approximate the integrated volatility $\int_{0}^{1} \sigma_t^2dt$ by a Riemann sum in all simulation studies. 
To achieve the optimal convergence rate for all estimators given in Section \ref{sec:estimator}, we set $k_n = \lfloor \sqrt{1/\Delta_n}\rfloor$ unless otherwise specified. 

\subsection{Simulation results}
We first compare the finite sample performance of our estimators with the above-mentioned estimators for jump activity index values $\beta = 0.5, 1, 1.5$. For our spot volatility estimators \eqref{est-spot++} and \eqref{est-spot-}, we take the relatively small value $u=\frac{(\log(n))^{-1/40}}{\sqrt{\text{BV}_n}}$, which consistently estimates $u^{\star}=\frac{(\log(n))^{-1/40}}{\sqrt{\int_0^1\sigma_t^2dt}}$.
Such a scheme was also used in \cite{JT2014} and \cite{LLL2018}.
For the leverage effect, we fix $\eta=0.3$ and vary $\rho=-0.6, -0.4, -0.2$. For each generated sample path, we calculate the relative bias $\frac{\widehat{\text{Lev}} - \text{Lev}}{\text{Lev}}$, where $\widehat{\text{Lev}}$ denotes the estimate from a generic leverage effect estimator and $\text{Lev}$ is the true leverage effect. The mean (M.), standard deviation (S.D.), and mean squared error (M.S.E.) of the relative biases are reported in Table \ref{com:lev}. From the table, we draw the following conclusions. Without jumps ($\gamma=0$), Lev-WM14 already performs poorly, with large M., S.D., and M.S.E.; this may be because it uses non-overlapping increments for spot volatility. Among the other three estimators, our leverage effect estimator almost always has the smallest absolute value of M., although it has a relatively larger S.D., which is consistent with our theoretical result that our estimator has a larger asymptotic variance. When jumps are present with $\gamma=0.2$, Lev-WM14 and Lev-AJ14 perform poorly, so we do not report their results in this scenario. Comparing our estimator with Lev-AFLWY17, we observe that our estimator has a smaller bias but larger S.D. when $\beta=0.5, 1$. For the larger value $\beta=1.5$, both the bias and S.D. of our estimator are the smallest, yielding the smallest M.S.E.; this shows that our estimator works better when jumps are intensive. 
The same experiment is conducted for volatility of volatility, but we use $n=2730\times 12$, which corresponds to 3-minute data over one year\footnote{We note that none of the volatility of volatility estimators works well for $n=2730$, suggesting that a larger amount of data is needed for estimating volatility of volatility than for estimating the leverage effect.}. The numerical results are presented in Table \ref{com:vov}, where we fix $\rho=-0.2$ and vary $\eta = 0.1, 0.2, 0.3$. ``Vov-BV09" does not work even in the absence of jumps. The conclusions for volatility of volatility are similar to those for the leverage effect. Without jumps or with inactive jumps, our estimator has smaller bias but larger S.D. When jumps are intensive, our estimator performs best in terms of all three measures, M., S.D., and M.S.E. 
Moreover, in most numerical experiments, the magnitude of M. is well controlled below $10\%$ for our proposed leverage effect estimator and volatility of volatility estimator. 

\begin{table}[!htbp]
	\centering
	\caption{The results of (M., S.D., M.S.E.) for 1000 relative biases obtained using different leverage effect estimators, with fixed $\eta=0.3$.}\label{com:lev}
	\vspace{0.2cm}
	{\small
		\begin{tabular}{c|c|c|c}
			\toprule[2pt]	
			& $\rho=-0.6$ & $\rho=-0.4$ & $\rho=-0.2$  \\
			\toprule[2pt]
			\multicolumn{4}{c}{$\gamma=0$}\\
			\midrule[1pt]
			Lev-WM14 & -0.120, 1.009, 1.032 & -0.031, 1.752, 3.072 & -0.738, 3.085, 10.06 \\
			\hline
			Lev-AJ14 & -0.099, 0.593, 0.362 & -0.039, 0.787, 0.621 & -0.084, 1.642,  2.706 \\
			\hline
			Lev-AFLWY17 & -0.079, 0.572, 0.334&  0.079, 0.762, 0.587 & -0.041, 1.839,  3.384  \\
			\hline
			Lev-our& -0.058, 0.621, 0.389 & 0.047, 0.824, 0.681 & 0.040, 1.883, 3.548  \\
			\midrule[1pt]
			\multicolumn{4}{c}{$\gamma=0.2, \beta=0.5$}\\
			\midrule[1pt]
			Lev-AFLWY17 & -0.123, 0.618, 0.397&  0.068, 0.963, 0.933 & 0.083, 1.875, 3.522  \\
			\hline
			Lev-our & -0.103, 0.636, 0.416 & 0.012, 0.981, 0.964 & 0.053, 1.899, 3.611 \\
			\midrule[1pt]
			\multicolumn{4}{c}{$\gamma=0.2, \beta=1$}\\
			\midrule[1pt]
			Lev-AFLWY17& -0.088, 0.605, 0.374 &  -0.119, 0.904, 0.832& -0.041, 2.269, 5.150  \\
			\hline
			Lev-our& -0.040, 0.616, 0.381 & -0.083, 0.936, 0.883 & -0.014, 2.132, 4.549
			\\
			\midrule[1pt]
			\multicolumn{4}{c}{$\gamma=0.2, \beta=1.5$}\\
			\midrule[1pt]
			Lev-AFLWY17& -0.130, 0.799, 0.656 & -0.082, 1.247, 1.563 & -0.120, 2.320, 5.398  \\
			\hline
			Lev-our& -0.077, 0.733, 0.543 &  -0.070, 1.212, 1.474 & -0.105, 2.310, 5.351
			\\
			\bottomrule[2pt]
	\end{tabular}}
\end{table}

\begin{table}[!htbp]
	\centering
	\caption{The results of (M., S.D., M.S.E.) for 1000 relative biases obtained using different volatility of volatility estimators, with fixed $\rho=-0.2$.}\label{com:vov}
	\vspace{0.2cm}
	{\small
		\begin{tabular}{c|c|c|c}
			\toprule[2pt]	
			& $\eta=0.1$ & $\eta=0.2$ & $\eta=0.3$  \\
			\toprule[2pt]
			\multicolumn{4}{c}{$\gamma=0$}\\
			\midrule[1pt]
			Vov-AJ14 &0.027, 1.057, 1.119 & -0.207, 0.406, 0.208 & -0.011, 0.209, 0.044 \\
			\hline
			Vov-V15&  0.162, 1.034, 1.097& -0.173, 0.415, 0.202 &-0.043, 0.214, 0.047 \\
			\hline
			Vov-BV09 & / &  / & /  \\
			\hline
			Vov-V15-thr &0.162, 1.034, 1.097&-0.173, 0.415, 0.202&-0.013, 0.216, 0.047 \\
			\hline
			Vov-our& 0.069, 1.106, 1.229& -0.100, 0.482, 0.243 &-0.001, 0.394, 0.155  \\
			\midrule[1pt]
			\multicolumn{4}{c}{$\gamma=0.2, \beta=0.5$}\\
			\midrule[1pt]
			Vov-V15-thr & 0.243, 1.293, 1.730&  0.136, 1.498, 2.263 &-0.020, 0.234, 0.055\\
			\hline
			Vov-our &0.181, 1.658, 2.781& 0.047, 1.188, 1.415& -0.007, 0.305, 0.093 \\
			\midrule[1pt]
			\multicolumn{4}{c}{$\gamma=0.2, \beta=1$}\\
			\midrule[1pt]
			Vov-V15-thr&  0.525, 1.198, 1.713 & -0.114, 0.527, 0.291 & -0.062, 0.325, 0.110\\
			\hline
			Vov-our& -0.035, 1.266, 1.604& -0.015, 0.508, 0.259 & -0.001, 0.332, 0.110
			\\
			\midrule[1pt]
			\multicolumn{4}{c}{$\gamma=0.2, \beta=1.5$}\\
			\midrule[1pt]
			Vov-V15&4.437, 3.589, 32.571  & 0.970, 0.853, 1.669 &  0.459, 0.589, 0.558 \\
			\hline
			Vov-our&1.053, 2.146,  5.716  &0.192, 0.735, 0.578  &  0.169, 0.547, 0.328
			\\
			\bottomrule[2pt]
	\end{tabular}}
\end{table}

Next, we verify the central limit theorems established in Theorems \ref{thm-lev-1} and \ref{thm-vov}, considering different values of $\kappa$, $u$, and $\beta$. We fix the parameters $\rho=-0.4$, $\eta=0.2$, and $b=0.5$, and first randomly generate a sample path of $\sigma^2$ in \eqref{sim:model}. 
Fixing this volatility path, we then generate 1000 sample paths of $X$ in \eqref{sim:model}. For each path, we calculate the estimates on the left-hand side of \eqref{thm-lev-clt} and report the mean (M.) and variance (Var.) in Table \ref{rob:lev}.\footnote{We recall that $\kappa_{opt}$ is defined as in \eqref{opt_kk} and $u^{\star} = \frac{(\log(n))^{-1/40}}{\sqrt{\int_0^1\sigma_t^2dt}}$ is used in the last simulation study. For the particular volatility path generated in this simulation, we have $\kappa_{opt} \approx 2$ and $u^{\star} \approx 7$.} We also report the theoretical variance (T-Var.) in \eqref{lev-var}\footnote{For \eqref{sim:model}, we have $h_2(t, \sigma_{t}^2,(\tilde{\sigma}_{t})^2, (\tilde{\sigma}'_{t})^2) = (\eta^2 \sigma_t^2)/3$. Thus, when $\eta$ is fixed, the theoretical variance depends only on the path of $\sigma^2$.} for evaluation. The same simulation study is conducted for volatility of volatility using the same parameters, except that we set $n=2730\times 12$. The results are displayed in Table \ref{rob:vov}.
\ignore{Besides, we also present the theoretical asymptotic variance by using the coefficients obtained in \cite{M2015} and \cite{AJ2014}, as given after Theorem \ref{thm-vov}, and denote it as T-Var-Vetter.} Tables \ref{rob:lev}--\ref{rob:vov} show that, for various jump activity indices and irrespective of the choice of $u$, all values of M. are close to 0 and the values of Var. are close to the corresponding theoretical values, T-Var. \ignore{Moreover, compared with T-Var-Vetter, our theoretical asymptotic variance estimates of Var. are closer to the corresponding sample variance Var., for most (If not all) of the cases.} The simulation results suggest that a data-driven choice of $u$ is not necessary for our estimators, and that a manual choice such as $u=1$ meets our requirements and can be used in practice. In addition, compared with $\kappa=1.5,1$, the smallest values of both Var. and T-Var. are obtained when $\kappa = \kappa_{opt}$ for estimating the leverage effect, which is consistent with our theory. 

\begin{table}[!htbp]
	\centering
	\caption{The results of (M., Var., T-Var.) for 1000 estimates obtained using our leverage estimator (the left-hand side of \eqref{thm-lev-clt}), for different values of $\kappa$ and $u$.}\label{rob:lev}
	\vspace{0.2cm}
	{\small
		\begin{tabular}{c|c|c|c}
			\toprule[2pt]	
			& $\kappa=\kappa_{opt}$ & $\kappa=1.5$ & $\kappa=1$  \\
			\toprule[2pt]
			\multicolumn{4}{c}{$\gamma=0$}\\
			\midrule[1pt]
			$u=u^{\star}$ &0.400e-2, 3.772e-5, 5.102e-5 & 0.305e-2, 4.568e-5, 5.390e-5 & 0.246e-2, 6.509e-5, 6.563e-5\\
			\hline
			$u=4$&0.336e-2, 3.846e-5, 4.800e-5&0.272e-2, 4.322e-5, 4.981e-5&0.209e-2, 5.454e-5, 5.949e-5 \\
			\hline
			$u=1$ & 0.370e-2, 3.469e-5, 4.759e-5 &  0.283e-2, 4.019e-5, 4.926e-5 & 0.229e-2, 5.545e-5, 5.866e-5 \\
			\midrule[1pt]
			\multicolumn{4}{c}{$\gamma=0.001, \beta=0.5$}\\
			\midrule[1pt]
			$u=u^{\star}$& 0.391e-2, 3.851e-5, 5.102e-5&  0.309e-2, 4.693e-5, 5.390e-5 &0.235e-2, 6.755e-5, 6.563e-5\\
			\hline
			$u=4$&0.343e-2, 3.112e-5, 4.800e-5&0.266e-2, 3.830e-5, 4.981e-5&0.198e-2, 5.442e-5, 5.949e-5 \\
			\hline
			$u=1$&0.340e-2, 4.806e-5, 4.759e-5& 0.280e-2, 5.744e-5, 4.926e-5& 0.218e-2, 7.275e-5, 5.866e-5 \\
			\midrule[1pt]
			\multicolumn{4}{c}{$\gamma=0.001, \beta=1$}\\
			\midrule[1pt]
			u=$u^{\star}$& 0.375e-2, 3.574e-5, 5.102e-5 & 0.273e-2, 4.391e-5, 5.390e-5 & 0.197e-2, 6.361e-5, 6.563e-5\\
			\hline
			$u=4$& 0.368e-2, 3.569e-5, 4.800e-5& 0.290e-2, 4.108e-5, 4.981e-5 & 0.206e-2, 5.587e-5, 5.949e-5\\
			\hline
			$u=1$& 0.403e-2, 3.513e-5 , 4.759e-5& 0.331e-2, 4.007e-5, 4.926e-5 & 0.266e-2, 5.312e-5, 5.866e-5\\
			\bottomrule[2pt]
	\end{tabular}}
\end{table}

\begin{table}[!htbp]
	\centering
	\caption{The results of mean (M., Var., T-Var.)\ignore{, theoretical variance obtained in \cite{M2015} (T-Var-Vetter)} for 1000 estimates obtained using our volatility of volatility estimator (the left-hand side of \eqref{thm-vov-clt}), for different values of $\kappa$ and $u$.}\label{rob:vov}
	\vspace{0.2cm}
	{\scriptsize
		\begin{tabular}{c|c|c|c}
			\toprule[2pt]	
			& $\kappa=\kappa_{opt}$ & $\kappa=1.5$ & $\kappa=1$  \\
			\toprule[2pt]
			\multicolumn{4}{c}{$\gamma=0$}\\
			\midrule[1pt]
			$u=u^{\star}$ &-0.168e-2, 9.894e-6, 9.936e-6
			&-0.106e-2, 2.219e-5, 2.260e-5
			& -0.141e-2, 5.867e-5, 6.092e-5
			\\
			\hline
			$u=4$& -0.153e-2, 1.044e-5, 9.569e-6
			& -0.116e-2, 2.098e-5, 2.013e-5
			&-0.145e-2, 5.336e-5, 5.360e-5
			\\
			\hline
			$u=1$ &-0.145e-2, 9.389e-6, 9.456e-6
			& -0.094e-2, 1.967e-5, 1.941e-5%
			&-0.126e-2, 4.973e-5, 5.147e-5
			\\
			\midrule[1pt]
			\multicolumn{4}{c}{$\gamma=0.001, \beta=0.5$}\\
			\midrule[1pt]
			$u=u^{\star}$& -0.150e-2, 9.771e-6, 9.936e-6
			& -0.092e-2, 2.223e-5, 2.260e-5
			&  -0.133e-2, 5.821e-5, 6.092e-5
			\\
			\hline
			$u=4$&-0.164e-2, 9.673e-6, 9.569e-6
			&-0.129e-2, 2.157e-5, 2.013e-5
			& -0.187e-2, 5.709e-5, 5.360e-5
			\\
			\hline
			$u=1$  & -0.075e-2, 1.092e-5, 9.456e-6
			&  0.018e-2, 2.493e-5,  1.941e-5
			& 0.106e-2, 6.657e-5, 5.147e-5
			\\
			\midrule[1pt]
			\multicolumn{4}{c}{$\gamma=0.001, \beta=1$}\\
			\midrule[1pt]
			$u=u^{\star}$& -0.154e-2, 9.998e-6, 9.936e-6
			& -0.116e-2, 2.338e-5, 2.260e-5
			&  -0.198e-2, 6.128e-5, 6.092e-5
			\\
			\hline
			$u=4$& -0.149e-2, 9.666e-6, 9.569e-6
			&  -0.111e-2, 2.112e-5 , 2.013e-5
			&-0.169e-2, 5.559e-5, 5.360e-5
			\\
			\hline
			$u=1$& -0.127e-2, 1.112e-5, 9.456e-6
			&  -0.078e-2, 2.178e-5, 1.941e-5
			&-0.091e-2, 5.404e-5, 5.147e-5
			\\
			\midrule[1pt]
			\multicolumn{4}{c}{$\gamma=0.001, \beta=1.5$}\\
			\midrule[1pt]
			$u=u^{\star}$& -0.175e-2, 9.881e-6, 9.936e-6
			&-0.127e-2, 2.200e-5, 2.260e-5
			&  -0.145e-2, 5.9346e-5, 6.092e-5
			\\
			\hline
			$u=4$& -0.145e-2, 9.445e-6, 9.569e-6
			& -0.111e-2, 1.961e-5, 2.013e-5
			& -0.158e-2, 5.086e-5 , 5.360e-5
			\\
			\hline
			$u=1$& -0.125e-2, 9.458e-6, 9.456e-6
			& -0.064e-2, 2.028e-5, 1.941e-5
			& -0.110e-2, 5.312e-5, 5.147e-5
			\\
			\bottomrule[2pt]
	\end{tabular}}
\end{table}

\begin{figure}[!htbp]
	\centering     
	\subfigure[$\gamma=0$]{\includegraphics[width=36mm]{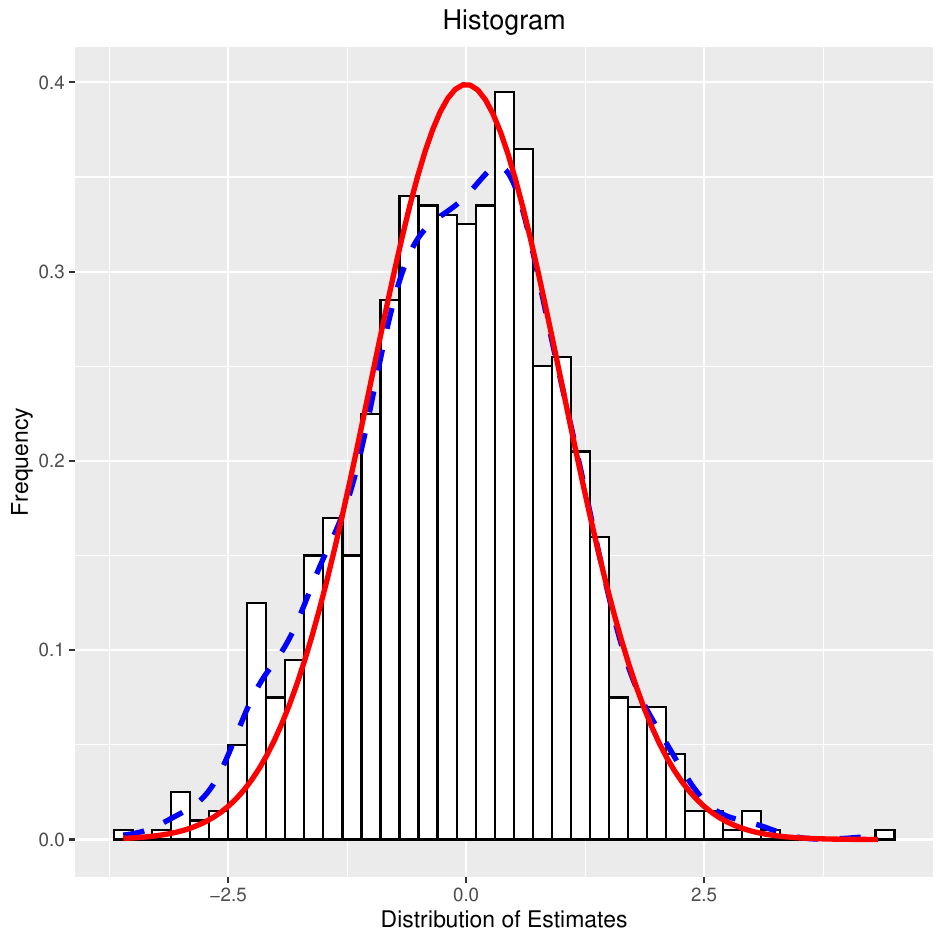}}
	\subfigure[$\gamma=0.01, \beta=0.5$]{\includegraphics[width=36mm]{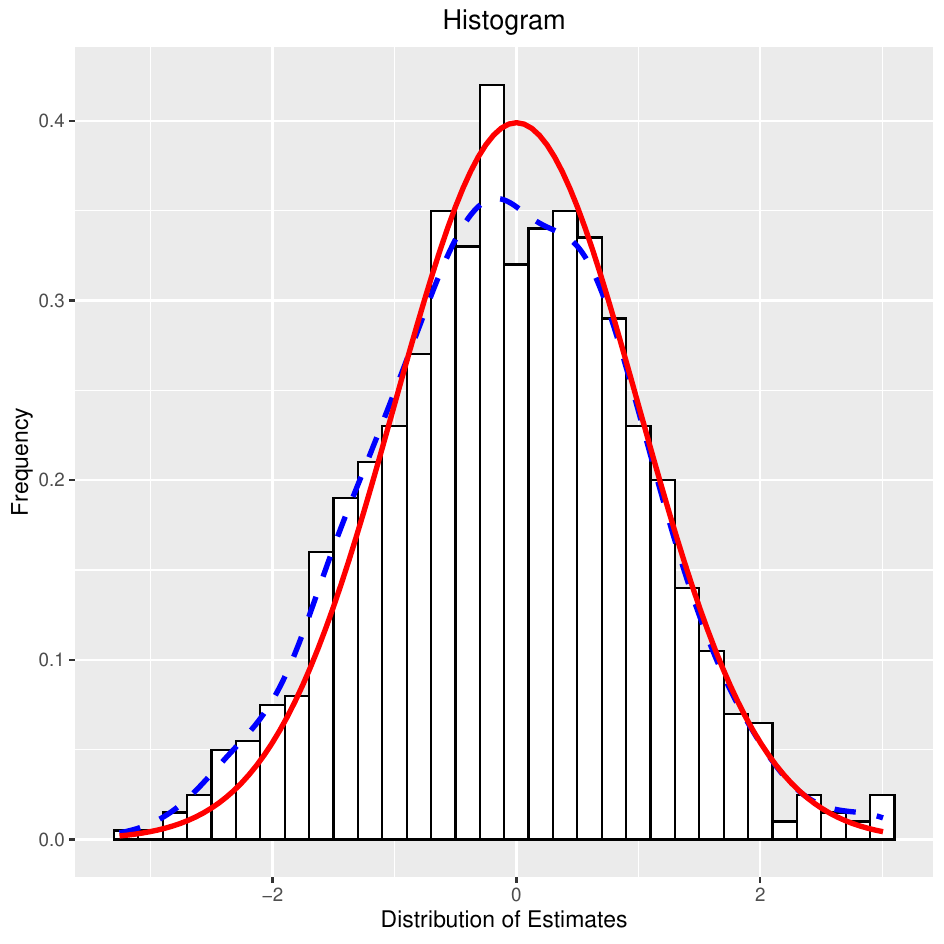}}
	\subfigure[$\gamma=0.01, \beta=1$]{\includegraphics[width=36mm]{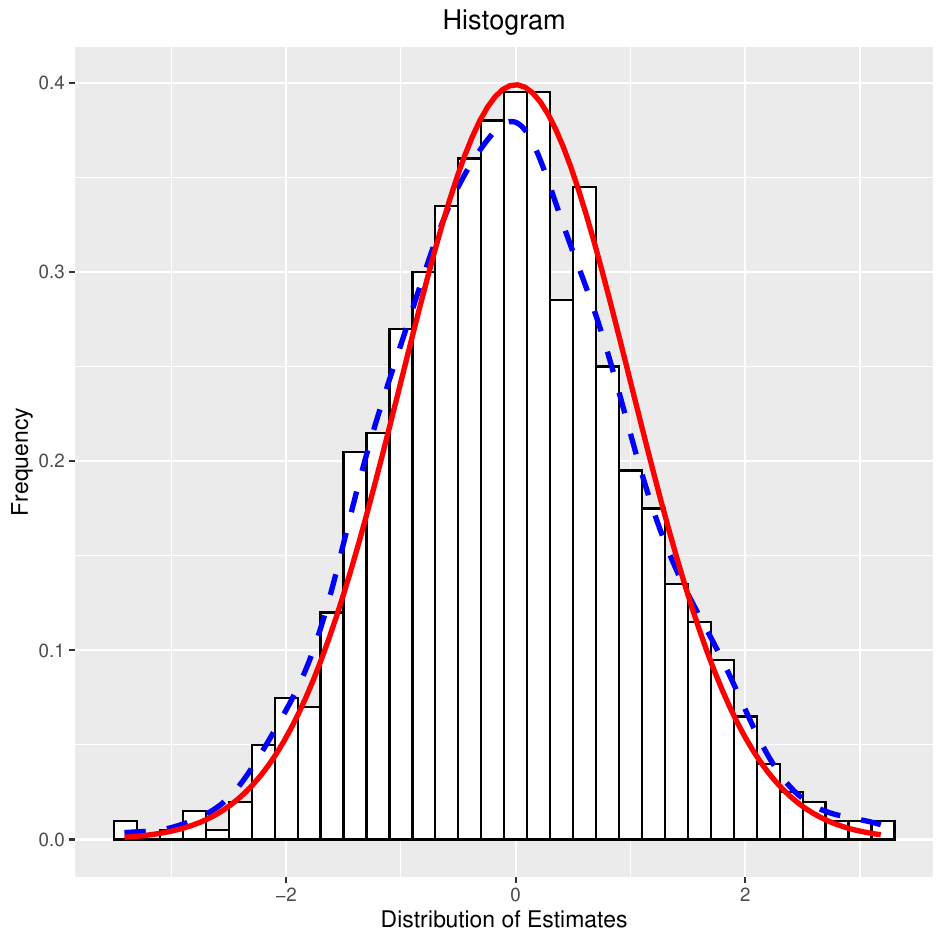}}
	\subfigure[$\gamma=0.01, \beta=1.5$]{\includegraphics[width=36mm]{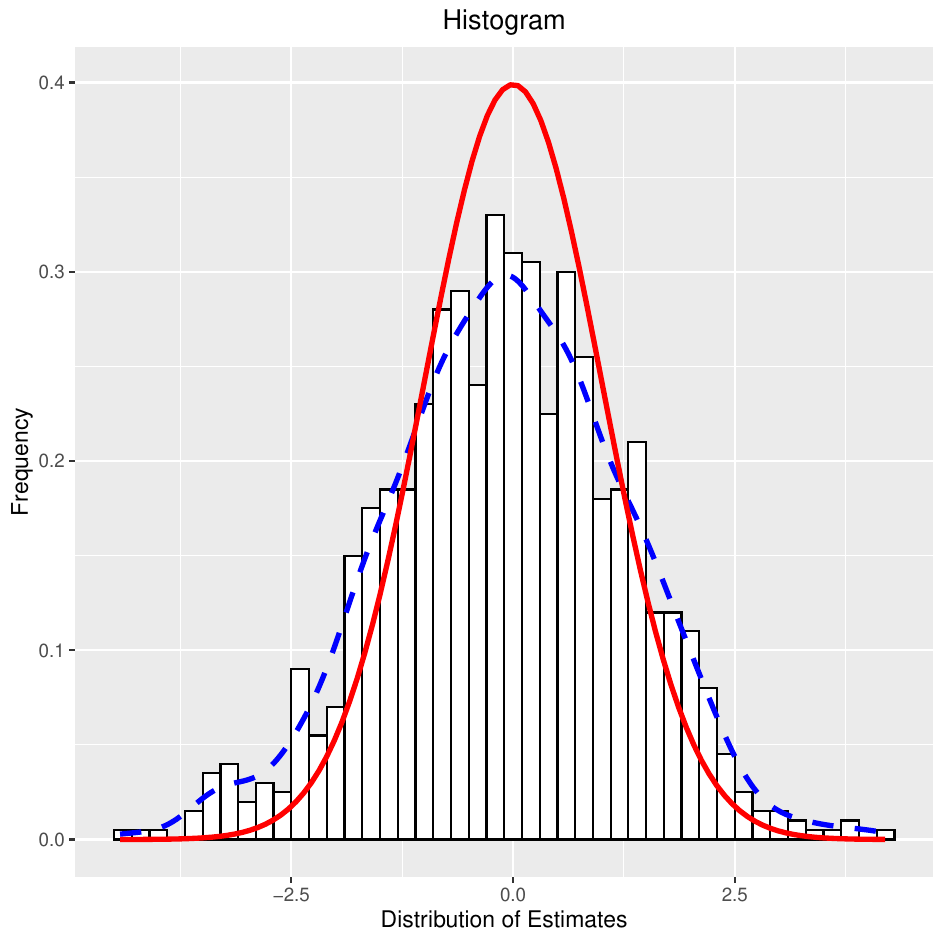}}
	\subfigure[$\gamma=0$]{\includegraphics[width=36mm]{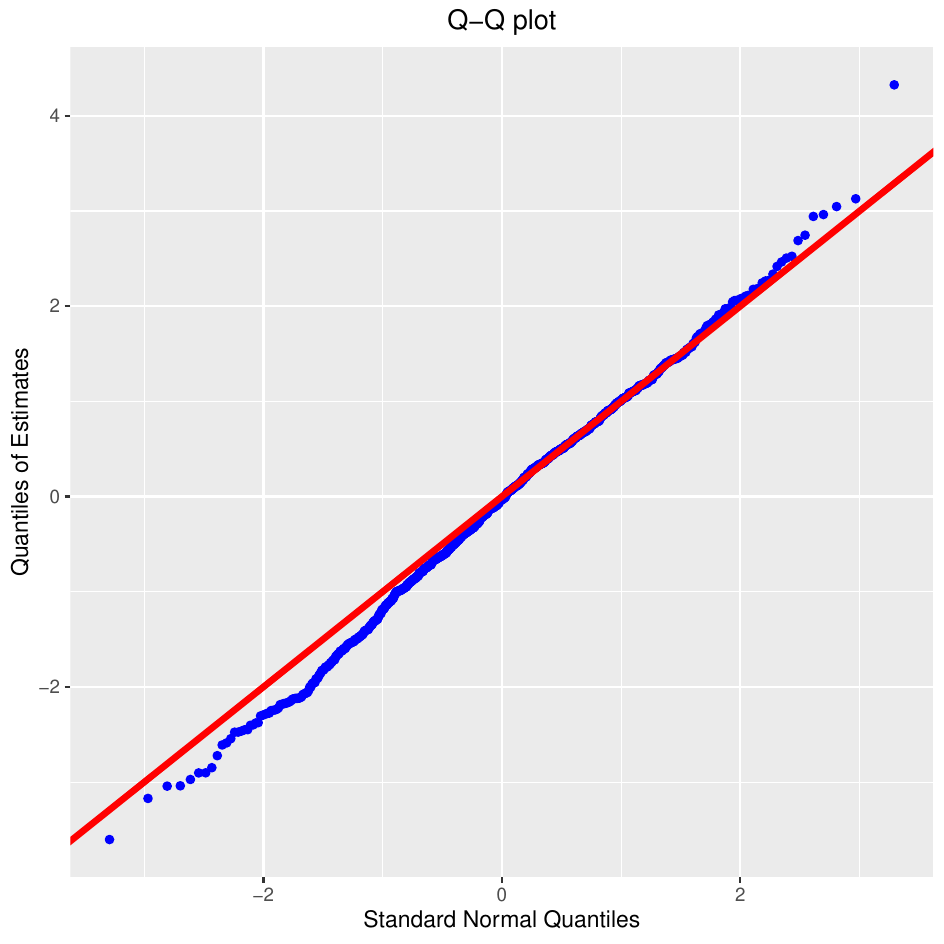}} 
	\subfigure[$\gamma=0.01, \beta=0.5$]{\includegraphics[width=36mm]{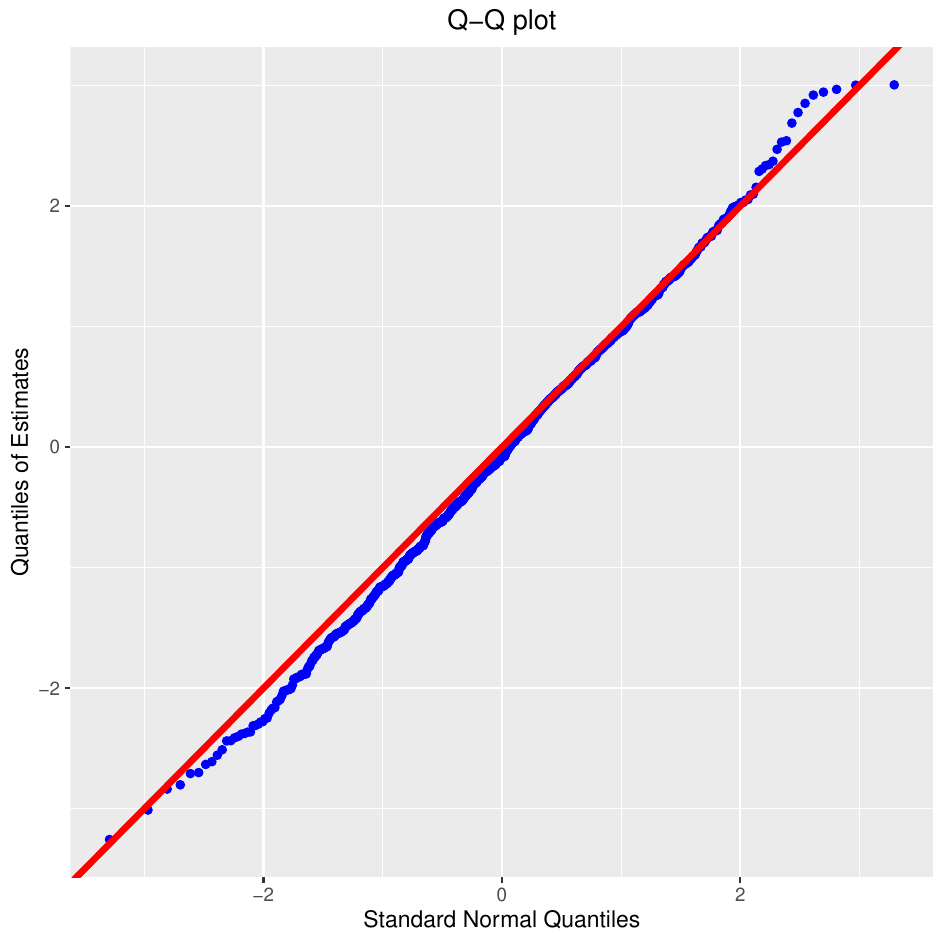}} 
	\subfigure[$\gamma=0.01, \beta=1$]{\includegraphics[width=36mm]{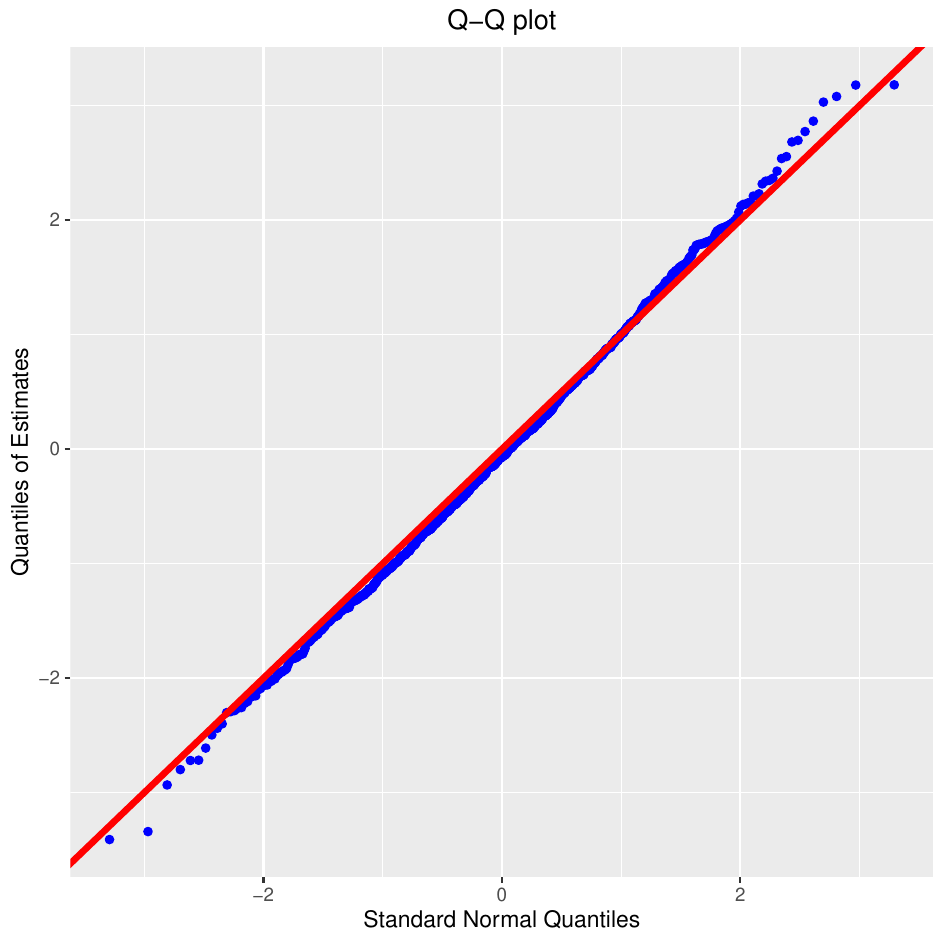}
	}
	\subfigure[$\gamma=0.01, \beta=1.5$]{\includegraphics[width=36mm]{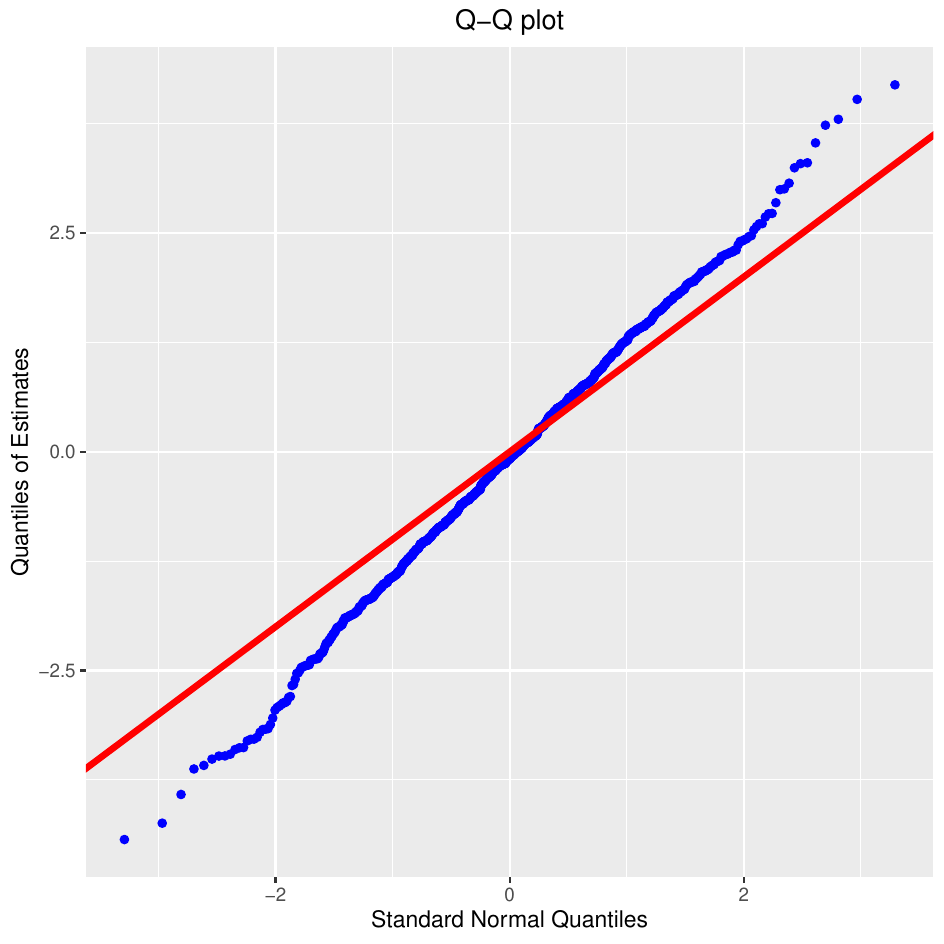}}  
	\caption{Histograms and Q-Q plots for the studentized statistics of the leverage effect. In the histograms, the red solid line is the density curve of the standard normal distribution, and the blue dotted line is the fitted density curve based on the estimates.}
	\label{fig:lev}
\end{figure}

\begin{figure}[!htbp]
	\centering     
	\subfigure[$\gamma=0$]{\includegraphics[width=30mm]{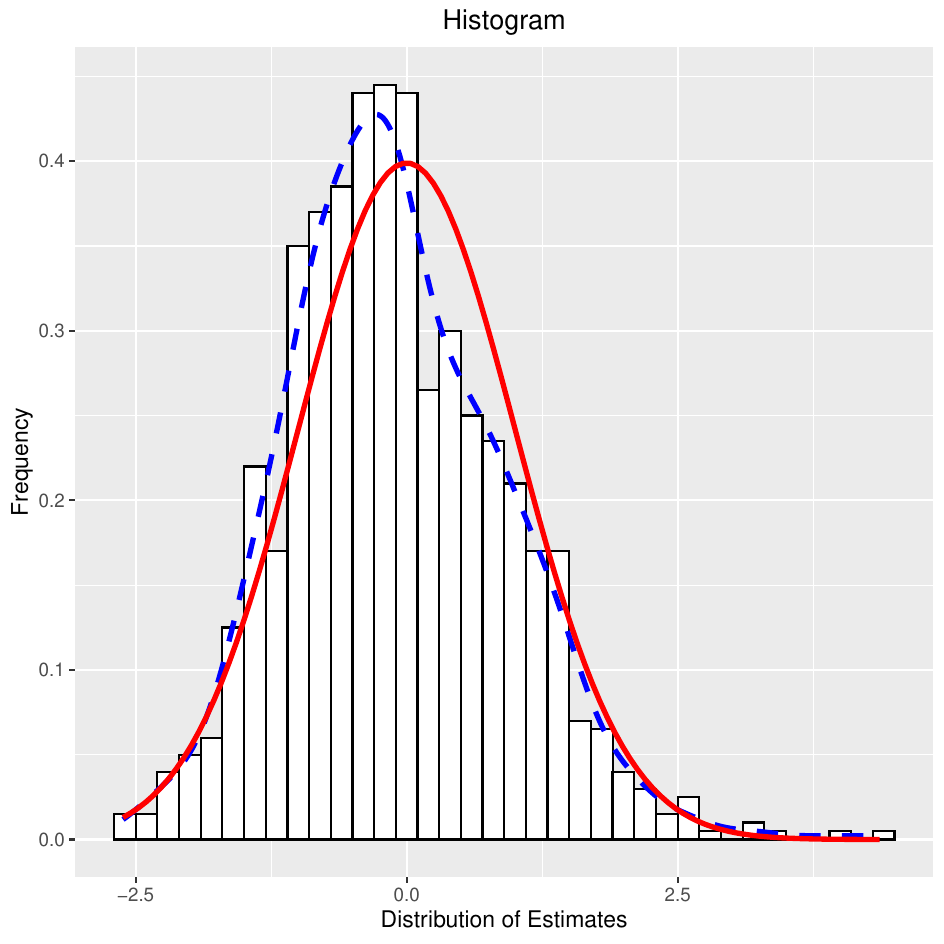}}
	\subfigure[$\gamma=0.01, \beta=0.5$]{\includegraphics[width=30mm]{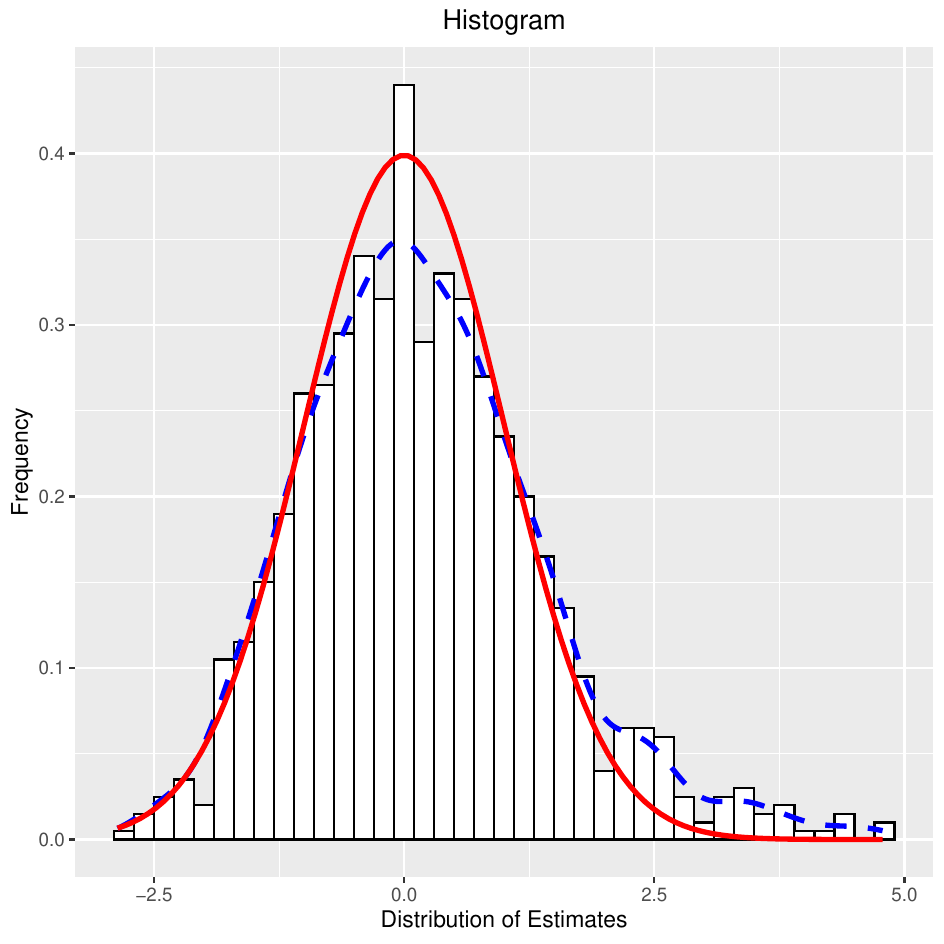}}
	\subfigure[$\gamma=0.01, \beta=1$]{\includegraphics[width=30mm]{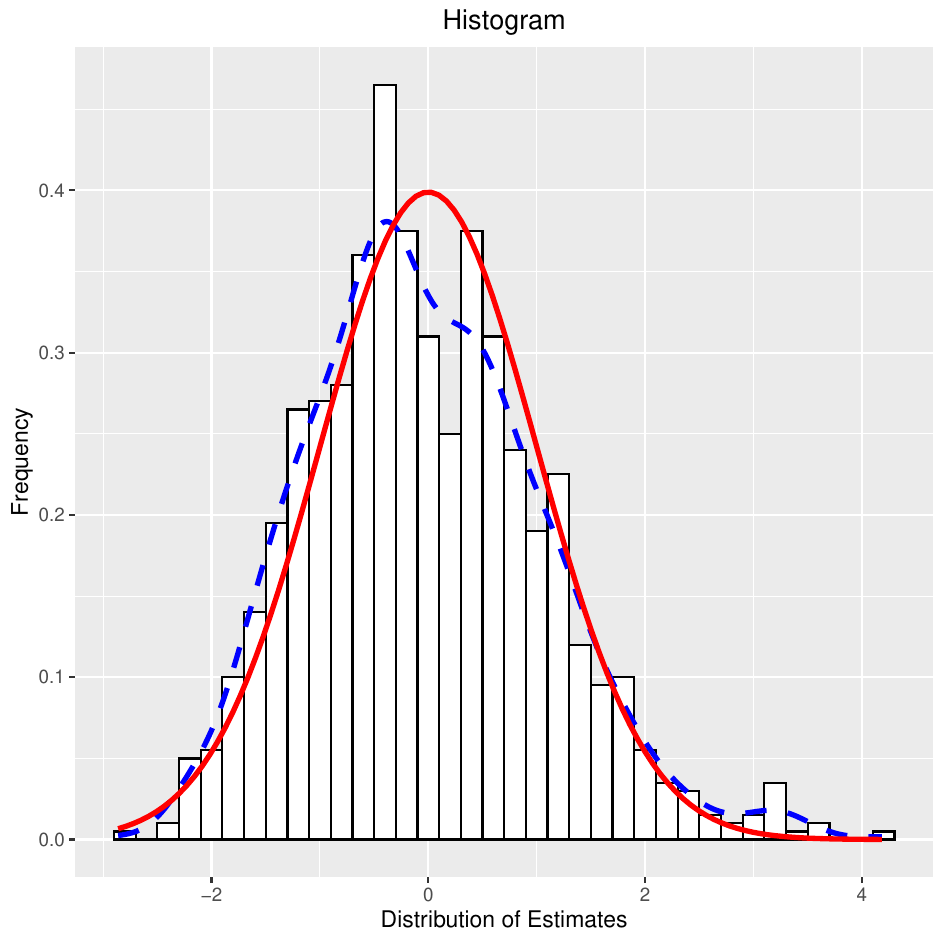}}
	\subfigure[$\gamma=0.01, \beta=1.5$]{\includegraphics[width=30mm]{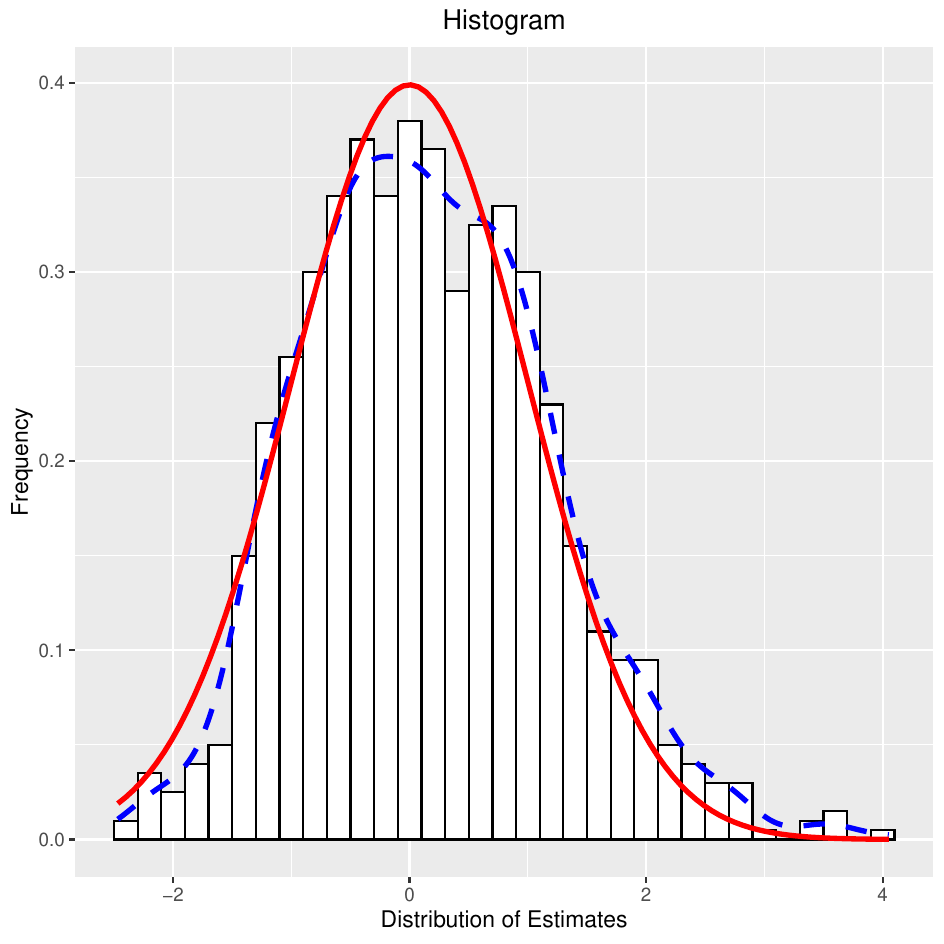}}
	\subfigure[$\gamma=0.01, \beta=1.8$]{\includegraphics[width=30mm]{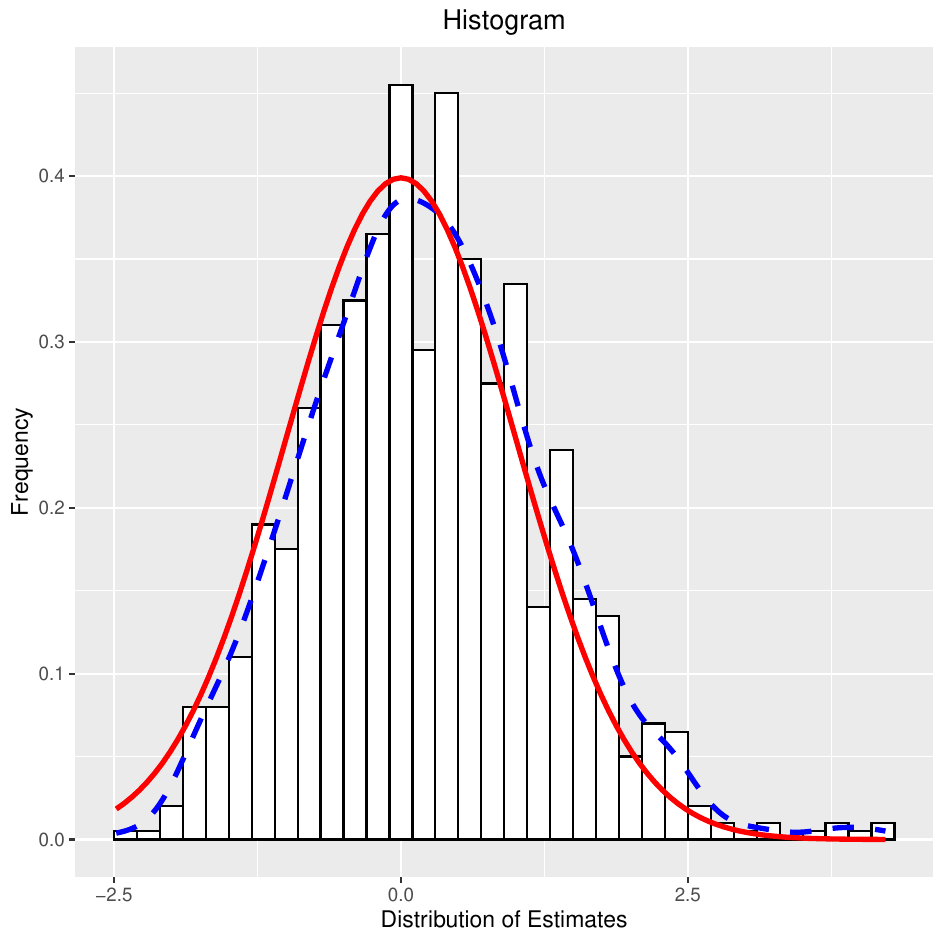}}
	\subfigure[$\gamma=0$]{\includegraphics[width=30mm]{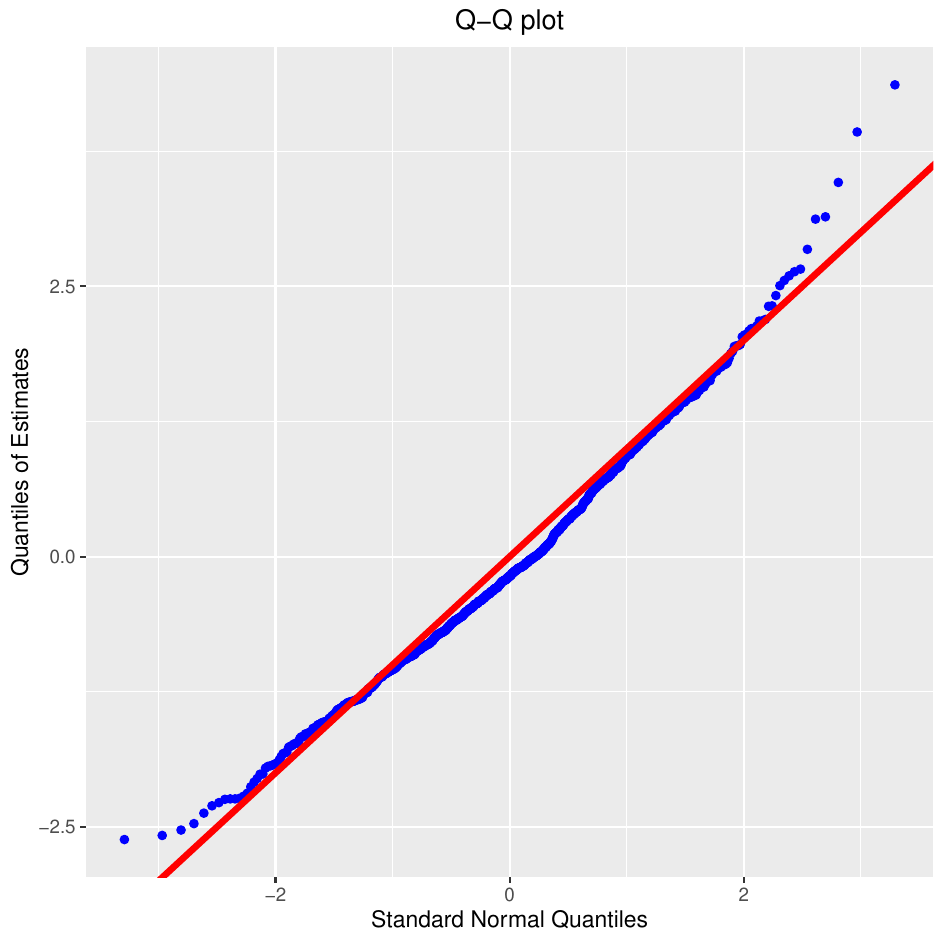}} 
	\subfigure[$\gamma=0.01, \beta=0.5$]{\includegraphics[width=30mm]{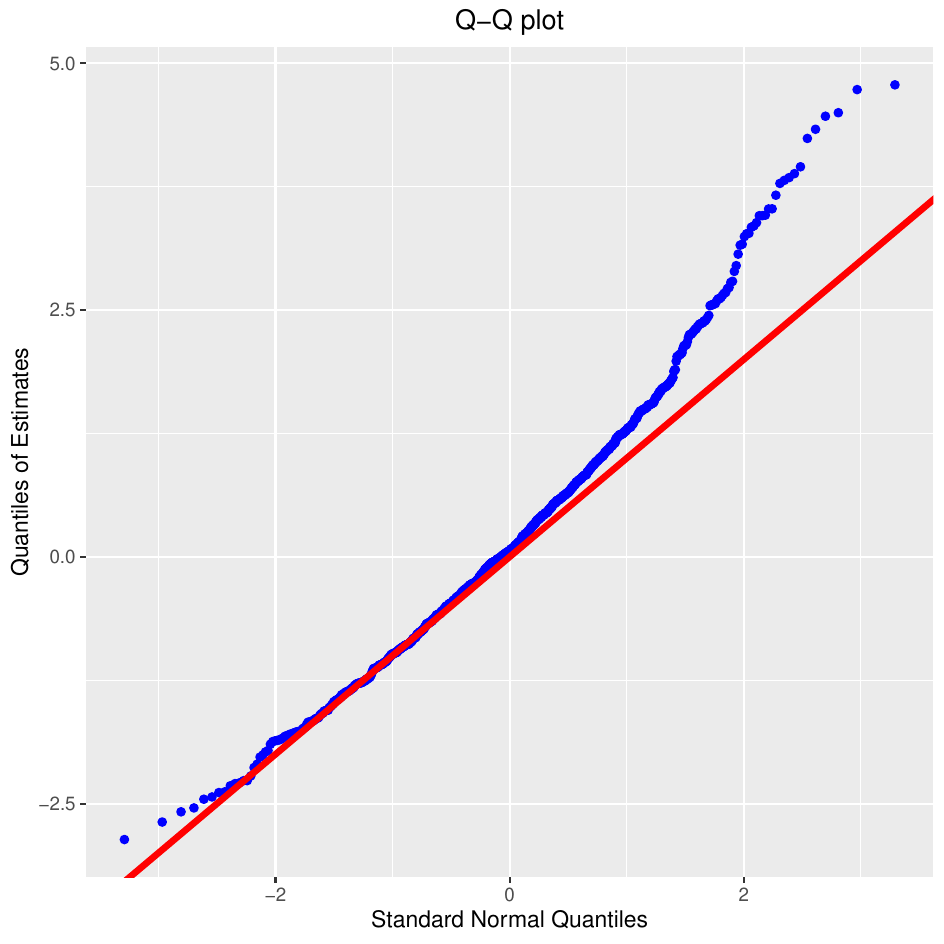}} 
	\subfigure[$\gamma=0.01, \beta=1$]{\includegraphics[width=30mm]{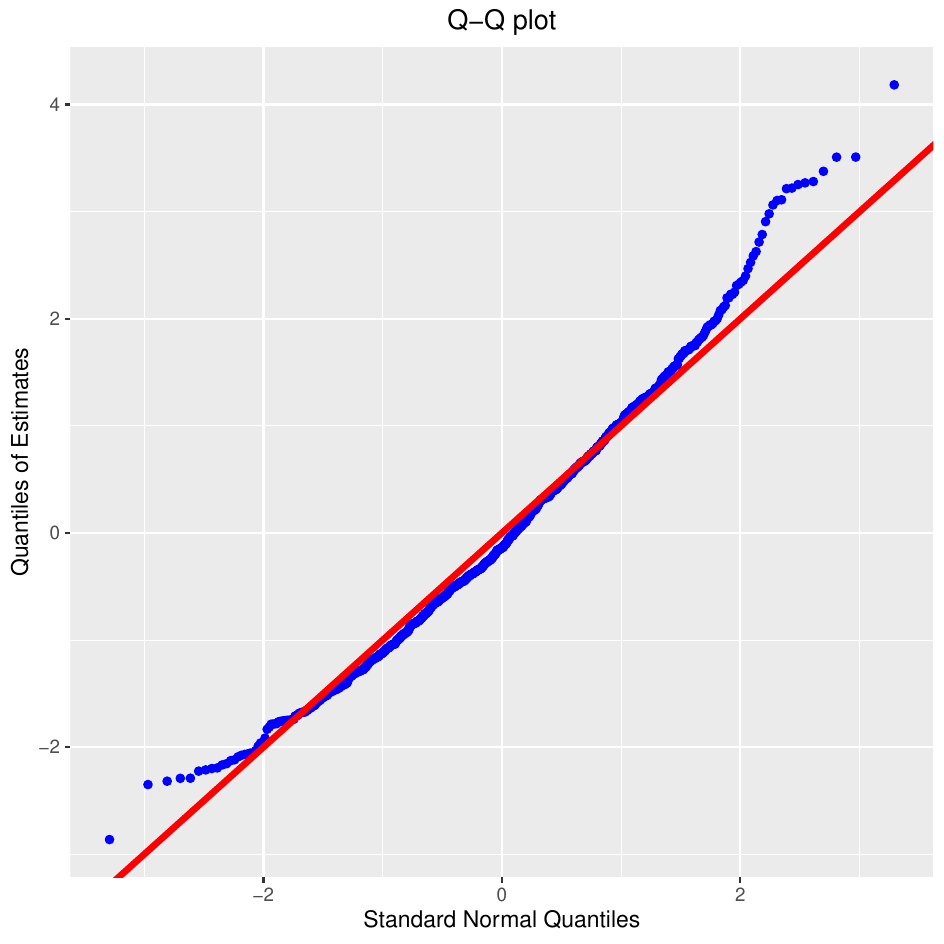}}
	\subfigure[$\gamma=0.01, \beta=1.5$]{\includegraphics[width=30mm]{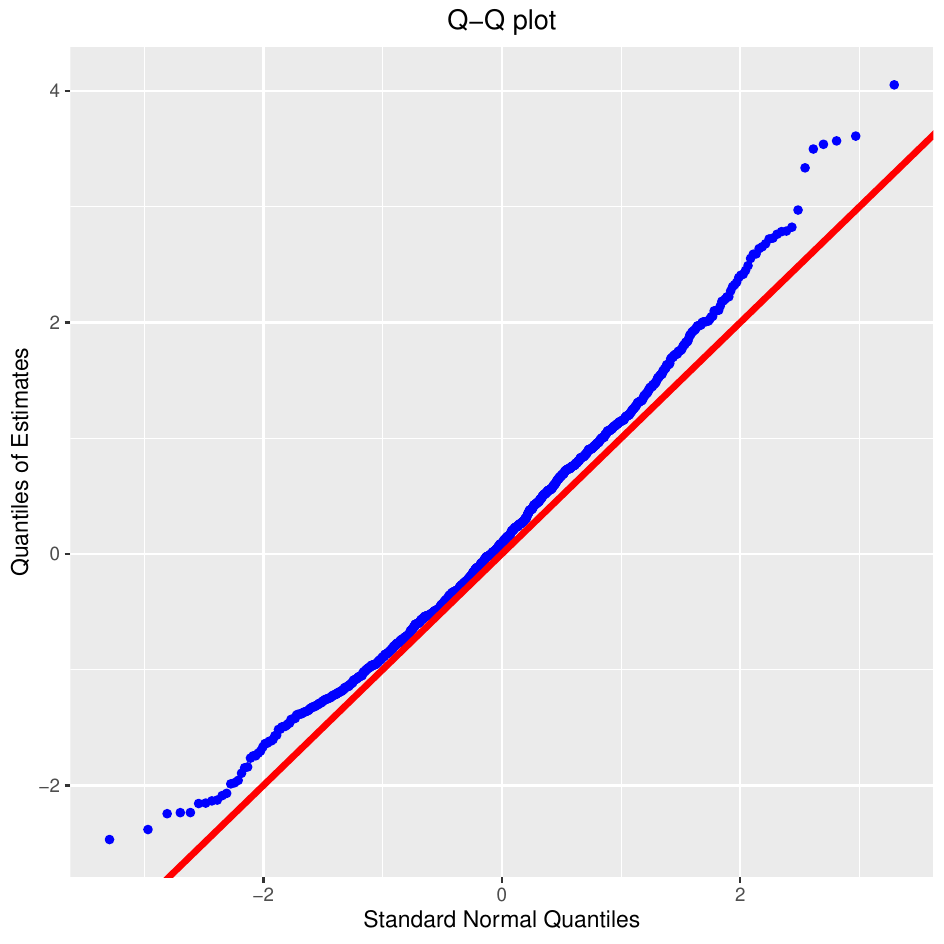}} 
	\subfigure[$\gamma=0.01, \beta=1.8$]{\includegraphics[width=30mm]{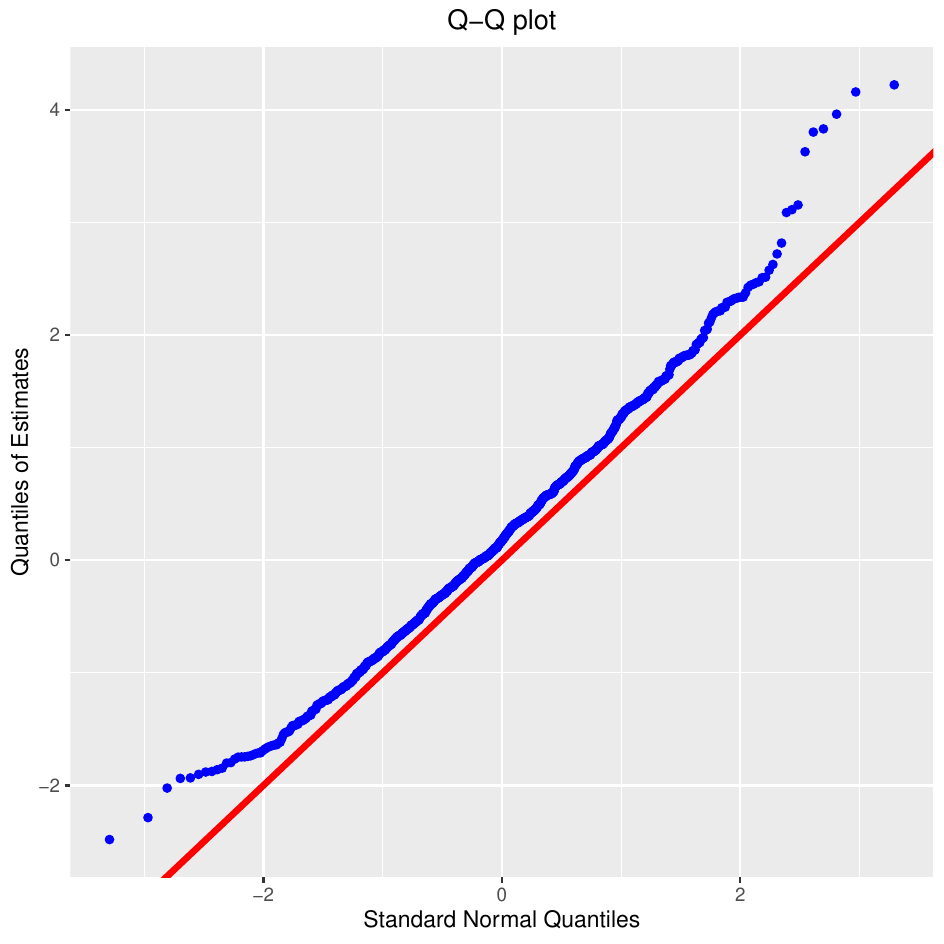}}  
	\caption{Histograms and Q-Q plots for the studentized statistics of volatility of volatility. In the histograms, the red solid line is the density curve of the standard normal distribution, and the blue dotted line is the fitted density curve based on the estimates.}
	\label{fig:vov}
\end{figure}

We then verify the feasible central limit theorems established in Corollary \ref{cor-spot-fea}. For the leverage effect, estimating the asymptotic variance requires estimating volatility of volatility, and the first experiment shows that a larger sample size is needed to ensure estimation accuracy\footnote{In practice, this can be achieved by using a relatively larger amount of historical data.}. Therefore, we take $n=2730\times12$. 
For the leverage effect, we fix $u=1$ and estimate $\kappa_{opt}$ in \eqref{opt_kk} by plugging in the corresponding estimators of the numerator and denominator, which are given in Section \ref{sec:vol-fun}, with the pre-specified value $\kappa=1$. We take $b=0.55$, since a small value of $k_n$ may lead to negative estimates of volatility of volatility. All other parameters remain the same as in the preceding simulation study. With a pre-generated volatility curve, we generate 1000 sample paths of $X$ and calculate the estimates on the left-hand side of \eqref{fea:lev}. The histograms and Q-Q plots for different jump intensity levels are presented in Figure \ref{fig:lev}, which shows that the distribution of the leverage effect estimates is close to the standard normal distribution. We also consider the case $\beta = 1.5$, where the theoretical requirement $\beta \leq 1$ is not satisfied. We observe that the mean remains close to 0 and the distribution appears symmetric, but the tail of the distribution is heavier than that of the standard normal distribution. 
For the volatility of volatility estimator, the performance is not satisfactory when $n=2730\times12$, perhaps because accurately estimating its asymptotic variance requires an even larger sample size. Using the true value of $Var(W|\mathcal{F})$, we present the histograms and Q-Q plots in Figure \ref{fig:vov}. When the condition $\beta<\frac{3}{2}$ in Theorem \ref{thm-vov} is satisfied, the distributions of the estimates are close to the standard normal distribution. When this condition is violated with $\beta=1.5, 1.8$, a pronounced positive bias emerges and increases with $\beta$. 

\ignore{
	\begin{figure}[htbp]
		\centering     
		\subfigure[$\gamma=0.01, \beta=1.5$]{\includegraphics[width=50mm]{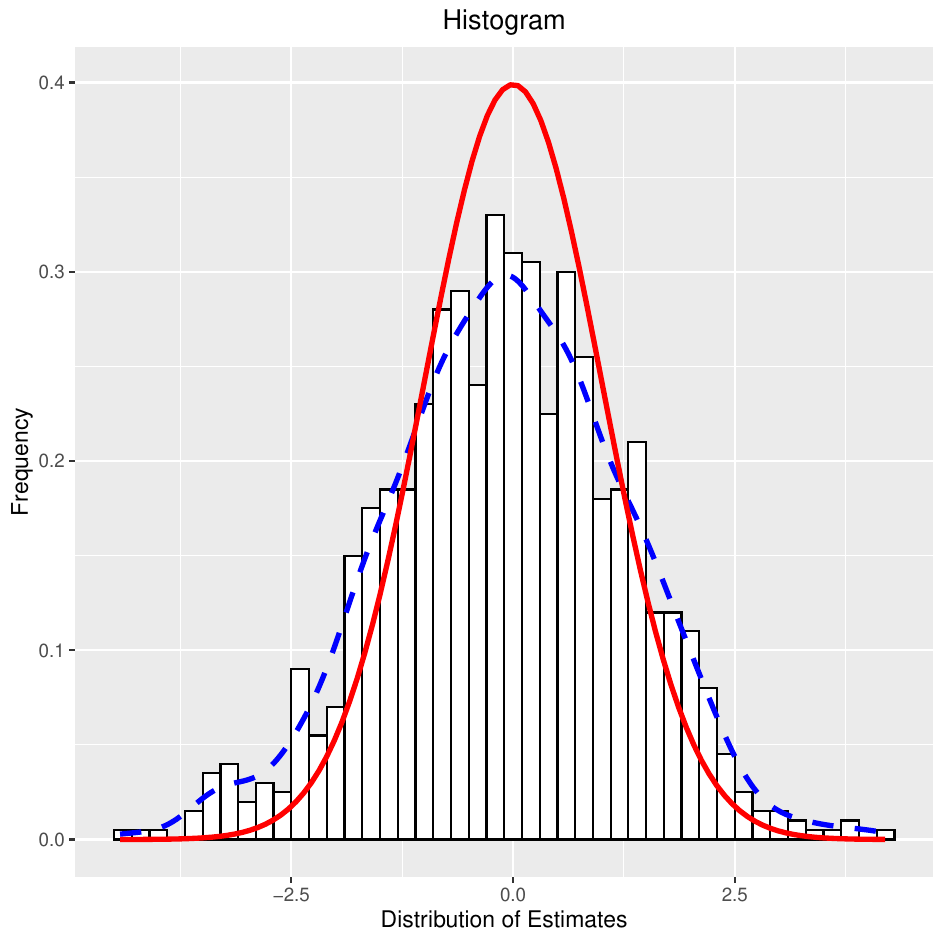}}
		\subfigure[$\gamma=0.01, \beta=1.5$]{\includegraphics[width=50mm]{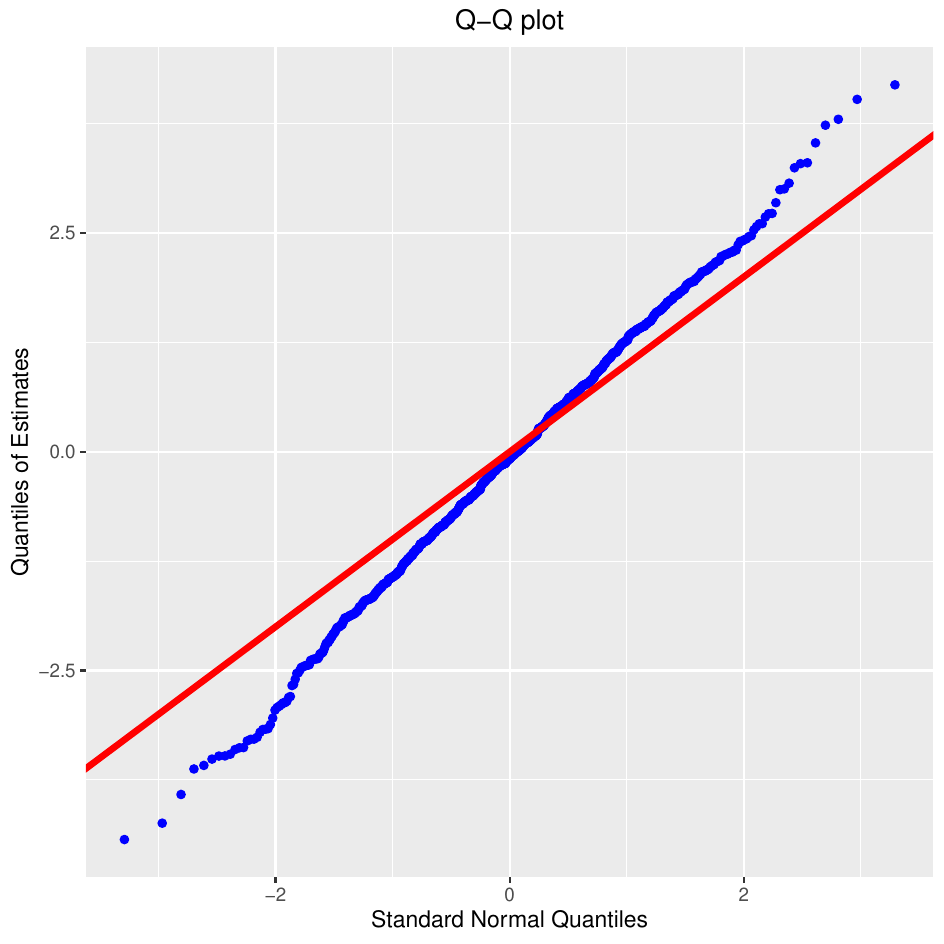}} 
		\caption{The histograms and Q-Q plots for the leverage effect estimates for the feasible central limit theorem: in the histograms, the red solid line is the density curve of the standard normal distribution, and the blue dotted line is the fitted density curve based on the estimates.}
		\label{fig:lev15}
	\end{figure}
}

\ignore{
	\begin{figure}[htbp]
		\centering     
		\subfigure[$\gamma=0.01, \beta=1.5$]{\includegraphics[width=55mm]{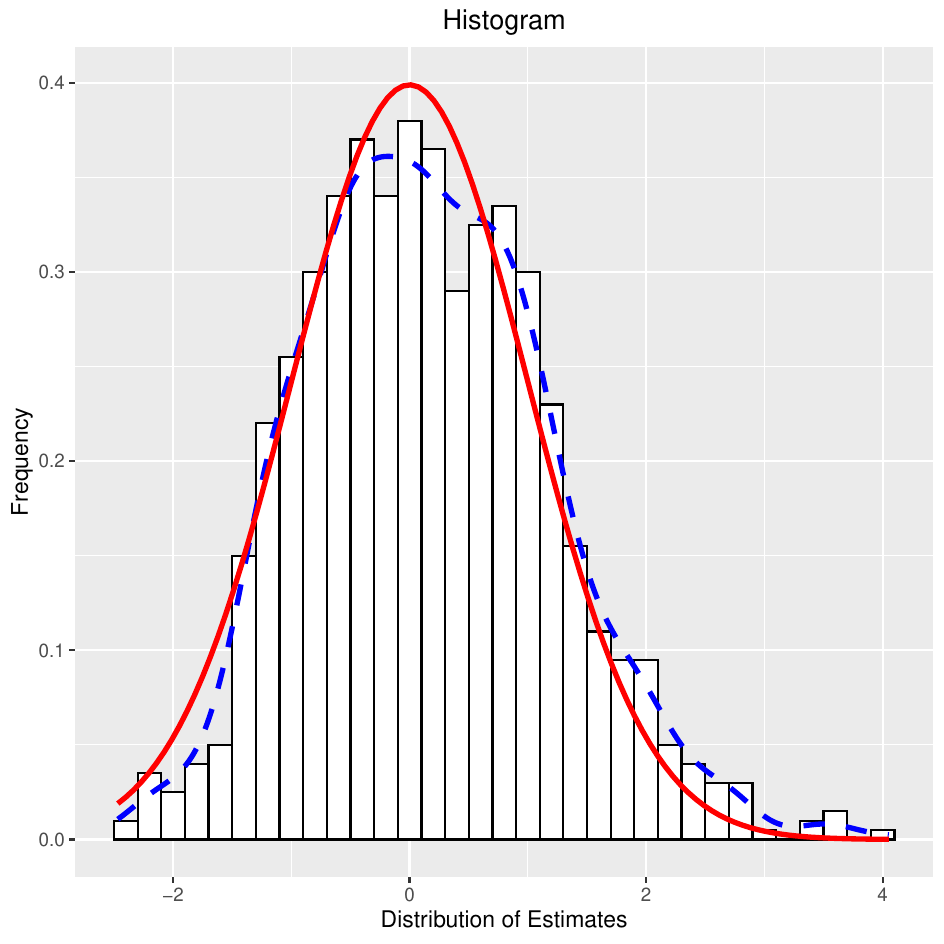}}
		\subfigure[$\gamma=0.01, \beta=1.8$]{\includegraphics[width=55mm]{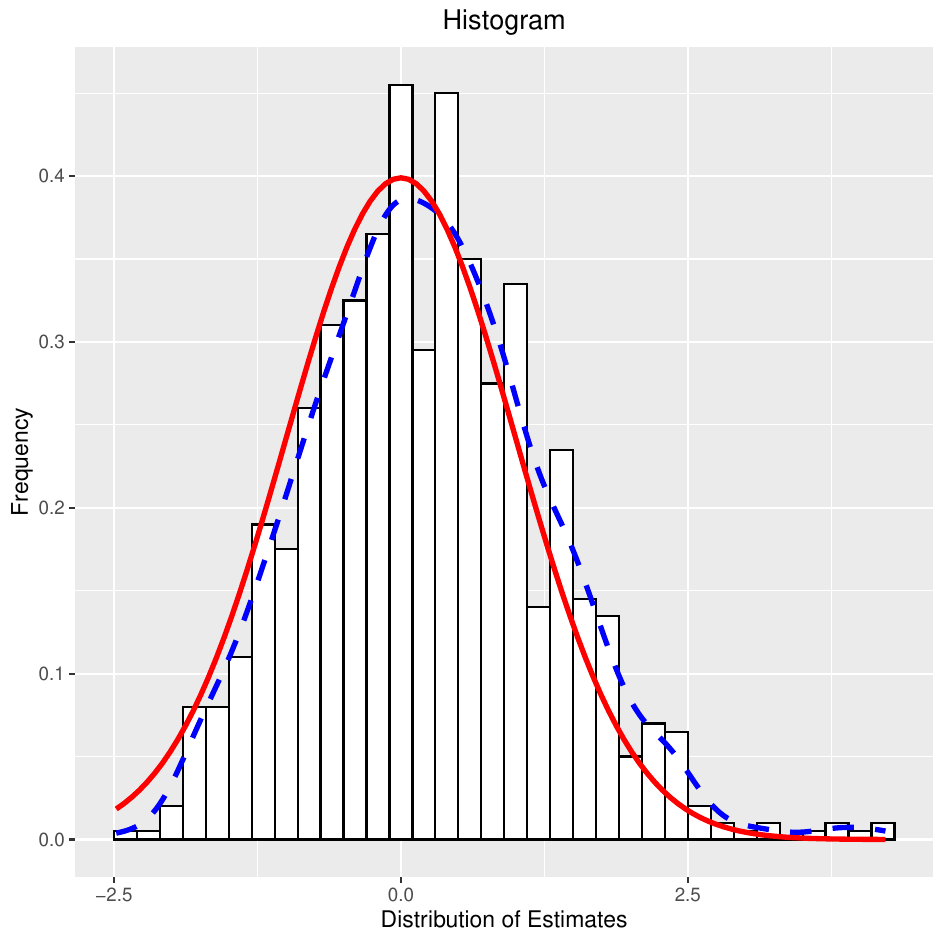}}
		\subfigure[$\gamma=0.01, \beta=1.5$]{\includegraphics[width=55mm]{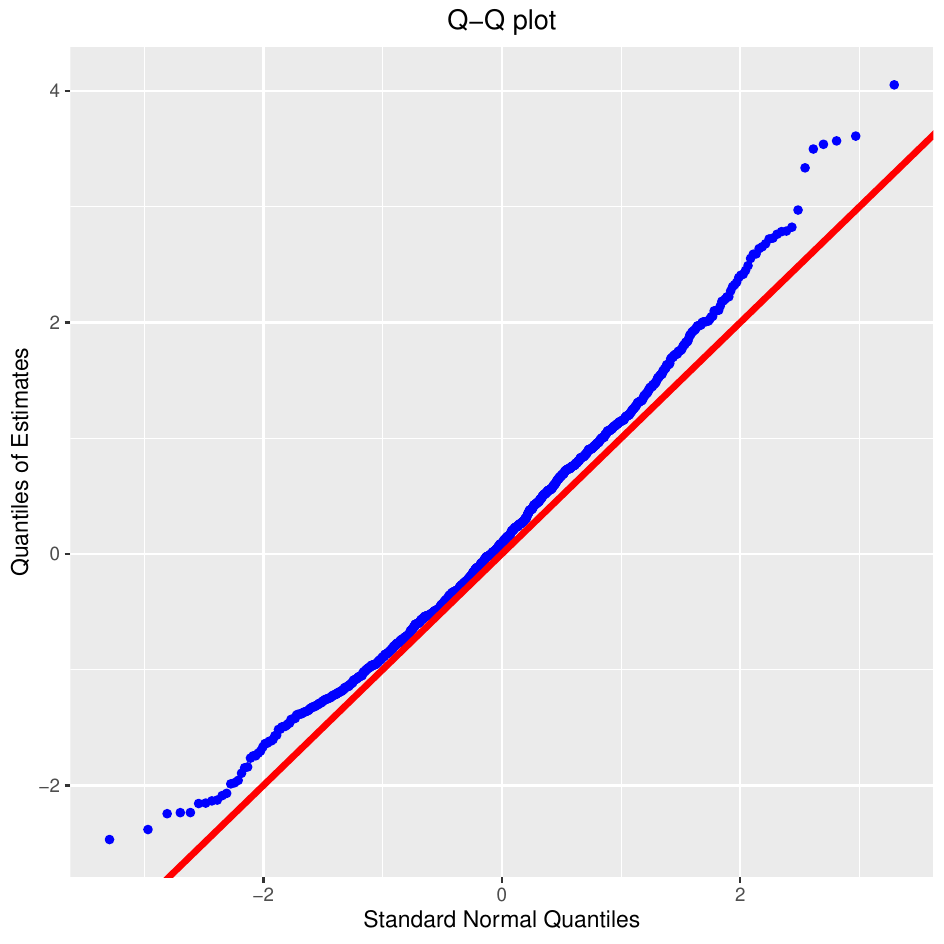}} 
		\subfigure[$\gamma=0.01, \beta=1.8$]{\includegraphics[width=55mm]{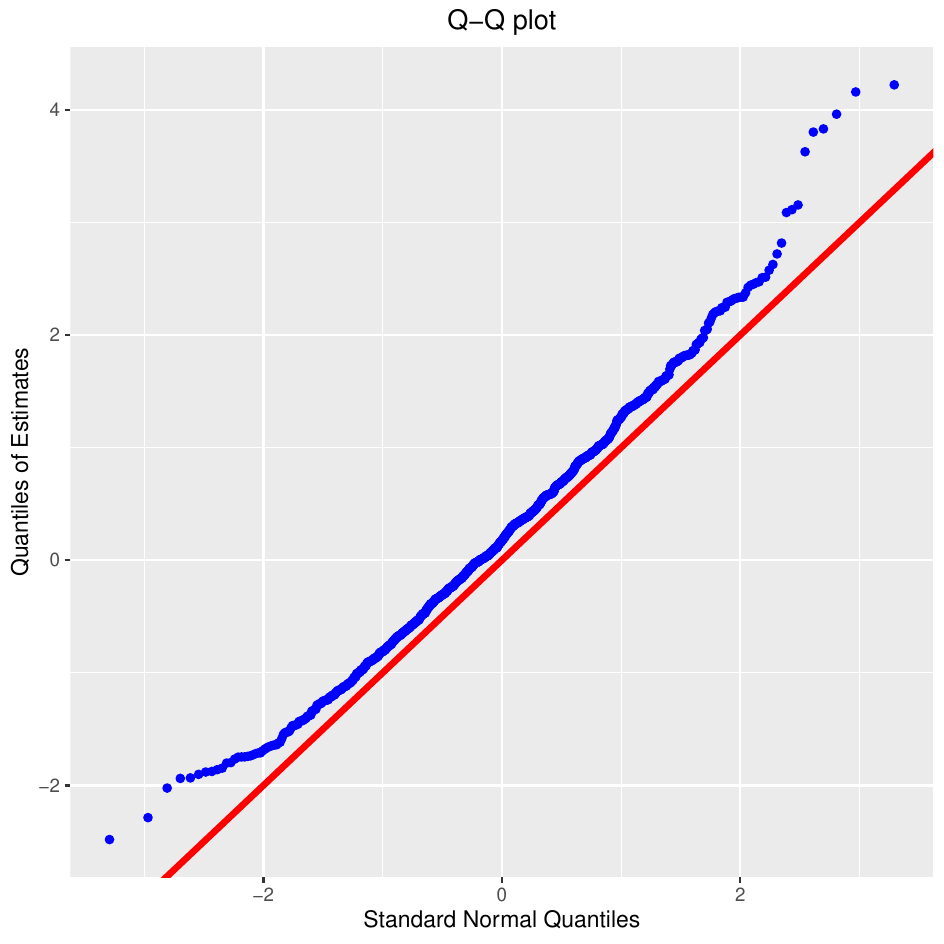}} 
		\caption{The histograms and Q-Q plots for the volatility of volatility estimates when the true asymptotic variance is used: in the histograms, the red solid line is the density curve of the standard normal distribution, and the blue dotted line is the fitted density curve based on the estimates.}
		\label{fig:vov18}
	\end{figure}
}
 
 \section{Empirical studies}\label{sec:emp}
 In this section, we apply our proposed leverage effect estimator and volatility of volatility estimator to real high-frequency financial data. Based on the feasible central limit theorems in Corollary \ref{cor-spot-fea}, we also conduct tests of zero leverage effect and zero volatility of volatility.
 
 The real high-frequency data used in this section are obtained from the 
 FirstRate database\footnote{\url{https://firstratedata.com}}. Instead of using tick-by-tick transaction prices directly, we use relatively sparse data to reduce the adverse effect of market microstructure noise. Specifically, we exclude all half-trading days, overnight returns, and weekend returns, and extract 3-minute log-return data during trading hours (from 9:30 a.m. to 4:00 p.m.) using the conventional previous-tick strategy (see, e.g., \cite{Z2011}), which uses the latest tick-by-tick price before each fixed time grid point as an approximation. We conduct our study over eight years, from January 3, 2011, to December 31, 2018, after the 2008 subprime mortgage crisis and before the COVID-19 pandemic. The subjects of our study are the SPDR S$\&$P 500 ETF (SPY), which tracks the S$\&$P 500 index, and four of its constituent stocks: Apple (APPL), Amazon (AMZN), Intel (INTC), and Microsoft (MSFT). 
 For the estimation of both the leverage effect and volatility of volatility, we fix $u=1$, $b=0.55$, and $\kappa=2$. 
 
 
 We first adopt the jump activity index (JAI) estimator proposed in \cite{JKLM2012} to estimate jump intensity in each month, assuming that the JAI is constant within each month. Based on the monthly estimates, the mean (M.) and standard deviation (S.D.) for each year are reported in Table \ref{emp:jai}. 
 The JAI estimator is defined as 
 \begin{align}
 	\widehat{\beta}_n (\omega, \alpha, \alpha', g) = \log \left( \frac{V(\omega, \alpha,g)_n}{V(\omega, \alpha',g)_n}  \right)\Big/\log\left(\frac{\alpha'}{\alpha}\right),
 \end{align}
 with $0<\alpha < \alpha'$, $0<\omega<1/2$, and $V(\omega, \alpha,g)_n = \sum_{i=1}^{n} g\left(\frac{\Delta_i^n X}{\alpha \Delta_n^{\omega}}\right)$, where $g(x)$ decreases to 0 as $|x|$ goes to 0. Following the empirical setting in \cite{JKLM2012}, we use $\omega = 1/5$, $\alpha=0.0013$, $\alpha'=2\alpha$, and 
 \begin{align*}
 	g(x) = 
 	\begin{cases}
 		&c^{-1} |x|^p, \quad ~~~~~~~ ~~~~~~~~~~~~~~~\text{if} \quad |x|\leq a,\\
 		&c^{-1}\left(a^{p}+\frac{pa^{p-1}}{2(b-a)}\left((b-a)^2- (|x|-b)^2\right)\right), \quad \text{if} \quad a<|x|<b, \\
 		& 1 \quad~~~~~~~~~~~~~~~~~~~~~~~~~~~~~~~ \text{if} \quad |x|\geq b,
 	\end{cases}
 \end{align*}
 with $a=6/5$, $b=7/5$, $p=5$, and $c=a^{p}+pa^{p-1}(b-a)/2$. Table \ref{emp:jai} shows that most JAI estimates are larger than 0.5, and some are even larger than 1, especially for SPY. These results indicate that jump activity is substantial in real high-frequency financial data, which motivates the treatment of jumps in this paper and makes our estimators attractive alternatives for empirical studies. 
 
 \begin{table*}[!htbp]
 	\centering
 	\tiny
 	\caption{The mean (M.) and standard deviation (S.D.) of the estimated jump activity index within each year, based on 12 monthly estimates from January 3, 2011, to December 31, 2018, for Apple (APPL), Amazon (AMZN), Intel (INTC), Microsoft (MSFT), and the SPDR S$\&$P 500 ETF (SPY).} \label{emp:jai}
 	\vspace{0.2cm}
 	\resizebox{\textwidth}{10mm}{
 		\begin{tabular}{|c|c|c|c|c|c|c|c|c|}
 			\toprule
 			&2011&2012&2013&2014&2015&2016&2017&2018\\
 			\hline
 			APPL&0.62, 0.16&0.59, 0.19&0.59, 0.14&0.72, 0.12&0.61, 0.16&0.72, 0.23&0.98, 0.20&0.61, 0.23\\
 			\hline
 			AMZN&0.38, 0.10&0.47, 0.08&0.52, 0.06& 0.49, 0.10 &0.50, 0.10&0.65, 0.27&0.88, 0.15&0.50, 0.22\\
 			\hline
 			INTC&0.67, 0.16&0.72, 0.10&0.67, 0.10&0.70, 0.14 &0.56, 0.09&0.68, 0.17&0.78, 0.07&0.45, 0.14\\
 			\hline
 			MSFT&0.65, 0.15&0.73, 0.08&0.70, 0.11&0.70, 0.13&0.58, 0.12&0.71, 0.22&0.99, 0.13&0.57, 0.23\\
 			\hline
 			SPY&0.86, 0.36&1.19, 0.21&1.41, 0.23&1.33, 0.30&1.13, 0.29&1.33, 0.41&1.95, 0.22&1.10, 0.51\\
 			\bottomrule
 		\end{tabular}
 	}
 \end{table*}
 
 We then use yearly data for volatility of volatility and report the volatility of volatility estimates (VoV.) and their estimated theoretical standard errors (T.S.D.), namely $n^{\frac{b-1}{2}}\sqrt{\widehat{Var(W|\mathcal{F})}}$, in Table \ref{emp:vov}. We also test the hypothesis in \eqref{vov_test} at the $5\%$ significance level using our proposed test statistic \eqref{vov_test_sta}. We reject the null hypothesis of zero volatility of volatility if $\widetilde{T}_n >1.64$; otherwise, if $\widetilde{T}_n \leq 1.64$, we do not reject $H_0$. The testing results from 2011 to 2018 are also included in Table \ref{emp:vov}. The table shows that SPY has the smallest values of both VoV. and T.S.D. in all years. For almost all investigated assets, these quantities attain their largest magnitudes in 2018 and are relatively smaller in 2017. The zero volatility of volatility test does not reject the null hypothesis in 2011 and 2015, but rejects it in 2012, 2013, and 2017 for all assets. Interestingly, in 2018, the test does not reject the null hypothesis for SPY, while all four constituent stocks favor the alternative hypothesis $H_1$. 
 
 \begin{table*}[!htbp]
 	\centering
 	\tiny
 	\caption{The yearly estimates of volatility of volatility (VoV.), its estimated theoretical standard error (T.S.D.), and the testing results for zero volatility of volatility at the $5\%$ significance level, for Apple (APPL), Amazon (AMZN), Intel (INTC), Microsoft (MSFT), and the SPDR S$\&$P 500 ETF (SPY), from January 3, 2011, to December 31, 2018. The notation $H_0$ means that the test does not reject the null hypothesis of zero volatility of volatility, and $H_1$ means that the test rejects the null hypothesis.}\label{emp:vov}
 	\vspace{0.2cm}
 	\resizebox{\textwidth}{10mm}{
 		\begin{tabular}{|c|c|c|c|c|c|c|c|c|}
 			\toprule
 			&2011&2012&2013&2014&2015&2016&2017&2018\\
 			\hline
 			APPL&0.105,0.065,$H_0$&0.065,0.026,$H_1$&0.034,0.013,$H_1$&0.041,0.022,$H_1$&0.187,0.174,$H_0$&0.017,0.008,$H_1$&0.012,0.006,$H_1$
 			&0.082,0.038,$H_1$\\
 			\hline
 			AMZN&0.275,0.170,$H_0$&0.036,0.013,$H_1$&0.041,0.022,$H_1$
 			&0.064,0.028,$H_1$&0.123,0.091,$H_0$&0.097,0.059,$H_0$&0.022,0.010,$H_1$&0.424,0.171,$H_1$\\
 			\hline
 			INTC&0.094,0.064,$H_0$&0.010,0.003,$H_1$&0.009,0.004,$H_1$
 			&0.043,0.033,$H_0$
 			&0.096,0.071,$H_0$&0.017,0.007,$H_1$&0.011,0.005,$H_1$
 			&0.145,0.069,$H_1$
 			\\
 			\hline
 			MSFT&0.092,0.070,$H_0$&0.005,0.002,$H_1$
 			&0.013,0.004,$H_1$&0.016,0.007,$H_1$
 			&0.123,0.116,$H_0$&0.026,0.013,$H_1$
 			&0.006,0.003,$H_1$&0.156,0.080,$H_1$\\
 			\hline
 			SPY&0.064,0.053,$H_0$&0.7e-3,0.3e-3,$H_1$&1.3e-3,0.6e-3,$H_1$
 			&4.1e-3,2.5e-3,$H_0$&0.025,0.024,$H_0$&4.2e-3,2.3e-3,$H_1$
 			&0.3e-3,0.1e-3,$H_1$&0.029,0.018,$H_0$\\
 			\bottomrule
 		\end{tabular}
 	}
 \end{table*}
 
 Similarly, for the leverage effect, we use yearly 3-minute log-return data and report the estimated leverage effect (LeV.) and its estimated theoretical standard error (T.S.D.), namely $\sqrt{\widehat{Var(U|\mathcal{F})}}/\sqrt{n}^{b \wedge (1-b)}$, in Table \ref{emp:lev}. Consider the following zero leverage effect testing problem:
 \begin{align}\label{lev_test}
 	H_0:  \mathcal{L}_{[0,T]}=0 \qquad \text{vs} \qquad  H_1: \mathcal{L}_{[0,T]} \neq 0,
 \end{align}
 By \eqref{fea:lev}, we use the test statistic
 \begin{align}\label{lev_test_sta}
 	\overline{T}_n := \sqrt{n}^{b\wedge(1-b)} \frac{ \widehat{\mathcal{L}}_{[0,T]}  }{\sqrt{\widehat{Var(U|\mathcal{F})}}}. 
 \end{align}
 At the $5\%$ significance level, we reject the null hypothesis of zero leverage effect if $|\overline{T}_n| >1.96$; otherwise, if $|\overline{T}_n| \leq 1.96$, we do not reject $H_0$. The testing results from 2011 to 2018 are also included in Table \ref{emp:lev}. The table shows that almost all leverage effect estimates are negative, which is consistent with the definition of the leverage effect. The magnitude of LeV. is relatively smaller for SPY than for the four individual stocks. For almost all investigated assets, LeV. and T.S.D. attain their largest magnitudes in 2011 and 2018, years in which the null hypothesis of zero leverage effect is also rejected. In 2015, the test does not reject $H_0$ for all assets. Interestingly, from 2011 to 2018, except for 2015, we reject the null hypothesis and obtain significantly negative leverage effect estimates for SPY, while the rejection rate for its four constituent stocks is around 50 percent over these years.
 
 \begin{table*}[!htbp]
 	\centering
 	\tiny
 	\caption{The yearly estimates of the leverage effect (LeV.), its estimated theoretical standard error (T.S.D.), and the testing results for zero leverage effect at the $5\%$ significance level, for Apple (APPL), Amazon (AMZN), Intel (INTC), Microsoft (MSFT), and the SPDR S$\&$P 500 ETF (SPY), from January 3, 2011, to December 31, 2018. The notation $H_0$ means that the test does not reject the null hypothesis of zero leverage effect, and $H_1$ means that the test rejects the null hypothesis.}\label{emp:lev}
 	\vspace{0.2cm}
 	\resizebox{\textwidth}{10mm}{
 		\begin{tabular}{|c|c|c|c|c|c|c|c|c|}
 			\toprule
 			&2011&2012&2013&2014&2015&2016&2017&2018\\
 			\hline
 			APPL&-0.028,0.011,$H_1$&-0.019,0.007,$H_1$&-0.012,0.005,$H_1$&-0.007,0.005,$H_0$&-0.031,0.021,$H_0$&-0.003,0.003,$H_0$
 			&-0.004,0.002,$H_0$
 			&-0.024,0.009,$H_1$\\
 			\hline
 			AMZN&-0.078,0.026,$H_1$&-0.002,0.005,$H_0$&-0.2e-3,6.4e-3,$H_0$
 			&-0.019,0.008,$H_1$&-0.025,0.014,$H_0$&-0.019,0.012,$H_0$&-0.004,0.003,$H_0$&-0.084,0.029,$H_1$\\
 			\hline
 			INTC&-0.026,0.012,$H_1$&-0.003,0.002,$H_0$&-0.002,0.002,$H_0$
 			&-0.007,0.006,$H_0$
 			&-0.016,0.012,$H_0$&-0.003,0.003,$H_0$&0.4e-3,
 			2.2e-3,$H_0$
 			&-0.044,0.014,$H_1$
 			\\
 			\hline
 			MSFT&-0.018,0.012,$H_0$&-0.5e-3,1.4e-3,$H_0$
 			&-0.8e-3,2.4e-3,$H_0$&-0.007,0.002,$H_1$
 			&-0.020,0.016,$H_0$&-0.010,0.004,$H_1$
 			&-1.9e-3,1.5e-3,$H_0$&-0.041,0.015,$H_1$\\
 			\hline
 			SPY&-0.027,0.009,$H_1$&-1.3e-3,0.3e-3,$H_1$&-1.7e-3,0.4e-3,$H_1$
 			&-4.4e-3,0.9e-3,$H_1$&-0.008,0.005,$H_0$&-0.003,0.001,$H_1$
 			&-0.3e-3,0.1e-3,$H_1$&-0.014,0.004,$H_1$\\
 			\bottomrule
 		\end{tabular}
 	}
 \end{table*}
 
\begin{figure}[!htbp]
 	\centering     
 	\includegraphics[width=160mm]{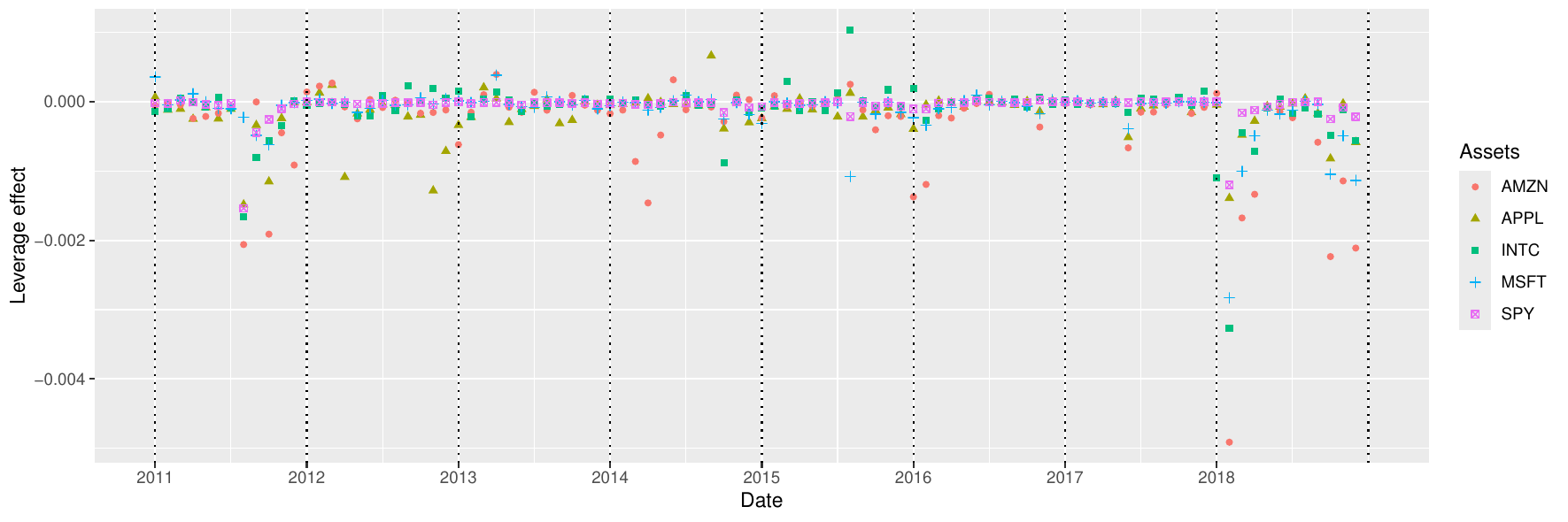}
 	\caption{The leverage effect estimates obtained using monthly data from January 3, 2011, to December 31, 2018, for Apple (APPL), Amazon (AMZN), Intel (INTC), Microsoft (MSFT), and the SPDR S$\&$P 500 ETF (SPY).}
 	\label{lev-fig}
 \end{figure}

 We also estimate the leverage effect using monthly data and report the estimates in Figure \ref{lev-fig}. Most estimates are negative and close to 0, especially for SPY. Repeating the above zero leverage effect testing procedure for all 96 months, the numbers of months in which the test fails to reject the null hypothesis are 78, 85, 82, 83, and 49 for APPL, AMZN, INTC, MSFT, and SPY, respectively. This suggests that a relatively larger sample size is preferable; using ultra-high-frequency data with market microstructure noise, such as tick-by-tick data, may partially address this issue. We leave this topic for future work. 
 
\section{Conclusion and future work}\label{sec:con}
We study statistical inference for the leverage effect and volatility of volatility using high-frequency data in the presence of jumps. Based on the empirical characteristic function of the increments, we propose consistent estimators and establish feasible central limit theorems under regular conditions through the consistent estimation of volatility functionals.

Several issues warrant further investigation. First, we do not consider market microstructure noise or irregular and endogenous observation times, all of which are common features of high-frequency data. Their effects on the estimation of the leverage effect and volatility of volatility deserve further study. Second, the effect of infinite-variation jumps on statistical inference for general volatility functionals remains to be explored. Third, although we propose a consistent estimator of the correlation-based leverage effect, we do not study its statistical inference. Moreover, we assume that $\mathcal{L}^{cor}_{[0,T]}$ is constant over $[0,T]$, which may be theoretically restrictive and empirically unrealistic. A rigorous testing procedure is therefore needed to assess this assumption, and theoretical results that do not rely on it should also be developed.

 
 

	\section{Appendix: Proofs}\label{sec:proofs}
\ignore{
\begin{center}
 \Large \textbf{Appendix: Proofs}
\end{center}
}
This section gives out all the proofs for the theoretical results, subsequently.
 Specifically, Section \ref{prof:pro1}, \ref{prof:thm1}, \ref{prof:thm2} include the proofs of Proposition \ref{pro-1}, Theorem \ref{thm-lev-1} and Theorem \ref{thm-vov}, respectively. Section \ref{prof:sec3_4-5} contains all the proofs for Section \ref{sec:vol-fun} and Section \ref{sec:fea-clt}, and Section \ref{prof:sec4} is for Section \ref{sec:diss}.

Throughout the proofs, we use $C>0$ to denote a generic constant and $\epsilon$ to denote a positive constant that may be arbitrarily close to 0. The shorthand $\E_{j}^n [\cdot]$ denotes the conditional expectation $\E [ \cdot | \mathcal{F}_{t_{j}^n} ] $ for $j=1,...,n$. By the localization procedure in \citet{JP2012}, we can equivalently assume that all coefficient processes of $X$ and $\sigma$ are bounded. Applying It$\hat{\text{o}}$'s lemma and It$\hat{\text{o}}$ isometry to \eqref{model-volvol}, we obtain
\begin{align}
	\E[(\sigma_{s}^2-\sigma^2_{t})^{2k}] \leq a_n (s-t)^{k},
\end{align}
for $k=1,2,...$. 
We introduce the following notations:
\begin{align*}
	\Delta_{j}^{n}X' &= \int_{t_{j-1}^n}^{t_j^n}\sigma_{t_{j-1}^n}dB_s + \int_{t_{j-1}^n}^{t_j^n}\int_{R} \gamma_{t_{j-1}^n} \cdot x (\mu-\nu)(ds,dx) = \sigma_{t_{j-1}^n} \Delta_{j}^n B +  \gamma_{t_{j-1}^n} \Delta_{j}^n L,\\
	S_i^n(u)&=  \frac{1}{k_n}\sum_{j \in I^n_{i+} } \cos{ \left(\frac{u\Delta_j^nX}{\sqrt{\Delta_n}} \right)},  \\
	A_i^n(u) &= \frac{1}{k_n}\sum_{j \in I^n_{i+} } \left( \cos \left(\frac{u\Delta_j^nX}{\sqrt{\Delta_n}} \right) - \cos{ \left(\frac{u\Delta_j^nX'}{\sqrt{\Delta_n}} \right)} \right), \\
	B_i^n(u) &= \frac{1}{k_n}\sum_{j \in I^n_{i+} }  \left( \cos  \left(\frac{u\Delta_j^nX'}{\sqrt{\Delta_n}} \right) - \E_{j-1}^{n} \left[ \cos \left(\frac{u\Delta_i^nX'}{\sqrt{\Delta_n}} \right) \right] \right), \\
	C_i^n(u) &=\frac{1}{k_n}\sum_{j \in I^n_{i+} }  \left( \E_{j-1}^{n} \left[ \cos\left(\frac{u\Delta_j^nX'}{\sqrt{\Delta_n}} \right) \right]\right)- f_{t_{i}^n,n}(u),
\end{align*}
where $f_{t,n}(u) = f^{(1)}_{t,n}(u) \cdot f^{(2)}_{t,n}(u)$ with 
$ f^{(1)}_{t,n}(u)= \exp{\left(- u^2\sigma^2_{t}/2 \right)}$, $ f^{(2)}_{t,n}(u)= \exp{\left(- u^2 b_{ t,n}(u)/2 \right)}$ and $ b_{t,n}(u)= 2C|\gamma_{t}|^{\beta}|u|^{\beta-2}\Delta_n^{1-\frac{\beta}{2}}$ for $t \in [0,T]$. 
Under Assumption \ref{asu-con} and with the boundedness of $\sigma$ and $\gamma$, we have, for $t,s\in[0,1]$,
\begin{align}\label{cont-sigma+gamma}
	\begin{split}
		\E\left[\left|f^{(1)}_{t,n}(u) - f^{(1)}_{s,n}(u)\right|\right]& \leq C\E[|\sigma_t-\sigma_s|]\leq C\left(\E[|\sigma_t-\sigma_s|^2]\right)^{1/2} \leq C|s-t|^{1/2}\\
		\E\left[\left|f^{(2)}_{t,n}(u) - f^{(2)}_{s,n}(u)\right|\right] &\leq C|u|^{\beta-2}\Delta_n^{1-\frac{\beta}{2}} \E[| |\gamma_t|^{\beta} -|\gamma_s|^{\beta} |] \leq C|u|^{\beta-2}\Delta_n^{1-\frac{\beta}{2}} \E[| |\gamma_t| -|\gamma_s| |]\\
		&\leq C|u|^{\beta-2}\Delta_n^{1-\frac{\beta}{2}} \E[|\gamma_t-\gamma_s |] \leq C|u|^{\beta-2}\Delta_n^{1-\frac{\beta}{2}} |s-t|^{1/2}.
	\end{split}
\end{align}
Moreover, throughout the proof, we suppress the dependence of the functions $h_1$ and $h_2$ on $\sigma^2, (\tilde{\sigma})^2, (\tilde{\sigma}')^2$ for simplicity; that is, we use
\begin{align*}
	h_1(u,t) = \frac{2(\exp{(-2 u^2\sigma_t^2)} - 2\exp{(- u^2\sigma_t^2)} + 1)}{u^4\exp{(- u^2\sigma_t^2 )}}, \quad 
	h_2(t) = \frac{4(\sigma_t^2 )(\tilde{\sigma}_t)^2+(\tilde{\sigma}'_t)^2)}{3}. 
\end{align*}

\subsection{Proof for Proposition \ref{pro-1} }\label{prof:pro1}
	\textbf{Proof of Proposition \ref{pro-1}:} Noting that $\widehat{\sigma}^2_{t} = \widehat{\sigma}^2_{t_{i+}^n}$ in \eqref{est-spot++} with $i=\lfloor t/\Delta_n \rfloor$, we write
	\begin{align*}
		&\widehat{\sigma}^2_{t} -\sigma^2_{t} - b_{ t}(u)\\
		&=\widehat{\sigma}^2_{t_{i+}^n} -\sigma^2_{t_{i+}^n} - b_{ t_{i+}^n,n}(u)+ (\sigma^2_{t_{i+}^n} -\sigma^2_{t}) + (b_{ t_{i+}^n,n}(u)- b_{ t}(u))\\
		&(\text{By conditions \eqref{cond-vov}.})\\
		&= \frac{-2}{u^2} \log\left( S_i^n(u) \vee \frac{1}{\sqrt{k_n}}\right) -\sigma^2_{t_{i+}^n} - b_{ t_{i+}^n,n}(u) + O_p(\sqrt{\Delta_n})\\
		& (\text{The equation holds almost surely, as given by Lemma 4 in \cite{LLL2018}}.)\\
		&= \frac{-2}{u^2} \log \left( S_i^n(u) \right) -\sigma^2_{t_{i+}^n} - b_{ t_{i+}^n,n}(u)+ O_p(\sqrt{\Delta_n})\\
		&=  \frac{-2}{u^2} \log{(1+\xi_{t_{i+}^n,n}(u))}+ O_p(\sqrt{\Delta_n}),
	\end{align*}
	with $\xi_{t_{i+}^n,n}(u)= \frac{  S_i^n(u) - f_{t_{i+}^n,n}(u) }{f_{t_{i+}^n,n}(u) }$.
	Since $|\log(1+x) - x | \leq Cx^2$, we have
	\begin{align}
		\left|\widehat{\sigma}^2_{t_{i+}^n} -\sigma^2_{t_{i+}^n} - b_{ t_{i+}^n,n}(u) - \frac{-2}{u^2} \xi_{t_{i+}^n,n}(u) \right| \leq \frac{C}{u^2} |\xi_{t_{i+}^n,n}(u)|^2.
	\end{align}
	We decompose 
	\begin{align}
		\frac{-2}{u^2}\xi_{t_{i+}^n,n}(u) =  \sum_{w=1}^{3}	\xi_{t_{i+}^n,n}^{(w)}(u),
	\end{align}
	with $\xi_{t_{i+}^n,n}^{(1)}(u) =\frac{-2}{u^2}\frac{A_i^n(u) }{f_{t_{i+}^n,n}(u)}$, $\xi_{t_{i+}^n,n}^{(2)}(u) = \frac{-2}{u^2}\frac{ B_i^n(u) }{f_{t_{i+}^n,n}(u)}$, and $\xi_{t_{i+}^n,n}^{(3)}(u)= \frac{-2}{u^2}\frac{ C_i^n(u) }{f_{t_{i+}^n,n}(u)}$. Then, to prove \eqref{pro1-res1}, \eqref{pro1-res2}, and \eqref{pro1-res3}, it is sufficient to show
	\begin{align}\label{xi-1}
		&\left( \sqrt{k_n} \wedge \frac{1}{\sqrt{k_n\Delta_n}}\right)	\xi_{t_{i+}^n,n}^{(1)}(u)  \longrightarrow^{p} 0, \\\label{xi-23}
		&\left( \sqrt{k_n}	\xi_{t_{i+}^n,n}^{(2)}(u),  \frac{1}{\sqrt{k_n\Delta_n}}	\xi_{t_{i+}^n,n}^{(3)}(u) \right) \longrightarrow^{L_s} \left(V_{t},V'_{t}\right), \\\label{xi^2}
		&\left(\sqrt{k_n} \wedge \frac{1}{\sqrt{k_n\Delta_n}}\right) \frac{u^2}{2}\left|\sum_{w=1}^{3}	\xi_{t_{i+}^n,n}^{(w)}(u)\right|^2 \leq \frac{Cu^2(k_n\sqrt{\Delta_n} \wedge 1)}{\sqrt{k_n\Delta_n}} \sum_{w=1}^{3} \left|	\xi_{t_{i+}^n,n}^{(w)}(u)\right|^2 \longrightarrow^{p} 0.
	\end{align}
	For \eqref{xi-1}, following the proofs of Lemmas 2 and 5 in \cite{LLL2018}, together with the boundedness of $\sigma, \gamma, \delta$ and condition \eqref{cond-vov}, we obtain
	\begin{align}\label{bound-A}
		\E[|\xi_{t_{i+}^n,n}^{(1)}(u)|] \leq
		\begin{cases}
			&C\frac{\sqrt{\Delta_n}}{u},\ \text{for}  \ 0<\max\{\beta,r\} < 1,\\
			&C\frac{\sqrt{\Delta_n}}{u} + C\frac{\Delta_n^{ \frac{5}{4p_1} - \frac{1}{4}}}{u}  + C\frac{\Delta_n^{ \frac{1}{p_2} - \frac{1}{2}}}{u},  \ \text{for}\ 1\leq \beta<p_1<2, 1\leq r<p_2<2.
		\end{cases}
	\end{align}
	If $\beta<2$ and $r<4/3$, we have  
	\begin{align*}
		\E\left[\left|\left(\sqrt{k_n} \wedge \frac{1}{\sqrt{k_n\Delta_n}}\right)	\xi_{t_{i+}^n,n}^{(1)}(u) \right|\right] \leq C\left( \frac{\Delta_n^{\frac{1}{4}} + \Delta_n^{\frac{5}{4(\beta+\epsilon)}-\frac{1}{2}} + \Delta_n^{\frac{1}{(r+\epsilon)}-\frac{3}{4}} }{u} \right) \rightarrow 0.
	\end{align*}
	For \eqref{xi-23}, 
	by Lemma S2.3 in \cite{LL2024}, we have 
	\begin{align*}
		\E_{j-1}^{n} \left[ \cos\left(\frac{u\Delta_j^nX'}{\sqrt{\Delta_n}} \right) \right] &= f^{(1)}_{t_{j-1}^n,n}(u) \cdot  \left( f^{(2)}_{t_{j-1}^n,n}(u) +o_p(u^4\sqrt{\Delta_n}) \right) \\
		&=  f^{(1)}_{t_{j-1}^n,n}(u) \cdot f^{(2)}_{t_{j-1}^n,n}(u) + o_p(u^4\sqrt{\Delta_n}).
	\end{align*}
	We write
	\begin{align}\label{xi-23-1}
		\begin{split}
			&\frac{1}{\sqrt{k_n\Delta_n}} \xi_{t_{i+}^n,n}^{(3)}(u) \\
			&= \sum_{j \in I^n_{i+} }  \frac{ -2\left( f^{(1)}_{t_{j-1}^n,n}(u) \cdot  f^{(2)}_{t_{j-1}^n,n}(u) - f^{(1)}_{t_{i+}^n,n}(u)f^{(2)}_{t_{i+}^n,n}(u)  \right) }{k_n \sqrt{k_n\Delta_n}u^2f^{(1)}_{t_{i+}^n,n}(u)f^{(2)}_{t_{i+}^n,n}(u)} + o_p\left(\frac{u^2}{\sqrt{k_n}}\right)\\
			&= \sum_{j \in I^n_{i+} }  \frac{ -2\left( f^{(1)}_{t_{j-1}^n,n}(u)  - f^{(1)}_{t_{i+}^n,n}(u)\right)  }{k_n \sqrt{k_n\Delta_n}u^2f^{(1)}_{t_{i+}^n,n}(u) }+\sum_{j \in I^n_{i+} }  \frac{ -2 f^{(1)}_{t_{j-1}^n,n}(u) \left(  f^{(2)}_{t_{j-1}^n,n}(u) - f^{(2)}_{t_{i+}^n,n}(u)  \right) }{k_n \sqrt{k_n\Delta_n}u^2f^{(1)}_{t_{i+}^n,n}(u)f^{(2)}_{t_{i+}^n,n}(u)} + o_p\left(\frac{u^2}{\sqrt{k_n}}\right).
		\end{split}
	\end{align}
	For $j \in I^n_{i+}$, we have 
	\begin{align*}
		&f^{(1)}_{t_{j-1}^n,n}(u) - f^{(1)}_{t_{i+}^n,n}(u) \\
		&\text{(By Taylor's expansion, the boundedness of $\sigma^2$ and condition \eqref{cond-vov}.)}\\
		&=  \exp{\left(- u^2\sigma^2_{t_{i+}^n}/2 \right)} \frac{-u^2}{2}(\sigma^2_{t_{j-1}^n}-\sigma^2_{t_{i+}^n})+ O_p(u^4k_n\Delta_n) \\ 
		&(\text{By It$\hat{\text{o}}$'s formula.})\\
		& =\exp{\left(- u^2\sigma^2_{t_{i+}^n}/2 \right)} \frac{-u^2}{2} \int_{t_{j-1}^n}^{t_j^n}\int_{t_{i+}^n}^{t_{j-1}^n} \left(2\sigma_u \tilde{\sigma}_udB_u+2\sigma_u \tilde{\sigma}'_udB'_u\right)ds \\
		&\quad + \exp{\left(- u^2\sigma^2_{t_{i+}^n}/2 \right)} \frac{-u^2}{2} \int_{t_{j-1}^n}^{t_j^n}\int_{t_{i+}^n}^{t_{j-1}^n}(2\sigma_u\tilde{b}_u + (\tilde{\sigma}_u)^2+(\tilde{\sigma}'_u)^2)duds +  O_p(u^4k_n\Delta_n) \\
		&(\text{Since $\sigma, \tilde{\sigma}, \tilde{\sigma}'$ are bounded.})\\
		&= \exp{\left(- u^2\sigma^2_{t_{i+}^n}/2 \right)} \frac{-u^2}{2} \int_{t_{j-1}^n}^{t_j^n}\int_{t_{i+}^n}^{t_{j-1}^n} \left(2\sigma_u \tilde{\sigma}_udB_u+2\sigma_u \tilde{\sigma}'_udB'_u\right)ds + O_p((u^2+u^4)k_n\Delta_n)\\
		&= \exp{\left(- u^2\sigma^2_{t_{i+}^n}/2 \right)} \frac{-u^2}{2} \left( \int_{t_{i+}^n}^{t_{j-1}^n} \left(2\sigma_u \tilde{\sigma}_u\right)dB_u +  \int_{t_{i+}^n}^{t_{j-1}^n} \left(2\sigma_u \tilde{\sigma}'_u\right) dB'_u \right)\\
		& \quad + \exp{\left(- u^2\sigma^2_{t_{i+}^n}/2 \right)} \frac{-u^2}{2\Delta_n}  \left(\int_{t_{j-1}^n}^{t_j^n}\int_{u}^{t_{j}^n} \left(2\sigma_u \tilde{\sigma}_u\right) dsdB_u + \int_{t_{j-1}^n}^{t_j^n}\int_{u}^{t_{j}^n} \left(2\sigma_u \tilde{\sigma}'_u\right) dsdB'_u  \right)  \\
		&(\text{Since $\sigma, \tilde{\sigma}, \tilde{\sigma}'$ are bounded.})\\
		&= \exp{\left(- u^2\sigma^2_{t_{i+}^n}/2 \right)} \frac{-u^2}{2} \left( \int_{t_{i+}^n}^{t_{j-1}^n} \left(2\sigma_u \tilde{\sigma}_u\right)dB_u +  \int_{t_{i+}^n}^{t_{j-1}^n} \left(2\sigma_u \tilde{\sigma}'_u\right) dB'_u \right) + O_p((u^2+u^4)k_n\Delta_n) \\
		&(\text{Using the boundedness of $\sigma,\tilde{\sigma},\tilde{\sigma}'$ and condition \eqref{cond-vov}.})\\
		&=  \exp{\left(- u^2\sigma^2_{t_{i+}^n}/2 \right)} \frac{-u^2}{2} \left( \int_{t_{i+}^n}^{t_{j-1}^n} \left(2\sigma_{t_{i+}^n} \tilde{\sigma}_{t_{i+}^n}\right)dB_u +  \int_{t_{i+}^n}^{t_{j-1}^n} \left(2\sigma_{t_{i+}^n} \tilde{\sigma}'_{t_{i+}^n}\right) dB'_u \right) + O_p((u^2+u^4)k_n\Delta_n).
	\end{align*}
	Since $\gamma$ is bounded, with condition \eqref{cond-vov}, we have
	\begin{align*}
		f^{(2)}_{t_{j-1}^n,n}(u) - f^{(2)}_{t_{i+}^n,n}(u) =O_p(u^{\beta} \Delta_n^{1-\beta/2} (k_n\Delta_n)^{1/2}). 
	\end{align*}
	Substituting the above results into \eqref{xi-23-1} yields 
	\begin{align*}
		\frac{1}{\sqrt{k_n\Delta_n}} \xi_{t_{i+}^n,n}^{(3)}(u) 
		& =\sum_{j \in I^n_{i+} } \xi_{t_{i+}^n,n}^{(3,j)}(u) + O_p(\sqrt{k_n\Delta_n}) + O_p \left( \frac{u^{\beta} \Delta_n^{1-\beta/2} (k_n\Delta_n)^{1/2}}{\sqrt{k_n\Delta_n}}\right) + o_p\left(\frac{u^2}{\sqrt{k_n}}\right),
	\end{align*}
	where
	\begin{align*}
		\xi_{t_{i+}^n,n}^{(3,j)}(u) =  \frac{ (i+k_n-j)\left(2\sigma_{t_{i+}^n} \tilde{\sigma}_{t_{i+}^n}\Delta_j^nB + 2\sigma_{t_{i+}^n} \tilde{\sigma}'_{t_{i+}^n} \Delta_j^nB'\right)    }{k_n \sqrt{k_n\Delta_n} }.
	\end{align*}	
	We write
	\begin{align*}
		\sqrt{k_n}	\xi_{t_{i+}^n,n}^{(2)}(u) = \sum_{j \in I^n_{i+} } \xi_{t_{i+}^n,n}^{(2,j)}(u),
	\end{align*}
	with 
	\begin{align*}
		\xi_{t_{i+}^n,n}^{(2,j)}(u)  = \frac{  -2 \left( \cos  \left(\frac{u\Delta_j^nX'}{\sqrt{\Delta_n}} \right) - \E_{j-1}^{n} \left[ \cos \left(\frac{u\Delta_j^nX'}{\sqrt{\Delta_n}} \right) \right] \right) }{\sqrt{k_n} u^2 f_{t_{i+}^n,n}(u)}. 
	\end{align*}
	Then, to prove \eqref{xi-23}, it remains to show 
	\begin{align*}
		\sum_{j \in I^n_{i+} } \left(  \xi_{t_{i+}^n,n}^{(2,j)}(u), \xi_{t_{i+}^n,n}^{(3,j)}(u) \right) \longrightarrow^{L_s} \left(V_{t},V'_{t}\right).
	\end{align*}
	Note that $\{\xi_{t_{i+}^n,n}^{(2,j)}(u), \mathcal{F}_{t_j^n}\}$ and $\{\xi_{t_{i+}^n,n}^{(3,j)}(u), \mathcal{F}_{t_j^n}\}$ are martingale difference arrays. Therefore, by Theorem 2.2.15 in \cite{JP2012}, it suffices to show
	\begin{align}\label{xi-23-var}
		\begin{split}
			&\sum_{j \in I^n_{i+} } \E_{j-1}^n \left[ \left(\xi_{t_{i+}^n,n}^{(2,j)}(u) \right)^2 \right] \longrightarrow^{p} \frac{2(\exp{(-2 u^2\sigma^2_{t} )} - 2\exp{(- u^2\sigma^2_{t} )} + 1)}{u^4\exp{(- u^2\sigma^2_{t} )}}, \\
			&\sum_{j \in I^n_{i+} } \E_{j-1}^n \left[ \left(\xi_{t_{i+}^n,n}^{(3,j)}(u) \right)^2 \right] \longrightarrow^{p} \frac{4(\sigma_{t} )^2((\tilde{\sigma}_{t})^2+(\tilde{\sigma}'_{t})^2)}{3}, \\
			&\sum_{j \in I^n_{i+} } \E_{j-1}^n \left[ \xi_{t_{i+}^n,n}^{(2,j)}(u) \cdot\xi_{t_{i+}^n,n}^{(3,j)}(u)   \right] \longrightarrow^{p} 0 ,
		\end{split}
	\end{align}
	and 
	\begin{align}\label{xi-23-4or}
		&\sum_{j \in I^n_{i+} } \E_{j-1}^n \left[ \left(\left(  \xi_{t_{i+}^n,n}^{(2,j)}(u)\right)^2+ \left(\xi_{t_{i+}^n,n}^{(3,j)}(u) \right)^2\right)^2\right] \leq 2
		\sum_{j \in I^n_{i+} } \E_{j-1}^n \left[ \left( \xi_{t_{i+}^n,n}^{(2,j)}(u)\right)^4+ \left(\xi_{t_{i+}^n,n}^{(3,j)}(u) \right)^4\right] \longrightarrow^{p} 0,\\   \label{xi-23-W}
		&\sum_{j \in I^n_{i+} } \E_{j-1}^n \left[ \left(  \xi_{t_{i+}^n,n}^{(2,j)}(u) + \xi_{t_{i+}^n,n}^{(3,j)}(u) \right)\cdot \left(W_{t_{i+k_n}^n} - W_{t_{i}^n} + B_{t_{i+k_n}^n} - B_{t_{i}^n} + B'_{t_{i+k_n}^n} - B'_{t_{i}^n} \right) \right]\longrightarrow^{p} 0, 
	\end{align}
	where $W$ is a bounded martingale orthogonal to both $B$ and $B'$.\\
	For \eqref{xi-23-var}, since $\gamma, \sigma$ are bounded, with condition \eqref{cond-vov}, applying Taylor's expansion yields 
	\begin{align}\label{appro-f12}
		f^{(1)}_{t_{j-1}^n,n}(u) = f^{(1)}_{t_{i+}^n,n}(u) + O_p(u^2\sqrt{k_n\Delta_n}), \qquad  f^{(2)}_{t_{j-1}^n,n}(u) = 1+ O_p(|u|^{\beta-2}\Delta_n^{1-\frac{\beta}{2}}).
	\end{align}
	Furthermore, we obtain
	\begin{align*}
		&\sum_{j \in I^n_{i+} } \E_{j-1}^n \left[ \left(\xi_{t_{i+}^n,n}^{(2,j)}(u) \right)^2 \right] \\
		&= \sum_{j \in I^n_{i+} } \frac{4}{k_nu^4(f_{t_{i+}^n,n}(u))^2} \left( \E_{j-1}^n \left[ \left( \cos  \left(\frac{u\Delta_j^nX'}{\sqrt{\Delta_n}} \right) \right)^2\right] - \left(\E_{j-1}^{n} \left[ \cos \left(\frac{u\Delta_j^nX'}{\sqrt{\Delta_n}} \right) \right]   \right)^2 \right)  \\
		&=  \sum_{j \in I^n_{i+} } \frac{4}{k_nu^4(f_{t_{i+}^n,n}(u))^2} \left( \E_{j-1}^n \left[\frac{  \cos  \left(\frac{2u\Delta_j^nX'}{\sqrt{\Delta_n}} \right) + 1 }{2}\right] - \left(  f^{(1)}_{t_{j-1}^n,n}(u) \cdot  f^{(2)}_{t_{j-1}^n,n}(u)  \right)^2 \right)  \\
		&=  \sum_{j \in I^n_{i+} } \frac{4}{k_nu^4(f_{t_{i+}^n,n}(u))^2} \left( \frac{1}{2} +\frac{1}{2} f^{(1)}_{t_{j-1}^n,n}(2u) \cdot  f^{(2)}_{t_{j-1}^n,n}(2u)- \left(  f^{(1)}_{t_{j-1}^n,n}(u) \cdot  f^{(2)}_{t_{j-1}^n,n}(u)  \right)^2 \right)  \\
		&=  \sum_{j \in I^n_{i+} } \frac{4}{k_nu^4(f^{(1)}_{t_{i+}^n,n}(u))^2} \left( \frac{1}{2} +\frac{1}{2} f^{(1)}_{t_{i+}^n,n}(2u) - \left(  f^{(1)}_{t_{i+}^n,n}(u)  \right)^2 + O_p\left(u^2\sqrt{k_n\Delta_n} + |u|^{\beta-2}\Delta_n^{1-\frac{\beta}{2}}\right) \right)  \\
		&\longrightarrow^{p} 
		\frac{2(\exp{(-2 u^2\sigma^2_{t} )} -2 \exp{(- u^2\sigma^2_{t} )} + 1)}{u^4\exp{(- u^2\sigma^2_{t} )}},
	\end{align*}
	and 
	\begin{align*}
		&\sum_{j \in I^n_{i+} } \E_{j-1}^n \left[ \left(\xi_{t_{i+}^n,n}^{(3,j)}(u) \right)^2 \right] = \sum_{j \in I^n_{i+} } \E_{j-1}^n \left[ \left(    \frac{ (i+k_n+1-j)\left(2\sigma_{t_{i+}^n} \tilde{\sigma}_{t_{i+}^n}\Delta_j^nB + 2\sigma_{t_{i+}^n} \tilde{\sigma}'_{t_{i+}^n} \Delta_j^nB'\right)    }{k_n \sqrt{k_n\Delta_n} }  \right)^2 \right] \\
		&\longrightarrow^{p} \frac{4(\sigma_{t} )^2((\tilde{\sigma}_{t})^2+(\tilde{\sigma}'_{t})^2)}{3},
	\end{align*}
	and
	\begin{align*}
		&\sum_{j \in I^n_{i+} } \E_{j-1}^n \left[ \xi_{t_{i+}^n,n}^{(3,j)}(u) \cdot\xi_{t_{i+}^n,n}^{(2,j)}(u)   \right]\\
		&= \sum_{j \in I^n_{i+} } \frac{-4((i+k_n-j))\sigma_{t_{i+}^n} \tilde{\sigma}_{t_{i+}^n} }{k_n^2 \sqrt{\Delta_n}u^2 f_{t_{i+}^n,n}(u) } \E_{j-1}^n \left[ \left(  \cos  \left(\frac{u(\sigma_{t_{j-1}^n} \Delta_j^n B + \gamma_{t_{j-1}^n} \Delta_j^n L)}{\sqrt{\Delta_n}} \right)  \right) \cdot \left(\Delta_j^nB \right)  \right] \\
		&\quad + \sum_{j \in I^n_{i+} } \frac{-4((i+k_n-j))\sigma_{t_{i+}^n} \tilde{\sigma}'_{t_{i+}^n} }{k_n^2 \sqrt{\Delta_n}u^2 f_{t_{i+}^n,n}(u) } \E_{j-1}^n \left[ \left(  \cos  \left(\frac{u(\sigma_{t_{j-1}^n} \Delta_j^n B + \gamma_{t_{j-1}^n} \Delta_j^n L)}{\sqrt{\Delta_n}} \right) \right) \cdot \left( \Delta_j^nB'\right)  \right] \\
		&(\text{Since $\sigma, \tilde{\sigma}, \tilde{\sigma}', \gamma$ are bounded.})\\
		&\leq \frac{C}{\sqrt{\Delta_n}u^2} \E_{j-1}^n \left[  \cos  \left(\frac{u(\sigma_{t_{j-1}^n} \Delta_j^n B )}{\sqrt{\Delta_n}} \right) \cos  \left(\frac{u( \gamma_{t_{j-1}^n} \Delta_j^n L)}{\sqrt{\Delta_n}} \right) \cdot \left(\Delta_j^nB \right) \right] \\
		&\quad- \frac{C}{\sqrt{\Delta_n}u^2} \E_{j-1}^n \left[  \sin \left(\frac{u(\sigma_{t_{j-1}^n} \Delta_j^n B )}{\sqrt{\Delta_n}} \right)  \sin \left(\frac{u(\gamma_{t_{j-1}^n} \Delta_j^n L )}{\sqrt{\Delta_n}} \right) \cdot \left(\Delta_j^nB \right) \right] \\
		&\quad + \frac{C}{\sqrt{\Delta_n}u^2} \E_{j-1}^n \left[  \cos  \left(\frac{u(\sigma_{t_{j-1}^n} \Delta_j^n B )}{\sqrt{\Delta_n}} \right) \cos  \left(\frac{u( \gamma_{t_{j-1}^n} \Delta_j^n L)}{\sqrt{\Delta_n}} \right) \cdot \left(\Delta_j^nB' \right) \right] \\
		&\quad- \frac{C}{\sqrt{\Delta_n}u^2} \E_{j-1}^n \left[  \sin \left(\frac{u(\sigma_{t_{j-1}^n} \Delta_j^n B )}{\sqrt{\Delta_n}} \right)  \sin \left(\frac{u(\gamma_{t_{j-1}^n} \Delta_j^n L )}{\sqrt{\Delta_n}} \right) \cdot \left(\Delta_j^nB' \right) \right] \\
		&(\text{Since $B, B', L$ are mutually independent, their distributions are symmetric around 0,}\\
		&\text{and the functions $\sin(x)$ and $x\cos{x}$ are odd.)}\\
		&\equiv 0. 
	\end{align*}
	For \eqref{xi-23-4or}, we have 
	\begin{align*}
		&\sum_{j \in I^n_{i+} } \E_{j-1}^n \left[ \left(  \xi_{t_{i+}^n,n}^{(2,j)}(u)\right)^4+ \left(\xi_{t_{i+}^n,n}^{(3,j)}(u) \right)^4\right] \\
		&= \sum_{j \in I^n_{i+} } \frac{16}{k_n^2 u^8(f_{t_{i+}^n,n}(u))^4} \E_{j-1}^n \left[ \left( \cos  \left(\frac{u\Delta_j^nX'}{\sqrt{\Delta_n}} \right) - \E_{j-1}^{n} \left[ \cos \left(\frac{u\Delta_j^nX'}{\sqrt{\Delta_n}} \right) \right] \right)^4 \right] \\
		&  \quad + \sum_{j \in I^n_{i+} }  \frac{ (i+k_n-j)^4}{k_n^6} \E_{j-1}^n \left[\left( \frac{ 2\sigma_{t_{i+}^n} \tilde{\sigma}_{t_{i+}^n} \Delta_j^nB + 2\sigma_{t_{i+}^n} \tilde{\sigma}'_{t_{i+}^n} \Delta_j^nB'}{\sqrt{\Delta_n}}\right)^4 \right]\\
		&(\text{Since $\sigma, \tilde{\sigma}, \tilde{\sigma}', \gamma$ and the cosine function are bounded. })\\
		&\leq  \frac{C}{k_n} \rightarrow 0. 
	\end{align*}
	The result \eqref{xi-23-W} follows from
	\begin{align*}
		&\sum_{j \in I^n_{i+} } \E_{j-1}^n \left[ \left(  \xi_{t_{i+}^n,n}^{(2,j)}(u) \right)\cdot \left(W_{t_{i+k_n}^n} - W_{t_{i}^n} + B_{t_{i+k_n}^n} - B_{t_{i}^n} + B'_{t_{i+k_n}^n} - B'_{t_{i}^n} \right) \right] \\
		& = \sum_{j \in I^n_{i+} } \E_{j-1}^n \left[ \left(  \xi_{t_{i+}^n,n}^{(2,j)}(u) \right)\cdot \left(\Delta_j^nW+ \Delta_j^n B + \Delta_j^n B'\right) \right] \\
		&= \sum_{j \in I^n_{i+} } \frac{-2}{\sqrt{k_n} u^2 f_{t_{i+}^n,n}(u)}  \left( \E_{j-1}^{n} \left[ \cos \left(\frac{u(\sigma_{t_{j-1}^n} \Delta_j^n B )}{\sqrt{\Delta_n}} \right) \cos  \left(\frac{u( \gamma_{t_{j-1}^n} \Delta_j^n L)}{\sqrt{\Delta_n}} \right) \cdot \left(\Delta_j^nW+ \Delta_j^n B + \Delta_j^n B'\right)\right] \right) \\
		& \quad + \sum_{j \in I^n_{i+} } \frac{2}{\sqrt{k_n} u^2 f_{t_{i+}^n,n}(u)}  \left( \E_{j-1}^{n} \left[  \sin  \left(\frac{u(\sigma_{t_{j-1}^n} \Delta_j^n B )}{\sqrt{\Delta_n}} \right) \sin  \left(\frac{u( \gamma_{t_{j-1}^n} \Delta_j^n L)}{\sqrt{\Delta_n}} \right) \cdot \left(\Delta_j^nW+ \Delta_j^n B + \Delta_j^n B'\right)\right] \right)\\
		&(\text{Since $B, B', L$ are mutually independent, their distributions are symmetric around 0,}\\
		&\text{and the functions $\sin(x)$ and $x\cos{x}$ are odd.)}\\
		&= \sum_{j \in I^n_{i+} } \frac{-2}{\sqrt{k_n} u^2 f_{t_{i+}^n,n}(u)}  \left( \E_{j-1}^{n} \left[ \cos \left(\frac{u(\sigma_{t_{j-1}^n} \Delta_j^n B )}{\sqrt{\Delta_n}} \right) \cos  \left(\frac{u( \gamma_{t_{j-1}^n} \Delta_j^n L)}{\sqrt{\Delta_n}} \right) \Delta_j^nW\right] \right) \\
		&(\text{By Holder's inequality and the boundedness of $\sigma, \gamma$.})\\
		&\leq C\frac{\sqrt{k_n}}{u^2} \left(  \E_{j-1}^{n} \left[ \left(\cos \left(\frac{u(\sigma_{t_{j-1}^n} \Delta_j^n B )}{\sqrt{\Delta_n}} \right) \right)^2 \right] \cdot \E_{j-1}^n \left[ \left( \cos  \left(\frac{u( \gamma_{t_{j-1}^n} \Delta_j^n L)}{\sqrt{\Delta_n}} \right) \Delta_j^nW \right)^2 \right] \right)^{1/2}\\
		&\leq C\sqrt{k_n\Delta_n} \longrightarrow^p 0, 
	\end{align*}
	and 
	\begin{align*}
		&\sum_{j \in I^n_{i+} } \E_{j-1}^n \left[ \left(  \xi_{t_{i+}^n,n}^{(3,j)}(u) \right)\cdot \left(W_{t_{i+k_n}^n} - W_{t_{i}^n} + B_{t_{i+k_n}^n} - B_{t_{i}^n} + B'_{t_{i+k_n}^n} - B'_{t_{i}^n} \right) \right] \\
		&= \sum_{j \in I^n_{i+} } \E_{j-1}^n \left[ \left(  \xi_{t_{i+}^n,n}^{(3,j)}(u) \right)\cdot \left(\Delta_j^nW+ \Delta_j^n B + \Delta_j^n B'\right) \right]\\	
		&=\sum_{j \in I^n_{i+} } \frac{ (i+k_n+1-j)  }{k_n \sqrt{k_n\Delta_n} } \cdot \E_{j-1}^n \left[\left(2\sigma_{t_{i+}^n} \tilde{\sigma}_{t_{i+}^n}\Delta_j^nB + 2\sigma_{t_{i+}^n} \tilde{\sigma}'_{t_{i+}^n} \Delta_j^nB'\right)   \left(\Delta_j^nW+ \Delta_j^n B + \Delta_j^n B'\right) \right]\\
		&= \sum_{j \in I^n_{i+} } \frac{ (i+k_n+1-j) \sqrt{\Delta_n} }{k_n \sqrt{k_n} }  \left(2\sigma_{t_{i+}^n} \tilde{\sigma}_{t_{i+}^n} + 2\sigma_{t_{i+}^n} \tilde{\sigma}'_{t_{i+}^n} \right)\leq C\sqrt{k_n\Delta_n} \rightarrow 0.
	\end{align*}
	This ends the proof of \eqref{xi-23}. \\
	For \eqref{xi^2}, since $\sigma, \gamma$ and the cosine function are bounded, together with the proof of \eqref{xi-1}, we have 
	\begin{align*}
		&\left(\sqrt{k_n} \wedge \frac{1}{\sqrt{k_n\Delta_n}}\right) \E \left[ \left|	\xi_{t_{i+}^n,n}^{(1)}(u)\right|^2\right] \leq C\E \left[ \left| \left(\sqrt{k_n} \wedge \frac{1}{\sqrt{k_n\Delta_n}}\right) 	\xi_{t_{i+}^n,n}^{(1)}(u)\right|\right]  \longrightarrow^p 0,
	\end{align*}
	and 
	\begin{align*}
		&\left(\sqrt{k_n} \wedge \frac{1}{\sqrt{k_n\Delta_n}}\right) \left( \E \left[ \left|	\xi_{t_{i+}^n,n}^{(2)}(u)\right|^2\right] + \E \left[ \left|	\xi_{t_{i+}^n,n}^{(3)}(u)\right|^2\right]  \right)\\
		& = \left( \frac{1}{k_n\sqrt{\Delta_n}}  \wedge 1 \right) \frac{1}{\sqrt{k_n}} \E \left[ \left| \sqrt{k_n}	\xi_{t_{i+}^n,n}^{(2)}(u)\right|^2 \right] +  (k_n\sqrt{\Delta_n} \wedge 1)\sqrt{k_n\Delta_n}  \E \left[ \frac{1}{\sqrt{k_n\Delta_n}} \left|	\xi_{t_{i+}^n,n}^{(3)}(u)\right|^2 \right] \\
		&(\text{From the proof of \eqref{xi-23} and since $\sigma, \tilde{\sigma}, \tilde{\sigma}'$ are bounded.})\\
		&\leq C\left(\frac{1}{\sqrt{k_n}} + \sqrt{k_n\Delta_n}\right) \rightarrow0.
	\end{align*}
	This completes the proof of Proposition \ref{pro-1}. \hfill $\square$
	
	\subsection{Proof for Theorem \ref{thm-lev-1}}\label{prof:thm1}
	For convenience in this part, we define, for $i=k_n+1,...,n-k_n$,
	\begin{align*}
		s_i^n
		&= \Delta_i^n X \cdot \left( \left( \widehat{\sigma}^2_{t_{i+}^n}  -(\sigma^2_{t_{i+}^n} + b_{ t_{i+}^n,n}(u) ) \right) - \left( \widehat{\sigma}^2_{t_{i-}^n}  -(\sigma^2_{t_{i-}^n} + b_{ t_{i-}^n,n}(u) ) \right) \right),
	\end{align*}
	and, for $i<j\leq i+2k_n$ with fixed $i$,
	\begin{align}\label{xi-2-3}
		\begin{split}
			&\xi_{t_{i+}^n,n}^{(2,j)}(u)  = \frac{  -2 \left( \cos  \left(\frac{u\Delta_j^nX'}{\sqrt{\Delta_n}} \right) - \E_{j-1}^{n} \left[ \cos \left(\frac{u\Delta_j^nX'}{\sqrt{\Delta_n}} \right) \right] \right) }{\sqrt{k_n} u^2 f_{t_{i+}^n,n}(u)},  \\
			&\xi_{t_{i+}^n,n}^{(3,j)}(u) =  \frac{ (i+k_n-j)\left(2\sigma_{t_{i+}^n} \tilde{\sigma}_{t_{i+}^n}\Delta_j^nB + 2\sigma_{t_{i+}^n} \tilde{\sigma}'_{t_{i+}^n} \Delta_j^nB'\right)    }{k_n \sqrt{k_n\Delta_n} }, \\
			&\xi_{t_{i-}^n,n}^{(2,j)}(u)  = \frac{  -2 \left( \cos  \left(\frac{u\Delta_j^nX'}{\sqrt{\Delta_n}} \right) - \E_{j-1}^{n} \left[ \cos \left(\frac{u\Delta_j^nX'}{\sqrt{\Delta_n}} \right) \right] \right) }{\sqrt{k_n} u^2 f_{t_{(i-k_n-1)}^n,n}(u)},  \\
			&\xi_{t_{i-}^n,n}^{(3,j)}(u) =  \frac{ (i-1-j)\left(2\sigma_{t_{(i-k_n-1)}^n} \tilde{\sigma}_{t_{(i-k_n-1)}^n}\Delta_j^nB + 2\sigma_{t_{(i-k_n-1)}^n} \tilde{\sigma}'_{t_{(i-k_n-1)}^n} \Delta_j^nB'\right)    }{k_n \sqrt{k_n\Delta_n} },
		\end{split}
	\end{align}
	and 
	\begin{align}
		\begin{split}
			&\xi_{t_{i\pm}^n,n}'(u) = \sqrt{n}^{b\wedge (1-b)} \left( \left(\widehat{\sigma}^2_{t_{i\pm}^n} - \sigma^2_{t_{i\pm}^n} - b_{ t_{i\pm}^n,n}(u) \right) - \frac{1}{\sqrt{k_n}} \sum_{j \in I^n_{i\pm} } \xi_{t_{i\pm}^n,n}^{(2,j)}(u) - \sqrt{k_n\Delta_n} \sum_{j \in I^n_{i\pm} } \xi_{t_{i\pm}^n,n}^{(3,j)}(u) \right). 
		\end{split}
	\end{align}
	Observe that $\widehat{\sigma}^2_{t_{i-}^n} = \widehat{\sigma}^2_{t_{(i-k_n-1)+}^n}$. From the proof of Proposition \ref{pro-1}, we have
	\begin{align}\label{spot-main}
		\begin{split}
			&\left( \sqrt{k_n} \wedge \frac{1}{\sqrt{k_n\Delta_n}}\right) \left(\widehat{\sigma}^2_{t_{i\pm}^n} -\sigma^2_{t_{i\pm}^n} - b_{ t_{\pm+}^n,n}(u) \right) \\
			&=   \left( \frac{1}{k_n\sqrt{\Delta_n}}  \wedge 1 \right) \sum_{j \in I^n_{i\pm} } \xi_{t_{i\pm}^n,n}^{(2,j)}(u) +  (k_n\sqrt{\Delta_n} \wedge 1) \sum_{j \in I^n_{i\pm} } \xi_{t_{i\pm}^n,n}^{(3,j)}(u) + O_p\left( h_n \right),
		\end{split}
	\end{align}
	with 
	\begin{align}\label{h_n}
		h_n = \left( \frac{\Delta_n^{ \frac{1}{4} \wedge \left(\frac{5}{4(\beta+\epsilon)}-\frac{1}{2}\right) \wedge \left(\frac{1}{(r+\epsilon)}-\frac{3}{4}\right) }}{u} + \sqrt{k_n\Delta_n} + \frac{1}{\sqrt{k_n}}+  \frac{ u^{\beta} \Delta_n^{1-\frac{\beta}{2} } (k_n\Delta_n)^{1/2}}{\sqrt{k_n\Delta_n}} \right).
	\end{align}
	Following the proofs of Lemmas 2 and 5 in \cite{LLL2018}, and applying Taylor's expansion together with the boundedness of $\sigma$ and the cosine function, we obtain, for $i,j=1,...,n$ and $k=1,2,\cdots$,
	\begin{align}\label{X-main}
		\begin{split}
			&\Delta_{j}^{n}X = \Delta_{j}^{n}X' + O_p\left( \Delta_n^{1 \wedge \left(\frac{5}{4(\beta+\epsilon)} + \frac{1}{4} \right)\wedge \frac{1}{(r+\epsilon)} }\right), \\
			&\Delta_{j}^{n}X = \sigma_{t_{j-1}^n} \Delta_{j}^{n}B + O_p\left(\Delta_n^{1\wedge \frac{1}{(\max\{\beta,r\}+\epsilon)}} \right) = \sigma_{t_{i}^n} \Delta_{j}^{n}B + O_p\left( \sqrt{|i-j+1|} \Delta_n+\Delta_n^{1\wedge \frac{1}{(\max\{\beta,r\}+\epsilon)}} \right),\\
			&\frac{\Delta_{j}^{n}X}{\sqrt{\Delta_n}} = \sigma_{t_{j-1}^n} \frac{\Delta_{j}^{n}B}{\sqrt{\Delta_n}} + O_p\left(\Delta_n^{(1 \wedge \frac{1}{(\max\{\beta,r\}+\epsilon)} )-\frac{1}{2}} \right), \\
			&\cos \left(\frac{\Delta_{j}^{n}X}{\sqrt{\Delta_n}} \right)= \cos \left(\sigma_{t_{j-1}^n} \frac{\Delta_{j}^{n}B}{\sqrt{\Delta_n}} \right)+ O_p\left(\Delta_n^{(1 \wedge \frac{1}{(\max\{\beta,r\}+\epsilon)} )-\frac{1}{2}} \right), \\
			&\frac{1}{k} \left( \cos \left(\frac{\Delta_{j}^{n}X}{\sqrt{\Delta_n}} \right) \right)^{k} = \frac{1}{k} \left( \cos \left(\sigma_{t_{j-1}^n} \frac{\Delta_{j}^{n}B}{\sqrt{\Delta_n}} \right) \right)^k + O_p\left(\Delta_n^{(1 \wedge \frac{1}{(\max\{\beta,r\}+\epsilon)} )-\frac{1}{2} } \right).
		\end{split}
	\end{align}
	
	Before proving Theorem \ref{thm-lev-1}, we provide the following lemmas and their proofs. 
	\begin{lem}\label{lem-vov-approx}
		Under Assumption \ref{asu-con} and \ref{asu-cha}, for any $i=k_n,...,n-k_n$ and $|j-i|\leq Ck_n$, we have
		\begin{align}
			f^{(1)}_{t_{j}^n,n}(u) - f^{(1)}_{t_{i}^n,n}(u) = O_p\left((k_n\Delta_n)^{\frac{3}{4}}\right),\\\label{lem3-res2}
			h_1(u,t_{j}^n)  - h_1(u,t_{i}^n)  = O_p\left((k_n\Delta_n)^{\frac{3}{4}}\right),\\  \label{lem3-res3}
			h_2(t_{j}^n) - h_2(t_{i}^n) =  O_p(\sqrt{k_n\Delta_n}).
		\end{align}
	\end{lem}
	\textbf{Proof of Lemma \ref{lem-vov-approx}:} Applying It$\hat{\text{o}}$'s lemma to \eqref{model-volvol}, we have 
	\begin{align*}
		&d \exp{\left(\frac{- u^2(\sigma^2_{s} - \sigma^2_{t_i^n} )}{2} \right)}\\
		&= \left(\frac{-u^2}{2}\exp{\left(\frac{- u^2( \sigma^2_{s} - \sigma^2_{t_{i}^n})}{2}\right)} \left(2\sigma_s\tilde{b}_s+(\tilde{\sigma}_s)^2 + (\tilde{\sigma}'_s)^2\right)  + \frac{u^4}{2} \exp{\left(\frac{- u^2( \sigma^2_{s} - \sigma^2_{t_{i}^n})}{2}\right)} ((\tilde{\sigma}_s)^2 + (\tilde{\sigma}'_s)^2) \sigma_s^2 \right)ds \\
		&\qquad -u^2\exp{\left(\frac{- u^2( \sigma^2_{s} - \sigma^2_{t_{i}^n})}{2}\right)} \sigma_s\tilde{\sigma}_sdB_s -u^2\exp{\left(\frac{- u^2( \sigma^2_{s} - \sigma^2_{t_{i}^n})}{2}\right)}  \sigma_s\tilde{\sigma}'_sdB'_s.
	\end{align*}
	Since $\tilde{b}, \sigma, \tilde{\sigma}, \tilde{\sigma}'$ are bounded, condition \eqref{cond-vov} and It$\hat{\text{o}}$'s isometry yield 
	\begin{align*}
		f^{(1)}_{t_{j}^n,n}(u) - f^{(1)}_{t_{i}^n,n}(u) &= \exp{\left(\frac{- u^2\sigma^2_{t_{i}^n}}{2}\right)} \left( \exp{\left(\frac{- u^2( \sigma^2_{t_{j}^n} - \sigma^2_{t_{i}^n})}{2}\right)} - 1\right)\\
		&= \exp{\left(\frac{- u^2\sigma^2_{t_{i}^n}}{2}\right)} \cdot \int_{t_{i}^n}^{t_{j}^n} d \exp{\left(\frac{- u^2(\sigma^2_{s} - \sigma^2_{t_i^n} )}{2} \right)} \\
		&= O_p\left((k_n\Delta_n)^{\frac{3}{4}}\right). 
	\end{align*}
	Furthermore, \eqref{lem3-res2} follows directly from the above conclusion.
	The result in \eqref{lem3-res3} follows directly from the boundedness of $\sigma, \tilde{\sigma}, \tilde{\sigma}'$ and condition \eqref{cond-vov}.
	\hfill $\square$
	
	\begin{lem}\label{lem-lev-1}
		Under the same assumptions and conditions as in Proposition \ref{pro-1}, 
		we have, for $i=k_n+1,...,n-k_n$, 
		\begin{align}\label{lem-lev-1-res1}
			\begin{split}
				\sqrt{k_n}\left( \widehat{\sigma}^2_{t_{i+}^n} -\sigma^2_{t_{i+}^n} - b_{ t_{i+}^n,n}(u), \widehat{\sigma}^2_{t_{i-}^n} -\sigma^2_{t_{i-}^n} - b_{ t_{i-}^n,n}(u) \right) &\longrightarrow^{L_s} \left( V_{i+}, V_{i-}\right), \qquad \text{if} \quad  \kappa = 0,\\
				\sqrt{k_n}\left( \widehat{\sigma}^2_{t_{i+}^n} -\sigma^2_{t_{i+}^n} - b_{ t_{i+}^n,n}(u), \widehat{\sigma}^2_{t_{i-}^n} -\sigma^2_{t_{i-}^n} - b_{ t_{i-}^n,n}(u) \right) &\longrightarrow^{L_s} \left(V_{i+} + \kappa V'_{i+}, V_{i-} + \kappa V'_{i-} \right),\\
				& \qquad \qquad \qquad \qquad   \text{if} \quad 0< \kappa < \infty,\\
				\frac{1}{\sqrt{k_n\Delta_n}}\left( \widehat{\sigma}^2_{t_{i+}^n} -\sigma^2_{t_{i+}^n} - b_{ t_{i+}^n,n}(u), \widehat{\sigma}^2_{t_{i-}^n} -\sigma^2_{t_{i-}^n} - b_{ t_{i-}^n,n}(u) \right) &\longrightarrow^{L_s} \left(V'_{i+},V'_{i-} \right) \qquad \text{if} \quad  \kappa = \infty,
			\end{split}
		\end{align}
		where $(V_{i+}, V'_{i+},V_{i-}, V'_{i-})$ is a vector of normal random variables defined on an extension of the probability space $(\Omega,\mathcal{F},\mathcal{F}_{0\leq t\leq T},\mathbb{P})$. Moreover, conditionally on the $\sigma$-field $\mathcal{F}$, it has zero mean, $\mathcal{F}$-conditional covariance $Cov(V_{i\pm},V'_{i\pm}|\mathcal{F})= \E[V_{i\pm}V'_{i\pm}|\mathcal{F}] = 0$, and $\mathcal{F}$-conditional variance 
		\begin{align}\label{pro1-var}
			\begin{split}
				&Var(V_{i\pm}|\mathcal{F}) = \E[(V_{i\pm})^2|\mathcal{F}] =  \frac{2(\exp{(-2 u^2\sigma^2_{t_{i\pm}^n} )} - 2\exp{(- u^2\sigma^2_{t_{i\pm}^n} )} + 1)}{u^4\exp{(- u^2\sigma^2_{t_{i\pm}^n} )}}, \\
				& Var(V'_{i\pm}|\mathcal{F}) = \E[(V'_{i\pm})^2|\mathcal{F}] =  \frac{4(\sigma_{t_{i\pm}^n} )^2((\tilde{\sigma}_{t_{i\pm}^n})^2+(\tilde{\sigma}'_{t_{i\pm}^n})^2)}{3}.
			\end{split}
		\end{align}
	\end{lem}
	\textbf{Proof of Lemma \ref{lem-lev-1}:} 
	By \eqref{spot-main}, conclusion \eqref{lem-lev-1-res1} can equivalently be proved by showing 
	\begin{align}\label{lem1-key}
		\sum_{j \in I^n_{i-} \cup I^n_{i+} } \left( \xi_{t_{i-}^n,n}^{(2,j)}(u), \xi_{t_{i-}^n,n}^{(3,j)}(u), \xi_{t_{i+}^n,n}^{(2,j)}(u),  \xi_{t_{i+}^n,n}^{(3,j)}(u) \right) \longrightarrow^{L_s} \left(V_{i-},V'_{i-}, V_{i+},V'_{i+}\right),
	\end{align}
	where we further define, for $w=2,3$, $\xi_{t_{i-}^n,n}^{(w,j)}(u) = 0$ if $j \in  I^n_{i+}$, and $\xi_{t_{i+}^n,n}^{(w,j)}(u) = 0$ if $j \in  I^n_{i-}$. 
	Noting that $\left\{\left( \xi_{t_{i-}^n,n}^{(2,j)}(u), \xi_{t_{i-}^n,n}^{(3,j)}(u), \xi_{t_{i+}^n,n}^{(2,j)}(u),  \xi_{t_{i+}^n,n}^{(3,j)}(u) \right), \mathcal{F}_{t_j^n}\right\}$ is a martingale difference array, by Theorem 2.2.15 in \cite{JP2012}, it suffices to show
	\begin{align}\label{lem-lev-1-1}
		&\sum_{j \in I^n_{i-} \cup I^n_{i+} } \E_{j-1}^n \left[ \left(\xi_{t_{i\pm}^n,n}^{(2,j)}(u) \right)^2 \right] \longrightarrow^{p}  \frac{2(\exp{(-2 u^2\sigma^2_{t_{i\pm}^n} )} - 2\exp{(- u^2\sigma^2_{t_{i\pm}^n} )} + 1)}{u^4\exp{(- u^2\sigma^2_{t_{i\pm}^n} )}}, \\\label{lem-lev-1-2}
		&\sum_{j \in I^n_{i-} \cup I^n_{i+} } \E_{j-1}^n \left[ \left(\xi_{t_{i\pm}^n,n}^{(3,j)}(u) \right)^2 \right] \longrightarrow^{p} \frac{4(\sigma_{t_{i\pm}^n} )^2((\tilde{\sigma}_{t_{i\pm}^n})^2+(\tilde{\sigma}'_{t_{i\pm}^n})^2)}{3}, \\\label{lem-lev-1-3}
		&\sum_{j \in I^n_{i-} \cup I^n_{i+} } \E_{j-1}^n \left[ \xi_{t_{i\pm}^n,n}^{(2,j)}(u) \cdot\xi_{t_{i\pm}^n,n}^{(3,j)}(u)  +  \xi_{t_{i-}^n,n}^{(2,j)}(u) \cdot\xi_{t_{i+}^n,n}^{(2,j)}(u) +  \xi_{t_{i-}^n,n}^{(3,j)}(u) \cdot\xi_{t_{i+}^n,n}^{(3,j)}(u) \right] \longrightarrow^{p} 0 ,\\ \label{lem-lev-1-5}
		&\sum_{j \in I^n_{i-} \cup I^n_{i+} } \E_{j-1}^n \left[ \left( \xi_{t_{i-}^n,n}^{(2,j)}(u)\right)^4+ \left(\xi_{t_{i-}^n,n}^{(3,j)}(u) \right)^4+ \left( \xi_{t_{i+}^n,n}^{(2,j)}(u)\right)^4+ \left(\xi_{t_{i+}^n,n}^{(3,j)}(u) \right)^4\right] \longrightarrow^{p} 0,
	\end{align}
	and 
	\begin{align}\label{lem-lev-1-4}
		\begin{split}
			&\sum_{j \in I^n_{i-} \cup I^n_{i+}  } \E_{j-1}^n \Bigg[ \left(  \xi_{t_{i-}^n,n}^{(2,j)}(u) + \xi_{t_{i-}^n,n}^{(3,j)}(u) + \xi_{t_{i+}^n,n}^{(2,j)}(u) + \xi_{t_{i+}^n,n}^{(3,j)}(u)\right) \\
			&\qquad \qquad \qquad \cdot \left(W_{t_{i+k_n}^n} - W_{t_{i-k_n-1}^n} + B_{t_{i+k_n}^n} - B_{t_{i-k_n-1}^n} + B'_{t_{i+k_n}^n} - B'_{t_{i-k_n-1}^n} \right) \Bigg]\longrightarrow^{p} 0,
		\end{split} 
	\end{align}
	where $W$ is a bounded martingale orthogonal to both $B$ and $B'$.
	For \eqref{lem-lev-1-1} and \eqref{lem-lev-1-2}, from the proof of Proposition \ref{pro-1}, we have
	\begin{align*}
		\sum_{j \in I^n_{i-} \cup I^n_{i+} } \E_{j-1}^n \left[ \left(\xi_{t_{i-}^n,n}^{(2,j)}(u) \right)^2 \right] 
		&\longrightarrow^p  \frac{2(\exp{(-2 u^2\sigma^2_{t_{i-}^n} )} - 2\exp{(- u^2\sigma^2_{t_{i-}^n} )} + 1)}{u^4\exp{(- u^2\sigma^2_{t_{i-}^n} )}}, \\
		\sum_{j \in I^n_{i-} \cup I^n_{i+} } \E_{j-1}^n \left[ \left(\xi_{t_{i-}^n,n}^{(3,j)}(u) \right)^2 \right]
		&\longrightarrow^{p}  \frac{4(\sigma_{t_{i-}^n} )^2((\tilde{\sigma}_{t_{i-}^n})^2+(\tilde{\sigma}'_{t_{i-}^n})^2)}{3}, 
	\end{align*}
	since Taylor's expansion, the boundedness of $\sigma$, and condition \eqref{cond-vov} yield
	\begin{align*}
		& \frac{2(\exp{(-2 u^2\sigma^2_{t_{(i-k_n-1)+}^n} )} - 2\exp{(- u^2\sigma^2_{t_{(i-k_n-1)+}^n} )} + 1)}{u^4\exp{(- u^2\sigma^2_{t_{(i-k_n-1)+}^n} )}} \\
		& = \frac{2(\exp{(-2 u^2\sigma^2_{t_{i-}^n} )} - 2\exp{(- u^2\sigma^2_{t_{i-}^n} )} + 1)}{u^4\exp{(- u^2\sigma^2_{t_{i-}^n} )}} + O_p(u^2\sqrt{k_n\Delta_n}),
	\end{align*}
	and 
	\begin{align*}
		& \frac{4(\sigma_{t_{(i-k_n-1)+}^n} )^2((\tilde{\sigma}_{t_{(i-k_n-1)+}^n})^2+(\tilde{\sigma}'_{t_{(i-k_n-1)+}^n})^2)}{3} = \frac{4(\sigma_{t_{i-}^n} )^2((\tilde{\sigma}_{t_{i-}^n})^2+(\tilde{\sigma}'_{t_{i-}^n})^2)}{3} + O_p(\sqrt{k_n\Delta_n}).
	\end{align*}
	The results for $\xi_{t_{i+}^n,n}^{(2,j)}(u)$ and $\xi_{t_{i+}^n,n}^{(3,j)}(u)$ can be found in Proposition \ref{pro-1}.
	For \eqref{lem-lev-1-3}, it is clear that $\E_{j-1}^n \left[ \xi_{t_{i+}^n,n}^{(w_1,j)}(u) \cdot\xi_{t_{i-}^n,n}^{(w_2,j)}(u) \right] \equiv 0$ for all $w_1, w_2 = \{2,3\}$ and $j \in I^n_{i-} \cup I^n_{i+}$. The results $\sum_{j \in I^n_{i-} \cup I^n_{i+} } \E_{j-1}^n \left[ \xi_{t_{i\pm}^n,n}^{(2,j)}(u) \cdot\xi_{t_{i\pm}^n,n}^{(3,j)}(u) \right] \longrightarrow^p 0 $, \eqref{lem-lev-1-4}, and \eqref{lem-lev-1-5} follow from the proof of Proposition \ref{pro-1}.
	
	This completes the proof of Lemma \ref{lem-lev-1}.       \hfill $\square$
	
	\begin{lem}\label{lem-lev-2}
		Under the same assumptions and conditions as in Theorem \ref{thm-lev-1}, for $i=k_n+1,...,n-k_n$, we have
		\begin{align}\label{lem-lev-2-1}
			\E_{i-k_n-1}^n \left[\left( \sqrt{n}^{b\wedge (1-b)} s^n_{i}  \right) \right] = O_p(\Delta_n^{1\wedge \frac{1}{(\max\{\beta,r\}+\epsilon)}} \cdot h_n).
		\end{align}
		For $j=k_n+1,...,n-k_n$ with $i<j$, if $j-i > 2k_n+1$, we have  
		\begin{align}\label{lem-lev-2-2}
			\E_{i-k_n-1}^n \left[ \left( \sqrt{n}^{b\wedge (1-b)} s^n_{i} \cdot \sqrt{n}^{b\wedge (1-b)} s^n_{j} \right) \right]= O_p(\Delta_n^{2 \wedge \frac{2}{(\max\{\beta,r\}+\epsilon)}}h_n^2),
		\end{align}
		if $j-i\leq 2k_n+1$, we have
		\begin{align}\label{lem-lev-2-3}
			\E_{i-k_n-1}^n \left[\left( \sqrt{n}^{b\wedge (1-b)} s^n_{i} \cdot \sqrt{n}^{b\wedge (1-b)} s^n_{j} \right)\right]= O_p(\Delta_n^{2 \wedge \frac{2}{(\max\{\beta,r\}+\epsilon)}}).
		\end{align}
	\end{lem}
	\textbf{Proof of Lemma \ref{lem-lev-2}:}  For \eqref{lem-lev-2-1}, we have 
	\begin{align*}
		&\E_{i-k_n-1}^n \left[ \sqrt{n}^{b\wedge (1-b)} s^n_{i} \right] \\
		&= \E_{i-k_n-1}^n \Bigg[ \Delta_{i+}^n X \cdot \left(  \sum_{j \in I^n_{i+} } \E_{j-1} \left[ \frac{1}{\sqrt{\kappa}}  \sqrt{n}^{0\wedge (1-2b)}  \xi_{t_{i+}^n,n}^{(2,j)}(u) + \sqrt{\kappa T} \sqrt{n}^{0\wedge (2b-1)}  \xi_{t_{i+}^n,n}^{(3,j)}(u) \right] \right) \\
		&\qquad \qquad \qquad \quad + \E\left[ \Delta_{i+}^nB \bigg| \left( \sigma_{t_{i+}^n} , \xi_{t_{i+}^n,n}'(u) \right) \right] \cdot \sigma_{t_{i+}^n}  \xi_{t_{i+}^n,n}'(u) \\ 
		&\qquad \qquad \qquad \quad + \left( \Delta_{i+}^n X - \sigma_{t_{i+}^n} \Delta_{i+}^nB  \right) \cdot \xi_{t_{i+}^n,n}'(u) \Bigg] \\ 
		&\quad - \E_{i-k_n-1}^n \Bigg[  \left(  \sqrt{n}^{b\wedge (1-b)} \left(\widehat{\sigma}^2_{t_{i-}^n} -\sigma^2_{t_{i-}^n} - b_{ t_{i-}^n,n}(u) \right)  \right) \cdot \sigma_{t_{i-}^n}  \E_{i-}^n \left[ \Delta_{i}^n B \right]\Bigg]\\
		&\quad - \E_{i-k_n-1}^n \Bigg[ \E \Bigg[ \left(  \sum_{j \in I^n_{i-} } \E_{j-1} \left[ \frac{1}{\sqrt{\kappa}}  \sqrt{n}^{0\wedge (1-2b)}  \xi_{t_{i-}^n,n}^{(2,j)}(u) + \sqrt{\kappa T} \sqrt{n}^{0\wedge (2b-1)}  \xi_{t_{i-}^n,n}^{(3,j)}(u) \right] \right)\\
		&\qquad \qquad \qquad \quad  \cdot  \Bigg|  \left( \Delta_{i}^n X, \sigma_{t_{i-}^n}, \Delta_{i}^nB  \right) \Bigg]  \left( \Delta_{i}^n X - \sigma_{t_{i-}^n} \Delta_{i}^nB  \right) \Bigg]\\
		&\quad - \E_{i-k_n-1}^n \Bigg[\left(  \xi_{t_{i-}^n,n}'(u)  \right) \cdot  \left( \Delta_{i}^n X - \sigma_{t_{i-}^n} \Delta_{i}^nB  \right) \Bigg]\\
		&(\text{By \eqref{spot-main} and  \eqref{X-main}.})\\
		& = O_p\left( \Delta_n^{1\wedge \frac{1}{(\max\{\beta,r\}+\epsilon)}} \cdot h_n \right).
	\end{align*}
	If $j-i>2k_n+1$, then there are no overlapping terms between $s^n_{i}$ and $s^n_{j}$, and successive conditioning gives
	\begin{align*}
		&\E_{i-k_n-1}^n \left[ \left( \sqrt{n}^{b\wedge (1-b)} s^n_{i} \cdot \sqrt{n}^{b\wedge (1-b)} s^n_j \right)\right]\\
		&= \E_{i-k_n-1}^n \left[ \sqrt{n}^{b\wedge (1-b)} s^n_{i} \cdot  \E\left[\sqrt{n}^{b\wedge (1-b)} s^n_j \Big| s_i^n \right] \right]\\
		&\text{(Following the proof of \eqref{lem-lev-2-1}.)}\\
		&= \E_{i-k_n-1}^n \left[ \sqrt{n}^{b\wedge (1-b)} s^n_{i} \cdot  \E\left[ \left( \left( \Delta_{j+}^n X - \sigma_{t_{j+}^n} \Delta_{j+}^nB  \right) \cdot \xi_{t_{j+}^n,n}'(u) -  \xi_{t_{j-}^n,n}'(u) \cdot  \left( \Delta_{j}^n X - \sigma_{t_{j-}^n} \Delta_{j}^nB  \right) \right) \Big| s_j^n \right] \right]\\
		& = \E_{i-k_n-1}^n \left[ \sqrt{n}^{b\wedge (1-b)} s^n_{i} \cdot  \left( \left( \Delta_{j+}^n X - \sigma_{t_{j+}^n} \Delta_{j+}^nB  \right) \cdot \xi_{t_{j+}^n,n}'(u) -  \xi_{t_{j-}^n,n}'(u) \cdot  \left( \Delta_{j}^n X - \sigma_{t_{j-}^n} \Delta_{j}^nB  \right) \right)  \right]\\
		&= \E_{i-k_n-1}^n \Big[ \E \left[\sqrt{n}^{b\wedge (1-b)} s^n_{i} \Big|\left( \Delta_{j+}^n X, \sigma_{t_{j+}^n}, \Delta_{j+}^nB, \xi_{t_{j+}^n,n}'(u), \xi_{t_{j-}^n,n}'(u), \Delta_{j}^n X, \sigma_{t_{j-}^n}, \Delta_{j}^nB \right) \right] \\
		& \quad \qquad \qquad \cdot  \left( \left( \Delta_{j+}^n X - \sigma_{t_{j+}^n} \Delta_{j+}^nB  \right) \cdot \xi_{t_{j+}^n,n}'(u) -  \xi_{t_{j-}^n,n}'(u) \cdot  \left( \Delta_{j}^n X - \sigma_{t_{j-}^n} \Delta_{j}^nB  \right) \right)  \Big]\\
		&\text{(Following the proof of \eqref{lem-lev-2-1}.)}\\
		&=  \E_{i-k_n-1}^n \Big[ \E \Big[ \left( \left( \Delta_{i+}^n X - \sigma_{t_{i+}^n} \Delta_{i+}^nB  \right) \cdot \xi_{t_{i+}^n,n}'(u) -  \xi_{t_{i-}^n,n}'(u) \cdot  \left( \Delta_{i}^n X - \sigma_{t_{i-}^n} \Delta_{i}^nB  \right) \right) \\
		&\qquad \qquad \qquad \ \ \Big|\left( \Delta_{j+}^n X, \sigma_{t_{j+}^n}, \Delta_{j+}^nB, \xi_{t_{j+}^n,n}'(u), \xi_{t_{j-}^n,n}'(u), \Delta_{j}^n X, \sigma_{t_{j-}^n}, \Delta_{j}^nB \right) \Big] \\
		& \qquad  \qquad \quad \cdot  \left( \left( \Delta_{j+}^n X - \sigma_{t_{j+}^n} \Delta_{j+}^nB  \right) \cdot \xi_{t_{j+}^n,n}'(u) -  \xi_{t_{j-}^n,n}'(u) \cdot  \left( \Delta_{j}^n X - \sigma_{t_{j-}^n} \Delta_{j}^nB  \right) \right)  \Big]\\
		&=  \E_{i-k_n-1}^n \Big[ \left( \left( \Delta_{i+}^n X - \sigma_{t_{i+}^n} \Delta_{i+}^nB  \right) \cdot \xi_{t_{i+}^n,n}'(u) -  \xi_{t_{i-}^n,n}'(u) \cdot  \left( \Delta_{i}^n X - \sigma_{t_{i-}^n} \Delta_{i}^nB  \right) \right) \\
		& \qquad  \qquad \quad \cdot  \left( \left( \Delta_{j+}^n X - \sigma_{t_{j+}^n} \Delta_{j+}^nB  \right) \cdot \xi_{t_{j+}^n,n}'(u) -  \xi_{t_{j-}^n,n}'(u) \cdot  \left( \Delta_{j}^n X - \sigma_{t_{j-}^n} \Delta_{j}^nB  \right) \right)  \Big]\\
		&(\text{By \eqref{spot-main} and  \eqref{X-main}.})\\
		&= O_p(\Delta_n^{2 \wedge \frac{2}{(\max\{\beta,r\}+\epsilon)}}h_n^2). 
	\end{align*}
	For $j-i \leq 2k_n+1$, we consider the special case $j=i+2$ for ease of presentation. Note that 
	\begin{align*}
		&\E_{i-k_n-1}^n \left[ \left( \sqrt{n}^{b\wedge (1-b)} s^n_{i} \cdot \sqrt{n}^{b\wedge (1-b)} s^n_{i+2} \right)\right]\\
		&=\E_{i-k_n-1}^n \Big[  \Delta_i^n X \cdot \sqrt{n}^{b\wedge (1-b)} \left( \left( \widehat{\sigma}^2_{t_{i+}^n}  -(\sigma^2_{t_{i+}^n} + b_{ t_{i+}^n,n}(u) ) \right) - \left( \widehat{\sigma}^2_{t_{i-}^n}  -(\sigma^2_{t_{i-}^n} + b_{ t_{i-}^n,n}(u) ) \right) \right) \\
		& \quad \cdot \Delta_{i+2}^n X \sqrt{n}^{b\wedge (1-b)} \left( \left( \widehat{\sigma}^2_{t_{(i+2)+}^n}  -(\sigma^2_{t_{(i+2)+}^n} + b_{ t_{(i+2)+}^n,n}(u) ) \right) - \left( \widehat{\sigma}^2_{t_{(i+2)-}^n}  -(\sigma^2_{t_{(i+2)-}^n} + b_{ t_{(i+2)-}^n,n}(u) ) \right) \right)\Big] \\
		&=-\E_{i-k_n-1}^n \left[  \Delta_i^n X\cdot \Delta_{i+2}^n X \cdot n^{b\wedge (1-b)} \left( \widehat{\sigma}^2_{t_{i+}^n}  -(\sigma^2_{t_{i+}^n} + b_{ t_{i+}^n,n}(u) ) \right) \cdot  \left( \widehat{\sigma}^2_{t_{(i+2)-}^n}  -(\sigma^2_{t_{(i+2)-}^n} + b_{ t_{(i+2)-}^n,n}(u) ) \right) \right]\\
		&\quad + \E_{i-k_n-1}^n \left[  \Delta_i^n X\cdot \Delta_{i+2}^n X \cdot n^{b\wedge (1-b)} \left( \widehat{\sigma}^2_{t_{i+}^n}  -(\sigma^2_{t_{i+}^n} + b_{ t_{i+}^n,n}(u) ) \right) \cdot  \left( \widehat{\sigma}^2_{t_{(i+2)+}^n}  -(\sigma^2_{t_{(i+2)+}^n} + b_{ t_{(i+2)+}^n,n}(u) ) \right) \right]\\
		&\quad -\E_{i-k_n-1}^n \left[  \Delta_i^n X\cdot \Delta_{i+2}^n X \cdot n^{b\wedge (1-b)} \left( \widehat{\sigma}^2_{t_{i-}^n}  -(\sigma^2_{t_{i-}^n} + b_{ t_{i-}^n,n}(u) ) \right) \cdot  \left( \widehat{\sigma}^2_{t_{(i+2)+}^n}  -(\sigma^2_{t_{(i+2)+}^n} + b_{ t_{(i+2)+}^n,n}(u) ) \right) \right]\\
		&\quad +\E_{i-k_n-1}^n \left[  \Delta_i^n X\cdot \Delta_{i+2}^n X \cdot n^{b\wedge (1-b)} \left( \widehat{\sigma}^2_{t_{i-}^n}  -(\sigma^2_{t_{i-}^n} + b_{ t_{i-}^n,n}(u) ) \right) \cdot  \left( \widehat{\sigma}^2_{t_{(i+2)-}^n}  -(\sigma^2_{t_{(i+2)-}^n} + b_{ t_{(i+2)-}^n,n}(u) ) \right) \right]\\
		&:=-A_i^n + B_i^n - C_i^n +D_i^n.
	\end{align*}
	For $A_i^n$, we define
	\begin{align*}
		\widehat{\sigma}^2_{t_{i+}^n,-(i+2)} &=  \frac{-2}{u^2} \log\left( \left( \frac{1}{k_n}\sum_{j \in I^n_{i+}, j\neq (i+2)} \cos{ \left(\frac{u\Delta_j^nX}{\sqrt{\Delta_n}} \right)} \right) \right),\\
		\widehat{\sigma}^2_{t_{(i+2)-}^n,-i} &=  \frac{-2}{u^2} \log\left( \left( \frac{1}{k_n}\sum_{j \in I^n_{(i+2)-}, j\neq i} \cos{ \left(\frac{u\Delta_j^nX}{\sqrt{\Delta_n}} \right)} \right) \right),
	\end{align*}
	and we decompose
	\begin{align*}
		&A_i^n =\\
		& \E_{i-k_n-1}^n \left[  \Delta_i^n X\cdot \Delta_{i+2}^n X \cdot n^{b\wedge (1-b)} \left( \widehat{\sigma}^2_{t_{i+}^n,-(i+2)}  -(\sigma^2_{t_{i+}^n} + b_{ t_{i+}^n,n}(u) ) \right) \cdot  \left( \widehat{\sigma}^2_{t_{(i+2)-}^n,-i}  -(\sigma^2_{t_{(i+2)-}^n} + b_{ t_{(i+2)-}^n,n}(u) ) \right) \right]\\
		&+ \E_{i-k_n-1}^n \left[  \Delta_i^n X\cdot \Delta_{i+2}^n X \cdot n^{b\wedge (1-b)} \left( \widehat{\sigma}^2_{t_{i+}^n,-(i+2)}  -(\sigma^2_{t_{i+}^n} + b_{ t_{i+}^n,n}(u) ) \right) \cdot  \left(\widehat{\sigma}^2_{t_{(i+2)-}^n}  -  \widehat{\sigma}^2_{t_{(i+2)-}^n,-i}   \right) \right]\\
		&  + \E_{i-k_n-1}^n \left[  \Delta_i^n X\cdot \Delta_{i+2}^n X \cdot n^{b\wedge (1-b)} \left( \widehat{\sigma}^2_{t_{i+}^n} - \widehat{\sigma}^2_{t_{i+}^n,-(i+2)}  \right) \cdot  \left( \widehat{\sigma}^2_{t_{(i+2)-}^n,-i}  -(\sigma^2_{t_{(i+2)-}^n} + b_{ t_{(i+2)-}^n,n}(u) ) \right) \right]\\
		&+ \E_{i-k_n-1}^n \left[  \Delta_i^n X\cdot \Delta_{i+2}^n X \cdot n^{b\wedge (1-b)} \left( \widehat{\sigma}^2_{t_{i+}^n} - \widehat{\sigma}^2_{t_{i+}^n,-(i+2)}  \right) \cdot  \left(\widehat{\sigma}^2_{t_{(i+2)-}^n}  -  \widehat{\sigma}^2_{t_{(i+2)-}^n,-i}   \right) \right]\\
		&:= A_i^n(1) + A_i^n(2) + A_i^n(3) + A_i^n(4).
	\end{align*}
	For $A_i^n(1)$, we have $A_i^n(1) = O_p(\Delta_n^{2 \wedge \frac{2}{(\max\{\beta,r\}+\epsilon)}})$ since
	\begin{align*}
		&A_i^n(1) \\
		& = \E_{i-k_n-1}^n \Big[ \E\left[ \Delta_i^n X \Big|  \left(\Delta_{i+2}^n X, \widehat{\sigma}^2_{t_{i+}^n,-(i+2)}, \sigma^2_{t_{i+}^n}, b_{ t_{i+}^n,n}(u), \widehat{\sigma}^2_{t_{(i+2)-}^n,-i}, \sigma^2_{t_{(i+2)-}^n}, b_{ t_{(i+2)-}^n,n}(u) \right) \right] \\
		& \qquad \qquad \quad \cdot n^{b\wedge (1-b)} \left( \widehat{\sigma}^2_{t_{i+}^n,-(i+2)}  -(\sigma^2_{t_{i+}^n} + b_{ t_{i+}^n,n}(u) \right) \cdot  \left( \widehat{\sigma}^2_{t_{(i+2)-}^n,-i}  -(\sigma^2_{t_{(i+2)-}^n} + b_{ t_{(i+2)-}^n,n}(u) ) \right) \Big]\\
		&(\text{Using a successive conditioning argument similar to that in the proof of \eqref{lem-lev-2-2}}.)\\
		&= \E_{i-k_n-1}^n \Big[ \left(\Delta_i^n X - \sigma_{t_i^n} \Delta_i^n B \right) \left(\Delta_i^n X - \sigma_{t_{i+2}^n} \Delta_{i+2}^n B \right)  \\
		& \qquad \qquad \quad \cdot n^{b\wedge (1-b)} \left( \widehat{\sigma}^2_{t_{i+}^n,-(i+2)}  -(\sigma^2_{t_{i+}^n} + b_{ t_{i+}^n,n}(u) ) \right) \cdot  \left( \widehat{\sigma}^2_{t_{(i+2)-}^n,-i}  -(\sigma^2_{t_{(i+2)-}^n} + b_{ t_{(i+2)-}^n,n}(u) ) \right) \Big]\\
		&\text{(By \eqref{X-main} and Lemma \ref{lem-lev-1}.)}\\
		&= O_p(\Delta_n^{2 \wedge \frac{2}{(\max\{\beta,r\}+\epsilon)}}).
	\end{align*}
	And 
	\begin{align*}
		&A_i^n(4)\\
		&=n^{b\wedge (1-b)} \E_{i-k_n-1}^n \left[  \Delta_i^n X\cdot \Delta_{i+2}^n X \cdot  \left( \widehat{\sigma}^2_{t_{i+}^n} - \widehat{\sigma}^2_{t_{i+}^n,-(i+2)}  \right) \cdot  \left(\widehat{\sigma}^2_{t_{(i+2)-}^n}  -  \widehat{\sigma}^2_{t_{(i+2)-}^n,-i}   \right) \right]\\
		& = \frac{4}{u^4}n^{b\wedge (1-b)} \E_{i-k_n-1}^n \Bigg[  \Delta_i^n X\cdot \Delta_{i+2}^n X \cdot  \log \left( 1+ \frac{\frac{1}{k_n} \cos{ \left(\frac{u\Delta_{i+2}^nX}{\sqrt{\Delta_n}} \right)} }{\left( \frac{1}{k_n}\sum_{j \in I^n_{i+}, j\neq (i+2)} \cos{ \left(\frac{u\Delta_j^nX}{\sqrt{\Delta_n}} \right)} \right) }  \right) \\
		& \qquad \qquad \qquad \quad \cdot \log \left( 1+ \frac{\frac{1}{k_n} \cos{ \left(\frac{u\Delta_i^nX}{\sqrt{\Delta_n}} \right)} }{\left( \frac{1}{k_n}\sum_{j \in I^n_{(i+2)-}, j\neq i} \cos{ \left(\frac{u\Delta_j^nX}{\sqrt{\Delta_n}} \right)} \right) }  \right)  \Bigg]\\
		&(\text{By Taylor's expansion of $\log(1+x)$.}) \\
		& = \frac{4}{u^4}n^{b\wedge (1-b)} \E_{i-k_n-1}^n \Bigg[ \E \Bigg[ \Delta_i^n X\cdot \Delta_{i+2}^n X \cdot  \sum_{k=1}^{\infty} \frac{(-1)^{k+1}}{k}\left( \frac{\frac{1}{k_n} \cos{ \left(\frac{u\Delta_{i+2}^nX}{\sqrt{\Delta_n}} \right)} }{\left( \frac{1}{k_n}\sum_{j \in I^n_{i+}, j\neq (i+2)} \cos{ \left(\frac{u\Delta_j^nX}{\sqrt{\Delta_n}} \right)} \right) }  \right)^k \\
		& \quad  \cdot \sum_{k=1}^{\infty} \frac{(-1)^{k+1}}{k} \left( \frac{\frac{1}{k_n} \cos{ \left(\frac{u\Delta_i^nX}{\sqrt{\Delta_n}} \right)} }{\left( \frac{1}{k_n}\sum_{j \in I^n_{(i+2)-}, j\neq i} \cos{ \left(\frac{u\Delta_j^nX}{\sqrt{\Delta_n}} \right)} \right) }  \right)^{k}  \Bigg] \Bigg| \{\Delta_j^n X: j \in I^n_{(i+2)-} \cup I^n_{i+} - \{i,i+2\}\} \Bigg] \Bigg]\\
		& = \frac{4}{u^4}n^{b\wedge (1-b)} \E_{i-k_n-1}^n \Bigg[ \E \Bigg[ \Delta_{i+2}^n X \cdot  \sum_{k=1}^{\infty} \frac{(-1)^{k+1}}{k}\left( \frac{\frac{1}{k_n} \cos{ \left(\frac{u\Delta_{i+2}^nX}{\sqrt{\Delta_n}} \right)} }{\left( \frac{1}{k_n}\sum_{j \in I^n_{i+}, j\neq (i+2)} \cos{ \left(\frac{u\Delta_j^nX}{\sqrt{\Delta_n}} \right)} \right) }  \right)^k \\
		& \quad \Bigg| \{\Delta_j^n X: j \in I^n_{(i+2)-} \cup I^n_{i+} - \{i+2\}\} \Bigg] \cdot  \Delta_i^n X \cdot \sum_{k=1}^{\infty} \frac{(-1)^{k+1}}{k} \left( \frac{\frac{1}{k_n} \cos{ \left(\frac{u\Delta_i^nX}{\sqrt{\Delta_n}} \right)} }{\left( \frac{1}{k_n}\sum_{j \in I^n_{(i+2)-}, j\neq i} \cos{ \left(\frac{u\Delta_j^nX}{\sqrt{\Delta_n}} \right)} \right) }  \right)^{k}   \Bigg]\\
		& \Bigg(R_i^n(1) = \Delta_{i+2}^n X \cdot  \sum_{k=1}^{\infty} \frac{(-1)^{k+1}}{k}\left( \frac{\frac{1}{k_n} \cos{ \left(\frac{u\Delta_{i+2}^nX}{\sqrt{\Delta_n}} \right)} }{\left( \frac{1}{k_n}\sum_{j \in I^n_{i+}, j\neq (i+2)} \cos{ \left(\frac{u\Delta_j^nX}{\sqrt{\Delta_n}} \right)} \right) }  \right)^k \\
		&\qquad \qquad \ - \sigma_{t_{i+2}^n} \Delta_{i+2}^n B \cdot  \sum_{k=1}^{\infty} \frac{(-1)^{k+1}}{k}\left( \frac{\frac{1}{k_n} \cos{ \left(\frac{u\sigma_{t_{i+2}^n} \Delta_{i+2}^n B}{\sqrt{\Delta_n}} \right)} }{\left( \frac{1}{k_n}\sum_{j \in I^n_{i+}, j\neq (i+2)} \cos{ \left(\frac{u\Delta_j^nX}{\sqrt{\Delta_n}} \right)} \right) }  \right)^k \Bigg)\\
		& = \frac{4}{u^4}n^{b\wedge (1-b)} \E_{i-k_n-1}^n \Bigg[ \E \Bigg[ \sigma_{t_{i+2}^n} \Delta_{i+2}^n B \cdot  \sum_{k=1}^{\infty} \frac{(-1)^{k+1}}{k}\left( \frac{\frac{1}{k_n} \cos{ \left(\frac{u\sigma_{t_{i+2}^n} \Delta_{i+2}^n B}{\sqrt{\Delta_n}} \right)} }{\left( \frac{1}{k_n}\sum_{j \in I^n_{i+}, j\neq (i+2)} \cos{ \left(\frac{u\Delta_j^nX}{\sqrt{\Delta_n}} \right)} \right) }  \right)^k + R_i^n(1) \\
		& \quad \Bigg| \{\Delta_j^n X: j \in I^n_{(i+2)-} \cup I^n_{i+} - \{i+2\}\} \Bigg] \cdot  \Delta_i^n X \cdot \sum_{k=1}^{\infty} \frac{(-1)^{k+1}}{k} \left( \frac{\frac{1}{k_n} \cos{ \left(\frac{u\Delta_i^nX}{\sqrt{\Delta_n}} \right)} }{\left( \frac{1}{k_n}\sum_{j \in I^n_{(i+2)-}, j\neq i} \cos{ \left(\frac{u\Delta_j^nX}{\sqrt{\Delta_n}} \right)} \right) }  \right)^{k}   \Bigg]\\
		&(\text{Since $x\cos^{k}(cx)$ is an odd function for any positive constant $c$ and $k\in\mathbb{N}^{+}$.})\\
		& = \frac{4}{u^4}n^{b\wedge (1-b)} \E_{i-k_n-1}^n \Bigg[ R_i^n(1) \cdot \E \Bigg[ \Delta_i^n X \cdot \sum_{k=1}^{\infty} \frac{(-1)^{k+1}}{k} \left( \frac{\frac{1}{k_n} \cos{ \left(\frac{u\Delta_i^nX}{\sqrt{\Delta_n}} \right)} }{\left( \frac{1}{k_n}\sum_{j \in I^n_{(i+2)-}, j\neq i} \cos{ \left(\frac{u\Delta_j^nX}{\sqrt{\Delta_n}} \right)} \right) }  \right)^{k}   \\
		& \quad \Bigg| \{\Delta_j^n X: j \in I^n_{(i+2)-} \cup I^n_{i+} - \{i\}\} \cup R_i^n(1) \Bigg] \Bigg]\\
		&\Bigg(R_i^n(2) = \Delta_{i}^n X \cdot  \sum_{k=1}^{\infty} \frac{(-1)^{k+1}}{k}\left( \frac{\frac{1}{k_n} \cos{ \left(\frac{u\Delta_{i}^nX}{\sqrt{\Delta_n}} \right)} }{\left( \frac{1}{k_n}\sum_{j \in I^n_{(i+2)-}, j\neq (i)} \cos{ \left(\frac{u\Delta_j^nX}{\sqrt{\Delta_n}} \right)} \right) }  \right)^k \\
		&\qquad \qquad \qquad \ - \sigma_{t_{i}^n} \Delta_{i}^n B \cdot  \sum_{k=1}^{\infty} \frac{(-1)^{k+1}}{k}\left( \frac{\frac{1}{k_n} \cos{ \left(\frac{u\sigma_{t_{i}^n} \Delta_{i}^n B}{\sqrt{\Delta_n}} \right)} }{\left( \frac{1}{k_n}\sum_{j \in I^n_{(i+2)-}, j\neq i} \cos{ \left(\frac{u\Delta_j^nX}{\sqrt{\Delta_n}} \right)} \right) }  \right)^k \Bigg)\\
		& = \frac{4}{u^4}n^{b\wedge (1-b)} \E_{i-k_n-1}^n \Bigg[ R_i^n(1) \cdot \E \Bigg[ \sigma_{t_{i}^n} \Delta_i^n B \cdot \sum_{k=1}^{\infty} \frac{(-1)^{k+1}}{k} \left( \frac{\frac{1}{k_n} \cos{ \left(\frac{u\sigma_{t_{i}^n} \Delta_i^n B }{\sqrt{\Delta_n}} \right)} }{\left( \frac{1}{k_n}\sum_{j \in I^n_{(i+2)-}, j\neq i} \cos{ \left(\frac{u\Delta_j^nX}{\sqrt{\Delta_n}} \right)} \right) }  \right)^{k}  + R_i^n(2) \\
		& \quad \Bigg| \{\Delta_j^n X: j \in I^n_{(i+2)-} \cup I^n_{i+} - \{i\}\} \cup R_i^n(1) \Bigg] \Bigg]\\
		&(\text{Since $x\cos^{k}(cx)$ is an odd function for any positive constant $c$ and $k\in\mathbb{N}^{+}$.})\\
		&= \frac{4}{u^2}  \cdot n^{b\wedge (1-b)} \E_{i-k_n-1}^n \left[R_i^n(1) \cdot R_i^n(2) \right]\\
		&(\text{By plugging in the results from \eqref{X-main}}.)\\
		&= \frac{4}{u^2}  \cdot \frac{n^{b\wedge (1-b)}}{n^{2b}} \cdot O_p(\Delta_n^{2 \wedge \frac{2}{(\max\{\beta,r\}+\epsilon)}}) = o_p(\Delta_n^{2 \wedge \frac{2}{ (\max\{\beta,r\}+\epsilon)}}).
	\end{align*}
	Similarly, we obtain $A_i^n(2),  A_i^n(3)= o_p(\Delta_n^{2 \wedge \frac{2}{(\max\{\beta,r\}+\epsilon)}})$, and furthermore $B_i^n,  C_i^n, D_i^n = O_p(\Delta_n^{2 \wedge \frac{2}{(\max\{\beta,r\}+\epsilon)}})$. We omit the detailed derivations to save space; they are available upon request. 
	The proof for the general case $j-i \leq 2k_n$ is the same as above. 
	
	This completes the proof of the lemma.  \hfill $\square$
	
	\textbf{Proof of Theorem \ref{thm-lev-1}:} We first prove the central limit theorem \eqref{thm-lev-clt}. Note that
	\begin{align*}
		&\widehat{\mathcal{L}}_{[0,T]} -  \mathcal{L}_{[0,T]} \\
		&= \sum_{i=k_n+1}^{n-k_n} s_i^n + \sum_{i=k_n+1}^{n-k_n} \left( \Delta_i^n X  \cdot  \left( (\sigma^2_{t_{i+}^n} + b_{ t_{i+}^n,n}(u) )- ( \sigma^2_{t_{i-}^n} + b_{ t_{i-}^n,n}(u) )\right) \right)  -  \mathcal{L}_{[0,T]}\\
		& (\text{Since $\gamma$ is bounded and with condition \eqref{cond-vov}.})\\
		&= \sum_{i=k_n+1}^{n-k_n} s_i^n  + \sum_{i=k_n+1}^{n-k_n} \left( \Delta_i^n X  \cdot   \left( \sigma^2_{t_{i+}^n} - \sigma^2_{t_{i-}^n} \right) \right)  -  \mathcal{L}_{[0,T]} + O_p(u^{\beta-2}\Delta_n^{\frac{2-\beta}{2}})\\
		&= \sum_{i=k_n+1}^{n-k_n} s_i^n  + \sum_{i=k_n+1}^{n-k_n} \left( \int_{t_{i-1}^n}^{t_i^n} \sigma_sdB_s  \cdot   \left( \sigma^2_{t_{i+}^n} - \sigma^2_{t_{i-}^n} \right) \right)  -  \mathcal{L}_{[0,T]}  \\
		&\quad + \sum_{i=k_n+1}^{n-k_n} \left( \left( \Delta_{i}^nX- \int_{t_{i-1}^n}^{t_i^n} \sigma_sdB_s \right) \cdot   \left( \sigma^2_{t_{i+}^n} - \sigma^2_{t_{i-}^n} \right) \right) + O_p(u^{\beta-2}\Delta_n^{\frac{2-\beta}{2}}) \\
		&(\text{By the proof of Lemma 2 in \cite{LLL2018}, the boundedness of $\sigma$ and the condition \eqref{cond-vov}}.)\\
		&= \sum_{i=k_n+1}^{n-k_n} s_i^n  + \sum_{i=k_n+1}^{n-k_n} \left( \int_{t_{i-1}^n}^{t_i^n} \sigma_sdB_s  \cdot   \left( \sigma^2_{t_{i+}^n} - \sigma^2_{t_{i-}^n} \right) \right)  -  \mathcal{L}_{[0,T]} + O_p\left( u^{\beta-2}\Delta_n^{\frac{2-\beta}{2}} + \Delta_n^{\frac{1}{\max\{\beta,r\}} - \frac{1}{2}} + \sqrt{\Delta_n} \right).
	\end{align*}
	When $\max\{\beta,r\}\leq 1$, we have 
	\begin{align*}
		\sqrt{n}^{b\wedge (1-b)}\left(u^{\beta-2}\Delta_n^{\frac{2-\beta}{2}} + \Delta_n^{\frac{1}{\max\{\beta,r\}} - \frac{1}{2}} + \sqrt{\Delta_n}\right) \rightarrow 0.
	\end{align*}
	By the proof of Theorem 3 (Step 3) in \cite{AFLWY2017}, we have 
	\begin{align*}
		\sqrt{n}^{b\wedge (1-b)} \left(\sum_{i=k_n+1}^{n-k_n} \left( \int_{t_{i-1}^n}^{t_i^n} \sigma_sdB_s  \cdot   \left( \sigma^2_{t_{i+}^n} - \sigma^2_{t_{i-}^n} \right) \right)  -  \mathcal{L}_{[0,T]}  \right)\longrightarrow^{p} 0.
	\end{align*}
	Thus, it remains to prove
	\begin{align}\label{lev-main}
		\sqrt{n}^{b\wedge (1-b)}	\sum_{i=k_n+1}^{n-k_n} s_i^n  \longrightarrow^{L_s} U.
	\end{align}
	Following the argument in \cite{JP2012} (Section 12.2.4) and \cite{AFLWY2017} (Step 4 in the proof of Theorem 3 therein), we split the sum over $i$ into big blocks of size $(\tilde{m}+2) k_n$, with $\tilde{m}$ satisfying
	\begin{align} \label{con-m}
		\tilde{m} \rightarrow  \infty, \quad \tilde{m}k_n\Delta_n \rightarrow 0.
	\end{align}
	We define $I(\tilde{m},n,l)= (l-1)(\tilde{m}+2)k_n + 1$ for $l=1,...,l_n(\tilde{m})$, where the total number of big blocks is $l_n(\tilde{m}) = \lfloor \frac{\lfloor T/\Delta_n\rfloor-1}{(\tilde{m}+2 )k_n} \rfloor$. In the $l$-th big block, for $s_i^n$ with $I(\tilde{m},n,l)+k_n + 1 \leq i\leq I(\tilde{m},n,l)  +(\tilde{m}+1)k_n$, the increments $\Delta_{j}^n X$ constituting $s_i^n$ lie within $I(\tilde{m},n,l)$, so $s_i^n$ is $\mathcal{F}_{I(\tilde{m},n,l+1) \Delta_n}$-measurable. To simplify notation, we use $(l,j)$ for the time point $(I(\tilde{m},n,l) + j) \Delta_n$ or its index $(I(\tilde{m},n,l) + j)$, and $\E^n_{(l,j)}[\cdot]$ for the conditional expectation $\E[\cdot | \mathcal{F}_{(l,j)}]$, with $l=1,...,l_n(\tilde{m})$ and $j$ any integer.
	We decompose $\sqrt{n}^{b\wedge (1-b)}	\sum_{i=k_n+1}^{n-k_n} s_i^n = Z(\tilde{m})^n  +  \tilde{Z}(\tilde{m})^n  + \tilde{Z}'(\tilde{m})^n$ with
	\begin{align*}
		&\xi(\tilde{m})_l^n =\sqrt{n}^{b\wedge (1-b)} \sum_{r=k_n+1}^{(\tilde{m}+1)k_n}  s^n_{(l,r)}, \quad \tilde{\xi}(\tilde{m})_l^n =\sqrt{n}^{b\wedge (1-b)} \sum_{r=-k_n}^{k_n} s^n_{( l,r)},\\
		&Z(\tilde{m})^n = \sum_{l=1}^{l_n(\tilde{m})} \xi(\tilde{m})_l^n, \quad \tilde{Z}(\tilde{m})^n = \sum_{l=2}^{l_n(\tilde{m})} \tilde{\xi}(\tilde{m})_l^n, \quad \tilde{Z}'(\tilde{m})^n =\sqrt{n}^{b\wedge (1-b)} \sum_{i=(l_n(\tilde{m})+1,-k_n)}^{n-k_n} s^n_{i}. 
	\end{align*}
	To prove \eqref{lev-main}, we will show that
	\begin{align}\label{lev-main-1}
		&\tilde{Z}(\tilde{m})^n  + \tilde{Z}'(\tilde{m})^n \longrightarrow^{p} 0,\\ \label{lev-main-2}
		&Z(\tilde{m})^n \longrightarrow^{L_s} U.
	\end{align}
	For \eqref{lev-main-1}, note that when $\tilde{m} > 2$, $\tilde{\xi}(\tilde{m})_l^n$ is $(l+1,-2k_n-1)$-measurable and there are no overlapping terms in the sequence $\{\tilde{\xi}(\tilde{m})_l^n: l=2,...,l_n(\tilde{m})\}$. By Lemma 4.1 in \cite{J2012}, $\tilde{Z}(\tilde{m})^n \longrightarrow^{p} 0$ can be proved by showing 
	\begin{align}
		\sum_{l=2}^{l_n(\tilde{m})} \E_{(l,-2k_n-1)}^n [\tilde{\xi}(\tilde{m})_l^n] \longrightarrow^p 0, \qquad 
		\sum_{l=2}^{l_n(\tilde{m})} \E_{(l,-2k_n-1)}^n [( \tilde{\xi}(\tilde{m})_l^n)^2]\longrightarrow^p 0. 
	\end{align}
	For $r=-k_n,...,k_n$ and $l=2,...,l_n(\tilde{m})$, Lemma \ref{lem-lev-2} gives
	\begin{align}\label{thm1-expec}
		\E_{(l,-2k_n-1)}^n \left[ \sqrt{n}^{b\wedge (1-b)} s^n_{( l,r)} \right]  = O_p\left( \Delta_n^{1\wedge \frac{1}{(\max\{\beta,r\}+\epsilon)}} \cdot h_n \right).
	\end{align}
	When $\max\{\beta,r\} \leq 1$, since $\tilde{m} \rightarrow \infty$, we have $l_n(\tilde{m}) k_n \cdot \Delta_n^{1\wedge \frac{1}{(\max\{\beta,r\}+\epsilon)}} h_n \rightarrow 0$.
	Thus, $\sum_{l=2}^{l_n(\tilde{m})} \E_{(l,-2k_n-1)}^n [\tilde{\xi}(\tilde{m})_l^n] \rightarrow^p 0$. Moreover,
	\begin{align*}
		\E_{(l,-k_n-1)}^n [( \tilde{\xi}(\tilde{m})_l^n)^2] &= \E_{(l,-k_n-1)}^n \left[\sum_{r=-k_n}^{k_n}  \left( \sqrt{n}^{b\wedge (1-b)} s^n_{( l,r)} \right)^2\right]\\
		&\quad + \E_{(l,-k_n-1)}^n \left[2 \sum_{r_1, r_2=-k_n}^{k_n} \sum_{r_1<r_2} \left( \sqrt{n}^{b\wedge (1-b)} s^n_{( l,r_1)} \cdot \sqrt{n}^{b\wedge (1-b)} s^n_{( l,r_2)} \right)\right]\\
		&(\text{By \eqref{spot-main}, \eqref{X-main}, \eqref{lem1-key} and Lemma \ref{lem-lev-2}.})\\
		&= O_p(k_n \Delta_n),
	\end{align*}
	as $\tilde{m} \longrightarrow \infty$, which implies $\sum_{l=2}^{l_n(\tilde{m})} \E_{(l,-k_n-1)}^n [( \tilde{\xi}(\tilde{m})_l^n)^2] \rightarrow^p 0$. 
	For $\tilde{Z}'(\tilde{m})^n \longrightarrow^{p} 0$, a similar argument gives
	\begin{align*}
		\E[\tilde{Z}'(\tilde{m})^n] = \tilde{m}k_n \cdot O_p\left(  \Delta_n^{1\wedge \frac{1}{(\max\{\beta,r\}+\epsilon)}} h_n \right)
	\end{align*}
	and 
	\begin{align*}
		\E[(\tilde{Z}'(\tilde{m})^n)^2] = O_p
		\left( (\tilde{m} k_n\Delta_n)^2\right).
	\end{align*}
	Since $\tilde{m} k_n\Delta_n \rightarrow 0$, Chebyshev's inequality implies convergence in probability. This proves \eqref{lev-main-1}. \\
	For \eqref{lev-main-2}, from the above proof of \eqref{lev-main-1}, if $\max\{\beta,r\} \leq 1$, since $k_n\Delta_n\rightarrow 0$ and $k_n\rightarrow \infty$, we have
	\begin{align*}
		\sum_{l=1}^{l_n(\tilde{m})} \E_{(l,0)}\left[ \xi(\tilde{m})_l^n\right] = O_p\left( n  \cdot\Delta_n^{1\wedge \frac{1}{(\max\{\beta,r\}+\epsilon)}} h_n \right) \longrightarrow^p 0,
	\end{align*}
	and $ \xi(\tilde{m})_l^n $ is $\mathcal{F}_{(l+1,0)}$-measurable, so $\left \{  \xi(\tilde{m})_l^n, \mathcal{F}_{(l+1,0)} \right \}$ behaves like a martingale difference array. By Theorem 2.2.15 in \cite{JP2012}, it remains to show
	\begin{align}\label{B-Ls-2-1}
		&\sum_{l=1}^{l_n(\tilde{m})} \E_{(l,0)}\left[ (\xi(\tilde{m})_l^n )^2 \right] \longrightarrow^p Var(U|\mathcal{F}),\\\label{B-Ls-2-2}
		&\sum_{l=1}^{l_n(\tilde{m})} \E_{(l,0)}\left[ (\xi(\tilde{m})_l^n )^4 \right] \longrightarrow^p 0,\\\label{B-Ls-2-3}
		&\sum_{l=1}^{l_n(\tilde{m})} \E_{(l,0)}\left[ \xi(\tilde{m})_l^n \cdot \left( \Delta_{l,\tilde{m}}^n B + \Delta_{l,\tilde{m}}^n B' + \Delta_{l,\tilde{m}}^n W \right) \right] \longrightarrow^p 0,
	\end{align} 
	where $\Delta_{l,\tilde{m}}^n M := M_{(l+1,0)} - M_{(l,0)}$ for $M=B,B'$ or $W$, with $W$ a bounded martingale orthogonal to both $B$ and $B'$. For \eqref{B-Ls-2-1}, we decompose
	\begin{align*}
		&\E_{(l,0)}\left[ (\xi(\tilde{m})_l^n )^2\right] \\ 
		&= \sum_{r=k_n+1}^{(\tilde{m}+1)k_n} \sum_{j=k_n+1}^{(\tilde{m}+1)k_n} \E_{(l,0)}[\sqrt{n}^{b\wedge (1-b)} s^n_{(l,r)} \cdot \sqrt{n}^{b\wedge (1-b)}  s^n_{(l,j)}]\\
		& = \sum_{r=k_n+1}^{(\tilde{m}+1)k_n}  \E_{(l,0)}[(\sqrt{n}^{b\wedge (1-b)} s^n_{(l,r)})^2 ] + n^{b\wedge (1-b)} \cdot  \sum_{r,j=k_n+1}^{(\tilde{m}+1)k_n} \sum_{|j-r|\leq 2k_n+1} \E_{(l,0)}[s^n_{(l,r)} s^n_{ (l,j)}]\\
		&\quad  + n^{b\wedge (1-b)}  \cdot \sum_{r,j=k_n+1}^{(\tilde{m}+1)k_n} \sum_{|j-r|>2k_n+1} \E_{(l,0)}[s^n_{(l,r)} s^n_{(l,j)}]\\
		&=:H(\tilde{m},1)_l^n + H(\tilde{m},2)_l^n  + H(\tilde{m},3)_l^n.
	\end{align*}
	By Lemma \ref{lem-lev-2}, when $\max\{\beta,r\} \leq 1$, we have
	\begin{align*}
		\sum_{l=1}^{l_n(\tilde{m})}  H(\tilde{m},2)_l^n + \sum_{l=1}^{l_n(\tilde{m})}  H(\tilde{m},3)_l^n  = O_p(\tilde{m} k_n \Delta_n) \longrightarrow^p 0,
	\end{align*}
	and
	\begin{align*}
		&\sum_{l=1}^{l_n(\tilde{m})} H(\tilde{m},1)_l^n =\sum_{l=1}^{l_n(\tilde{m})}   \sum_{r=k_n+1}^{(\tilde{m}+1)k_n}  \E_{(l,0)}[(\sqrt{n}^{b\wedge (1-b)} s^n_{(l,r)})^2 ]\\
		&= \sum_{l=1}^{l_n(\tilde{m})}   \sum_{r=k_n+1}^{(\tilde{m}+1)k_n}  \E_{(l,0)}[(\sqrt{n}^{b\wedge (1-b)} s^n_{(l,r)})^2 ] \\
		& = \sum_{l=1}^{l_n(\tilde{m})}   \sum_{r=k_n+1}^{(\tilde{m}+1)k_n} \E_{(l,0)} \Bigg[ (\Delta_i^n X)^2 \cdot \Bigg(\sqrt{n}^{b\wedge (1-b)}  \left( \widehat{\sigma}^2_{t_{i+}^n}  -(\sigma^2_{t_{i+}^n} + b_{ t_{i+}^n,n}(u) ) \right)\\
		& \qquad \qquad \qquad \qquad \quad -  \left(\sqrt{n}^{b\wedge (1-b)} \left( \widehat{\sigma}^2_{t_{i-}^n}  -(\sigma^2_{t_{i-}^n} + b_{ t_{i-}^n,n}(u) ) \right)\right)^2 \Bigg] \\
		&\text{(By \eqref{X-main} and the results of Lemma \ref{lem-lev-1}.)} \\
		& = \sum_{l=1}^{l_n(\tilde{m})}  \sum_{r=k_n+1}^{(\tilde{m}+1)k_n} \Bigg( \sigma_{(l,0)}^2 \E_{(l,0)} [ (\Delta_i^n B)^2 \cdot  \Bigg(\sqrt{n}^{b\wedge (1-b)}  \left( \widehat{\sigma}^2_{t_{i+}^n}  -(\sigma^2_{t_{i+}^n} + b_{ t_{i+}^n,n}(u) ) \right)\\
		& \qquad \qquad \quad -  \sqrt{n}^{b\wedge (1-b)} \left( \widehat{\sigma}^2_{t_{i-}^n}  -(\sigma^2_{t_{i-}^n} + b_{ t_{i-}^n,n}(u) ) \right)\Bigg)^2 \Bigg]  + O_p\left( \sqrt{\tilde{m}k_n} \Delta_n^{\frac{3}{2}}+ \sqrt{\Delta_n} \Delta_n^{1 \wedge \frac{1}{p}} \right) \Bigg)\\
		&(\text{With condition \eqref{con-m} and from the proof of Lemma \ref{lem-lev-1}}.) \\
		& = \sum_{l=1}^{l_n(\tilde{m})}  \sum_{r=k_n+1}^{(\tilde{m}+1)k_n} \Bigg( \sigma_{(l,0)}^2 \E_{(l,0)} [ (\Delta_i^n B)^2 \cdot \Bigg(\frac{1}{\sqrt{\kappa}} \sqrt{n}^{0\wedge (1-2b)}  \left( \sum_{j \in I^n_{i+} }  \xi_{t_{i+}^n,n}^{(2,j)}(u) - \sum_{j \in I^n_{i-} }  \xi_{t_{i-}^n,n}^{(2,j)}(u) \right)\\
		& \qquad \qquad \quad - \sqrt{\kappa T} \sqrt{n}^{0\wedge (2b-1)} \left( \sum_{j \in I^n_{i+} }  \xi_{t_{i+}^n,n}^{(3,j)}(u) - \sum_{j \in I^n_{i-} }  \xi_{t_{i-}^n,n}^{(3,j)}(u) \right) \Bigg)^2 \Bigg] \Bigg) + o_p\left( 1 \right) \\
		&(\text{Since $(\Delta_i^n B)^2- \Delta_n \longrightarrow^p 0$, and by Lemma \ref{lem-lev-1} and Proposition 2.5 in}\\
		&\ \  \text{\cite{PV2010}.})\\
		&= \sum_{l=1}^{l_n(\tilde{m})}  \sum_{r=k_n+1}^{(\tilde{m}+1)k_n} \Bigg( \Delta_n\cdot  \sigma_{(l,0)}^2\cdot \E_{(l,0)} \Bigg[ \Bigg( \frac{1}{\kappa} \Bigg( \frac{2(\exp{(-2 u^2\sigma^2_{t_{i+}^n} )} - 2\exp{(- u^2\sigma^2_{t_{i+}^n} )} + 1)}{u^4\exp{(- u^2\sigma^2_{t_{i+}^n} )}} \\
		& \qquad +   \frac{2(\exp{(-2 u^2\sigma^2_{t_{i-}^n} )} - 2\exp{(- u^2\sigma^2_{t_{i-}^n} )} + 1)}{u^4\exp{(- u^2\sigma^2_{t_{i-}^n} )}}\Bigg)  \cdot  1_{\{b \in (0,1/2]\}}  + \kappa T \cdot \Bigg(\frac{4(\sigma_{t_{i+}^n} )^2((\tilde{\sigma}_{t_{i+}^n})^2+(\tilde{\sigma}'_{t_{i+}^n})^2)}{3}   \\
		& \qquad  + \frac{4(\sigma_{t_{i-}^n} )^2((\tilde{\sigma}_{t_{i-}^n})^2+(\tilde{\sigma}'_{t_{i-}^n})^2)}{3} \Bigg) \cdot 1_{\{b\in [1/2,1)\}}  \Bigg) \Bigg]  \Bigg)+o_p(1) \\
		&(\text{Using the boundedness of $\sigma,\tilde{\sigma},\tilde{\sigma}'$ and the condition \eqref{cond-vov}.})\\
		&= \sum_{l=1}^{l_n(\tilde{m})}  \sum_{r=k_n+1}^{(\tilde{m}+1)k_n} \Bigg( \Delta_n\cdot  \sigma_{(l,0)}^2\cdot  \Bigg( \frac{2}{\kappa} \Bigg( \frac{2(\exp{(-2 u^2\sigma^2_{(l,0)} )} - 2\exp{(- u^2\sigma^2_{(l,0)} )} + 1)}{u^4\exp{(- u^2\sigma^2_{(l,0)} )}} \Bigg)  \cdot  1_{\{b \in (0,1/2]\}}  \\
		& \qquad + 2\kappa T \cdot \Bigg(\frac{4(\sigma_{(l,0)} )^2((\tilde{\sigma}_{(l,0)}^2+(\tilde{\sigma}'_{(l,0)})^2)}{3} \Bigg) \cdot 1_{\{b\in [1/2,1)\}}  \Bigg) \Bigg) +o_p(1) \\
		&  \longrightarrow^{p} Var(U|\mathcal{F}).
	\end{align*}
	This proves \eqref{B-Ls-2-1}. For \eqref{B-Ls-2-2}, we decompose
	\begin{align*}
		&\sum_{l=1}^{l_n(\tilde{m})} \E_{(l,0)}\left[ \left( \sum_{r=k_n+1}^{(\tilde{m}+1)k_n} \left( \sqrt{n}^{b\wedge (1-b)}s^n_{(l,r)}\right) \right)^4 \right] \\
		& \leq  C \sum_{l=1}^{l_n(\tilde{m})} \E_{(l,0)}\left[  \sum_{r_1} \left( \sqrt{n}^{b\wedge (1-b)}s^n_{(l,r_1)}\right)^4 \right] 
		\\
		&\quad +C \sum_{l=1}^{l_n(\tilde{m})} \E_{(l,0)}\left[  \sum_{r_1}\sum_{r_2} \left( \left( \sqrt{n}^{b\wedge (1-b)}s^n_{(l,r_1)}\right)^3 \left( \sqrt{n}^{b\wedge (1-b)}s^n_{(l,r_2)}\right)  \right) \right] \\
		&\quad +C \sum_{l=1}^{l_n(\tilde{m})} \E_{(l,0)}\left[  \sum_{r_1}\sum_{r_2} \left( \left( \sqrt{n}^{b\wedge (1-b)}s^n_{(l,r_1)}\right)^2 \left( \sqrt{n}^{b\wedge (1-b)}s^n_{(l,r_2)}\right)^2  \right) \right] \\
		&\quad +C \sum_{l=1}^{l_n(\tilde{m})} \E_{(l,0)}\left[  \sum_{r_1}\sum_{r_2}\sum_{r_3} \left( \left( \sqrt{n}^{b\wedge (1-b)}s^n_{(l,r_1)}\right)^2 \left( \sqrt{n}^{b\wedge (1-b)}s^n_{(l,r_2)}\right) \left( \sqrt{n}^{b\wedge (1-b)}s^n_{(l,r_3)}\right) \right) \right] \\
		&\quad +C \sum_{l=1}^{l_n(\tilde{m})} \E_{(l,0)}\left[  \sum_{r_1}\sum_{r_2}\sum_{r_3}\sum_{r_4} \left( \left( \sqrt{n}^{b\wedge (1-b)} \right)^4 \left(s^n_{(l,r_1)} s^n_{(l,r_2)}s^n_{(l,r_3)}s^n_{(l,r_4)}\right) \right) \right] \\
		&=:S_i^n (1) + S_i^n (2) + S_i^n (3) +S_i^n (4) +S_i^n (5), 
	\end{align*} 
	where $r_1 < r_2 < r_3 < r_4$ and $r_1, r_2, r_3, r_4 =k_n+1,...,(\tilde{m}+1)k_n$. Following the proof of \eqref{B-Ls-2-1}, we obtain
	\begin{align*}
		\E_{(l,0)}\left[ \left( \sqrt{n}^{b\wedge (1-b)}s^n_{(l,r)}\right)^2 \right] = O_p(\Delta_n), \ \E_{(l,0)}\left[ \left( \sqrt{n}^{b\wedge (1-b)}s^n_{(l,r)}\right)^4 \right] = O_p(\Delta_n^2),
	\end{align*}
	Thus, $S_i^n (1) = O_p(\Delta_n)$. Since
	\begin{align*}
		&\sum_{r_1}\sum_{r_2} \E_{(l,0)}\left[  \left( \left( \sqrt{n}^{b\wedge (1-b)}s^n_{(l,r_1)}\right)^3 \left( \sqrt{n}^{b\wedge (1-b)}s^n_{(l,r_2)}\right)  \right) \right] \\
		&= \sum_{r_1}\sum_{r_2-r_1\leq 2k_n} \E_{(l,0)}\left[  \left( \left( a^n_{(l,r_1)}\right)^3 \left( \sqrt{n}^{b\wedge (1-b)}s^n_{(l,r_2)}\right)  \right) \right]  \\
		&\quad + \sum_{r_1}\sum_{r_2-r_1>2k_n} \E_{(l,0)}\left[  \left( \left( \sqrt{n}^{b\wedge (1-b)}s^n_{(l,r_1)}\right)^3 \right] \E_{(l,0)}\left[  \left( \sqrt{n}^{b\wedge (1-b)}s^n_{(l,r_2)}\right)  \right) \right] \\
		&\text{(By applying Holder's inequality and using the results in Lemma \ref{lem-lev-2}.)}\\
		&\leq C\tilde{m}k_n\cdot k_n\cdot \Delta_n^2+ C (\tilde{m}k_n)^2 \cdot \Delta_n^{\frac{3}{2}} \cdot  \Delta_n h_n,
	\end{align*}
	Thus, $S_i^n(2) = O_p\left( \tilde{m}k_n\Delta_n +  \tilde{m}k_n\Delta_n \sqrt{\Delta_n} h_n \right)$. 
	Similarly, decomposing the summation into the cases $r_{k+1} - r_{k} \leq 2k_n$ and $r_{k+1} - r_{k} > 2k_n$ for $k=1,2,3$, and using successive conditioning and Holder's inequality, we obtain
	\begin{align*}
		&S_i^n(3) = O_p\left( \tilde{m}k_n\Delta_n \right), \qquad S_i^n(4) = O_p\left( (\tilde{m}k_n\Delta_n)^2 h_n^2 + \tilde{m} (k_n\Delta_n)^2 \right), \\
		&S_i^n(5) = O_p\left( (\tilde{m} k_n\Delta_n)^3h_n^4 + (\tilde{m} k_n\Delta_n)^2(k_n\Delta_n) h_n  +(\tilde{m} k_n\Delta_n)(k_n\Delta_n)^2h_n^2  \right).
	\end{align*}
	This proves \eqref{B-Ls-2-2}.
	For \eqref{B-Ls-2-3}, with $M=B+B'+W$, an argument similar to the proof of \eqref{lem-lev-2-1} gives, for $j=i-k_n,...,n$,
	\begin{align*}
		&\E_{i-k_n-1}^n \left[ \sqrt{n}^{b\wedge (1-b)} s^n_{i}\cdot \sum_{j=i-k_n}^{n} \Delta_j^nM \right] \\
		&= \E_{i-k_n-1}^n \left[ \sqrt{n}^{b\wedge (1-b)} s^n_{i}\cdot \sum_{j=i-k_n}^{i+k_n+1} \Delta_j^nM \right] \\
		&=  \E_{i-k_n-1}^n \left[ \Delta_i^n X \cdot \left( \Delta_i^n M +  \sum_{j=i+1}^{i+k_n+1} \Delta_j^nM \right) \cdot \left( \sqrt{n}^{b\wedge (1-b)} \left( \widehat{\sigma}^2_{t_{i+}^n}  -(\sigma^2_{t_{i+}^n} + b_{ t_{i+}^n,n}(u) ) \right) \right) \right] \\
		&\quad -  \E_{i-k_n-1}^n \left[ \Delta_i^n X \cdot \left( \Delta_i^n M + \sum_{j=i-k_n}^{i-1} \Delta_j^nM \right)\cdot  \left( \sqrt{n}^{b\wedge (1-b)} \left( \widehat{\sigma}^2_{t_{i-}^n}  -(\sigma^2_{t_{i-}^n} + b_{ t_{i-}^n,n}(u) ) \right) \right) \right] \\
		&= \E_{i-k_n-1}^n \Bigg[ \Delta_{i+}^n X \Delta_i^n M \cdot \left(  \sum_{j \in I^n_{i+} } \E_{j-1} \left[ \frac{1}{\sqrt{\kappa}}  \sqrt{n}^{0\wedge (1-2b)}  \xi_{t_{i+}^n,n}^{(2,j)}(u) + \sqrt{\kappa T} \sqrt{n}^{0\wedge (2b-1)}  \xi_{t_{i+}^n,n}^{(3,j)}(u) \right] \right) \\
		&\qquad \qquad \qquad \quad + \E\left[ \Delta_{i+}^nB \Delta_i^n M \bigg| \left( \sigma_{t_{i+}^n} , \xi_{t_{i+}^n,n}'(u) \right) \right] \cdot \sigma_{t_{i+}^n}  \xi_{t_{i+}^n,n}'(u) \Bigg] \\ 
		&\qquad \qquad \qquad \quad + \left( \Delta_{i+}^n X - \sigma_{t_{i+}^n} \Delta_{i+}^nB  \right) \Delta_i^n M \cdot \xi_{t_{i+}^n,n}'(u) \Bigg] \\ 
		&+ \E_{i-k_n-1}^n \Bigg[ \Delta_{i+}^n X  \cdot \left(  \sum_{j \in I^n_{i+} } \E_{j-1} \left[ \left( \frac{1}{\sqrt{\kappa}}  \sqrt{n}^{0\wedge (1-2b)}  \xi_{t_{i+}^n,n}^{(2,j)}(u) + \sqrt{\kappa T} \sqrt{n}^{0\wedge (2b-1)}  \xi_{t_{i+}^n,n}^{(3,j)}(u) \right) \Delta_j^n M \right] \right) \\
		&\qquad \qquad \qquad \quad + \E\left[ \Delta_{i+}^nB \bigg| \left( \sigma_{t_{i+}^n} , \xi_{t_{i+}^n,n}'(u) \right) \right] \cdot \sigma_{t_{i+}^n}  \xi_{t_{i+}^n,n}'(u) \sum_{j=i+1}^{i+k_n+1} \Delta_j^n M \Bigg] \\ 
		&\qquad \qquad \qquad \quad + \left( \Delta_{i+}^n X - \sigma_{t_{i+}^n} \Delta_{i+}^nB  \right) \cdot \xi_{t_{i+}^n,n}'(u)\cdot  \sum_{j=i+1}^{i+k_n+1} \Delta_j^n M   \Bigg] \\
		&\quad - \E_{i-k_n-1}^n \Bigg[ \E \Bigg[ \left(  \sum_{j \in I^n_{i-} } \E_{j-1} \left[ \left( \frac{1}{\sqrt{\kappa}}  \sqrt{n}^{0\wedge (1-2b)}  \xi_{t_{i-}^n,n}^{(2,j)}(u) + \sqrt{\kappa T} \sqrt{n}^{0\wedge (2b-1)}  \xi_{t_{i-}^n,n}^{(3,j)}(u) \right)  \right] \right)\\
		&\qquad \qquad \qquad \quad  \cdot  \Bigg|  \left( \Delta_{i}^n X, \Delta_{i}^nM  \right) \Bigg]  \left( \Delta_{i}^n X  \Delta_{i}^nM \right) \Bigg]\\
		&\quad - \E_{i-k_n-1}^n \Bigg[\left(  \xi_{t_{i-}^n,n}'(u)  \right) \cdot \Delta_i^n M \cdot  \left( \Delta_{i}^n X - \sigma_{t_{i-}^n} \Delta_{i}^nB  \right) \Bigg]\\
		&\quad - \E_{i-k_n-1}^n \Bigg[\left(  \xi_{t_{i-}^n,n}'(u)  \right) \cdot \Delta_i^n M \sigma_{t_{i-}^n} \Delta_{i}^nB \Bigg]\\
		&\quad - \E_{i-k_n-1}^n \Bigg[  \left(  \sqrt{n}^{b\wedge (1-b)} \left(\widehat{\sigma}^2_{t_{i-}^n} -\sigma^2_{t_{i-}^n} - b_{ t_{i-}^n,n}(u) \right)  \sum_{j=i-k_n}^{i-1} \Delta_i^n M \right) \cdot \sigma_{t_{i-}^n}  \E_{i-}^n \left[ \Delta_{i}^n B \right]\Bigg]\\
		&\quad - \E_{i-k_n-1}^n \Bigg[ \E \Bigg[ \left(  \sum_{j \in I^n_{i-} } \E_{j-1} \left[ \left( \frac{1}{\sqrt{\kappa}}  \sqrt{n}^{0\wedge (1-2b)}  \xi_{t_{i-}^n,n}^{(2,j)}(u) + \sqrt{\kappa T} \sqrt{n}^{0\wedge (2b-1)}  \xi_{t_{i-}^n,n}^{(3,j)}(u) \right) \cdot \Delta_j^n M \right] \right)\\
		&\qquad \qquad \qquad \quad  \cdot  \Bigg|  \left( \Delta_{i}^n X, \sigma_{t_{i-}^n}, \Delta_{i}^nB  \right) \Bigg]  \left( \Delta_{i}^n X - \sigma_{t_{i-}^n} \Delta_{i}^nB  \right) \Bigg]\\
		&\quad - \E_{i-k_n-1}^n \Bigg[\left(  \xi_{t_{i-}^n,n}'(u)  \right) \cdot \sum_{j=i-k_n}^{i-1} \Delta_j^n M \cdot  \left( \Delta_{i}^n X - \sigma_{t_{i-}^n} \Delta_{i}^nB  \right) \Bigg]\\
		&(\text{By \eqref{spot-main}, \eqref{X-main}, the proof of \eqref{xi-23-W}, $\E[\Delta_i^n X] = O_p(\Delta_n)$, and Holder's inequality.})\\
		& = O_p\left( \Delta_n \cdot (h_n + \sqrt{k_n\Delta_n}) \right).
	\end{align*}
	Thus
	\begin{align*}
		&\sum_{l=1}^{l_n(\tilde{m})} \E_{(l,0)}\left[ \xi(\tilde{m})_l^n \cdot \left( \Delta_{l,\tilde{m}}^n B + \Delta_{l,\tilde{m}}^n B' + \Delta_{l,\tilde{m}}^n W \right) \right] \\
		& = \sum_{l=1}^{l_n(\tilde{m})} \E_{(l,0)}\left[  \sqrt{n}^{b\wedge (1-b)} \sum_{r=k_n+1}^{(\tilde{m}+1)k_n}  s^n_{(l,r)}  \cdot  \left( \sum_{r=1}^{(\tilde{m}+2)k_n} \left( \Delta_{(l,r)}^n B + \Delta_{(l,r)}^n B' + \Delta_{(l,r)}^n W \right)  \right) \right] \\
		&= O_p\left(h_n+\sqrt{k_n\Delta_n}\right),
	\end{align*}
	which implies \eqref{B-Ls-2-3}. This proves \eqref{lev-main-2} and \eqref{thm-lev-clt}.
	
	The consistency result \eqref{thm-lev-con} is implied by \eqref{thm-lev-clt}. Moreover, if we fix $b=\frac{1}{2}$, by repeating the above proof steps, we find that it requires
	\begin{align*}
		\frac{1}{\sqrt{n}^{b \wedge (1-b)}}\left( n  \cdot\Delta_n^{1\wedge \frac{1}{(\max\{\beta,r\}+\epsilon)}} h_n \right) \rightarrow 0,
	\end{align*}
	which holds when $\max\{\beta,r\}\leq \frac{4}{3}$. 
	
	This completes the proof of Theorem \ref{thm-lev-1}. \hfill $\square$
	
	\subsection{Proof for Theorem \ref{thm-vov}}\label{prof:thm2}
	For $i=k_n+1,...,n-k_n$, we define
	\begin{align}\label{a_i^n}
		\begin{split}
			a_i^n &=  \frac{1}{\sqrt{k_n\Delta_n}} \Bigg( \frac{3\Delta_n}{2} \left( \frac{1}{\sqrt{k_n\Delta_n}}\left( \widehat{\sigma}^2_{t_{i+}^n}  -(\sigma^2_{t_{i+}^n} + b_{ t_{i+}^n,n}(u) ) \right) - \frac{1}{\sqrt{k_n\Delta_n}}\left( \widehat{\sigma}^2_{t_{i-}^n}  -(\sigma^2_{t_{i-}^n} + b_{ t_{i-}^n,n}(u) ) \right) \right)^2 \\
			&\qquad \qquad \qquad \qquad - \frac{3}{k_n^2} h_1(u,\sigma_{t_{i+}^n}^2)- 4(\sigma_{t_{i+}^n}^2 )((\tilde{\sigma}_{t_{i+}^n})^2+(\tilde{\sigma}'_{t_{i+}^n})^2)\Delta_n  \Bigg)\\
			&=: a_i^n(1) + a_i^n (2) + a_i^n(3),\\
			e_i^n &= \frac{1}{\sqrt{k_n\Delta_n}} \frac{-3}{ k_n^2}  h'_1(u, \sigma_{t_{i-k_n-1}^n}^2 ) \Bigg(  \frac{1}{\sqrt{k_n^2\Delta_n}} \sum_{j \in I^n_{i+} } \xi_{t_{i+}^n,n}^{(2,j)}(u) + \sum_{j \in I^n_{i+} } \xi_{t_{i+}^n,n}^{(3,j)}(u) \Bigg),
		\end{split}
	\end{align}
	with
	\begin{align*}
		a_i^n(1)&= \frac{1}{\sqrt{k_n\Delta_n}} \Bigg( \frac{3\Delta_n}{2} \Bigg( \Bigg(  \frac{1}{\sqrt{k_n^2\Delta_n}} \sum_{j \in I^n_{i+} } \xi_{t_{i+}^n,n}^{(2,j)}(u) + \sum_{j \in I^n_{i+} } \xi_{t_{i+}^n,n}^{(3,j)}(u) \Bigg) \\
		&\qquad \qquad \quad \qquad - \Bigg( \frac{1}{\sqrt{k_n^2 \Delta_n}} \sum_{j \in I^n_{i-} } \xi_{t_{i-}^n,n}^{(2,j)}(u) + \sum_{j \in I^n_{i-} } \xi_{t_{i-}^n,n}^{(3,j)}(u) \Bigg) \Bigg)^2 \\
		&\qquad \qquad \quad - \frac{3}{k_n^2\Delta_n} h_1(u,t_{i+}^n,\sigma_{t_{i+}^n}^2) \Delta_n - 4(\sigma_{t_{i+}^n}^2 )((\tilde{\sigma}_{t_{i+}^n})^2+(\tilde{\sigma}'_{t_{i+}^n})^2)\Delta_n  \Bigg) \\
		a_i^n(2)&= \frac{3\Delta_n}{\sqrt{k_n\Delta_n}} \Bigg( \Bigg(  \frac{1}{\sqrt{k_n^2\Delta_n}} \sum_{j \in I^n_{i+} } \xi_{t_{i+}^n,n}^{(2,j)}(u) + \sum_{j \in I^n_{i+} } \xi_{t_{i+}^n,n}^{(3,j)}(u) \Bigg) \\
		&\qquad \qquad \quad \qquad - \Bigg( \frac{1}{\sqrt{k_n^2 \Delta_n}} \sum_{j \in I^n_{i-} } \xi_{t_{i-}^n,n}^{(2,j)}(u) + \sum_{j \in I^n_{i-} } \xi_{t_{i-}^n,n}^{(3,j)}(u) \Bigg) \Bigg) \left( \xi_{t_{i+}^n,n}''(u)-\xi_{t_{i-}^n,n}''(u)  \right)\\
		a_i^n(3)& =  \frac{3\Delta_n}{\sqrt{k_n\Delta_n}}  \left( \xi_{t_{i+}^n,n}''(u)-\xi_{t_{i-}^n,n}''(u)  \right)^2,
	\end{align*}
	where
	\begin{align}\label{xi'}
		\begin{split}
			&\xi_{t_{i\pm}^n,n}''(u) = \frac{1}{\sqrt{k_n\Delta_n}} \left(\widehat{\sigma}^2_{t_{i\pm}^n} -\sigma^2_{t_{i+}^n} - b_{ t_{i\pm}^n,n}(u) \right) - \frac{1}{\sqrt{k_n^2\Delta_n}} \sum_{j \in I^n_{i\pm} } \xi_{t_{i\pm}^n,n}^{(2,j)}(u) - \sum_{j \in I^n_{i\pm} } \xi_{t_{i\pm}^n,n}^{(3,j)}(u),
		\end{split}
	\end{align}
	and $\xi_{t_{i\pm}^n,n}^{(2,j)}(u)$ and $\xi_{t_{i\pm}^n,n}^{(3,j)}(u)$ are defined as in \eqref{xi-2-3}, for $i<j\leq i+2k_n$ with fixed $i$. From \eqref{spot-main}, we have 
	\begin{align}
		\xi_{t_{i\pm}^n,n}''(u)= O_p\left( h_n \right). 
	\end{align}
	By \eqref{appro-f12} and Lemmas 2 and 6 in \cite{LLL2018}, together with the boundedness of $\sigma, \tilde{\sigma}, \tilde{\sigma}'$ and condition \eqref{cond-vov}, we obtain 
	\begin{align}\label{xi-xi'}
		\begin{split}
			\xi_{t_{i\pm}^n,n}^{(2,j)}(u) &= \xi_{i,n}^{'(2,j)}(u) + O_p\left( \frac{1}{\sqrt{k_n}} \left( \sqrt{k_n\Delta_n} + \Delta_n^{1-\frac{\beta}{2}} + \Delta_n^{\frac{1}{(\beta+\epsilon)} - \frac{1}{2}} \right) \right)\\
			\xi_{t_{i+}^n,n}^{(3,j)}(u) &= \xi_{i,n}^{'(3,j)}(u) + O_p\left( \sqrt{\Delta_n} \right),
		\end{split}
	\end{align}
	where
	\begin{align*}
		\xi_{i,n}^{'(2,j)}(u)&= \frac{  -2 \left( \cos  \left(\frac{u\sigma_{t_{(i-k_n-1)}^n}\Delta_j^nB}{\sqrt{\Delta_n}} \right) - \E_{j-1}^{n} \left[ \cos \left(\frac{u\sigma_{t_{(i-k_n-1)}^n}\Delta_j^nB}{\sqrt{\Delta_n}} \right) \right] \right) }{\sqrt{k_n} u^2 f^{(1)}_{t_{(i-k_n-1)}^n,n}(u)}, \\
		\xi_{i,n}^{'(3,j)}(u) &= \frac{ (i+k_n-j)\left(2\sigma_{t_{(i-k_n-1)}^n} \tilde{\sigma}_{t_{(i-k_n-1)}^n}\Delta_j^nB + 2\sigma_{t_{(i-k_n-1)}^n} \tilde{\sigma}'_{t_{(i-k_n-1)}^n} \Delta_j^nB'\right)    }{k_n \sqrt{k_n\Delta_n} }.
	\end{align*}
	Following the calculations in Lemma 7 of \cite{LLL2018}, we obtain
	\begin{align}\label{E-xi4}
		\begin{split}
			&\E_{i-k_n-1}^{n} \left[(\xi_{i,n}^{'(2,j)}(u) )^2\right] = \frac{1}{k_n} h_1(u,t_{i+}^n,\sigma_{t_{(i-k_n-1)}^n}^2), \qquad \E_{i-k_n-1}^{n} \left[(\xi_{i,n}^{'(2,j)}(u) )^4\right] 
			= \frac{1}{k_n^2} h_3(u,\sigma_{t_{(i-k_n-1)}^n}^2),\\
			&\E_{i-k_n-1}^{n} \left[(\xi_{i,n}^{'(2,j)}(u) )^2\frac{\left(2\sigma_{t_{(i-k_n-1)}^n} \tilde{\sigma}_{t_{(i-k_n-1)}^n}\Delta_j^nB + 2\sigma_{t_{(i-k_n-1)}^n} \tilde{\sigma}'_{t_{(i-k_n-1)}^n} \Delta_j^nB'\right)^2}{\Delta_n}   \right]\\
			&:=  \frac{1}{k_n} h_4(u,\sigma_{t_{(i-k_n-1)}^n}^2) = O_p\left( \frac{1}{k_n}\right),
		\end{split}
	\end{align}
	with
	\begin{align*}
		h_3(u,\sigma_{t_{(i-k_n-1)}^n}^2):= \frac{2 f^{(1)}_{t_{(i-k_n-1)}^n,n}(4u) - 16f^{(1)}_{t_{(i-k_n-1)}^n,n}(\sqrt{10}u) + 48f^{(1)}_{t_{(i-k_n-1)}^n,n}(\sqrt{6}u) -  40 f^{(1)}_{t_{(i-k_n-1)}^n,n}(2u) + 2}{ u^8 f^{(1)}_{t_{(i-k_n-1)}^n,n}(2u)},
	\end{align*}
	since
	\begin{align*}
		&\E_{i-k_n-1}^{n} \left[\left( \cos  \left(\frac{u\sigma_{t_{(i-k_n-1)}^n}\Delta_j^nB}{\sqrt{\Delta_n}} \right) - \E_{j-1}^{n} \left[ \cos \left(\frac{u\sigma_{t_{(i-k_n-1)}^n}\Delta_j^nB}{\sqrt{\Delta_n}} \right) \right] \right)^4 \right]\\
		&=  \E_{i-k_n-1}^{n} \left[\left( \cos  \left(\frac{u\sigma_{t_{(i-k_n-1)}^n}\Delta_j^nB}{\sqrt{\Delta_n}} \right) \right)^4 \right] - 4 \cdot f^{(1)}_{t_{(i-k_n-1)}^n,n}(u) \cdot  \E_{i-k_n-1}^{n} \left[\left( \cos  \left(\frac{u\sigma_{t_{(i-k_n-1)}^n}\Delta_j^nB}{\sqrt{\Delta_n}} \right) \right)^3 \right] \\
		&\quad +  6\cdot f^{(1)}_{t_{(i-k_n-1)}^n,n}(\sqrt{2}u) \cdot  \E_{i-k_n-1}^{n} \left[\left( \cos  \left(\frac{u\sigma_{t_{(i-k_n-1)}^n}\Delta_j^nB}{\sqrt{\Delta_n}} \right) \right)^2 \right] - 3\cdot f^{(1)}_{t_{(i-k_n-1)}^n,n}(2u)\\
		&\text{(Using $(\cos{x})^2 = \frac{\cos{2x}+1}{2}$, $(\cos{x})^3 =\frac{\cos{3x} + 3\cos{x}}{4}$ and $(\cos{x})^4 = \frac{\cos{4x} + 4\cos{2x} + 1}{8}$.)}\\
		& = \frac{1}{8} f^{(1)}_{t_{(i-k_n-1)}^n,n}(4u) - f^{(1)}_{t_{(i-k_n-1)}^n,n}(\sqrt{10}u) + 3f^{(1)}_{t_{(i-k_n-1)}^n,n}(\sqrt{6}u) - \frac{5}{2} f^{(1)}_{t_{(i-k_n-1)}^n,n}(2u) + \frac{1}{8},
	\end{align*}
	and 
	\begin{align*}
		&\E_{i-k_n-1}^{n} \left[(\xi_{i,n}^{'(2,j)}(u) )^2\frac{\left(2\sigma_{t_{(i-k_n-1)}^n} \tilde{\sigma}_{t_{(i-k_n-1)}^n}\Delta_j^nB + 2\sigma_{t_{(i-k_n-1)}^n} \tilde{\sigma}'_{t_{(i-k_n-1)}^n} \Delta_j^nB'\right)^2}{\Delta_n}   \right]\\
		&\text{(By Holder's inequality.)}\\
		&\leq \left(\E_{i-k_n-1}^{n} \left[(\xi_{i,n}^{'(2,j)}(u) )^4 \right] \E_{i-k_n-1}^{n} \left[\frac{\left(2\sigma_{t_{(i-k_n-1)}^n} \tilde{\sigma}_{t_{(i-k_n-1)}^n}\Delta_j^nB + 2\sigma_{t_{(i-k_n-1)}^n} \tilde{\sigma}'_{t_{(i-k_n-1)}^n} \Delta_j^nB'\right)^2}{\Delta_n}   \right] \right)^{1/2}\\
		&\leq \frac{1}{k_n} \left(27 h_3(u,\sigma_{t_{(i-k_n-1)}^n}^2) (h_2(\sigma_{t_{(i-k_n-1)}^n}^2))^2 \right)^{1/2}. 
	\end{align*}
	Furthermore, following the proof of Proposition \ref{pro-1}, we obtain
	\begin{align}\label{2-moment}
		&\E_{i-k_n-1}^n  \Bigg[\Bigg(   \sum_{j \in I^n_{i+} } \left( \frac{1}{\sqrt{k_n^2\Delta_n}}\xi_{i,n}^{'(2,j)}(u) + \xi_{i,n}^{'(3,j)}(u) \right) \Bigg)^2 \Bigg] =\frac{1}{k_n^2\Delta_n}h_1(u, t_{(i-k_n-1)}^n) +h_2(t_{(i-k_n-1)}^n),
	\end{align}
	and 
	\begin{align}\label{4-moment}
		\begin{split}
			&\E_{i-k_n-1}^n  \Bigg[\Bigg(   \sum_{j \in I^n_{i+} } \left( \frac{1}{\sqrt{k_n^2\Delta_n}}\xi_{i,n}^{'(2,j)}(u) + \xi_{i,n}^{'(3,j)}(u) \right) \Bigg)^4 \Bigg] \\
			&\text{(Noting that the functions $x(\cos{(cx)})^{k}$ and $x^{k}\cos{(cx)}$ are odd for any constant $c$ and $k\in \mathbb{N}^{+}$.)}\\
			&= \E_{i-k_n-1}^n  \Bigg[  \sum_{j \in I^n_{i+} } \Bigg(  \frac{1}{\sqrt{k_n^2\Delta_n}}\xi_{i,n}^{'(2,j)}(u)\Bigg)^4 + 12\sum_{j_1,j_2 \in I^n_{i+} } \sum_{j_1\neq j_2} \Bigg(\Bigg(  \frac{1}{\sqrt{k_n^2\Delta_n}}\xi_{i,n}^{'(2,j_1)}(u)\Bigg )^2 \Bigg(  \frac{1}{\sqrt{k_n^2\Delta_n}}\xi_{i,n}^{'(2,j_2)}(u)\Bigg )^2   \Bigg)   \Bigg] \\
			& \quad +\E_{i-k_n-1}^n  \Bigg[ 6\sum_{j_1 \in I^n_{i+}}\sum_{j_2 \in I^n_{i+} } \Bigg(\Bigg(  \frac{1}{\sqrt{k_n^2\Delta_n}} \xi_{i,n}^{'(2,j_1)}(u)\Bigg)^2 \Bigg( \xi_{i,n}^{'(3,j_2)}(u)\Bigg)^2 \Bigg)  \Bigg]\\
			&\quad  +  \E_{i-k_n-1}^n  \Bigg[  \sum_{j \in I^n_{i+} } \Bigg(\Bigg( \xi_{i,n}^{'(3,j)}(u)\Bigg)^4 + 12\sum_{j_1,j_2 \in I^n_{i+} }\sum_{j_1\neq j_2} \Bigg(\Bigg( \xi_{i,n}^{'(3,j_1)}(u)\Bigg)^2 \Bigg( \xi_{i,n}^{'(3,j_2)}(u)\Bigg)^2 \Bigg)  \Bigg] \\
			&= \frac{12}{k_n^4\Delta_n^2}(h_1(u,t_{(i-k_n-1)}^n))^2  + \frac{6}{k_n^2\Delta_n}h_1(u,t_{(i-k_n-1)}^n)h_2(t_{(i-k_n-1)}^n)  + 12 (h_2(t_{(i-k_n-1)}^n) )^2+ O_p\left(\frac{1}{k_n}\right).
		\end{split}
	\end{align}
	The same results can be proved similarly when $j \in I^n_{i-}$.
	
	Before proving Theorem \ref{thm-vov}, we provide the following lemmas and their proofs. 
	
	\begin{lem}\label{lem-vov-2}
		Under the same assumptions and conditions as in Theorem \ref{thm-vov}, for $i=k_n+1,...,n-k_n$ and $m=3,4,...$, we have  
		\begin{align}\label{lem-vov-2-1}
			\E_{i-k_n-1}^n \left[a^n_{i}  \right] = O_p\left(\frac{\sqrt{\Delta_n}}{\sqrt{k_n}}\left(u^2 (k_n\Delta_n)^{\frac{3}{4}}+ |u|^{\beta-2}\Delta_n^{1-\frac{\beta}{2}} + h_n^2 \right) \right), 
		\end{align}
		and 
		\begin{align}\label{lem-vov-2-4}
			\begin{split}
				\E_{i-k_n-1}^n \left[\left( a^n_{i} \right)^2\right] &=\frac{\Delta_n}{k_n} \left( \frac{117}{2} \frac{(h_1(u,t_{i-k_n-1}^n))^2}{k_n^4\Delta_n^2}+ 36 \frac{h_1(u,t_{i-k_n-1}^n) h_2(t_{i-k_n-1}^n)}{k_n^2\Delta_n} + \frac{117}{2} (h_2(t_{i-k_n-1}^n))^2 \right) \\
				&\quad + O_p\left(\frac{\Delta_n h_n}{k_n}\right),
			\end{split}
		\end{align}
		and
		\begin{align}\label{lem-vov-2-4'}
			&	  \E_{i-k_n-1}^n \left[\left( a^n_{i} \right)^m\right] = O_p\left( \frac{\Delta_n^{m/2}}{k_n^{m/2}}\right), \  \E_{i-k_n-1}^n \left[e^n_{i} \right] \equiv 0, \ \E_{i-k_n-1}^n \left[\left( e^n_{i} \right)^m\right] = O_p\left( \frac{\Delta_n^{m/2}}{k_n^{m/2}}\right), \\\label{lem-vov-e1}
			&\E_{i-k_n-1}^n \left[\left( e^n_{i} \right)^2\right] = \frac{\Delta_n}{k_n} \cdot \frac{9 (h'_1(u, \sigma_{t_{i-k_n-1}^n}^2 ) )^2}{(k_n^2 \Delta_n)^2} \left(  \frac{h_1(u,t_{i-k_n-1}^n)}{k_n^2\Delta_n} + h_2(t_{i-k_n-1}^n) \right)+  O_p\left(\frac{\Delta_n h_n}{k_n}\right).
		\end{align}
		Furthermore, for $j=k_n+1,...,n-k_n$ with $i<j$, we have
		\begin{align}\label{lem-vov-2-6}
			\E_{i-k_n-1}^n \left[ ( a^n_{i} + e^n_{i} ) \cdot (a^n_{j} + e^n_{j})\right] \leq C\frac{\Delta_n}{k_n},
		\end{align}
		and 
		\begin{align}\label{lem-vov-2-3}
			\begin{split}
				\sum_{j=i+1}^{i+k_n}\E_{i-k_n-1}^n \left[\left( a^n_{i} \cdot a^n_{j} \right)\right]&=  \left( \frac{9}{2} \frac{(h_1(u,t_{i-k_n-1}^n))^2}{k_n^4\Delta_n^2}+ \frac{197}{40}\frac{h_1(u,t_{i-k_n-1}^n) h_2(t_{i-k_n-1}^n)}{k_n^2\Delta_n} + \frac{153}{35} (h_2(t_{i-k_n-1}^n))^2 \right)  \Delta_n \\
				&\quad +O_p(\Delta_n^{\frac{3}{2}} +\Delta_n h_n),
			\end{split}
		\end{align}
		and
		\begin{align}\label{lem-vov-2-5}
			\begin{split}
				\sum_{j=i+k_n+1}^{i+2k_n} \E_{i-k_n-1}^n \left[\left( a^n_{i} \cdot a^n_{j} \right)\right]=& \left( \frac{9}{4} \frac{(h_1(u,t_{i-k_n-1}^n))^2}{k_n^4\Delta_n^2}+ \frac{63}{16} \frac{h_1(u,t_{i-k_n-1}^n) h_2(t_{i-k_n-1}^n)}{k_n^2\Delta_n} + \frac{45}{28} (h_2(t_{i-k_n-1}^n))^2 \right)  \Delta_n \\
				&+O_p(\Delta_n^{\frac{3}{2}} +\Delta_n h_n),
			\end{split}
		\end{align}
		and 
		\begin{align}\label{a_e_i^n}
			\begin{split}
				&\sum_{j=i+1}^{i+k_n}\E_{i-k_n-1}^n \left[\left( a^n_{i} \cdot e^n_{j} \right)\right]= O_p( \Delta_n h_n), \ \sum_{j=i+1}^{i+k_n}\E_{i-k_n-1}^n \left[\left( e^n_{i} \cdot a^n_{j} \right)\right]= O_p( \Delta_n h_n),\\
				&\sum_{j=i+1}^{i+k_n}\E_{i-k_n-1}^n \left[\left( e^n_{i} \cdot e^n_{j} \right)\right]= \frac{9(h'_1(u,t_{i-k_n-1}^n))^2}{k_n^4\Delta_n^2} \left( \frac{1}{2}\frac{h_1(u,t_{i-k_n-1}^n) }{k_n^2\Delta_n} + \frac{1}{4} h_2(t_{i-k_n-1}^n) \right)  \Delta_n +O_p(\Delta_n\sqrt{k_n\Delta_n}),\\
				&\sum_{j=i+k_n+1}^{i+2k_n}\E_{i-k_n-1}^n \left[\left( e^n_{i} \cdot a^n_{j} \right)\right] =O_p( \Delta_n h_n), \ 	\sum_{j=i+k_n+1}^{i+2k_n} \left( \E_{i-k_n-1}^n \left[\left( a^n_{i} \cdot e^n_{j} \right)\right] + \E_{i-k_n-1}^n \left[\left( e^n_{i} \cdot e^n_{j} \right)\right] \right) \equiv 0,
			\end{split}
		\end{align}
		and, if $j-i > 2k_n$, we have  
		\begin{align}\label{lem-vov-2-2}
			\E_{i-k_n-1}^n \left[ ( a^n_{i} + e^n_{i} ) \cdot (a^n_{j} + e^n_{j})\right] = O_p\left(\frac{\Delta_n}{k_n}\left(u^2 (k_n\Delta_n)^{\frac{3}{4}}+ |u|^{\beta-2}\Delta_n^{1-\frac{\beta}{2}} + h_n^2\right)^2\right).
		\end{align}
	\end{lem}
	\textbf{Proof of Lemma \ref{lem-vov-2}:} For \eqref{lem-vov-2-1}, we write $a_i^n =a_i^n(1) + a_i^n (2) + a_i^n(3)$.
	By the proof of Lemma \ref{lem-lev-1} and the results in Lemma \ref{lem-vov-approx}, we have 
	\begin{align}\label{E-a_i(1)}
		\E_{i-k_n-1}^n \left[\left( a^n_{i}(1)  \right) \right] = O_p\left(\frac{\sqrt{\Delta_n}}{\sqrt{k_n}}\left(u^2 (k_n\Delta_n)^{\frac{3}{4}}+ |u|^{\beta-2}\Delta_n^{1-\frac{\beta}{2}}\right)\right),
	\end{align}
	and
	\begin{align*}
		&\E_{i-k_n-1}^n \left[\left( a^n_{i}(2)  \right) \right] \\
		&=\frac{3\Delta_n}{\sqrt{k_n\Delta_n}}  \E_{i-k_n-1}^n \Bigg[ \E\Bigg[\Bigg( \Bigg(  \frac{1}{\sqrt{k_n^2\Delta_n}} \sum_{j \in I^n_{i+} } \xi_{t_{i+}^n,n}^{(2,j)}(u) + \sum_{j \in I^n_{i+} } \xi_{t_{i+}^n,n}^{(3,j)}(u) \Bigg) \\
		&\qquad - \Bigg( \frac{1}{\sqrt{k_n^2 \Delta_n}} \sum_{j \in I^n_{i-} } \xi_{t_{i-}^n,n}^{(2,j)}(u) + \sum_{j \in I^n_{i-} } \xi_{t_{i-}^n,n}^{(3,j)}(u) \Bigg) \Bigg) \Bigg| \left( \xi_{t_{i+}^n,n}''(u), \xi_{t_{i-}^n,n}''(u)  \right)  \Bigg] \left( \xi_{t_{i+}^n,n}''(u)-\xi_{t_{i-}^n,n}''(u)  \right) \Bigg]\\
		&\equiv 0.
	\end{align*}
	We have $\E_{i-k_n-1}^n \left[\left( a^n_{i}(3)  \right) \right] = O_p(\frac{h_n^2\sqrt{\Delta_n}}{\sqrt{k_n}})$ since $\xi_{t_{i\pm}^n,n}''(u)= O_p\left( h_n \right)$. These results, together with \eqref{spot-main}, yield \eqref{lem-vov-2-1}. \\
	For \eqref{lem-vov-2-4}, from the proof of \eqref{lem-vov-2-1}, we have
	\begin{align}\label{res:a2a3}
		a^n_{i}(3) = O_p\left(\frac{h_n^2\sqrt{\Delta_n}}{\sqrt{k_n}}\right), \qquad a^n_{i}(2)= O_p\left( \frac{\sqrt{\Delta_n}h_n}{\sqrt{k_n}} \right),
	\end{align}
	since Holder's inequality and \eqref{E-a_i(1)} give
	\begin{align*}
		&\E_{i-k_n-1}^n \left[\left| a^n_{i}(2) \right| \right] \\
		&\leq  \frac{C\sqrt{\Delta_n}}{\sqrt{k_n}} \Bigg(\E_{i-k_n-1}^n  \Bigg[ \Bigg( \Bigg(  \frac{1}{\sqrt{k_n^2\Delta_n}} \sum_{j \in I^n_{i+} } \xi_{t_{i+}^n,n}^{(2,j)}(u) + \sum_{j \in I^n_{i+} } \xi_{t_{i+}^n,n}^{(3,j)}(u) \Bigg) \\
		&\qquad \qquad \quad - \Bigg( \frac{1}{\sqrt{k_n^2 \Delta_n}} \sum_{j \in I^n_{i-} } \xi_{t_{i-}^n,n}^{(2,j)}(u) + \sum_{j \in I^n_{i-} } \xi_{t_{i-}^n,n}^{(3,j)}(u) \Bigg) \Bigg)^2 \Bigg] \cdot \E_{i-k_n-1}^n  \left[ \left( \xi_{t_{i+}^n,n}''(u)-\xi_{t_{i-}^n,n}''(u)  \right)^2 \right] \Bigg)^{1/2}\\
		&\leq C\frac{\sqrt{\Delta_n}h_n}{\sqrt{k_n}}.  
	\end{align*}
	Moreover, we have 
	\begin{align*}
		&\E_{i-k_n-1}^n \left[\left( a^n_{i}(1) \right)^2\right]\\
		& = \frac{\Delta_n}{k_n} \cdot  \E_{i-k_n-1}^n  \Bigg[\Bigg( \frac{3}{2} \Bigg(  \frac{1}{\sqrt{k_n^2\Delta_n}} \sum_{j \in I^n_{i+} } \xi_{t_{i+}^n,n}^{(2,j)}(u) + \sum_{j \in I^n_{i+} } \xi_{t_{i+}^n,n}^{(3,j)}(u) - \frac{1}{\sqrt{k_n^2 \Delta_n}} \sum_{j \in I^n_{i-} } \xi_{t_{i-}^n,n}^{(2,j)}(u) - \sum_{j \in I^n_{i-} } \xi_{t_{i-}^n,n}^{(3,j)}(u)  \Bigg)^2 \\
		&\qquad \qquad \qquad - \frac{3}{k_n^2\Delta_n} h_1(u, t_{i+}^n,\sigma_{t_{i+}^n}^2) - 4(\sigma_{t_{i+}^n}^2 )((\tilde{\sigma}_{t_{i+}^n})^2+(\tilde{\sigma}'_{t_{i+}^n})^2) \Bigg)^2 \Bigg]\\
		&\text{(By \eqref{xi-xi'}.)}\\
		&= \frac{\Delta_n}{k_n} \cdot  \E_{i-k_n-1}^n  \Bigg[\Bigg( \frac{3}{2} \Bigg(  \sum_{j \in I^n_{i+} } \left( \frac{1}{\sqrt{k_n^2\Delta_n}}\xi_{i,n}^{'(2,j)}(u) + \xi_{i,n}^{'(3,j)}(u) \right)-\sum_{j \in I^n_{i-} }  \left(\frac{1}{\sqrt{k_n^2 \Delta_n}} \xi_{i,n}^{'(2,j)}(u) + \xi_{t_{i-}^n,n}^{(3,j)}(u)\right)  \Bigg)^2 \\
		&\qquad \qquad \qquad - \frac{3}{k_n^2\Delta_n} h_1(u, \sigma_{t_{i-k_n-1}^n}^2) - 4(\sigma_{t_{i-k_n-1}^n}^2 )((\tilde{\sigma}_{t_{i-k_n-1}^n})^2+(\tilde{\sigma}'_{t_{i-k_n-1}^n})^2) \Bigg)^2 \Bigg]  +O_p\left(\frac{\Delta_n^{\frac{3}{2}}}{k_n}\right)\\
		&\text{(Following the calculation in the proof of Lemma \ref{lem-lev-1}.)}\\
		&=  \frac{9\Delta_n}{4k_n} \cdot  \E_{i-k_n-1}^n  \Bigg[\Bigg(   \sum_{j \in I^n_{i+} } \left( \frac{1}{\sqrt{k_n^2\Delta_n}}\xi_{i,n}^{'(2,j)}(u) + \xi_{i,n}^{'(3,j)}(u) \right)-\sum_{j \in I^n_{i-} }  \left(\frac{1}{\sqrt{k_n^2 \Delta_n}} \xi_{i,n}^{'(2,j)}(u) + \xi_{t_{i-}^n,n}^{(3,j)}(u)\right)  \Bigg)^4\Bigg] \\
		&\quad - \frac{\Delta_n}{k_n}\Bigg( \frac{3}{k_n^2\Delta_n} h_1(u, \sigma_{t_{i-k_n-1}^n}^2) + 4(\sigma_{t_{i-k_n-1}^n}^2 )((\tilde{\sigma}_{t_{i-k_n-1}^n})^2+(\tilde{\sigma}'_{t_{i-k_n-1}^n})^2) \Bigg)^2   +O_p\left(\frac{\Delta_n^{\frac{3}{2}}}{k_n}\right)\\
		&=  \frac{9\Delta_n}{4k_n} \cdot \E_{i-k_n-1}^n  \Bigg[\Bigg(   \sum_{j \in I^n_{i+} } \left( \frac{1}{\sqrt{k_n^2\Delta_n}}\xi_{i,n}^{'(2,j)}(u) + \xi_{i,n}^{'(3,j)}(u) \right) \Bigg)^4 + \Bigg(\sum_{j \in I^n_{i-} }  \left(\frac{1}{\sqrt{k_n^2 \Delta_n}} \xi_{i,n}^{'(2,j)}(u) + \xi_{t_{i-}^n,n}^{(3,j)}(u)\right)  \Bigg)^4\Bigg] \\
		&+  \frac{27\Delta_n}{2k_n} \cdot  \E_{i-k_n-1}^n  \Bigg[\Bigg(   \sum_{j \in I^n_{i+} } \left( \frac{1}{\sqrt{k_n^2\Delta_n}}\xi_{i,n}^{'(2,j)}(u) + \xi_{i,n}^{'(3,j)}(u) \right) \Bigg)^2 \cdot  \Bigg(\sum_{j \in I^n_{i-} }  \left(\frac{1}{\sqrt{k_n^2 \Delta_n}} \xi_{i,n}^{'(2,j)}(u) + \xi_{t_{i-}^n,n}^{(3,j)}(u)\right)  \Bigg)^2\Bigg] \\
		&- \frac{\Delta_n}{k_n}\Bigg( \frac{3}{k_n^2\Delta_n} h_1(u, \sigma_{t_{i-k_n-1}^n}^2) + 4(\sigma_{t_{i-k_n-1}^n}^2 )((\tilde{\sigma}_{t_{i-k_n-1}^n})^2+(\tilde{\sigma}'_{t_{i-k_n-1}^n})^2) \Bigg)^2   +O_p\left(\frac{\Delta_n^{\frac{3}{2}}}{k_n}\right)\\
		&\text{(Using \eqref{2-moment} and \eqref{4-moment}.)}\\
		&=\frac{\Delta_n}{k_n} \left( \frac{117}{2} \frac{(h_1)^2}{k_n^4\Delta_n^2}+ 36 \frac{h_1 h_2}{k_n^2\Delta_n} + \frac{117}{2} (h_2)^2 \right) + O_p\left(\frac{\Delta_n^{\frac{3}{2}}}{k_n}\right).
	\end{align*}
	The above results directly yield the conclusion for $\E_{i-k_n-1}^n \left[\left( a^n_{i} \right)^2\right]$ in \eqref{lem-vov-2-4}.  
	Similarly, we obtain \eqref{lem-vov-2-4'}. The results in \eqref{lem-vov-e1} for $e_i^n$ follow directly from the proof of Proposition \ref{pro-1}. The result in \eqref{lem-vov-2-6} follows directly from Holder's inequality, \eqref{lem-vov-2-4}, and \eqref{lem-vov-e1}. \\
	For \eqref{lem-vov-2-3}, let $k:=j-i$ with $k\in\{2,...,k_n-1\}$. From the proof of \eqref{lem-vov-2-4}, we have 
	\begin{align*}
		\E_{i-k_n-1}^n \left[\left( a^n_{i} \cdot a^n_{i+k} \right)\right] &= \sum_{s_1=1}^{3} \sum_{s_2=1}^{3} \E_{i-k_n-1}^n \left[\left( a^n_{i}(s_1) \cdot a^n_{i+k}(s_2) \right)\right]\\
		&= \E_{i-k_n-1}^n \left[\left( a^n_{i}(1) \cdot a^n_{i+k}(1) \right)\right] + O_p\left(\frac{\Delta_n}{k_n} h_n\right). 
	\end{align*}
	For $\E_{i-k_n-1}^n \left[\left( a^n_{i}(1) \cdot a^n_{i+k}(1) \right)\right]$, we define 
	\begin{align*}
		&I_i^n(1)= \{i-k_n,...,i+k-k_n-1\}, \ I_i^n(2)=\{i+k-k_n,...,i-1\},\  I_i^n(3)=\{i\}, \\ &I_i^n(4)=\{i+1,...,i+k-1\},\ I_i^n(5)=\{i+k\}, \ I_i^n(6)=\{i+k+1,...,i+k_n\}, \\
		& I_i^n(7)=\{i+k_n+1, i+k+k_n\},
	\end{align*}
	and 
	\begin{align*}
		&A_{i,n}^{+}(j)= \left( \frac{1}{\sqrt{k_n^2\Delta_n}}\xi_{i,n}^{'(2,j)}(u) + \xi_{i,n}^{'(3,j)}(u) \right), \quad A_{i,n}^{-}(j)= \left(\frac{1}{\sqrt{k_n^2 \Delta_n}} \xi_{i,n}^{'(2,j)}(u) + \xi_{t_{i-}^n,n}^{(3,j)}(u)\right) \\
		&B_{i,n}^{+}(j) = \left( \frac{1}{\sqrt{k_n^2\Delta_n}}\xi_{i+k,n}^{'(2,j)}(u) + \xi_{i+k,n}^{'(3,j)}(u) \right), \quad B_{i,n}^{-}(j) =  \left(\frac{1}{\sqrt{k_n^2 \Delta_n}} \xi_{i+k,n}^{'(2,j)}(u) + \xi_{t_{{(i+k)}-}^n,n}^{(3,j)}(u)\right).
	\end{align*}
	Following the proof of \eqref{lem-vov-2-4}, we have 
	\begin{align*}
		&\E_{i-k_n-1}^n \left[\left( a^n_{i}(1) \cdot a^n_{i+k}(1) \right)\right] \\
		&=  \frac{9\Delta_n}{4k_n} \cdot  \E_{i-k_n-1}^n  \Bigg[\Bigg(   \sum_{m=4,5,6} \sum_{j \in I^n_{i}(m) } \left( A_{i,n}^{+}(j)\right)-\sum_{m=1,2}\sum_{j \in I^n_{i}(m) }  \left(A_{i,n}^{-}(j) \right)  \Bigg)^2\\
		&\qquad \qquad \qquad \qquad \quad   \cdot \Bigg(   \sum_{m=6,7}\sum_{j \in I^n_{i}(m) } \left( B_{i,n}^{+}(j) \right)  -\sum_{m=2,3,4}\sum_{j \in I^n_{i}(m) }  \left(B_{i,n}^{-}(j) \right)  \Bigg)^2 \Bigg] \\
		&\quad - \frac{\Delta_n}{k_n}\Bigg( \frac{3}{k_n^2\Delta_n} h_1(u, \sigma_{t_{i-k_n-1}^n}^2) + 4(\sigma_{t_{i-k_n-1}^n}^2 )((\tilde{\sigma}_{t_{i-k_n-1}^n})^2+(\tilde{\sigma}'_{t_{i-k_n-1}^n})^2) \Bigg)^2   +O_p\left(\frac{\Delta_n^{\frac{3}{2}}}{k_n}\right)\\
		&=\frac{9\Delta_n}{4k_n} \cdot  \E_{i-k_n-1}^n  \Bigg[\sum_{j_1,j_2\in I_i^n(2)} \sum_{j_1\neq j_2}( (A_{i,n}^{-}(j_1))^2 (B_{i,n}^{-}(j_2))^2) + \sum_{j\in I_i^n(2)} ( (A_{i,n}^{-}(j))^2 (B_{i,n}^{-}(j))^2)  \\
		&\qquad +  2 \sum_{j_1,j_2\in I_i^n(2)} \sum_{j_1\neq j_2} ( (A_{i,n}^{-}(j_1))(A_{i,n}^{-}(j_2))) (B_{i,n}^{-}(j_2)) (B_{i,n}^{-}(j_2)) \\
		&\qquad  +  \left(\sum_{j \in I_i^n(2)} (A_{i,n}^{-}(j))^2 \right)  \cdot \left(\sum_{j \in I_i^n(4)} (B_{i,n}^{-}(j))^2 + \sum_{j \in I_i^n(6)} (B_{i,n}^{+}(j))^2\right)  \\
		&\qquad +  4 \left(\sum_{j \in I_i^n(2)} (A_{i,n}^{-}(j) B_{i,n}^{-}(j) ) \right)  \cdot \left( \sum_{j \in I_i^n(6)} (A_{i,n}^{+}(j)B_{i,n}^{+}(j))-\sum_{j \in I_i^n(4)} (A_{i,n}^{+}(j) B_{i,n}^{-}(j)) \right) \\
		& \qquad + \left( \sum_{j \in I_i^n(4) \cup I_i^n(6)} (A_{i,n}^{+}(j))^2 \right) \cdot \left( \sum_{j \in I_i^n(2)} (B_{i,n}^{-}(j))^2  \right)\\
		& \qquad  + \left( \sum_{j_1,j_2 \in I_i^n(4)} \sum_{j_1\neq j_2} ((A_{i,n}^{+}(j_1))^2 (B_{i,n}^{-}(j_2))^2) \right) + \left( \sum_{j \in I_i^n(4)} ((A_{i,n}^{+}(j))^2 (B_{i,n}^{-}(j))^2) \right) \\
		&\qquad + 2 \left( \sum_{j_1,j_2 \in I_i^n(4)} \sum_{j_1\neq j_2} ((A_{i,n}^{+}(j_1))(A_{i,n}^{+}(j_2)) (B_{i,n}^{-}(j_1)) (B_{i,n}^{-}(j_2))) \right) \\
		& \qquad + \left( \sum_{j_1,j_2 \in I_i^n(6)} \sum_{j_1\neq j_2} ((A_{i,n}^{+}(j_1))^2 (B_{i,n}^{+}(j_2))^2) \right) + \left( \sum_{j \in I_i^n(6)} ((A_{i,n}^{+}(j))^2 (B_{i,n}^{+}(j))^2) \right) \\
		&\qquad + 2 \left( \sum_{j_1,j_2 \in I_i^n(6)} \sum_{j_1\neq j_2} ((A_{i,n}^{+}(j_1))(A_{i,n}^{+}(j_2)) (B_{i,n}^{+}(j_1)) (B_{i,n}^{+}(j_2))) \right) \\
		& \qquad + \left( \sum_{j \in I_i^n(4) } (A_{i,n}^{+}(j))^2 \right) \cdot \left( \sum_{j \in I_i^n(6)} (B_{i,n}^{+}(j))^2  \right) + \left( \sum_{j \in I_i^n(6)} (A_{i,n}^{+}(j))^2 \right) \cdot \left( \sum_{j \in I_i^n(4)} (B_{i,n}^{-}(j))^2  \right)\\
		&\qquad - 4\left( \sum_{j \in I_i^n(4) } (A_{i,n}^{+}(j) B_{i,n}^{-}(j) ) \right) \cdot \left( \sum_{j \in I_i^n(6)} (A_{i,n}^{+}(j)  B_{i,n}^{+}(j))  \right)  \\
		& \qquad + \left( \sum_{j \in I_i^n(4) \cup I_i^n(6) } (A_{i,n}^{+}(j))^2 + \sum_{j \in I_i^n(2) } (A_{i,n}^{-}(j))^2 \right) \cdot \left( \sum_{j \in I_i^n(3) } (B_{i,n}^{-}(j))^2 + \sum_{j \in I_i^n(7) } (B_{i,n}^{+}(j))^2 \right)  \\
		& \qquad + \left( \sum_{j \in I_i^n(5) } (A_{i,n}^{+}(j))^2 + \sum_{j \in I_i^n(1) } (A_{i,n}^{-}(j))^2 \right) \cdot \left( \sum_{m=2,3,4}\sum_{j \in I_i^n(m) } (B_{i,n}^{-}(j))^2 + \sum_{m=6,7} \sum_{j \in I_i^n(m) } (B_{i,n}^{+}(j))^2 \right)  \Bigg]  \\
		&\quad - \frac{\Delta_n}{k_n}\Bigg( \frac{3}{k_n^2\Delta_n} h_1(u, \sigma_{t_{i-k_n-1}^n}^2) + 4(\sigma_{t_{i-k_n-1}^n}^2 )((\tilde{\sigma}_{t_{i-k_n-1}^n})^2+(\tilde{\sigma}'_{t_{i-k_n-1}^n})^2) \Bigg)^2   +O_p\left(\frac{\Delta_n^{\frac{3}{2}}}{k_n}\right)\\
		&\text{(Using \eqref{E-xi4}. For convenience, we use the shorthand notation} \\
		&\text{ $h_1, h_2, h_3, h_4$ for  $h_1(u, t_{i-k_n-1}^n), h_2(t_{i-k_n-1}^n), h_3(u, t_{i-k_n-1}^n), h_4(u, t_{i-k_n-1}^n)$ hereafter.)}\\
		&=\frac{9\Delta_n}{4k_n} \cdot \Bigg(\\
		&\qquad \Big( \Big(\frac{(k_n-k)h_1}{k_n(k_n^2\Delta_n)} + \frac{3h_2}{k_n^3} \sum_{s=1}^{k_n-k}s^2\Big)\Big(\frac{(k_n-k)h_1}{k_n(k_n^2\Delta_n)} + \frac{3h_2}{k_n^3} \sum_{s=k+1}^{k_n}s^2\Big) \\
		&\qquad \qquad - \sum_{s=1}^{k_n-k}\Big(\frac{h_1}{k_n(k_n^2\Delta_n)} + \frac{3h_2s^2}{k_n^3} \Big)\Big(\frac{h_1}{k_n(k_n^2\Delta_n)} + \frac{3h_2(k+s)^2}{k_n^3}\Big)  \Big)\\
		&\qquad + \Big( \frac{(k_n-k)h_3}{k_n^2(k_n^2\Delta_n)^2} + \frac{h_4}{k_n^4(k_n^2\Delta_n)} \sum_{s=k+1}^{k_n} s^2 + \frac{4h_4}{k_n^4(k_n^2\Delta_n)} \sum_{s=1}^{k_n-k} ((k+s)s) \\
		&\qquad \qquad + \frac{h_4}{k_n^4(k_n^2\Delta_n)} \sum_{s=1}^{k_n-k} s^2 + \frac{27h_2^2}{k_n^6} \sum_{s=1}^{k_n-k} ((k+s)^2s^2)   \Big) \\
		&\qquad + 2\Big( \Big(\sum_{s=1}^{k_n-k} \Big( \frac{h_1}{k_n(k_n^2\Delta_n)} + \frac{3h_2(s(k+s))}{k_n^3}\Big) \Big)^2 - \sum_{s=1}^{k_n-k} \Big(  \Big( \frac{h_1}{k_n(k_n^2\Delta_n)} + \frac{3h_2(s(k+s))}{k_n^3}\Big)^2 \Big) \Big) \\
		& \qquad + \Big(\frac{(k_n-k)h_1}{k_n(k_n^2\Delta_n)} + \frac{3h_2}{k_n^3} \sum_{s=1}^{k_n-k}s^2\Big) \Big(\frac{(k_n-1)h_1}{k_n(k_n^2\Delta_n)} + \frac{3h_2}{k_n^3} \Big(\Big(\sum_{s=1}^{k_n}s^2\Big) - k^2\Big)\Big)  \\
		&+ 4\Big(  \Big(\frac{(k_n-k)h_1}{k_n(k_n^2\Delta_n)} + \frac{3h_2}{k_n^3} \sum_{s=1}^{k_n-k}((k+s)s)\Big)  \Big(\frac{(k_n-2k)h_1}{k_n(k_n^2\Delta_n)} + \frac{3h_2}{k_n^3} \Big(\sum_{s=1}^{k_n-k}((k+s)s)-\sum_{s=1}^{k-1}((k_n-k+1+s)s)  \Big)\Big)  \Big) \\
		& \qquad + \Big( \Big(\frac{(k_n-1)h_1}{k_n(k_n^2\Delta_n)} + \frac{3h_2}{k_n^3}\Big(\Big(\sum_{s=1}^{k_n}s^2\Big)-(k_n-k+1)^2\Big) \Big) \Big(\frac{(k_n-k)h_1}{k_n(k_n^2\Delta_n)} + \frac{3h_2}{k_n^3} \sum_{s=k+1}^{k_n}s^2\Big) \\
		&\qquad \qquad - \sum_{j=i+1}^{i+k_n}\Big(\frac{h_1}{k_n(k_n^2\Delta_n)} + \frac{3h_2(i+k_n+1-j)^2}{k_n^3} \Big)\Big(\frac{h_1}{k_n(k_n^2\Delta_n)} + \frac{3h_2(i+k-j)^2}{k_n^3}\Big)  \Big)\\
		& \qquad + \Big( \Big(\frac{(k-1)h_1}{k_n(k_n^2\Delta_n)} + \frac{3h_2}{k_n^3} \sum_{s=k_n-k+2}^{k_n}s^2\Big)\Big(\frac{(k+1)h_1}{k_n(k_n^2\Delta_n)} + \frac{3h_2}{k_n^3} \sum_{s=1}^{k-1}s^2\Big)  \\
		&\qquad \qquad - \sum_{s=1}^{k-1}\Big(\frac{h_1}{k_n(k_n^2\Delta_n)} + \frac{3h_2s^2}{k_n^3} \Big)\Big(\frac{h_1}{k_n(k_n^2\Delta_n)} + \frac{3h_2(k_n+1-k+s)^2}{k_n^3}\Big)  \Big)\\
		&\qquad + \Big( \frac{(k-1)h_3}{k_n^2(k_n^2\Delta_n)^2} + \frac{h_4}{k_n^4(k_n^2\Delta_n)} \sum_{s=1}^{k-1} s^2 + \frac{4h_4}{k_n^4(k_n^2\Delta_n)} \sum_{s=1}^{k-1} ((k_n-k+1+s)s) \\
		&\qquad \qquad + \frac{h_4}{k_n^4(k_n^2\Delta_n)} \sum_{s=1}^{k-1} (k_n+1-k+s)^2 + \frac{27h_2^2}{k_n^6} \sum_{s=1}^{k-1} ((k_n+1-k+s)^2s^2)   \Big) \\
		&\qquad + 2\Big( \Big(\sum_{s=1}^{k-1} \Big( \frac{h_1}{k_n(k_n^2\Delta_n)} + \frac{3h_2(s+k_n+1-k)s}{k_n^3}\Big) \Big)^2 \\
		&\qquad \qquad  - \sum_{s=1}^{k-1} \Big(  \Big( \frac{h_1}{k_n(k_n^2\Delta_n)} + \frac{3h_2(s+k_n+1-k)s}{k_n^3}\Big)^2 \Big) \Big) \\
		& \qquad + \Big( \Big(\frac{(k_n-k)h_1}{k_n(k_n^2\Delta_n)} + \frac{3h_2}{k_n^3} \sum_{s=1}^{k_n-k}s^2\Big)\Big(\frac{(k_n-k)h_1}{k_n(k_n^2\Delta_n)} + \frac{3h_2}{k_n^3} \sum_{s=1}^{k_n-k}(k+s)^2\Big)  \\
		&\qquad \qquad - \sum_{j=1}^{k_n-k}\Big(\frac{h_1}{k_n(k_n^2\Delta_n)} + \frac{3h_2s^2}{k_n^3} \Big)\Big(\frac{h_1}{k_n(k_n^2\Delta_n)} + \frac{3h_2(k+s)^2}{k_n^3}\Big)  \Big)\\
		&\qquad + \Big( \frac{(k_n-k)h_3}{k_n^2(k_n^2\Delta_n)^2} + \frac{h_4}{k_n^4(k_n^2\Delta_n)} \sum_{s=1}^{k_n-k} (k+s)^2 + \frac{4h_4}{k_n^4(k_n^2\Delta_n)} \sum_{s=1}^{k_n-k} ((k+s)s) \\
		&\qquad \qquad + \frac{h_4}{k_n^4(k_n^2\Delta_n)} \sum_{s=1}^{k_n-k} s^2 + \frac{27h_2^2}{k_n^6} \sum_{s=1}^{k_n-k} ((k+s)^2s^2)   \Big) \\
		&\qquad + 2\Big( \Big(\sum_{s=1}^{k_n-k} \Big( \frac{h_1}{k_n(k_n^2\Delta_n)} + \frac{3h_2(k+s)s}{k_n^3}\Big) \Big)^2 - \sum_{s=1}^{k_n-k} \Big(  \Big( \frac{h_1}{k_n(k_n^2\Delta_n)} + \frac{3h_2(k+s)s}{k_n^3}\Big)^2 \Big) \Big) \\
		&\qquad + \Big( \frac{(k-1)h_1}{k_n(k_n^2\Delta_n)} + \frac{3h_2}{k_n^3}\Big(\sum_{s=k_n-k+2}^{k_n} s^2 \Big)\Big) \Big( \frac{(k_n-k)h_1}{k_n(k_n^2\Delta_n)} + \frac{3h_2}{k_n^3}\Big(\sum_{s=k+1}^{k_n} s^2 \Big)\Big)\\
		&\qquad + \Big( \frac{(k-1)h_1}{k_n(k_n^2\Delta_n)} + \frac{3h_2}{k_n^3}\Big(\sum_{s=1}^{k-1} s^2 \Big)\Big) \Big( \frac{(k_n-k)h_1}{k_n(k_n^2\Delta_n)} + \frac{3h_2}{k_n^3}\Big(\sum_{s=1}^{k_n-k} s^2 \Big)\Big)\\
		&\qquad - 4  \Big( \frac{(k-1)h_1}{k_n(k_n^2\Delta_n)} + \frac{3h_2}{k_n^3}\Big(\sum_{s=1}^{k-1} (k_n+1-k+s)s \Big)\Big) \Big( \frac{(k_n-k)h_1}{k_n(k_n^2\Delta_n)} + \frac{3h_2}{k_n^3}\Big(\sum_{s=1}^{k_n-k} (k+s)s \Big)\Big)\\
		&+ \Big( \frac{(2k_n-k-1)h_1}{k_n(k_n^2\Delta_n)} + \frac{3h_2}{k_n^3}\Big(\sum_{s=1}^{k_n-k} (s^2) +\sum_{s=1}^{k_n} (s^2) - (k_n-k+1)^2 \Big)\Big) \Big( \frac{(k+1)h_1}{k_n(k_n^2\Delta_n)} + \frac{3h_2}{k_n^3}\Big(\Big(\sum_{s=1}^{k}s^2 \Big) + k^2\Big)\Big)\\
		&\qquad + \Big( \frac{(k+1)h_1}{k_n(k_n^2\Delta_n)} + \frac{3h_2}{k_n^3}\Big(\Big(\sum_{s=k_n+1-k}^{k_n} s^2\Big) + (k_n-k+1)^2\Big)\Big) \Bigg( \frac{2}{k_n^2\Delta_n} h_1 + 2 h_2 \Bigg)\\
		&\quad \Bigg)- \frac{\Delta_n}{k_n}\Bigg( \frac{3h_1}{k_n^2\Delta_n} + 3h_2 \Bigg)^2   +O_p\left(\frac{\Delta_n^{\frac{3}{2}}}{k_n}\right). 
	\end{align*}
	Applying Cauchy's inequality and Holder's inequality together with \eqref{lem-vov-2-4}, we have
	\begin{align*} 
		\E_{i-k_n-1}^n[a_i a_{i+1} + a_i a_{i+k_n}] \leq C\sqrt{\E_{i-k_n-1}^n[a_i^2] \cdot \E_{i-k_n-1}^n[a_{i+1}^2+a_{i+k_n}^2]} \leq  C\frac{\Delta_n}{k_n}.
	\end{align*}
	Summing the above results yields
	\begin{align*}
		&\sum_{j=i+1}^{i+k_n} \E_{i-k_n-1}^n \left[\left( a^n_{i} \cdot a^n_{j} \right)\right]=\\
		&=\sum_{k=2}^{k_n-1} \Bigg( \frac{9\Delta_n}{4k_n} \cdot  \Bigg(\\
		&\qquad  \Big( \Big(\frac{(k_n-k)h_1}{k_n(k_n^2\Delta_n)} + \frac{3h_2}{k_n^3} \sum_{s=1}^{k_n-k}s^2\Big)\Big(\frac{(k_n-k)h_1}{k_n(k_n^2\Delta_n)} + \frac{3h_2}{k_n^3} \sum_{s=k+1}^{k_n}s^2\Big) \\
		&\qquad +2 \Big(\sum_{s=1}^{k_n-k} \Big( \frac{h_1}{k_n(k_n^2\Delta_n)} + \frac{3h_2(s(k+s))}{k_n^3}\Big) \Big)^2  \\
		& \qquad + \Big(\frac{(k_n-k)h_1}{k_n(k_n^2\Delta_n)} + \frac{3h_2}{k_n^3} \sum_{s=1}^{k_n-k}s^2\Big) \Big(\frac{(k_n-1)h_1}{k_n(k_n^2\Delta_n)} + \frac{3h_2}{k_n^3} \Big(\Big(\sum_{s=1}^{k_n}s^2\Big) - k^2\Big)\Big)   \\
		&+ 4\Big(  \Big(\frac{(k_n-k)h_1}{k_n(k_n^2\Delta_n)} + \frac{3h_2}{k_n^3} \sum_{s=1}^{k_n-k}((k+s)s)\Big)  \Big(\frac{(k_n-2k)h_1}{k_n(k_n^2\Delta_n)} + \frac{3h_2}{k_n^3} \Big(\sum_{s=1}^{k_n-k}(k+s)s-\sum_{s=1}^{k-1}(k_n-k+1+s)s \Big)\Big)  \Big) \\
		& \qquad + \Big( \Big(\frac{(k_n-1)h_1}{k_n(k_n^2\Delta_n)} + \frac{3h_2}{k_n^3} \Big(\sum_{s=1}^{k_n}s^2\Big)-(k_n-k+1)^2\Big)\Big(\frac{(k_n-k)h_1}{k_n(k_n^2\Delta_n)} + \frac{3h_2}{k_n^3} \sum_{s=k+1}^{k_n}s^2\Big)  \\
		& \qquad + \Big( \Big(\frac{(k-1)h_1}{k_n(k_n^2\Delta_n)} + \frac{3h_2}{k_n^3} \sum_{s=k_n-k+2}^{k_n}s^2\Big)\Big(\frac{(k+1)h_1}{k_n(k_n^2\Delta_n)} + \frac{3h_2}{k_n^3} \sum_{s=1}^{k-1}s^2\Big)  \\
		&\qquad + 2 \Big(\sum_{s=1}^{k-1} \Big( \frac{h_1}{k_n(k_n^2\Delta_n)} + \frac{3h_2(s+k_n+1-k)s}{k_n^3}\Big) \Big)^2  \\
		& \qquad + \Big( \Big(\frac{(k_n-k)h_1}{k_n(k_n^2\Delta_n)} + \frac{3h_2}{k_n^3} \sum_{s=1}^{k_n-k}s^2\Big)\Big(\frac{(k_n-k)h_1}{k_n(k_n^2\Delta_n)} + \frac{3h_2}{k_n^3} \sum_{s=1}^{k_n-k}(k+s)^2\Big)   \\
		&\qquad + 2 \Big(\sum_{s=1}^{k_n-k} \Big( \frac{h_1}{k_n(k_n^2\Delta_n)} + \frac{3h_2(k+s)s}{k_n^3}\Big) \Big)^2 \\
		&\qquad  + \Big( \frac{(k-1)h_1}{k_n(k_n^2\Delta_n)} + \frac{3h_2}{k_n^3}\Big(\sum_{j=k_n-k+2}^{k_n} s^2 \Big)\Big) \Big( \frac{(k_n-k)h_1}{k_n(k_n^2\Delta_n)} + \frac{3h_2}{k_n^3}\Big(\sum_{j=k+1}^{k_n} s^2 \Big)\Big)\\
		&\qquad + \Big( \frac{(k-1)h_1}{k_n(k_n^2\Delta_n)} + \frac{3h_2}{k_n^3}\Big(\sum_{s=1}^{k-1} s^2 \Big)\Big) \Big( \frac{(k_n-k)h_1}{k_n(k_n^2\Delta_n)} + \frac{3h_2}{k_n^3}\Big(\sum_{s=1}^{k_n-k} s^2 \Big)\Big)\\
		&\qquad - 4  \Big( \frac{(k-1)h_1}{k_n(k_n^2\Delta_n)} + \frac{3h_2}{k_n^3}\Big(\sum_{s=1}^{k-1} (k_n+1-k+s)s \Big)\Big) \Big( \frac{(k_n-k)h_1}{k_n(k_n^2\Delta_n)} + \frac{3h_2}{k_n^3}\Big(\sum_{s=1}^{k_n-k} (k+s)s \Big)\Big)\\
		&+ \Big( \frac{(2k_n-k-1)h_1}{k_n(k_n^2\Delta_n)} + \frac{3h_2}{k_n^3}\Big(\sum_{s=1}^{k_n-k} (s^2) +\sum_{s=1}^{k_n} (s^2) - (k_n-k+1)^2 \Big)\Big) \Big( \frac{(k+1)h_1}{k_n(k_n^2\Delta_n)} + \frac{3h_2}{k_n^3}\Big(\sum_{s=1}^{k} (s^2) + k^2\Big)\Big)\\
		&\qquad + \Big( \frac{(k+1)h_1}{k_n(k_n^2\Delta_n)} + \frac{3h_2}{k_n^3}\Big(\sum_{s=k_n+1-k}^{k_n} (s^2) + (k_n-k+1)^2\Big)\Big) \Bigg( \frac{2}{k_n^2\Delta_n} h_1 + 2 h_2 \Bigg) \Bigg)\\
		&\qquad - \frac{\Delta_n}{k_n}\Bigg( \frac{3h_1}{k_n^2\Delta_n} + 3h_2 \Bigg)^2   +O_p\left(\frac{\Delta_n^{\frac{3}{2}}}{k_n} + \frac{\Delta_n }{k_n} h_n \right) \Bigg) + O_p\left(\frac{\Delta_n}{k_n}\right)\\
		&= \frac{9\Delta_n}{4}\Bigg( \Big(\frac{1}{3} \frac{h_1^2}{(k_n^2\Delta_n)^2} + \frac{13}{20} \frac{h_1}{k_n^2\Delta_n}h_2 + \frac{17}{70} h_2^2 \Big) + 2\Big(\frac{1}{3} \frac{h_1^2}{(k_n^2\Delta_n)^2} + \frac{11}{20} \frac{h_1}{k_n^2\Delta_n}h_2 + \frac{33}{140} h_2^2 \Big) \\
		&\quad + \Big(\frac{1}{2} \frac{h_1^2}{(k_n^2\Delta_n)^2} + \frac{3}{4} \frac{h_1}{k_n^2\Delta_n}h_2 + \frac{1}{4} h_2^2 \Big)  +4  \Big(\frac{1}{6} \frac{h_1^2}{(k_n^2\Delta_n)^2} + \frac{7}{20} \frac{h_1}{k_n^2\Delta_n}h_2 + \frac{103}{560} h_2^2 \Big) \\
		&\quad + \Big(\frac{1}{2} \frac{h_1^2}{(k_n^2\Delta_n)^2} + \frac{5}{4} \frac{h_1}{k_n^2\Delta_n}h_2 + \frac{3}{4} h_2^2 \Big)  +  \Big(\frac{1}{3} \frac{h_1^2}{(k_n^2\Delta_n)^2} + \frac{13}{20} \frac{h_1}{k_n^2\Delta_n}h_2 + \frac{17}{70} h_2^2 \Big) \\
		&\quad + 2 \Big(\frac{1}{3} \frac{h_1^2}{(k_n^2\Delta_n)^2} + \frac{1}{4} \frac{h_1}{k_n^2\Delta_n}h_2 + \frac{33}{140} h_2^2 \Big) + \Big(\frac{1}{3} \frac{h_1^2}{(k_n^2\Delta_n)^2} + \frac{13}{20} \frac{h_1}{k_n^2\Delta_n}h_2 + \frac{17}{70} h_2^2 \Big) \\
		&\quad +  2 \Big(\frac{1}{3} \frac{h_1^2}{(k_n^2\Delta_n)^2} + \frac{11}{20} \frac{h_1}{k_n^2\Delta_n}h_2 + \frac{33}{140} h_2^2 \Big) + \Big(\frac{1}{6} \frac{h_1^2}{(k_n^2\Delta_n)^2} + \frac{3}{5} \frac{h_1}{k_n^2\Delta_n}h_2 + \frac{71}{140} h_2^2 \Big) \\
		&\quad + \Big(\frac{1}{6} \frac{h_1^2}{(k_n^2\Delta_n)^2} + \frac{1}{10} \frac{h_1}{k_n^2\Delta_n}h_2 + \frac{1}{140} h_2^2 \Big) - 4 \Big(\frac{1}{6} \frac{h_1^2}{(k_n^2\Delta_n)^2} + \frac{1}{5} \frac{h_1}{k_n^2\Delta_n}h_2 + \frac{29}{560} h_2^2 \Big) \\
		&\quad + \Big(\frac{2}{3} \frac{h_1^2}{(k_n^2\Delta_n)^2} + \frac{17}{20} \frac{h_1}{k_n^2\Delta_n}h_2 + \frac{9}{35} h_2^2 \Big) +\Big(2 \frac{h_1}{k_n^2\Delta_n} + 2 h_2 \Big)\Big( \frac{1}{2} \frac{h_1}{k_n^2\Delta_n}+ \frac{3}{4}h_2 \Big) \Bigg) \\
		&\quad -  \Delta_n \Bigg( \frac{3h_1}{k_n^2\Delta_n} + 3h_2 \Bigg)^2 +O_p\left(\Delta_n^{\frac{3}{2}} +\Delta_n h_n + \frac{\Delta_n}{k_n}\right)\\
		&=\left( \frac{9}{2} \frac{h_1^2}{(k_n^2\Delta_n)^2} + \frac{197}{40} \frac{h_1}{k_n^2\Delta_n}h_2 + \frac{153}{35} h_2^2   \right)\Delta_n +O_p(\Delta_n^{\frac{3}{2}} +\Delta_n h_n). 
	\end{align*}
	For \eqref{lem-vov-2-5}, the proof of \eqref{lem-vov-2-3} shows that it is enough to calculate $\E_{i-k_n-1}^n \left[\left( a^n_{i}(1) \cdot a^n_{i+k}(1) \right)\right]$ for $k=k_n+1,...,2k_n$ and sum the results, since the remaining terms are negligible. Similarly, we define 
	\begin{align*}
		&I_i^n(1)= \{i-k_n,...,i-1\}, \ I_i^n(2)=\{i+1,...,i+k-k_n\},\  I_i^n(3)=\{i+k-k_n+1,...,i+k_n\}, \\ &I_i^n(4)=\{i+k_n+1,...,i+k-1\},\ I_i^n(5)=\{i+k+1,...,i+k+k_n\},
	\end{align*}
	and 
	\begin{align*}
		&A_{i,n}^{+}(j)= \left( \frac{1}{\sqrt{k_n^2\Delta_n}}\xi_{i,n}^{'(2,j)}(u) + \xi_{i,n}^{'(3,j)}(u) \right), \quad A_{i,n}^{-}(j)= \left(\frac{1}{\sqrt{k_n^2 \Delta_n}} \xi_{i,n}^{'(2,j)}(u) + \xi_{t_{i-}^n,n}^{(3,j)}(u)\right), \\
		&B_{i,n}^{+}(j) = \left( \frac{1}{\sqrt{k_n^2\Delta_n}}\xi_{i+k,n}^{'(2,j)}(u) + \xi_{i+k,n}^{'(3,j)}(u) \right), \quad B_{i,n}^{-}(j) =  \left(\frac{1}{\sqrt{k_n^2 \Delta_n}} \xi_{i+k,n}^{'(2,j)}(u) + \xi_{t_{{(i+k)}-}^n,n}^{(3,j)}(u)\right).
	\end{align*}
	Following the proof of \eqref{lem-vov-2-3}, we have 
	\begin{align*}
		&\E_{i-k_n-1}^n \left[\left( a^n_{i}(1) \cdot a^n_{i+k}(1) \right)\right] \\
		&=  \frac{9\Delta_n}{4k_n} \cdot  \E_{i-k_n-1}^n  \Bigg[\Bigg(   \sum_{m=2,3} \sum_{j \in I^n_{i}(m) } \left( A_{i,n}^{+}(j)\right)-\sum_{j \in I^n_{i}(1) }  \left(A_{i,n}^{-}(j) \right)  \Bigg)^2\\
		&\qquad \qquad \qquad \qquad \quad   \cdot \Bigg(   \sum_{j \in I^n_{i}(5) } \left( B_{i,n}^{+}(j) \right)  -\sum_{m=3,4}\sum_{j \in I^n_{i}(m) }  \left(B_{i,n}^{-}(j) \right)  \Bigg)^2 \Bigg] \\
		&\quad - \frac{\Delta_n}{k_n}\Bigg( \frac{3}{k_n^2\Delta_n} h_1(u, \sigma_{t_{i-k_n-1}^n}^2) + 4(\sigma_{t_{i-k_n-1}^n}^2 )((\tilde{\sigma}_{t_{i-k_n-1}^n})^2+(\tilde{\sigma}'_{t_{i-k_n-1}^n})^2) \Bigg)^2   +O_p\left(\frac{\Delta_n^{\frac{3}{2}}}{k_n}\right)\\
		&=\frac{9\Delta_n}{4k_n} \cdot  \E_{i-k_n-1}^n  \Bigg[\sum_{j_1,j_2\in I_i^n(3)} \sum_{j_1\neq j_2}( (A_{i,n}^{+}(j_1))^2 (B_{i,n}^{-}(j_2))^2) + \sum_{j\in I_i^n(3)} ( (A_{i,n}^{+}(j))^2 (B_{i,n}^{-}(j))^2)  \\
		&\qquad +  2 \sum_{j_1,j_2\in I_i^n(3)} \sum_{j_1\neq j_2} ( (A_{i,n}^{+}(j_1))(A_{i,n}^{+}(j_2))) (B_{i,n}^{-}(j_1)) (B_{i,n}^{-}(j_2)) \\
		&\qquad  +  \left(\sum_{j \in I_i^n(3)} (A_{i,n}^{+}(j))^2 \right)  \cdot \left(\sum_{j \in I_i^n(5)} (B_{i,n}^{+}(j))^2 + \sum_{j \in I_i^n(4)} (B_{i,n}^{-}(j))^2\right)  \\
		& \qquad + \left( \sum_{j \in I_i^n(2) } (A_{i,n}^{+}(j))^2 + \sum_{j \in I_i^n(1) } (A_{i,n}^{-}(j))^2 \right) \cdot \left(   \sum_{j \in I^n_{i}(5) } \left( B_{i,n}^{+}(j) \right)^2  +\sum_{m=3,4}\sum_{j \in I^n_{i}(m) }  \left(B_{i,n}^{-}(j) \right)^2  \right)  \Bigg]  \\
		&\quad - \frac{\Delta_n}{k_n}\Bigg( \frac{3}{k_n^2\Delta_n} h_1(u, \sigma_{t_{i-k_n-1}^n}^2) + 4(\sigma_{t_{i-k_n-1}^n}^2 )((\tilde{\sigma}_{t_{i-k_n-1}^n})^2+(\tilde{\sigma}'_{t_{i-k_n-1}^n})^2) \Bigg)^2   +O_p\left(\frac{\Delta_n^{\frac{3}{2}}}{k_n}\right)\\
		&=\frac{9\Delta_n}{4k_n} \cdot \Bigg( \Big( \Big(\frac{(2k_n-k)h_1}{k_n(k_n^2\Delta_n)} + \frac{3h_2}{k_n^3} \sum_{s=1}^{2k_n-k}(2k_n-k-s+1)^2\Big)\Big(\frac{(2k_n-k)h_1}{k_n(k_n^2\Delta_n)} + \frac{3h_2}{k_n^3} \sum_{s=1}^{2k_n-k}(k_n-s)^2\Big) \\
		&\qquad \qquad - \sum_{s=1}^{2k_n-k}\Big(\frac{h_1}{k_n(k_n^2\Delta_n)} + \frac{3h_2(2k_n-k-s+1)^2}{k_n^3} \Big)\Big(\frac{h_1}{k_n(k_n^2\Delta_n)} + \frac{3h_2(k_n-s)^2}{k_n^3}\Big)  \Big)\\
		&\qquad + \Big( \frac{(2k_n-k)h_3}{k_n^2(k_n^2\Delta_n)^2} + \frac{h_4}{k_n^4(k_n^2\Delta_n)} \sum_{s=1}^{2k_n-k} (k_n-s)^2 + \frac{4h_4}{k_n^4(k_n^2\Delta_n)} \sum_{s=1}^{2k_n-k} ((k_n-s)(2k_n-k-s+1)) \\
		&\qquad \qquad + \frac{h_4}{k_n^4(k_n^2\Delta_n)} \sum_{s=1}^{2k_n-k} (2k_n-k-s+1)^2+ \frac{27h_2^2}{k_n^6} \sum_{s=1}^{2k_n-k} ((k_n-s)^2(2k_n-k-s+1)^2)   \Big) \\
		&\qquad + 2\Big( \Big(\sum_{s=1}^{2k_n-k} \Big( \frac{h_1}{k_n(k_n^2\Delta_n)} + \frac{3h_2((k_n-s)(2k_n-k-s+1))}{k_n^3}\Big) \Big)^2 \\
		&\qquad - \sum_{s=1}^{2k_n-k} \Big(  \Big( \frac{h_1}{k_n(k_n^2\Delta_n)} + \frac{3h_2((k_n-s)(2k_n-k-s+1))}{k_n^3}\Big)^2 \Big) \Big) \\
		& \qquad + \Big(\frac{(2k_n-k)h_1}{k_n(k_n^2\Delta_n)} + \frac{3h_2}{k_n^3} \sum_{s=1}^{2k_n-k}(2k_n-k-s+1)^2\Big)  \Big(\frac{(2k_n-1)h_1}{k_n(k_n^2\Delta_n)} + \frac{3h_2}{k_n^3} \Big(\sum_{s=1}^{k_n}s^2+ \sum_{s=1}^{k_n-1}s^2 \Big)\Big)  \\
		&\qquad + \Big( \frac{kh_1}{k_n(k_n^2\Delta_n)} + \frac{3h_2}{k_n^3}\Big( \sum_{s=1}^{k-k_n}(k_n+1-s)^2 + \sum_{s=1}^{k_n}(k_n+1-s)^2 \Big)\Big) \Bigg( \frac{2}{k_n^2\Delta_n} h_1 + 2 h_2 \Bigg)  \Bigg)\\
		&\quad  - \frac{\Delta_n}{k_n}\Bigg( \frac{3h_1}{k_n^2\Delta_n} + 3h_2 \Bigg)^2   +O_p\left(\frac{\Delta_n^{\frac{3}{2}}}{k_n}\right). 
	\end{align*}
	Summing up these results for $k=k_n+1,...,2k_n-1$ yields
	\begin{align*}
		&\sum_{j=i+k_n+1}^{i+2k_n-1} \E_{i-k_n-1}^n \left[\left( a^n_{i} \cdot a^n_{j} \right)\right]\\
		&= \frac{9\Delta_n}{4}\Bigg( \Big(\frac{1}{3} \frac{h_1^2}{(k_n^2\Delta_n)^2} + \frac{13}{20} \frac{h_1}{k_n^2\Delta_n}h_2 + \frac{17}{70} h_2^2 \Big) + 2\Big(\frac{1}{3} \frac{h_1^2}{(k_n^2\Delta_n)^2} + \frac{11}{20} \frac{h_1}{k_n^2\Delta_n}h_2 + \frac{33}{140} h_2^2 \Big) \\
		&\quad + \Big( \frac{h_1^2}{(k_n^2\Delta_n)^2} + \frac{3}{2} \frac{h_1}{k_n^2\Delta_n}h_2 + \frac{1}{2} h_2^2 \Big) +\Big(2 \frac{h_1}{k_n^2\Delta_n} + 2 h_2 \Big)\Big( \frac{3}{2} \frac{h_1}{k_n^2\Delta_n}+ \frac{7}{4}h_2 \Big) \Bigg) \\
		&\quad -  \Delta_n \Bigg( \frac{3h_1}{k_n^2\Delta_n} + 3h_2 \Bigg)^2 +O_p\left(\Delta_n^{\frac{3}{2}} +\Delta_n h_n + \frac{\Delta_n}{k_n}\right)\\
		&=\left( \frac{9}{4} \frac{h_1^2}{(k_n^2\Delta_n)^2} + \frac{63}{16} \frac{h_1}{k_n^2\Delta_n}h_2 + \frac{45}{28} h_2^2   \right)\Delta_n +O_p(\Delta_n^{\frac{3}{2}} +\Delta_n h_n),
	\end{align*}
	where we recall that $h_n$ is defined in \eqref{h_n}.
	After calculations similar to those above and an application of Holder's inequality with \eqref{E-xi4}, we obtain \eqref{a_e_i^n}.
	The result in \eqref{lem-vov-2-2} follows directly from successive conditioning, as used to prove \eqref{lem-lev-2-2} in Lemma \ref{lem-lev-2}, together with \eqref{lem-vov-2-1}, since there are no overlapping terms between $a_i^n+e_i^n$ and $a_j^n+e_j^n$.
	
	This completes the proof of Lemma \ref{lem-vov-2}.
	\hfill $\square$
	
	\textbf{Proof of Theorem  \ref{thm-vov}: }
	We focus on the proof of the central limit theorem \eqref{thm-vov-clt} for $b=\frac{1}{2}$ and show that 
	\begin{align}\label{thm2:b12}
		\frac{1}{\sqrt{k_n\Delta_n}} \left(\widehat{VoV}_{[0,T]} - VoV_{[0,T]} \right) \longrightarrow^{L_s} \frac{1}{\sqrt{\kappa T}}\cdot W,
	\end{align}
	with
	\begin{align*}	
		&\text{Var}\left(\frac{1}{\sqrt{\kappa T} } \cdot W|\mathcal{F}\right)\\
		&= 9\frac{(h'_1(u,t,\sigma_{t}^2))^2(h_1(u,t,\sigma_{t}^2))}{\kappa^6T^3} +  \frac{9}{2} \frac{(h'_1(u,t,\sigma_{t}^2))^2h_2(t,\sigma_{t}^2,(\tilde{\sigma}_{t})^2, (\tilde{\sigma}'_{t})^2)}{\kappa^4T^2}  + \frac{27}{2} \frac{(h_1(u,t,\sigma_{t}^2))^2}{\kappa^4T^2} \\
		&\quad + \frac{709}{40} \frac{h_1(u,t,\sigma_{t}^2)}{\kappa^2T}h_2(t,\sigma_{t}^2,(\tilde{\sigma}_{t})^2, (\tilde{\sigma}'_{t})^2) + \frac{837}{70} (h_2(t, \sigma_{t}^2,(\tilde{\sigma}_{t})^2, (\tilde{\sigma}'_{t})^2))^2. 
	\end{align*}
	We decompose 
	\begin{align*}
		&\frac{1}{\sqrt{k_n\Delta_n}} \left(\widehat{VoV}_{[0,T]} - VoV_{[0,T]} \right) \\
		&=\frac{1}{\sqrt{k_n\Delta_n}}\sum_{i=k_n+1}^{n-k_n} \left(\frac{3}{2k_n} \left( \widehat{\sigma}^2_{t_{i+}^n} - \widehat{\sigma}^2_{t_{i-}^n} \right)^2 - \frac{3}{k_n^2} h_1(u,t_{i+}^n,\widehat{\sigma}_{t_{i+}^n}^2) - 4(\sigma_{t_{i+}^n}^2 )((\tilde{\sigma}_{t_{i+}^n})^2+(\tilde{\sigma}'_{t_{i+}^n})^2)\Delta_n \right)\\
		&\quad + \frac{1}{\sqrt{k_n\Delta_n}} \left(\sum_{i=k_n+1}^{n-k_n} \left( 4(\sigma_{t_{i+}^n}^2 )((\tilde{\sigma}_{t_{i+}^n})^2+(\tilde{\sigma}'_{t_{i+}^n})^2) \Delta_n \right) - \int_{0}^{T} 4(\sigma_{t}^2 )((\tilde{\sigma}_{t})^2+(\tilde{\sigma}'_{t})^2) dt \right)\\
		& =  \frac{1}{\sqrt{k_n\Delta_n}} \sum_{i=k_n+1}^{n-k_n} \Big( \frac{3}{2k_n} \left( \left( \widehat{\sigma}^2_{t_{i+}^n}  -(\sigma^2_{t_{i+}^n} + b_{ t_{i+}^n,n}(u) ) \right) - \left( \widehat{\sigma}^2_{t_{i-}^n}  -(\sigma^2_{t_{i-}^n} + b_{ t_{i-}^n,n}(u) ) \right) \right)^2 \\
		&\qquad \qquad \qquad \qquad - \frac{3}{k_n^2\Delta_n} h_1(u,t_{i+}^n, \sigma_{t_{i+}^n}^2)\Delta_n - 4(\sigma_{t_{i+}^n}^2 )((\tilde{\sigma}_{t_{i+}^n})^2+(\tilde{\sigma}'_{t_{i+}^n})^2)\Delta_n  \Big)\\
		&\quad + \frac{1}{\sqrt{k_n\Delta_n}}  \sum_{i=k_n+1}^{n-k_n} \Big( \frac{3}{k_n} \left( \left( \widehat{\sigma}^2_{t_{i+}^n}  -(\sigma^2_{t_{i+}^n} + b_{ t_{i+}^n,n}(u) ) \right) - \left( \widehat{\sigma}^2_{t_{i-}^n}  -(\sigma^2_{t_{i-}^n} + b_{ t_{i-}^n,n}(u) ) \right) \right) \\
		& \qquad \qquad \quad \ \cdot  \left( (\sigma^2_{t_{i+}^n} + b_{ t_{i+}^n,n}(u) )   - (\sigma^2_{t_{i-}^n} + b_{ t_{i-}^n,n}(u) )  \right) \Big) \\
		& \quad +\frac{1}{\sqrt{k_n\Delta_n}} \sum_{i=k_n+1}^{n-k_n}  \frac{3}{2k_n} \left( (\sigma^2_{t_{i+}^n} + b_{ t_{i+}^n,n}(u) )   - (\sigma^2_{t_{i-}^n} + b_{ t_{i-}^n,n}(u) )  \right)^2 \\
		&\quad + \frac{1}{\sqrt{k_n\Delta_n}} \left(\sum_{i=k_n+1}^{n-k_n} \left( 4(\sigma_{t_{i+}^n}^2 )((\tilde{\sigma}_{t_{i+}^n})^2+(\tilde{\sigma}'_{t_{i+}^n})^2) \Delta_n \right) - \int_{0}^{T} 4(\sigma_{t}^2 )((\tilde{\sigma}_{t})^2+(\tilde{\sigma}'_{t})^2) dt \right)\\
		&\quad + \frac{1}{\sqrt{k_n\Delta_n}} \left(\frac{3}{k_n^2\Delta_n}  \sum_{i=k_n+1}^{n-k_n} \left(h_1(u,t_{i+}^n, \sigma_{t_{i+}^n}^2) - h_1(u,t_{i+}^n, \widehat{\sigma}_{t_{i+}^n}^2) \right)\Delta_n  \right)  \\
		&:= A^n + B^n + C^n+D^n+E^n.
	\end{align*}
	Using the boundedness of $\sigma,\tilde{\sigma},\tilde{\sigma}'$ and the conditions \eqref{cond-vov}, we obtain 
	\begin{align*}
		C^n = O_p\left( \frac{1}{\sqrt{k_n}} \cdot \frac{1}{\sqrt{k_n^2 \Delta_n}}  \right), \quad D^n = O_p \left( \sqrt{k_n\Delta_n} + \frac{1}{\sqrt{k_n}} \right),
	\end{align*}
	Thus, $C^n+D^n\longrightarrow^p 0$.
	For $B^n$, we define 
	\begin{align*}
		B'^n =  \frac{1}{\sqrt{k_n\Delta_n}}  \sum_{i=k_n+1}^{n-k_n} \Big( \frac{3}{k_n} \left( \left( \widehat{\sigma}^2_{t_{i+}^n}  -(\sigma^2_{t_{i+}^n} + b_{ t_{i+}^n,n}(u) ) \right) - \left( \widehat{\sigma}^2_{t_{i-}^n}  -(\sigma^2_{t_{i-}^n} + b_{ t_{i-}^n,n}(u) ) \right) \right) \left( (\sigma^2_{t_{i+}^n} - \sigma^2_{t_{i-}^n} ) \right) \Big).
	\end{align*}
	Condition \eqref{cond-vov} and Lemma \ref{lem-lev-1} imply $B^n - B'^n= O_p\left( \frac{ \Delta_n^{\frac{1-\beta}{2}}}{k_n} \right)$. Following the proof of Lemma \ref{lem-lev-2} (directly replacing $\Delta_{i}^n X$ in $s_i^n$ with $\Delta_{i}^n \sigma^2$), we obtain $\E[B'^n] = O_p(\frac{1}{k_n})$ and $\E[(B'^n)^2 ] = O_p(\frac{1}{k_n^2})$. These results and Chebyshev's inequality imply that $B^n  \longrightarrow^{p} 0$. For $A_n+E^n$, recall that $a_i^n, e_i^n$ are defined as in \eqref{a_i^n}; applying Taylor's expansion and using condition \eqref{cond-vov} and Lemma \ref{lem-vov-approx}, we obtain 
	\begin{align*}
		\frac{1}{\sqrt{k_n\Delta_n}} \left(\frac{3}{k_n^2\Delta_n}  \left(h_1(u,t_{i+}^n, \sigma_{t_{i+}^n}^2) - h_1(u,t_{i+}^n, \widehat{\sigma}_{t_{i+}^n}^2) \right)\Delta_n  \right) = e_i^n + O_p\left( \frac{\sqrt{k_n\Delta_n}+h_n}{k_n^2} + \frac{|u|^{\beta-2}\Delta_n^{1-\frac{\beta}{2}}}{\sqrt{k_n\Delta_n}k_n^2} \right). 
	\end{align*}
	Thus, to prove \eqref{thm2:b12}, it remains to show
	\begin{align}\label{thm-vov-main}
		\sum_{i=k_n+1}^{n-k_n} (a_i^n + e_i^n)   \longrightarrow^{L_s} \frac{1}{\sqrt{\kappa T} } \cdot W.
	\end{align}
	As in the proof of Theorem \ref{thm-lev-1}, we split the sum over $i$ into big blocks of size $(\tilde{m}+2) k_n$, with $\tilde{m}$ satisfying
	\begin{align} \label{con-m}
		\tilde{m} \rightarrow  \infty, \quad \tilde{m}k_n\Delta_n \rightarrow 0.
	\end{align}
	We define $I(\tilde{m},n,l)= (l-1)(\tilde{m}+2)k_n + 1$ for $l=1,...,l_n(\tilde{m})$, where the total number of big blocks is $l_n(\tilde{m}) = \lfloor \frac{\lfloor T/\Delta_n\rfloor-1}{(\tilde{m}+2 )k_n} \rfloor$. In the $l$-th big block, for $t_i^n$ with $I(\tilde{m},n,l)+k_n + 1 \leq i\leq I(\tilde{m},n,l)  +(\tilde{m}+1)k_n$, the increments $\Delta_{j}^n X$ constituting $a_i^n, e_i^n$ lie within $I(\tilde{m},n,l)$, so $(a_i^n,e_i^n)$ is $\mathcal{F}_{I(\tilde{m},n,l+1) \Delta_n}$-measurable. For simplicity, we use $(l,j)$ for the time point $(I(\tilde{m},n,l) + j) \Delta_n$ or its index $(I(\tilde{m},n,l) + j)$, and $\E^n_{(l,j)}[\cdot]$ for the conditional expectation $\E[\cdot | \mathcal{F}_{(l,j)}]$, with $l=1,...,l_n(\tilde{m})$ and $j$ any integer.
	We decompose $\sum_{i=k_n+1}^{n-k_n} (a_i^n+e_i^n) = W(\tilde{m})^n  +  \tilde{W}(\tilde{m})^n  + \tilde{W}'(\tilde{m})^n$ with
	\begin{align*}
		&\zeta(\tilde{m})_l^n =\sum_{r=k_n+1}^{(\tilde{m}+1)k_n}  (a^n_{(l,r)} + e^n_{(l,r)}), \quad \tilde{\zeta}(\tilde{m})_l^n = \sum_{r=-k_n}^{k_n} (a^n_{(l,r)}+e^n_{(l,r)}),\\
		&W(\tilde{m})^n = \sum_{l=1}^{l_n(\tilde{m})} \zeta(\tilde{m})_l^n, \quad \tilde{W}(\tilde{m})^n = \sum_{l=2}^{l_n(\tilde{m})} \tilde{\zeta}(\tilde{m})_l^n, \quad \tilde{W}'(\tilde{m})^n = \sum_{i=(l_n(\tilde{m})+1,-k_n)}^{n-k_n} (a^n_{i}+e^n_{i}). 
	\end{align*}
	To prove the first result in \eqref{thm-vov-main}, it suffices to show that 
	\begin{align}\label{vov-main-1}
		&\tilde{W}(\tilde{m})^n  + \tilde{W}'(\tilde{m})^n \longrightarrow^{p} 0,\\ \label{vov-main-2}
		&W(\tilde{m})^n \longrightarrow^{L_s} \frac{1}{\sqrt{\kappa T} } \cdot W.
	\end{align}
	For \eqref{vov-main-1}, note that whenever $\tilde{m} > 2$, $\tilde{\zeta}(\tilde{m})_l^n$ is $(l+1,-2k_n-1)$-measurable and there are no overlapping terms in the sequence $\{\tilde{\zeta}(\tilde{m})_l^n: l=2,...,l_n(\tilde{m})\}$. By Lemma 4.1 in \cite{J2012}, $\tilde{W}(\tilde{m})^n \longrightarrow^{p} 0$ can be proved by showing 
	\begin{align}
		\sum_{l=2}^{l_n(\tilde{m})} \E_{(l,-2k_n-1)}^n [\tilde{\zeta}(\tilde{m})_l^n] \longrightarrow^p 0, \qquad 
		\sum_{l=2}^{l_n(\tilde{m})} \E_{(l,-2k_n-1)}^n [( \tilde{\zeta}(\tilde{m})_l^n)^2] \longrightarrow^p 0. 
	\end{align}
	For $r=-k_n,...,k_n$ and $l=2,...,l_n(\tilde{m})$, Lemma \ref{lem-vov-2} gives
	\begin{align}\label{thm1-expec}
		\E_{(l,-2k_n-1)}^n \left[ a^n_{( l,r)} + e^n_{(l,r)}\right]  = O_p\left(\frac{\sqrt{\Delta_n}}{\sqrt{k_n}}\left(u^2 (k_n\Delta_n)^{\frac{3}{4}}+ |u|^{\beta-2}\Delta_n^{1-\frac{\beta}{2}} + h_n^2\right)\right).
	\end{align}
	When $\beta<3/2$ and $r<1$, since $\tilde{m} \rightarrow \infty$ and $n^{1/2}/k_n =O(1)$, we have 
	\begin{align*}
		l_n(\tilde{m}) k_n \cdot \left(\frac{\sqrt{\Delta_n}}{\sqrt{k_n}}\left(u^2 (k_n\Delta_n)^{\frac{3}{4}}+ |u|^{\beta-2}\Delta_n^{1-\frac{\beta}{2}} + h_n^2\right)\right)  \rightarrow 0,
	\end{align*}
	Thus, $\sum_{l=2}^{l_n(\tilde{m})} \E_{(l,-2k_n-1)}^n [\tilde{\zeta}(\tilde{m})_l^n] \longrightarrow^p 0$. Moreover,
	\begin{align*}
		&\E_{(l,-k_n-1)}^n [( \tilde{\zeta}(\tilde{m})_l^n )^2] \\
		&= \E_{(l,-k_n-1)}^n \left[\sum_{r=-k_n}^{k_n}  \left( a^n_{( l,r)} +e^n_{(l,r)} \right)^2\right]\\
		&\quad + \E_{(l,-k_n-1)}^n \left[2 \sum_{r_1, r_2=-k_n}^{k_n} \sum_{r_1<r_2<r_1+2k_n+1} \left( (a^n_{( l,r_1)} + e^n_{(l,r_1)} ) \cdot (a^n_{( l,r_2)} + e^n_{(l,r_2)}) \right)\right]\\
		&\quad + \E_{(l,-k_n-1)}^n \left[2 \sum_{r_1, r_2=-k_n}^{k_n} \sum_{r_1+2k_n+1<r_2} \left( (a^n_{( l,r_1)} + e^n_{(l,r_1)})\cdot ( a^n_{( l,r_2)} + e^n_{(l,r_2)} ) \right)\right]\\
		&(\text{By Lemma \ref{lem-vov-2}.})\\
		&= O_p(k_n \Delta_n),
	\end{align*}
	as $\tilde{m} \rightarrow \infty$, which implies $\sum_{l=2}^{l_n(\tilde{m})} \E_{(l,-k_n-1)}^n [( \tilde{\zeta}(\tilde{m})_l^n)^2] \longrightarrow^p 0$.
	For $\tilde{W}'(\tilde{m})^n \longrightarrow^{p} 0$, a similar argument gives
	\begin{align*}
		\E[\tilde{W}'(\tilde{m})^n] = \tilde{m}k_n \cdot O_p \left(\frac{\sqrt{\Delta_n}}{\sqrt{k_n}}\left(u^2 (k_n\Delta_n)^{\frac{3}{4}}+ |u|^{\beta-2}\Delta_n^{1-\frac{\beta}{2}} + h_n^2\right)\right),
	\end{align*}
	and 
	\begin{align*}
		\E[(\tilde{W}'(\tilde{m})^n)^2] = o_p
		\left( \tilde{m} k_n \Delta_n \right),
	\end{align*}
	Since $\tilde{m} k_n\Delta_n \rightarrow 0$, Chebyshev's inequality implies convergence in probability. This proves \eqref{vov-main-1}. \\
	For \eqref{vov-main-2}, from the above proof of \eqref{vov-main-1}, if $\beta <3/2$ and $r<1$, since $k_n\Delta_n\rightarrow 0$ and $n^{1/2}/k_n = O(1)$, we have
	\begin{align*}
		\sum_{l=1}^{l_n(\tilde{m})} \E_{(l,0)}\left[ \zeta(\tilde{m})_l^n\right] =n  \cdot O_p \left(\frac{\sqrt{\Delta_n}}{\sqrt{k_n}}\left(u^2 (k_n\Delta_n)^{\frac{3}{4}}+ |u|^{\beta-2}\Delta_n^{1-\frac{\beta}{2}} + h_n^2\right)\right)  \longrightarrow^p 0,
	\end{align*}
	and $\zeta(\tilde{m})_l^n$ is $\mathcal{F}_{(l+1,0)}$-measurable, so $\left \{  \zeta(\tilde{m})_l^n, \mathcal{F}_{(l+1,0)} \right \}$ behaves like a martingale difference array. By Theorem 2.2.15 in \cite{JP2012}, it remains to show
	\begin{align}\label{V-Ls-2-1}
		&\sum_{l=1}^{l_n(\tilde{m})} \E_{(l,0)}\left[ (\zeta(\tilde{m})_l^n )^2 \right] \longrightarrow^p \text{Var}\left(\frac{1}{\sqrt{\kappa T} } \cdot W|\mathcal{F}\right),\\\label{V-Ls-2-2}
		&\sum_{l=1}^{l_n(\tilde{m})} \E_{(l,0)}\left[ (\zeta(\tilde{m})_l^n )^4 \right] \longrightarrow^p 0,\\\label{V-Ls-2-3}
		&\sum_{l=1}^{l_n(\tilde{m})} \E_{(l,0)}\left[ \zeta(\tilde{m})_l^n \cdot \left( \Delta_{l,\tilde{m}}^n M \right) \right] \longrightarrow^p 0,
	\end{align} 
	where $\Delta_{l,\tilde{m}}^n M := M_{(l+1,0)} - M_{(l,0)}$ for $M=B,B'$ or $W$, with $W$ a bounded martingale orthogonal to both $B$ and $B'$.
	For \eqref{V-Ls-2-1}, we decompose
	\begin{align*}
		&\E_{(l,0)}\left[ (\zeta(\tilde{m})_l^n )^2\right] \\ 
		&= \sum_{r=k_n+1}^{(\tilde{m}+1)k_n} \sum_{j=k_n+1}^{(\tilde{m}+1)k_n} \E_{(l,0)}[(a^n_{(l,r)} + e^n_{(l,r)})\cdot (a^n_{(l,j)}+ e^n_{(l,j)})]\\
		& = \sum_{r=k_n+1}^{(\tilde{m}+1)k_n}  \E_{(l,0)}[(a^n_{(l,r)} + e^n_{(l,r)})^2 ] + \sum_{r,j=k_n+1}^{(\tilde{m}+1)k_n} \sum_{|j-r|\leq 2k_n+1} \E_{(l,0)}[(a^n_{(l,r)} + e^n_{(l,r)})\cdot (a^n_{(l,j)}+ e^n_{(l,j)})] \\
		&\quad +  \sum_{r,j=k_n+1}^{(\tilde{m}+1)k_n} \sum_{|j-r|>2k_n+1} \E_{(l,0)}[(a^n_{(l,r)} + e^n_{(l,r)})\cdot (a^n_{(l,j)}+ e^n_{(l,j)})]\\
		&:=H(\tilde{m},1)_l^n + H(\tilde{m},2)_l^n  + H(\tilde{m},3)_l^n.
	\end{align*}
	By Lemma \ref{lem-vov-2}, when $\beta<3/2$, we have
	\begin{align*}
		\sum_{l=1}^{l_n(\tilde{m})}  H(\tilde{m},1)_l^n + \sum_{l=1}^{l_n(\tilde{m})}  H(\tilde{m},3)_l^n  = O_p\left(\tilde{m} k_n \Delta_n \sqrt{k_n\Delta_n} + \frac{1}{k_n}\right),
	\end{align*}
	and
	\begin{align}
		\begin{split}
			\sum_{l=1}^{l_n(\tilde{m})} H(\tilde{m},2)_l^n  = \sum_{l=1}^{l_n(\tilde{m})} (H(u,(l,0)) \cdot  \tilde{m}k_n\Delta_n) + O_p\left( \sqrt{\tilde{m}k_n\Delta_n} + h_n \right) \longrightarrow^{p} \frac{ Var(W|\mathcal{F})}{\kappa T}.
		\end{split}
	\end{align}
	We prove \eqref{V-Ls-2-2} and \eqref{V-Ls-2-3} only for $a_i^n$ for illustration; the same results for $e_i^n$ can be obtained similarly. 
	For \eqref{V-Ls-2-2}, we decompose
	\begin{align*}
		&\sum_{l=1}^{l_n(\tilde{m})} \E_{(l,0)}\left[ \left( \sum_{r=k_n+1}^{(\tilde{m}+1)k_n} \left( a^n_{(l,r)}\right) \right)^4 \right] \\
		& \leq C \sum_{l=1}^{l_n(\tilde{m})} \E_{(l,0)}\left[  \sum_{r_1} \left( a^n_{(l,r_1)}\right)^4 \right]  +C \sum_{l=1}^{l_n(\tilde{m})} \E_{(l,0)}\left[  \sum_{r_1}\sum_{r_2} \left( \left( a^n_{(l,r_1)}\right)^3 \left( a^n_{(l,r_2)}\right)  \right) \right] \\
		&\quad +C \sum_{l=1}^{l_n(\tilde{m})} \E_{(l,0)}\left[  \sum_{r_1}\sum_{r_2} \left( \left( a^n_{(l,r_1)}\right)^2 \left( a^n_{(l,r_2)}\right)^2  \right) \right] \\
		&\quad +C \sum_{l=1}^{l_n(\tilde{m})} \E_{(l,0)}\left[  \sum_{r_1}\sum_{r_2}\sum_{r_3} \left( \left( a^n_{(l,r_1)}\right)^2 \left( a^n_{(l,r_2)}\right) \left( a^n_{(l,r_3)}\right) \right) \right] \\
		&\quad +C \sum_{l=1}^{l_n(\tilde{m})} \E_{(l,0)}\left[  \sum_{r_1}\sum_{r_2}\sum_{r_3}\sum_{r_4} \left( \left(a^n_{(l,r_1)} a^n_{(l,r_2)}a^n_{(l,r_3)}a^n_{(l,r_4)}\right) \right) \right] \\
		&=:S_i^n (1) + S_i^n (2) + S_i^n (3) +S_i^n (4) +S_i^n (5), 
	\end{align*} 
	where $r_1 < r_2 < r_3 < r_4$ and $r_1, r_2, r_3, r_4 =k_n+1,...,(\tilde{m}+1)k_n$. 
	By Lemma \ref{lem-vov-2}, we have $S_i^n (1) = O_p\left(\frac{\Delta_n}{k_n^2}\right)$.
	Since
	\begin{align*}
		&\sum_{r_1}\sum_{r_2} \E_{(l,0)}\left[  \left( \left( a^n_{(l,r_1)}\right)^3 \left( a^n_{(l,r_2)}\right)  \right) \right] \\
		&= \sum_{r_1}\sum_{r_2-r_1\leq 2k_n} \E_{(l,0)}\left[  \left( \left( a^n_{(l,r_1)}\right)^3 \left( a^n_{(l,r_2)}\right)  \right) \right]  + \sum_{r_1}\sum_{r_2-r_1>2k_n} \E_{(l,0)}\left[  \left( \left( a^n_{(l,r_1)}\right)^3 \right] \E_{(l,0)}\left[  \left( a^n_{(l,r_2)}\right)  \right) \right] \\
		&\text{(Using the results in Lemma \ref{lem-vov-2} and Holder's inequality.)}\\
		&\leq C\tilde{m}k_n\cdot k_n\cdot \frac{\Delta_n^2}{k_n^2} + C (\tilde{m}k_n)^2 \cdot \frac{\Delta_n^{\frac{3}{2}}}{k_n^{\frac{3}{2}}} \cdot  \left(\frac{\sqrt{\Delta_n}}{\sqrt{k_n}}\left(u^2 (k_n\Delta_n)^{\frac{3}{4}}+ |u|^{\beta-2}\Delta_n^{1-\frac{\beta}{2}} + h_n^2\right)\right),
	\end{align*}
	Thus,
	\begin{align*}
		S_i^n(2) = O_p\left( \frac{\tilde{m}k_n\Delta_n}{k_n^2}\left(u^2 (k_n\Delta_n)^{\frac{3}{4}}+ |u|^{\beta-2}\Delta_n^{1-\frac{\beta}{2}} + h_n^2\right) \right). 
	\end{align*}
	Similarly, decomposing the summation into the cases $r_{k+1} - r_{k} \leq 2k_n$ and $r_{k+1} - r_{k} > 2k_n$ for $k=1,2,3$, we obtain
	\begin{align*}
		&S_i^n(3) = O_p\left( \frac{\tilde{m}k_n\Delta_n}{k_n^2}\right), \qquad S_i^n(4) = O_p\left( \frac{\tilde{m}k_n\Delta_n}{k_n^2}\right), \\
		&S_i^n(5) = O_p\left( (\tilde{m} k_n\Delta_n)^2(k_n\Delta_n)^{\frac{1}{2}} + (\tilde{m} k_n\Delta_n)(k_n\Delta_n)^{\frac{3}{4}}  +k_n\Delta_n  \right).
	\end{align*}
	This proves \eqref{V-Ls-2-2}.
	For \eqref{V-Ls-2-3}, with $M=B+ B'+W$, as in the proof of \eqref{lem-vov-2-1}, we decompose
	\begin{align*}
		&\E_{(l,0)}\left[ \sum_{r=k_n+1}^{(\tilde{m}+1)k_n}  a^n_{(l,r)} \cdot \left( \Delta_{l,\tilde{m}}^n M \right) \right] \\
		& = \E_{(l,0)}\left[ \sum_{r=k_n+1}^{(\tilde{m}+1)k_n}  \left(a^n_{(l,r)}(1) + a^n_{(l,r)}(2)  + a^n_{(l,r)}(3) \right) \cdot \left( \Delta_{l,\tilde{m}}^n M \right) \right],
	\end{align*}
	where $a^n_{(l,r)}(1), a^n_{(l,r)}(2),a^n_{(l,r)}(3)$ are defined in the proof of \eqref{lem-vov-2-1}. We have 
	\begin{align*}
		&\E_{(l,0)} \left[  \left( a^n_{(l,r)}(3) \cdot \Delta_{l,\tilde{m}}^n M \right)  \right] \\
		&= \E_{(l,0)} \Bigg[   \Bigg(  \frac{1}{\sqrt{k_n\Delta_n}} \Bigg( \frac{3\Delta_n}{2} \Bigg( \Bigg(  \frac{1}{\sqrt{k_n^2\Delta_n}} \sum_{j \in I^n_{i+} } \xi_{t_{i+}^n,n}^{(2,j)}(u) + \sum_{j \in I^n_{i+} } \xi_{t_{i+}^n,n}^{(3,j)}(u) \Bigg) \\
		&\qquad \qquad \quad \qquad - \Bigg( \frac{1}{\sqrt{k_n^2 \Delta_n}} \sum_{j \in I^n_{i-} } \xi_{t_{i-}^n,n}^{(2,j)}(u) + \sum_{j \in I^n_{i-} } \xi_{t_{i-}^n,n}^{(3,j)}(u) \Bigg) \Bigg)^2 \\
		&\qquad \qquad \quad - \frac{3}{k_n^2\Delta_n} h_1(u, \sigma_{t_{i+}^n}^2) \Delta_n - 4(\sigma_{t_{i+}^n}^2 )((\tilde{\sigma}_{t_{i+}^n})^2+(\tilde{\sigma}'_{t_{i+}^n})^2)\Delta_n  \Bigg) \cdot \Delta_{l,\tilde{m}}^n M \Bigg)  \Bigg] \\
		& = \E_{(l,0)} \Bigg[  \frac{\Delta_n}{k_n} \Bigg( \Bigg(  \frac{1}{\sqrt{k_n^2\Delta_n}} \sum_{j \in I^n_{i+} } \xi_{t_{i+}^n,n}^{(2,j)}(u) + \sum_{j \in I^n_{i+} } \xi_{t_{i+}^n,n}^{(3,j)}(u) \Bigg) \\
		&\qquad \qquad \quad \qquad - \Bigg( \frac{1}{\sqrt{k_n^2 \Delta_n}} \sum_{j \in I^n_{i-} } \xi_{t_{i-}^n,n}^{(2,j)}(u) + \sum_{j \in I^n_{i-} } \xi_{t_{i-}^n,n}^{(3,j)}(u) \Bigg) \Bigg)^2 \cdot \Delta_{l,\tilde{m}}^n M \Bigg)  \Bigg]\\
		&= O_p\left( \Delta_n \frac{\sqrt{\Delta_n}}{k_n} \right), 
	\end{align*}
	since
	\begin{align*}
		&\E_{(l,0)} \Bigg[ \Bigg( \Bigg( \sum_{j \in I^n_{i+} } \xi_{t_{i+}^n,n}^{(2,j)}(u) \Bigg)^2 \cdot \Delta_{l,\tilde{m}}^n M \Bigg)  \Bigg]\\
		&=\sum_{j \in I^n_{i+} } \E_{j-1}^n \left[ \left(  \xi_{t_{i+}^n,n}^{(2,j)}(u) \right)^2 \cdot \left(W_{t_{i+k_n}^n} - W_{t_{i}^n} + B_{t_{i+k_n}^n} - B_{t_{i}^n} + B'_{t_{i+k_n}^n} - B'_{t_{i}^n} \right) \right] \\
		& = \sum_{j \in I^n_{i+} } \E_{j-1}^n \left[ \left(  \xi_{t_{i+}^n,n}^{(2,j)}(u) \right)^2 \cdot \left(\Delta_j^nW+ \Delta_j^n B + \Delta_j^n B'\right) \right] \\
		&= \sum_{j \in I^n_{i+} } \frac{2}{k_nu^4 (f_{t_{i+}^n,n}(u) )^2}  \left( \E_{j-1}^{n} \left[ \cos \left(\frac{2u(\sigma_{t_{j-1}^n} \Delta_j^n B )}{\sqrt{\Delta_n}} \right) \cos  \left(\frac{2u( \gamma_{t_{j-1}^n} \Delta_j^n L)}{\sqrt{\Delta_n}} \right) \cdot \left(\Delta_j^nW+ \Delta_j^n B + \Delta_j^n B'\right)\right] \right) \\
		& + \sum_{j \in I^n_{i+} } \frac{2}{\sqrt{k_n} u^2 (f_{t_{i+}^n,n}(u) )^2}  \left( \E_{j-1}^{n} \left[  \sin  \left(\frac{2u(\sigma_{t_{j-1}^n} \Delta_j^n B )}{\sqrt{\Delta_n}} \right) \sin  \left(\frac{2u( \gamma_{t_{j-1}^n} \Delta_j^n L)}{\sqrt{\Delta_n}} \right) \cdot \left(\Delta_j^nW+ \Delta_j^n B + \Delta_j^n B'\right)\right] \right)\\
		&(\text{Similar to the proof of \eqref{xi-23-W}.})\\
		&\leq C\sqrt{\Delta_n}, 
	\end{align*}
	and 
	\begin{align*}
		&\E_{(l,0)} \Bigg[ \Bigg( \Bigg( \sum_{j \in I^n_{i+} } \xi_{t_{i+}^n,n}^{(2,j)}(u) \Bigg)^2 \cdot \Delta_{l,\tilde{m}}^n M \Bigg)  \Bigg]\\
		&= \sum_{j \in I^n_{i+} } \E_{j-1}^n \left[ \left(  \xi_{t_{i+}^n,n}^{(3,j)}(u) \right)^2 \cdot \left(W_{t_{i+k_n}^n} - W_{t_{i}^n} + B_{t_{i+k_n}^n} - B_{t_{i}^n} + B'_{t_{i+k_n}^n} - B'_{t_{i}^n} \right) \right] \\
		&= \sum_{j \in I^n_{i+} } \E_{j-1}^n \left[ \left(  \xi_{t_{i+}^n,n}^{(3,j)}(u) \right)^2 \cdot \left(\Delta_j^nW+ \Delta_j^n B + \Delta_j^n B'\right) \right]\\	
		&=\sum_{j \in I^n_{i+} } \frac{ (i+k_n+1-j)^2  }{k_n^2 k_n\Delta_n } \cdot \E_{j-1}^n \left[\left(2\sigma_{t_{i+}^n} \tilde{\sigma}_{t_{i+}^n}\Delta_j^nB + 2\sigma_{t_{i+}^n} \tilde{\sigma}'_{t_{i+}^n} \Delta_j^nB'\right)^2 \left(\Delta_j^nW+ \Delta_j^n B + \Delta_j^n B'\right) \right]\\
		&\equiv 0,
	\end{align*}
	and 
	\begin{align*}
		&\E_{(l,0)} \Bigg[ \Bigg( \sum_{j \in I^n_{i+} } \xi_{t_{i+}^n,n}^{(2,j)}(u) \Bigg) \Bigg( \sum_{j \in I^n_{i+} } \xi_{t_{i+}^n,n}^{(3,j)}(u) \Bigg) \cdot \Delta_{l,\tilde{m}}^n M  \Bigg]\\
		&= \sum_{j \in I^n_{i+} } \E_{j-1}^n \left[ \left(  \xi_{t_{i+}^n,n}^{(2,j)}(u) \right)  \left(  \xi_{t_{i+}^n,n}^{(3,j)}(u) \right) \cdot \left(W_{t_{i+k_n}^n} - W_{t_{i}^n} + B_{t_{i+k_n}^n} - B_{t_{i}^n} + B'_{t_{i+k_n}^n} - B'_{t_{i}^n} \right) \right] \\
		&= \sum_{j \in I^n_{i+} } \E_{j-1}^n \left[ \left(  \xi_{t_{i+}^n,n}^{(2,j)}(u) \right)  \left(  \xi_{t_{i+}^n,n}^{(3,j)}(u) \right) \cdot \left(\Delta_j^nW+ \Delta_j^n B + \Delta_j^n B'\right) \right] \\
		&=\sum_{j \in I^n_{i+} } \frac{  -2(i+k_n+1-j)  }{k_n \sqrt{k_n\Delta_n} \sqrt{k_n} u^2 f_{t_{i+}^n,n}(u) } \cdot \E_{j-1}^n \Bigg[ \left( \cos  \left(\frac{u\Delta_j^nX'}{\sqrt{\Delta_n}} \right) - \E_{j-1}^{n} \left[ \cos \left(\frac{u\Delta_j^nX'}{\sqrt{\Delta_n}} \right) \right] \right) \\
		&\qquad \qquad \cdot \left(2\sigma_{t_{i+}^n} \tilde{\sigma}_{t_{i+}^n}\Delta_j^nB + 2\sigma_{t_{i+}^n} \tilde{\sigma}'_{t_{i+}^n} \Delta_j^nB'\right)   \left(\Delta_j^nW+ \Delta_j^n B + \Delta_j^n B'\right) \Bigg]\\
		&\leq C\sqrt{\Delta_n},
	\end{align*}
	and similar results can be obtained for the other terms. 
	Moreover,
	\begin{align*}
		&\E_{(l,0)} \left[  \left( a^n_{(l,r)}(2) \cdot \Delta_{l,\tilde{m}}^n M \right)  \right] \\
		& = \E_{(l,0)} \Bigg[ \E\Bigg[ \Bigg( \frac{3\Delta_n}{\sqrt{k_n\Delta_n}} \Bigg( \Bigg(  \frac{1}{\sqrt{k_n^2\Delta_n}} \sum_{j \in I^n_{i+} } \xi_{t_{i+}^n,n}^{(2,j)}(u) + \sum_{j \in I^n_{i+} } \xi_{t_{i+}^n,n}^{(3,j)}(u) \Bigg) \\
		&\qquad \qquad  - \Bigg( \frac{1}{\sqrt{k_n^2 \Delta_n}} \sum_{j \in I^n_{i-} } \xi_{t_{i-}^n,n}^{(2,j)}(u) + \sum_{j \in I^n_{i-} } \xi_{t_{i-}^n,n}^{(3,j)}(u) \Bigg) \Bigg)  \cdot \Delta_{l,\tilde{m}}^n M \Bigg) \Bigg|  (\xi_{t_{i+}^n,n}''(u), \xi_{t_{i-}^n,n}''(u)) \Bigg] \cdot \\
		& \qquad \cdot \left( \xi_{t_{i+}^n,n}''(u)-\xi_{t_{i-}^n,n}''(u)  \right)  \Bigg]\\
		&\text{(Using the results in the proof of \eqref{xi-23-W}, $\xi_{t_{i\pm}^n,n}''(u) =O_p(h_n)$, and Holder's inequality.)}\\
		&= O_p(\Delta_n h_n),
	\end{align*}
	and 
	\begin{align*}
		&\E_{(l,0)} \left[  \left( a^n_{(l,r)}(3) \cdot \Delta_{l,\tilde{m}}^n M \right)  \right] \\
		&= \E_{(l,0)} \left[  \left( 	\frac{3\Delta_n}{\sqrt{k_n\Delta_n}}  \left( \xi_{t_{i+}^n,n}''(u)-\xi_{t_{i-}^n,n}''(u)  \right)^2 \cdot \Delta_{l,\tilde{m}}^n M \right)  \right] \\
		&\text{(Using $\xi_{t_{i\pm}^n,n}''(u)=O_p( h_n)$ and Holder's inequality.)}\\
		&= O_p(\Delta_n \sqrt{\tilde{m}}(h_n)^2).
	\end{align*}
	With these results, we obtain 
	\begin{align*}
		& \sum_{l=1}^{l_n(\tilde{m})} \E_{(l,0)}\left[ \zeta(\tilde{m})_l^n \cdot \left( \Delta_{l,\tilde{m}}^n M \right) \right] \\
		& = \sum_{l=1}^{l_n(\tilde{m})} \E_{(l,0)}\left[ \sum_{r=k_n+1}^{(\tilde{m}+1)k_n}  a^n_{(l,r)} \cdot \sum_{r=1}^{2\tilde{m}k_n}  \left( \Delta_{l,\tilde{m}}^n M \right) \right]  \\
		&= O_p\left(  h_n + \sqrt{\tilde{m}} (h_n)^2\right),
	\end{align*}
	which implies \eqref{V-Ls-2-3}. This proves \eqref{vov-main-2}; hence \eqref{thm-vov-main} and conclusion \eqref{thm2:b12} hold.
	The result for the case $\frac{1}{2}<b<1$ follows directly from the above result with $\kappa=\infty$. Plugging $k_n = \lfloor \kappa n^{b} \rfloor$ into \eqref{thm2:b12} completes the proof of \eqref{thm-vov-clt}.
	
	For the consistency result \eqref{thm-vov-con}, the above proof shows that it requires
	\begin{align*}
		\sqrt{k_n\Delta_n}\cdot n  \cdot \left(\frac{\sqrt{\Delta_n}}{\sqrt{k_n}}\left(u^2 (k_n\Delta_n)^{\frac{3}{4}}+ |u|^{\beta-2}\Delta_n^{1-\frac{\beta}{2}} + h_n^2\right)\right)  \longrightarrow 0,
	\end{align*}
	which always holds when $\beta<2$ and $r<4/3$. 
	
	This completes the proof of Theorem \ref{thm-vov}. 
	\hfill $\square$
	
	\subsection{Proofs for Section \ref{sec:vol-fun} and \ref{sec:fea-clt}}\label{prof:sec3_4-5}
	
	\textbf{Proof of Theorem \ref{thm:vovfunc}:} 
	By Theorem 1 in \cite{LLL2018}, we have $\widehat{\sigma}_t^2 \longrightarrow^{p} \sigma_t^2$, and hence $g(\widehat{\sigma}_t^2) \longrightarrow^{p} g(\sigma_t^2)$. Since $\widehat{\sigma}_t^2$ and $\sigma_t^2$ are bounded almost surely, and 
	\begin{align*}
		&\E\left[ \left| \widehat{I}(g) - I(g) \right|\right] \\
		&= \E \left[ \left| \sum_{i=0}^{n-k_n} \left(   \int_{t_{i}^n}^{t_{i+1}^n} \left( g( \widehat{\sigma}^2_{t}) - g(\sigma_t^2) \right)dt  \right) -  \int_{t_{n-k_n+1}^n}^{T} g(\sigma_t^2)dt \right|\right]\\
		& \leq  \sum_{i=0}^{n-k_n} \left(   \int_{t_{i}^n}^{t_{i+1}^n} \E \left[ \left|  \left( g( \widehat{\sigma}^2_t) - g(\sigma_t^2) \right) \right|\right]dt  \right)  + Ck_n\Delta_n, 
	\end{align*}
	the conclusion follows from the dominated convergence theorem and Chebyshev's theorem. This completes the proof. 
	\hfill $\square$
	
	\textbf{Proof of Theorem \ref{thm-uv-hat}:} We focus on the case $b=\frac{1}{2}$; the case $b>\frac{1}{2}$ follows directly. We first show that
	\begin{align}\label{proof:H3}
		\widehat{H_n^3} =\frac{n}{k_n^2}  \sum_{i=k_n+1}^{n-k_n}   \left( \widehat{\sigma}^2_{t_{i+}^n} - \widehat{\sigma}^2_{t_{i-}^n} \right)^4 \longrightarrow^{p}  \int_{0}^{T} \left( \frac{135}{2} \frac{h_1^2}{\kappa^4T^2}+ 38 \frac{h_1 h_2}{\kappa^2T} + \frac{135}{2} (h_2)^2 \right) dt.
	\end{align}
	We decompose 
	\begin{align*}
		&\frac{n}{k_n^2}  \sum_{i=k_n+1}^{n-k_n}   \left( \widehat{\sigma}^2_{t_{i+}^n} - \widehat{\sigma}^2_{t_{i-}^n} \right)^4 \\
		& = \frac{n}{k_n^2}  \sum_{i=k_n+1}^{n-k_n} \left( \left( \widehat{\sigma}^2_{t_{i+}^n}  -(\sigma^2_{t_{i+}^n} + b_{ t_{i+}^n,n}(u) ) \right) - \left( \widehat{\sigma}^2_{t_{i-}^n}  -(\sigma^2_{t_{i-}^n} + b_{ t_{i-}^n,n}(u) ) \right) \right)^4 \\
		&\quad - \frac{n}{k_n^2}  \sum_{i=k_n+1}^{n-k_n} \Big( 4 \left( \left( \widehat{\sigma}^2_{t_{i+}^n}  -(\sigma^2_{t_{i+}^n} + b_{ t_{i+}^n,n}(u) ) \right) - \left( \widehat{\sigma}^2_{t_{i-}^n}  -(\sigma^2_{t_{i-}^n} + b_{ t_{i-}^n,n}(u) ) \right) \right)^3 \\
		& \qquad \qquad \quad \ \cdot  \left( (\sigma^2_{t_{i+}^n} + b_{ t_{i+}^n,n}(u) )   - (\sigma^2_{t_{i-}^n} + b_{ t_{i-}^n,n}(u) )  \right) \Big) \\
		&\quad - \frac{n}{k_n^2}  \sum_{i=k_n+1}^{n-k_n} \Big( 4 \left( \left( \widehat{\sigma}^2_{t_{i+}^n}  -(\sigma^2_{t_{i+}^n} + b_{ t_{i+}^n,n}(u) ) \right) - \left( \widehat{\sigma}^2_{t_{i-}^n}  -(\sigma^2_{t_{i-}^n} + b_{ t_{i-}^n,n}(u) ) \right) \right) \\
		& \qquad \qquad \quad \ \cdot  \left( (\sigma^2_{t_{i+}^n} + b_{ t_{i+}^n,n}(u) )   - (\sigma^2_{t_{i-}^n} + b_{ t_{i-}^n,n}(u) )  \right)^3 \Big) \\
		&\quad + \frac{n}{k_n^2}  \sum_{i=k_n+1}^{n-k_n} \Big( 6 \left( \left( \widehat{\sigma}^2_{t_{i+}^n}  -(\sigma^2_{t_{i+}^n} + b_{ t_{i+}^n,n}(u) ) \right) - \left( \widehat{\sigma}^2_{t_{i-}^n}  -(\sigma^2_{t_{i-}^n} + b_{ t_{i-}^n,n}(u) ) \right) \right)^2 \\
		& \qquad \qquad \quad \ \cdot  \left( (\sigma^2_{t_{i+}^n} + b_{ t_{i+}^n,n}(u) )   - (\sigma^2_{t_{i-}^n} + b_{ t_{i-}^n,n}(u) )  \right)^2 \Big) \\
		& \quad +\frac{n}{k_n^2} \sum_{i=k_n+1}^{n-k_n}  \left( (\sigma^2_{t_{i+}^n} + b_{ t_{i+}^n,n}(u) )   - (\sigma^2_{t_{i-}^n} + b_{ t_{i-}^n,n}(u) )  \right)^4\\
		&:= P_i^{n,1} + P_i^{n,2} +P_i^{n,3} +P_i^{n,4} +P_i^{n,5}.
	\end{align*}
	For $P_i^{n,1}$, we rewrite
	\begin{align*}
		P_i^{n,1} &= \frac{4k_n}{9}\sum_{i=k_n+1}^{n-k_n} \Bigg( \left(\frac{3}{2} \sqrt{\frac{\Delta_n}{k_n}}\left( \frac{\left( \widehat{\sigma}^2_{t_{i+}^n}  -(\sigma^2_{t_{i+}^n} + b_{ t_{i+}^n,n}(u) ) \right)}{\sqrt{k_n\Delta_n}} - \frac{\left( \widehat{\sigma}^2_{t_{i-}^n}  -(\sigma^2_{t_{i-}^n} + b_{ t_{i-}^n,n}(u) ) \right)}{\sqrt{k_n\Delta_n}} \right)^2\right)^2 \\
		& \qquad \quad - \E_{i-k_n-1}^n \left[ \left(\frac{3}{2} \sqrt{\frac{\Delta_n}{k_n}}\left( \frac{\left( \widehat{\sigma}^2_{t_{i+}^n}  -(\sigma^2_{t_{i+}^n} + b_{ t_{i+}^n,n}(u) ) \right)}{\sqrt{k_n\Delta_n}} - \frac{\left( \widehat{\sigma}^2_{t_{i-}^n}  -(\sigma^2_{t_{i-}^n} + b_{ t_{i-}^n,n}(u) ) \right)}{\sqrt{k_n\Delta_n}} \right)^2\right)^2 \right]  \Bigg)\\
		& \quad +\frac{4k_n}{9}\sum_{i=k_n+1}^{n-k_n} \E_{i-k_n-1}^n \left[ \left(\frac{3}{2} \sqrt{\frac{\Delta_n}{k_n}}\left( \frac{\left( \widehat{\sigma}^2_{t_{i+}^n}  -(\sigma^2_{t_{i+}^n} + b_{ t_{i+}^n,n}(u) ) \right)}{\sqrt{k_n\Delta_n}} - \frac{\left( \widehat{\sigma}^2_{t_{i-}^n}  -(\sigma^2_{t_{i-}^n} + b_{ t_{i-}^n,n}(u) ) \right)}{\sqrt{k_n\Delta_n}} \right)^2\right)^2 \right]\\
		&:= \frac{4k_n}{9} \sum_{j=k_n+1}^{n-k_n} p_i^{n,1}(j) + \frac{4k_n}{9} \sum_{j=k_n+1}^{n-k_n} \overline{p_i^{n,1}}(j). 
	\end{align*}
	Since
	\begin{align*}
		\E\left[\frac{4k_n}{9} \sum_{j=k_n+1}^{n-k_n} p_i^{n,1}(j) \right] =\frac{4k_n}{9} \sum_{j=k_n+1}^{n-k_n} \E\left[ \E_{i-k_n-1}\left[ p_i^{n,1}(j) \right] \right] \equiv 0,
	\end{align*}
	applying Holder's inequality with Lemma \ref{lem-vov-2}, we obtain
	\begin{align*}
		\E\left[ \left(\frac{4k_n}{9} \sum_{j=k_n+1}^{n-k_n} p_i^{n,1}(j) \right)^2\right] = \frac{16k_n^2}{81}  \sum_{j_1,j_2=k_n+1}^{n-k_n} \sum_{|j_2-j_1|\leq k_n+1}^{n-k_n} \E \left[ p_i^{n,1}(j_1) p_i^{n,1}(j_2)\right] \leq C k_n\Delta_n,
	\end{align*}
	Thus, $\frac{4k_n}{9} \sum_{j=k_n+1}^{n-k_n} p_i^{n,1}(j) \longrightarrow^{p} 0$. 
	Moreover,
	\begin{align*}
		& \frac{4k_n}{9} \sum_{j=k_n+1}^{n-k_n} \overline{p_i^{n,1}}(j) \\
		& =\frac{4k_n}{9} \sum_{j=k_n+1}^{n-k_n}  \E_{i-k_n-1}^n \left[ \left(a_i^n +\frac{1}{\sqrt{k_n\Delta_n}}  \left( \frac{3}{k_n^2\Delta_n} h_1\Delta_n+3h_2\Delta_n  \right) \right)^2 \right]\\
		&\text{(Using the results in Lemma \ref{lem-vov-2}.)}\\
		&= \frac{4\Delta_n}{9} \sum_{j=k_n+1}^{n-k_n} \E_{i-k_n-1}^n \left[  \Bigg( \left( \frac{117}{2} \frac{h_1^2}{k_n^4\Delta_n^2}+ 36 \frac{h_1 h_2}{k_n^2\Delta_n} + \frac{117}{2} (h_2)^2 \right) +   \left( \frac{3}{k_n^2\Delta_n} h_1+3h_2  \right)^2 \Bigg) \right]+o_p(1)\\
		&\longrightarrow^{p} \int_{0}^{T} \left( 30 \frac{h_1^2}{\kappa^4T^2}+ 24 \frac{h_1 h_2}{\kappa^2T} + 30 (h_2)^2 \right) dt.
	\end{align*}
	Thus
	\begin{align*}
		P_i^{n,1} \longrightarrow^p \int_{0}^{T} \left( 30 \frac{h_1^2}{\kappa^4T^2}+ 24\frac{h_1 h_2}{\kappa^2T} + 30(h_2)^2 \right) dt. 
	\end{align*}
	For $ P_i^{n,2}$, Holder's inequality with Lemma \ref{lem-vov-2} gives
	\begin{align*}
		&\E_{i-k_n+1}\Big[ \Big|\left( \left( \widehat{\sigma}^2_{t_{i+}^n}  -(\sigma^2_{t_{i+}^n} + b_{ t_{i+}^n,n}(u) ) \right) - \left( \widehat{\sigma}^2_{t_{i-}^n}  -(\sigma^2_{t_{i-}^n} + b_{ t_{i-}^n,n}(u) ) \right) \right)^3 \\
		& \qquad  \cdot  \left( (\sigma^2_{t_{i+}^n} + b_{ t_{i+}^n,n}(u) )   - (\sigma^2_{t_{i-}^n} + b_{ t_{i-}^n,n}(u) )  \right) \Big) \Big| \Big] \\
		&\leq \left( \E_{i-k_n+1}\Big[ \left( \left( \widehat{\sigma}^2_{t_{i+}^n}  -(\sigma^2_{t_{i+}^n} + b_{ t_{i+}^n,n}(u) ) \right) - \left( \widehat{\sigma}^2_{t_{i-}^n}  -(\sigma^2_{t_{i-}^n} + b_{ t_{i-}^n,n}(u) ) \right) \right)^6 \Big] \right)^{1/2}\\
		& \qquad  \cdot \left(\E_{i-k_n+1}\Big[  \left( (\sigma^2_{t_{i+}^n} + b_{ t_{i+}^n,n}(u) )   - (\sigma^2_{t_{i-}^n} + b_{ t_{i-}^n,n}(u) )  \right) \Big)^2 \Big] \right)^{1/2}\\
		&=k_n^{9/4}\Delta_n^{3/4} \left(\E_{i-k_n-1}^n \left[ \left(\frac{3}{2} \sqrt{\frac{\Delta_n}{k_n}}\left( \frac{\left( \widehat{\sigma}^2_{t_{i+}^n}  -(\sigma^2_{t_{i+}^n} + b_{ t_{i+}^n,n}(u) ) \right)}{\sqrt{k_n\Delta_n}} - \frac{\left( \widehat{\sigma}^2_{t_{i-}^n}  -(\sigma^2_{t_{i-}^n} + b_{ t_{i-}^n,n}(u) ) \right)}{\sqrt{k_n\Delta_n}} \right)^2\right)^3 \right]\right)^{1/2}\\
		& \qquad  \cdot \left(\E_{i-k_n+1}\Big[  \left( (\sigma^2_{t_{i+}^n} + b_{ t_{i+}^n,n}(u) )   - (\sigma^2_{t_{i-}^n} + b_{ t_{i-}^n,n}(u) )  \right) \Big)^2 \Big] \right)^{1/2}\\
		&=k_n^{9/4}\Delta_n^{3/4} \left(\E_{i-k_n-1}^n \left[ \left(a_i^n +\frac{1}{\sqrt{k_n\Delta_n}}  \left( \frac{3}{k_n^2\Delta_n} h_1\Delta_n+3h_2\Delta_n  \right) \right)^3 \right]\right)^{1/2}\\
		& \qquad  \cdot \left(\E_{i-k_n+1}\Big[  \left( (\sigma^2_{t_{i+}^n} + b_{ t_{i+}^n,n}(u) )   - (\sigma^2_{t_{i-}^n} + b_{ t_{i-}^n,n}(u) )  \right) \Big)^2 \Big] \right)^{1/2}\\
		&\leq Ck_n^{3/2}\Delta_n^2.
	\end{align*}
	Furthermore, $\E[|P_i^{n,2}|] \leq C/\sqrt{k_n}$ and $ P_i^{n,2} \longrightarrow^p 0$. Similarly, we can prove that 
	\begin{align*}
		P_i^{n,3}= O_p\left( \frac{\Delta_n^{1/4}}{k_n^{7/4}} \right), \ P_i^{n,4}= O_p\left( \frac{\Delta_n^{1/2}}{k_n^{3/2}} \right), \ P_i^{n,5}=O_p\left( \frac{1}{k_n^{2}} \right),
	\end{align*}
	Thus, $P_i^{n,3} +P_i^{n,4} +P_i^{n,5} \longrightarrow^{p} 0$.
	The above results together imply conclusion \eqref{proof:H3}.\\
	A direct application of Theorem \ref{thm:vovfunc} yields
	\begin{align}\label{proof:H1}
		\widehat{H_n^1} \longrightarrow^p \int_{0}^{T}  \frac{(h_1(u,t, \sigma^2_t) )^2}{\kappa^4T^2} dt,
	\end{align}
	and 
	\begin{align}\label{Uhat-1}
		\Delta_n \sum_{i=0}^{n-k_n}  (\widehat{\sigma}^2_{t_{i+}^n} h_1(u,t_{i+}^n, \widehat{\sigma}^2_{t_{i+}^n}) ) \longrightarrow^{p} \int_{0}^{T}  \sigma^2_{t} h_1(u,t,\sigma^2_{t}) dt.
	\end{align}
	Now we prove 
	\begin{align}\label{proof:H2}
		\widehat{H_n^2} \longrightarrow^p \int_{0}^{T} \frac{h_1(u,t,\sigma_{t}^2)}{\kappa^2T} h_2(t,\sigma_{t}^2,(\tilde{\sigma}_{t})^2, (\tilde{\sigma}'_{t})^2) dt.
	\end{align}
	We decompose 
	\begin{align*}
		&\sum_{i=k_n+1}^{n-k_n} \left( \frac{h_1(u,t_{i+}^n, \widehat{\sigma}^2_{t_{i+}^n}) }{k_n^2\Delta_n} \cdot \frac{ 1 }{2k_n} \left( \widehat{\sigma}^2_{t_{i+}^n} - \widehat{\sigma}^2_{t_{i-}^n} \right)^2 \right)\\
		&= \sum_{i=k_n+1}^{n-k_n} \left( \frac{h_1(u,t_{i-k_n-1}^n, \sigma^2_{t_{i-k_n-1}^n}) }{k_n^2\Delta_n} \cdot \frac{ 1 }{2k_n} \left( \left( \widehat{\sigma}^2_{t_{i+}^n}  -(\sigma^2_{t_{i+}^n} + b_{ t_{i+}^n,n}(u) ) \right) - \left( \widehat{\sigma}^2_{t_{i-}^n}  -(\sigma^2_{t_{i-}^n} + b_{ t_{i-}^n,n}(u) ) \right) \right)^2  \right)\\
		&\quad + \sum_{i=k_n+1}^{n-k_n} \Bigg( \frac{h_1(u,t_{i+}^n, \widehat{\sigma}^2_{t_{i+}^n}) - h_1(u,t_{i-k_n-1}^n, \sigma^2_{t_{i-k_n-1}^n})  }{k_n^2\Delta_n} \\
		&\qquad \qquad \cdot \frac{ 1 }{2k_n} \left( \left( \widehat{\sigma}^2_{t_{i+}^n}  -(\sigma^2_{t_{i+}^n} + b_{ t_{i+}^n,n}(u) ) \right) - \left( \widehat{\sigma}^2_{t_{i-}^n}  -(\sigma^2_{t_{i-}^n} + b_{ t_{i-}^n,n}(u) ) \right) \right)^2  \Bigg)\\
		&\quad + \sum_{i=k_n+1}^{n-k_n} \Bigg( \frac{h_1(u,t_{i+}^n, \widehat{\sigma}^2_{t_{i+}^n}) }{k_n^2\Delta_n} \cdot \frac{ 1 }{k_n}  \Big( \left( \left( \widehat{\sigma}^2_{t_{i+}^n}  -(\sigma^2_{t_{i+}^n} + b_{ t_{i+}^n,n}(u) ) \right) - \left( \widehat{\sigma}^2_{t_{i-}^n}  -(\sigma^2_{t_{i-}^n} + b_{ t_{i-}^n,n}(u) ) \right) \right) \\
		& \qquad \qquad \quad \ \cdot  \left( (\sigma^2_{t_{i+}^n} + b_{ t_{i+}^n,n}(u) )   - (\sigma^2_{t_{i-}^n} + b_{ t_{i-}^n,n}(u) )  \right) \Big) \Bigg)\\  
		&+\quad \sum_{i=k_n+1}^{n-k_n} \left( \frac{h_1(u,t_{i+}^n, \widehat{\sigma}^2_{t_{i+}^n}) }{k_n^2\Delta_n} \cdot \frac{ 1 }{2k_n} \left( (\sigma^2_{t_{i+}^n} + b_{ t_{i+}^n,n}(u) )   - (\sigma^2_{t_{i-}^n} + b_{ t_{i-}^n,n}(u) )  \right)^2 \right)\\
		&:= Q_i^{n,1} + Q_i^{n,2} +Q_i^{n,3} +  Q_i^{n,4}. 
	\end{align*}
	For $Q_i^{n,1}$, we write
	\begin{align*}
		&Q_i^{n,1}=\\
		& \frac{\sqrt{k_n\Delta_n}}{3}\sum_{i=k_n+1}^{n-k_n} \Bigg( \frac{h_1(u,t_{i-k_n-1}^n) }{k_n^2\Delta_n} \left(\frac{3}{2} \sqrt{\frac{\Delta_n}{k_n}}\left( \frac{\left( \widehat{\sigma}^2_{t_{i+}^n}  -(\sigma^2_{t_{i+}^n} + b_{ t_{i+}^n,n}(u) ) \right)}{\sqrt{k_n\Delta_n}} - \frac{\left( \widehat{\sigma}^2_{t_{i-}^n}  -(\sigma^2_{t_{i-}^n} + b_{ t_{i-}^n,n}(u) ) \right)}{\sqrt{k_n\Delta_n}} \right)^2\right) \\
		& - \E_{i-k_n-1}^n \left[  \frac{h_1(u,t_{i-k_n-1}^n) }{k_n^2\Delta_n} \left(\frac{3}{2} \sqrt{\frac{\Delta_n}{k_n}}\left( \frac{\left( \widehat{\sigma}^2_{t_{i+}^n}  -(\sigma^2_{t_{i+}^n} + b_{ t_{i+}^n,n}(u) ) \right)}{\sqrt{k_n\Delta_n}} - \frac{\left( \widehat{\sigma}^2_{t_{i-}^n}  -(\sigma^2_{t_{i-}^n} + b_{ t_{i-}^n,n}(u) ) \right)}{\sqrt{k_n\Delta_n}} \right)^2\right)  \right]  \Bigg)\\
		& +\frac{\sqrt{k_n\Delta_n}}{3}\sum_{i=k_n+1}^{n-k_n} \E_{i-k_n-1}^n \Bigg[ \frac{h_1(u,t_{i-k_n-1}^n) }{k_n^2\Delta_n} \Bigg(\frac{3}{2} \sqrt{\frac{\Delta_n}{k_n}}\Big( \frac{\left( \widehat{\sigma}^2_{t_{i+}^n}  -(\sigma^2_{t_{i+}^n} + b_{ t_{i+}^n,n}(u) ) \right)}{\sqrt{k_n\Delta_n}} \\
		&\qquad \qquad \qquad \qquad \qquad \qquad - \frac{\left( \widehat{\sigma}^2_{t_{i-}^n}  -(\sigma^2_{t_{i-}^n} + b_{ t_{i-}^n,n}(u) ) \right)}{\sqrt{k_n\Delta_n}} \Big)^2\Bigg)  \Bigg]\\
		&:= \frac{\sqrt{k_n\Delta_n}}{3} \sum_{j=k_n+1}^{n-k_n} q_i^{n,1}(j) + \frac{\sqrt{k_n\Delta_n}}{3} \sum_{j=k_n+1}^{n-k_n} \overline{q_i^{n,1}}(j). 
	\end{align*}
	Since
	\begin{align*}
		\E\left[\frac{\sqrt{k_n\Delta_n}}{3} \sum_{j=k_n+1}^{n-k_n} q_i^{n,1}(j) \right] =\frac{\sqrt{k_n\Delta_n}}{3} \sum_{j=k_n+1}^{n-k_n} \E\left[ \E_{i-k_n-1}\left[ q_i^{n,1}(j) \right] \right] \equiv 0,
	\end{align*}
	applying Holder's inequality with Lemma \ref{lem-vov-2}, we obtain
	\begin{align*}
		\E\left[ \left(\frac{\sqrt{k_n\Delta_n}}{3} \sum_{j=k_n+1}^{n-k_n} q_i^{n,1}(j) \right)^2\right] = \frac{k_n\Delta_n}{9}  \sum_{j_1,j_2=k_n+1}^{n-k_n} \sum_{|j_2-j_1|\leq k_n+1}^{n-k_n} \E \left[ q_i^{n,1}(j_1) q_i^{n,1}(j_2)\right] \leq \frac{C}{k_n},
	\end{align*}
	Thus, $\frac{\sqrt{k_n\Delta_n}}{3} \sum_{j=k_n+1}^{n-k_n} q_i^{n,1}(j)  \longrightarrow^{p} 0$. 
	Moreover,
	\begin{align*}
		&\frac{\sqrt{k_n\Delta_n}}{3}  \sum_{j=k_n+1}^{n-k_n} \overline{p_i^{n,1}}(j) \\
		& = \frac{\sqrt{k_n\Delta_n}}{3}  \sum_{j=k_n+1}^{n-k_n}  \E_{i-k_n-1}^n \left[ \frac{h_1(u,t_{i-k_n-1}^n, \sigma^2_{t_{i-k_n-1}^n}) }{k_n^2\Delta_n} \left(a_i^n +\frac{1}{\sqrt{k_n\Delta_n}}  \left( \frac{3}{k_n^2\Delta_n} h_1\Delta_n+3h_2\Delta_n  \right) \right) \right]\\
		&\text{(Using the results in Lemma \ref{lem-vov-2}.)}\\
		&= \Delta_n\sum_{j=k_n+1}^{n-k_n} \E_{i-k_n-1}^n \left[ \frac{h_1}{k_n^2\Delta_n}  \left( \frac{1}{k_n^2\Delta_n} h_1+ h_2 \right)  \right]+o_p(1)\\
		&\longrightarrow^{p} \int_{0}^{T} \left(  \frac{h_1^2}{\kappa^4T^2}+ \frac{h_1 h_2}{\kappa^2T} \right) dt.
	\end{align*}
	Thus
	\begin{align*}
		Q_i^{n,1} \longrightarrow^p \int_{0}^{T} \left(  \frac{h_1^2}{\kappa^4T^2}+ \frac{h_1 h_2}{\kappa^2T} \right) dt.
	\end{align*}
	For $ Q_i^{n,2}$, Holder's inequality with Lemma \ref{lem-vov-2} gives
	\begin{align*}
		&\E_{i-k_n+1}\Big[ \Big|\frac{h_1(u,t_{i+}^n, \widehat{\sigma}^2_{t_{i+}^n}) - h_1(u,t_{i-k_n-1}^n, \sigma^2_{t_{i-k_n-1}^n})  }{k_n^2\Delta_n} \\
		&\qquad \qquad \cdot \frac{ 1 }{2k_n} \left( \left( \widehat{\sigma}^2_{t_{i+}^n}  -(\sigma^2_{t_{i+}^n} + b_{ t_{i+}^n,n}(u) ) \right) - \left( \widehat{\sigma}^2_{t_{i-}^n}  -(\sigma^2_{t_{i-}^n} + b_{ t_{i-}^n,n}(u) ) \right) \right)^2 \Big| \Big] \\
		&\leq \left( \E_{i-k_n+1}\Big[ \left( \frac{h_1(u,t_{i+}^n, \widehat{\sigma}^2_{t_{i+}^n}) - h_1(u,t_{i-k_n-1}^n, \sigma^2_{t_{i-k_n-1}^n})  }{k_n^2\Delta_n} \right)^2 \Big] \right)^{1/2}\\
		&\quad \cdot \frac{\sqrt{k_n\Delta_n}}{3} \left(\E_{i-k_n-1}^n \left[ \left(\frac{3}{2} \sqrt{\frac{\Delta_n}{k_n}}\left( \frac{\left( \widehat{\sigma}^2_{t_{i+}^n}  -(\sigma^2_{t_{i+}^n} + b_{ t_{i+}^n,n}(u) ) \right)}{\sqrt{k_n\Delta_n}} - \frac{\left( \widehat{\sigma}^2_{t_{i-}^n}  -(\sigma^2_{t_{i-}^n} + b_{ t_{i-}^n,n}(u) ) \right)}{\sqrt{k_n\Delta_n}} \right)^2\right)^2 \right]\right)^{1/2}\\
		& = \left( \E_{i-k_n+1}\Big[ \left( \frac{h_1(u,t_{i+}^n, \widehat{\sigma}^2_{t_{i+}^n}) - h_1(u,t_{i-k_n-1}^n, \sigma^2_{t_{i-k_n-1}^n})  }{k_n^2\Delta_n} \right)^2 \Big] \right)^{1/2}\\
		&\quad \cdot \frac{\sqrt{k_n\Delta_n}}{3}  \left(\E_{i-k_n-1}^n \left[ \left(a_i^n +\frac{1}{\sqrt{k_n\Delta_n}}  \left( \frac{3}{k_n^2\Delta_n} h_1\Delta_n+ 3h_2\Delta_n  \right) \right)^2 \right]\right)^{1/2}\\
		&= o_p(\Delta_n).
	\end{align*}
	Furthermore, $\E[|Q_i^{n,2}|] = o_p(1)$ and $ Q_i^{n,2} \longrightarrow^p 0$. Similarly, we can prove that 
	\begin{align*}
		Q_i^{n,3}= o_p\left( \sqrt{\Delta_n} \right), \ Q_i^{n,4}= O_p\left( \sqrt{k_n\Delta_n} \right).
	\end{align*}
	Thus, $Q_i^{n,3} +Q_i^{n,4} \longrightarrow^{p} 0$.
	Combining the above results with \eqref{proof:H1} yields \eqref{proof:H2}. Similarly, we obtain
	\begin{align}\label{Uhat-2}
		\sum_{i=k_n+1}^{n-k_n} \frac{\widehat{\sigma}^2_{t_{i+}^n} }{3} \left( \frac{3}{2k_n}  \left( \widehat{\sigma}^2_{t_{i+}^n} - \widehat{\sigma}^2_{t_{i-}^n} \right)^2 - \frac{3}{k_n^2} h_1(u,\widehat{\sigma}_{t_{i+}^n}^2) \right) \longrightarrow^{p}  \int_{0}^{T} \sigma_{t}^2 h_2(t, \sigma_{t}^2,(\tilde{\sigma}_{t})^2, (\tilde{\sigma}'_{t})^2)dt,
	\end{align}
	and 
	\begin{align}\label{H'_1}
		\widehat{{H'_n}^1} \longrightarrow^{p}& \int_{0}^{T}\frac{(h'_1(u,t,\sigma_{t}^2))^2(h_1(u,t,\sigma_{t}^2))}{\kappa^6T^3} dt,\\ \label{H'_2} 	 
		\widehat{{H'_n}^2}\longrightarrow^{p}&\int_{0}^{T}\frac{(h'_1(u,t,\sigma_{t}^2))^2h_2(t,\sigma_{t}^2,(\tilde{\sigma}_{t})^2, (\tilde{\sigma}'_{t})^2)}{\kappa^4T^2}dt. 
	\end{align}
	The conclusions follow from \eqref{proof:H3}--\eqref{H'_2}. 
	
	This completes the proof of Theorem \ref{thm-uv-hat}.  
	\hfill $\square$
	
	\textbf{Proof of Theorem \ref{thm-vv-hat}:} For $\widehat{Var(V_{t}|\mathcal{F}) }$, by the mean value theorem, there exists $\overline{\sigma_{t}^2}$ between $\sigma_{t}^2$ and $\widehat{\sigma}_{t}^2$ such that 
	\begin{align*}
		&\E[|\widehat{Var(V_{t}|\mathcal{F}) } - Var(V_{t}|\mathcal{F}) | ]\\
		&=\E[| h_1(u,t, \widehat{\sigma}_{t}^2) - h_1(u,t, \sigma_{t}^2)| ]\\
		&= \E[ |h'_1(u,t, \overline{\sigma_{t}^2}) (\widehat{\sigma}_{t}^2- \sigma_{t}^2)| ]\\
		&\text{(By Holder's inequality.)}\\
		&\leq ( \E[ |h'_1(u,t, \overline{\sigma_{t}^2})|^2] \cdot \E[| (\widehat{\sigma}_{t}^2- \sigma_{t})|^2 ] )^{1/2}\\
		&\text{(Since $\sigma_{t}^2$ and $\widehat{\sigma}_{t}^2$ are bounded, together with Theorem 1 in \cite{LLL2018}.)}\\
		&=o_p(1),
	\end{align*}
	Thus,
	\begin{align}
		\widehat{Var(V_{t}|\mathcal{F})} \longrightarrow^{p} Var(V_{t}|\mathcal{F}). 
	\end{align}
	
	For $\widehat{Var(V'_{t}|\mathcal{F})}$, following the proof of Theorem \ref{thm-vov}, we obtain 
	\begin{align*}
		\widehat{Var(V'_{t}|\mathcal{F})} - \frac{1}{m_n\Delta_n} \int_{ \lfloor t/\Delta_n \rfloor\Delta_n }^{(\lfloor t/\Delta_n \rfloor+m_n)\Delta_n} h_2(t, \sigma_{t}^2,(\tilde{\sigma}_{t})^2, (\tilde{\sigma}'_{t})^2) dt \longrightarrow^p 0.
	\end{align*}
	Furthermore,
	\begin{align*}
		\E\left[ \left|\frac{1}{m_n\Delta_n} \int_{ \lfloor t/\Delta_n \rfloor\Delta_n }^{(\lfloor t/\Delta_n \rfloor+m_n)\Delta_n}  h_2(t, \sigma_{t}^2,(\tilde{\sigma}_{t})^2, (\tilde{\sigma}'_{t})^2) dt- h_2(t, \sigma_{t}^2,(\tilde{\sigma}_{t})^2, (\tilde{\sigma}'_{t})^2)\right| \right] \leq C\sqrt{m_n\Delta_n}.
	\end{align*}
	The above results together yield 
	\begin{align}
		\widehat{Var(V'_{t}|\mathcal{F})} \longrightarrow^{p} Var(V'_{t}|\mathcal{F}).
	\end{align}
	This completes the proof. 
	\hfill $\square$
	
	\textbf{Proof of Corollary \ref{cor-spot-fea}: } By Proposition 2.5 in \cite{PV2010}, the conclusions follow directly from Theorems \ref{thm-uv-hat} and \ref{thm-vv-hat} in Section \ref{sec:vol-fun}, together with Proposition \ref{pro-1}, Theorem \ref{thm-lev-1}, and Theorem \ref{thm-vov}, respectively.
	\hfill $\square$
	
	\subsection{Proofs for Section \ref{sec:diss}}\label{prof:sec4}
	\textbf{Proof of Theorem \ref{thm:lev-cor}:} By Theorem 2.7 in \cite{V2000}, the conclusion follows directly from Theorem \ref{thm-lev-1}, Theorem \ref{thm-vov}, and Theorem 1 in \cite{JT2014}. \hfill $\square$
	
	\textbf{Proof of Theorem \ref{thm:lev-fun}:} The conclusion can be proved similarly to Theorem \ref{thm-lev-1}. We define, for $i=k_n+1,...,n-k_n$,
	\begin{align*}
		\tilde{s}_i^n&= \left( \Delta_i^n X \cdot \left( F(\widehat{\sigma}^2_{t_{i+}^n} )-  F(\widehat{\sigma}^2_{t_{i-}^n}) \right) - \Delta_i^n X  \cdot  \left( F(\sigma^2_{t_{i+}^n} + b_{ t_{i+}^n,n}(u) )- F( \sigma^2_{t_{i-}^n} + b_{ t_{i-}^n,n}(u) )\right) \right).
	\end{align*}
	By the mean value theorem, there exist $\eta_{i+}^n$ between $\widehat{\sigma}^2_{t_{i+}^n}$ and $(\sigma^2_{t_{i+}^n} + b_{ t_{i+}^n,n}(u) )$, and $\eta_{i-}^n$ between $\widehat{\sigma}^2_{t_{i-}^n}$ and $(\sigma^2_{t_{i-}^n} + b_{ t_{i-}^n,n}(u) )$, such that 
	\begin{align*}
		\tilde{s}_i^n& = \Delta_i^n X \cdot F'(\sigma^2_{t_{i+}^n} + b_{ t_{i+}^n,n}(u) ) \cdot \left( \widehat{\sigma}^2_{t_{i+}^n}  -(\sigma^2_{t_{i+}^n} + b_{ t_{i+}^n,n}(u) ) \right) \\
		&\quad - \Delta_i^n X \cdot F'( \sigma^2_{t_{i-}^n} + b_{ t_{i-}^n,n}(u) ) \cdot \left( \widehat{\sigma}^2_{t_{i-}^n}  -(\sigma^2_{t_{i-}^n} + b_{ t_{i-}^n,n}(u) ) \right)\\
		&\quad +  \Delta_i^n X \cdot \frac{F''(\eta_{i+}^n )}{2} \cdot \left( \widehat{\sigma}^2_{t_{i+}^n}  -(\sigma^2_{t_{i+}^n} + b_{ t_{i+}^n,n}(u) ) \right)^2\\
		&\quad -  \Delta_i^n X \cdot \frac{F''(\eta_{i-}^{n} )}{2} \cdot \left( \widehat{\sigma}^2_{t_{i+}^n}  -(\sigma^2_{t_{i+}^n} + b_{ t_{i+}^n,n}(u) ) \right)^2\\
		&\text{(Since $F$ is twice continuously differentiable, and using the boundedness of $\sigma$ and}\\
		& \text{ the condition \eqref{cond-vov}, $b_{t,n}(u) = O_p(\Delta_n^{1-\frac{\beta}{2}})$, \eqref{X-main} and Lemma \ref{lem-lev-1}.)}\\
		&= F'( \sigma^2_{t_{i-k_n-1}^n} ) \cdot s_i^n \\
		&\quad + O_p\left( \Delta_n^{1\wedge \frac{1}{(\max\{\beta,r\}+\epsilon)}} \cdot \left(\left( \frac{1}{\sqrt{k_n}} \vee \sqrt{k_n\Delta_n} \right)^2 + \left( \frac{1}{\sqrt{k_n}} \vee \sqrt{k_n\Delta_n} \right)\cdot (\sqrt{k_n\Delta_n} + \Delta_n^{1-\frac{\beta}{2}}) \right) \right), 
	\end{align*}
	where $s_i^n$ is defined in the proof of Theorem \ref{thm-lev-1}. 
	We decompose
	\begin{align*}
		&\widehat{\mathcal{L}}^{func}_{[0,T]} -  \mathcal{L}^{func}_{[0,T]} \\
		&= \sum_{i=k_n+1}^{n-k_n} \tilde{s}_i^{n} + \sum_{i=k_n+1}^{n-k_n} \left( \Delta_i^n X  \cdot  \left( F(\sigma^2_{t_{i+}^n} + b_{ t_{i+}^n,n}(u) )- F( \sigma^2_{t_{i-}^n} + b_{ t_{i-}^n,n}(u) )\right) \right)  -  \mathcal{L}_{[0,T]}\\
		& (\text{Since $\gamma$ is bounded and with the condition \eqref{cond-vov}.})\\
		&= \sum_{i=k_n+1}^{n-k_n} \tilde{s}_i^n  + \sum_{i=k_n+1}^{n-k_n} \left( \Delta_i^n X  \cdot   \left( F(\sigma^2_{t_{i+}^n}) -F(\sigma^2_{t_{i-}^n}) \right) \right)  -  \mathcal{L}^{func}_{[0,T]} + O_p(u^{\beta-2}\Delta_n^{\frac{1+\alpha-\beta}{2}})\\
		&= \sum_{i=k_n+1}^{n-k_n} \tilde{s}_i^n  + \sum_{i=k_n+1}^{n-k_n} \left( \int_{t_{i-1}^n}^{t_i^n} \sigma_sdB_s  \cdot   \left( F(\sigma^2_{t_{i+}^n}) - F(\sigma^2_{t_{i-}^n}) \right) \right)  -  \mathcal{L}^{func}_{[0,T]}  \\
		&\quad + \sum_{i=k_n+1}^{n-k_n} \left( \left( \Delta_{i}^nX- \int_{t_{i-1}^n}^{t_i^n} \sigma_sdB_s \right) \cdot   \left( F(\sigma^2_{t_{i+}^n}) -F( \sigma^2_{t_{i-}^n}) \right) \right) + O_p(u^{\beta-2}\Delta_n^{\frac{1+\alpha-\beta}{2}}) \\
		&(\text{By the proof of Lemma 2 in \cite{LLL2018}, the boundedness of $\sigma$ and the condition \eqref{cond-vov}.})\\
		&= \sum_{i=k_n+1}^{n-k_n}(F'( \sigma^2_{t_{i-k_n-1}^n} ) \cdot s_i^n)  + \sum_{i=k_n+1}^{n-k_n} \left( \int_{t_{i-1}^n}^{t_i^n} \sigma_sdB_s  \cdot   \left( F(\sigma^2_{t_{i+}^n}) - F(\sigma^2_{t_{i-}^n}) \right) \right)  -  \mathcal{L}^{func}_{[0,T]} \\
		&\quad + O_p\left( u^{\beta-2}\Delta_n^{\frac{1+\alpha-\beta}{2}} + \Delta_n^{\frac{1}{\max\{\beta,r\}} - \frac{1}{2}} + \sqrt{\Delta_n} \right)\\
		&\quad + O_p\left( \Delta_n^{1\wedge \frac{1}{(\max\{\beta,r\}+\epsilon)}} \cdot \left(\left( \frac{1}{\sqrt{k_n}} \vee \sqrt{k_n\Delta_n} \right)^2 + \left( \frac{1}{\sqrt{k_n}} \vee \sqrt{k_n\Delta_n} \right)\cdot (\sqrt{k_n\Delta_n} + \Delta_n^{1-\frac{\beta}{2}}) \right) \right).
	\end{align*}
	When $\alpha=1$ and $\max\{\beta,r\}\leq 1$, we have
	\begin{align*}
		\sqrt{n}^{b\wedge (1-b)} \cdot \left( \widehat{\mathcal{L}}^{func}_{[0,T]} -  \mathcal{L}^{func}_{[0,T]} \right) = \sqrt{n}^{b\wedge (1-b)} \cdot  \sum_{i=k_n+1}^{n-k_n}(F'( \sigma^2_{t_{i-k_n-1}^n} ) \cdot s_i^n)  + o_p(1).
	\end{align*}
	Following the proof of Theorem \ref{thm-lev-1} and Lemma \ref{lem-lev-2}, the conclusion is obtained under the same conditions. 
	
	This completes the proof of Theorem \ref{thm:lev-fun}. \hfill $\square$

\ignore{
\hspace{-0.25in}
\bibliographystyle{model2-names}
\bibliography{liu}

	\bigskip
	
	\hspace{-0.25in}
	\bibliographystyle{model2-names}
	\bibliography{liu}

@article{CT2024,
	author = {Chong, Carsten H and Todorov, Viktor},
	journal = {Journal of Econometrics},
	number = {1},
	pages = {105669},
	publisher = {Elsevier},
	title = {Volatility of volatility and leverage effect from options},
	volume = {240},
	year = {2024}}

@book{V2000,
	author = {Van der Vaart, Aad W},
	publisher = {Cambridge University Press},
	title = {Asymptotic Statistics},
	year = {2000}}

@article{V2011,
	author = {Veraart, Almut ED},
	journal = {Advances in Statistical Analysis},
	number = {3},
	pages = {253--291},
	publisher = {Springer},
	title = {How precise is the finite sample approximation of the asymptotic distribution of realised variation measures in the presence of jumps?},
	volume = {95},
	year = {2011}}

@article{BV2009,
	author = {Barndorff-Nielsen, Ole E and Veraart, Almut},
	journal = {CREATES research paper},
	title = {Stochastic volatility of volatility in continuous time},
	year = {2009}}

@article{LLZ2022,
	author = {Li, Yingying and Liu, Guangying and Zhang, Zhiyuan},
	journal = {Journal of Econometrics},
	number = {2},
	pages = {422--451},
	publisher = {Elsevier},
	title = {Volatility of volatility: Estimation and tests based on noisy high frequency data with jumps},
	volume = {229},
	year = {2022}}

@article{JR2014,
	author = {Jacod, Jean and Reiss, Markus},
	journal = {Annals of Statistics},
	pages = {1131-1144},
	title = {A remark on the rates of convergence for integrated volatility estimation in the presence of jumps},
	volume = {42},
	year = {2014}}

@article{BSW2006,
	author = {Barndorff-Nielsen, Ole E and Shephard, Neil and Winkel, Matthias},
	journal = {Stochastic processes and their applications},
	number = {5},
	pages = {796--806},
	publisher = {Elsevier},
	title = {Limit theorems for multipower variation in the presence of jumps},
	volume = {116},
	year = {2006}}

@article{M2011,
	author = {Mancini, Cecilia},
	journal = {Stochastic Processes and their Applications},
	number = {4},
	pages = {845--855},
	publisher = {Elsevier},
	title = {The speed of convergence of the threshold estimator of integrated variance},
	volume = {121},
	year = {2011}}

@incollection{W2006,
	author = {Woerner, Jeannette HC},
	booktitle = {Stochastic Finance},
	pages = {343--364},
	publisher = {Boston, MA: Springer US},
	title = {Power and multipower variation: {I}nference for high frequency data},
	year = {2006}}

@article{C1982,
	author = {Andrew A. Christie},
	journal = {Journal of Financial Economics},
	pages = {407-432},
	title = {The stochastic behavior of common stock variances: {V}alue, leverage and interest rate effects},
	volume = {10},
	year = {1982}}

@article{M2015,
	author = {Vetter, Mathias},
	journal = {Bernoulli},
	pages = {2393-2418},
	title = {Estimation of integrated volatility of volatility with applications to goodness-of-fit testing},
	volume = {21},
	year = {2015}}

@article{PV2010,
	author = {M. Podolskij AND M. Vetter},
	journal = {Statistica Neerlandica},
	pages = {329-351},
	title = {Understanding limit theorems for semimartingales: {A} short survey},
	volume = {64},
	year = {2010}}

@inproceedings{J2012,
	author = {Jacod, Jean},
	booktitle = {Statistical Methods for Stochastic Differential Equations},
	editor = {A.L. Matthieu Kessler and M. Sorensen},
	journal = {Statistical methods for stochastic differential equations},
	pages = {191--310},
	publisher = {Taylor and Francis},
	title = {Statistics and high frequency data},
	volume = {124},
	year = {2012}}

@article{KX2017,
	author = {Kalnina, Ilze and Xiu, Dacheng},
	journal = {Journal of the American Statistical Association},
	number = {517},
	pages = {384--396},
	publisher = {Taylor \& Francis},
	title = {Nonparametric estimation of the leverage effect: {A} trade-off between robustness and efficiency},
	volume = {112},
	year = {2017}}

@article{AFLWY2017,
	author = {A{\"\i}t-Sahalia, Yacine and Fan, Jianqing and Laeven, Roger JA and Wang, Christina Dan and Yang, Xiye},
	journal = {Journal of the American Statistical Association},
	number = {520},
	pages = {1744--1758},
	publisher = {Taylor \& Francis},
	title = {Estimation of the continuous and discontinuous leverage effects},
	volume = {112},
	year = {2017}}

@article{LL2024,
	author = {Liu, Qiang and Liu, Zhi},
	journal = {Econometrics Journal},
	pages = {278-298},
	publisher = {Oxford University Press},
	title = {Estimating spot volatility under infinite variation jumps with dependent market microstructure noise},
	volume = {27},
	year = {2024}}

@article{LLL2018,
	author = {Liu, Qiang and Liu, Yiqi and Liu, Zhi},
	journal = {Stochastic Processes and their Applications},
	number = {6},
	pages = {1958--1987},
	publisher = {Elsevier},
	title = {Estimating spot volatility in the presence of infinite variation jumps},
	volume = {128},
	year = {2018}}

@article{WM2014,
	author = {Wang, Christina D and Mykland, Per A},
	journal = {Journal of the American Statistical Association},
	number = {505},
	pages = {197--215},
	publisher = {Taylor \& Francis},
	title = {The estimation of leverage effect with high-frequency data},
	volume = {109},
	year = {2014}}

@article{ABDL2003,
	author = {T. G. Andersen and T. Bollerslev and F. Diebold and P. Labys},
	journal = {Econometrica},
	pages = {579-625},
	title = {Modeling and forecasting realized volatility},
	volume = {71 (3)},
	year = {2003}}

@article{BN2004,
	author = {O. E. Barndorff-Nielsen and N. Shephard},
	journal = {Journal of Financial Econometrics},
	pages = {1-37},
	title = {Power and Bipower Variation with Stochastic Volatility and Jumps},
	volume = {2 (1)},
	year = {2004}}

@article{MR2011,
	author = {Mancini, C. and Ren{\`o}, R.},
	journal = {Journal of Econometrics},
	pages = {77-92},
	title = {Threshold estimation of Markov models with jumps and interest rate modeling},
	volume = {160},
	year = {2011}}

@article{D2010,
	author = {Kristensen, D.},
	journal = {Econometric Theory},
	pages = {60-93},
	title = {Nonparametric filtering of the realised spot volatility: {A} kernel-based approach},
	volume = {26},
	year = {2010}}

@article{DS1994,
	author = {Delbaen, F. and Schachermayer, W.},
	journal = {Mathematische Annalen},
	pages = {463-520},
	title = {A general version of the fundamental theorem of asset pricing},
	volume = {300},
	year = {1994}}

@article{J2008,
	author = {Jacod, J.},
	journal = {Stochastic Processes and their Application},
	pages = {517-559},
	title = {Asymptotic properties of realized power variations and related functionals of semimartingales},
	volume = {118 (4)},
	year = {2008}}

@article{JKLM2012,
	author = {Jing, B. Y and Kong, X. B. and Liu, Z. and Mykland, P. A.},
	journal = {Journal of Econometrics},
	pages = {213-223},
	title = {On the jump activity index for semimartingales},
	volume = {166 (2)},
	year = {2012}}

@article{JR2013,
	author = {Jacod, J. and Rosenbaum, M.},
	journal = {Annals of Statistics},
	pages = {1462-1484},
	title = {Quarticity and other functionals of volatility: {E}fficient estimation},
	volume = {41},
	year = {2013}}

@book{JP2012,
	author = {Jacod, J. and Protter, P},
	publisher = {Springer-Verlag Berlin Heidelberg},
	title = {Discretization of Processes},
	year = {2012}}

@book{JS2003,
	author = {Jacod, J. and Shiryayev, A. V.},
	publisher = {Springer-Verlag Berlin Heidelberg New York},
	title = {Limit theorems for Stochastic Processes},
	year = {2003}}

@article{JT2014,
	author = {Jean Jacod and Viktor Todorov},
	journal = {Annals of Statistics},
	number = {3},
	pages = {1029--1069},
	title = {Efficient Estimation of Integrated Volatility in Presence of Infinite Variation Jumps},
	volume = {42},
	year = {2014}}

@article{Mancini2009,
	author = {Mancini, C.},
	journal = {Scandinavian Journal of Statistics},
	number = {2},
	pages = {270-296},
	title = {Nonparametric threshold estimation for models with stochastic diffusion coefficient and jumps},
	volume = {36},
	year = {2009}}

@article{YJ2009,
	author = {Y. A\"{i}t-Sahalia and J. Jacod},
	journal = {Annals of Statistics},
	pages = {2202-2244},
	title = {Estimating the degree of activity of jumps in high frequency data},
	volume = {37},
	year = {2009}}

@article{YP2014,
	author = {Zu, Y. and Boswijk, P.},
	journal = {Journal of Econometrics},
	pages = {117-135},
	title = {Estimating spot volatility with high-frequency financial data.},
	volume = {181},
	year = {2014}}

@article{Z2011,
	author = {Zhang, L.},
	journal = {Journal of Econometrics},
	number = {1},
	pages = {33-47},
	title = {Estimating Covariation: Epps Effect, Microstructure Noise},
	volume = {160},
	year = {2011}}

@book{AJ2014,
	author = {Y. A\"{i}t-Sahalia and J. Jacod},
	publisher = {Princeton University Press, New Jersey},
	title = {High-Frequency Financial Econometrics},
	year = {2014}}
\end{document}